%% file: main.tex
\documentclass[11pt]{article}

\usepackage[margin=1in]{geometry}
\usepackage{amsthm,amsmath,amsfonts,amssymb}
\usepackage{natbib}
\usepackage[colorlinks,citecolor=blue,urlcolor=blue,hypertexnames=false]{hyperref}
\usepackage{graphicx}

\usepackage{algorithm}
\usepackage{algorithmic}
\usepackage{array}
\usepackage{bm}
\usepackage{bbm}
\usepackage{color}
\usepackage{float}
\usepackage{booktabs}
\usepackage{bigints}
\usepackage{extarrows}
\usepackage{multirow}
\usepackage{subcaption}
\usepackage[bbgreekl]{mathbbol}
\usepackage{mathtools}
\usepackage{mathrsfs}
\usepackage[section]{placeins}
\usepackage{verbatim}
\usepackage{enumitem}
\usepackage{xcolor}
\usepackage{optidef}

\usepackage{mdframed}

\DeclareSymbolFontAlphabet{\mathbbm}{bbold}
\DeclareSymbolFontAlphabet{\mathbb}{AMSb}%

\theoremstyle{plain}
\newtheorem{theorem}{Theorem}[section]
\newtheorem{lemma}{Lemma}[section]

\newtheorem{corollary}{Corollary}[section]
\newtheorem{definition}{Definition}
\newtheorem{remark}{Remark}[section]

\numberwithin{equation}{section}
\numberwithin{theorem}{section}
\numberwithin{lemma}{section}
\numberwithin{proposition}{section}
\numberwithin{corollary}{section}
\numberwithin{definition}{section}
\numberwithin{remark}{section}
\numberwithin{table}{section}
\numberwithin{figure}{section}

\newenvironment{block}[1]
{\par\medskip\noindent\textbf{#1.}\ }
{\par\medskip}

\newcommand{\mE}{\mathbb{E}}
\newcommand{\mP}{\mathbb{P}}

\newcommand{\tr}{\operatorname{tr}}
\newcommand{\calA}{\mathcal{A}}
\newcommand{\calB}{\mathcal{B}}
\newcommand{\calD}{\mathcal{D}}
\newcommand{\calK}{\mathcal{K}}

\newcommand{\calG}{\mathcal{G}}
\newcommand{\calH}{\mathcal{H}}
\newcommand{\calS}{\mathcal{S}}
\newcommand{\calV}{\mathcal{V}}
\newcommand{\calT}{\mathcal{T}}
\newcommand{\calU}{\mathcal{U}}
\newcommand{\calQ}{\mathcal{Q}}

\newcommand{\SC}{T_{\rm sc}}
\newcommand{\MC}{T_{\rm mc}}
\newcommand{\SNR}{\operatorname{SNR}}

\newcommand{\bz}{\mathbf{z}}

\allowdisplaybreaks

\begin{document}

\title{Adaptable High-Dimensional Change Point Detection via Ridge Regularization}
\author{
Haoran Li\thanks{Corresponding author. Email: \texttt{hzl0152@auburn.edu}}
~~~and ~~~Haotian Xu\thanks{Email: \texttt{haotian.xu@auburn.edu}}\\[0.5em]
{\small Department of Mathematics and Statistics, Auburn University}
}
\date{}

\maketitle

\begin{abstract}
We study the problem of detecting multiple change points in the mean vectors of an independent sequence of high-dimensional observations. Targeting at detecting dense alternatives in the regime where the dimension is comparable to the sample size, we propose a family of ridge-regularized CUSUM statistics built upon the adaptable ridge-regularized Hotelling's $T^2$ test of Li et al. (Ann. Statist. \textbf{48} (2020) 1815--1847). Ridge regularization provides stable covariance normalization while allowing the tests to adapt to the underlying population covariance structure. To accommodate multiple change points without prior knowledge of their number or locations, we scan over a multi-scale collection of adjacent segments. We further enhance adaptability by aggregating information across a collection of ridge regularization parameters through the maximum of the corresponding test statistics. Under mild conditions, we establish the limiting distributions of the proposed statistics under the null hypothesis and a class of local alternatives. Extensive simulation studies demonstrate their finite-sample performance.
\end{abstract}




\input{1_intro}
\input{2_method}
\input{3_theory}

\input{5_simulation}

\input{8_discussion}

\section*{Supplementary Material}
The Supplementary Material includes additional simulation results and detailed proofs of the main theoretical results in the paper. 

\bibliographystyle{plainnat}
\bibliography{reference}

\newpage

\input{Supp_Material}

\end{document}

%% file: 1_intro.tex
\section{Introduction}\label{sec:introduction}
Change point detection is a classical problem in statistics with applications across a wide range of disciplines, including econometrics, signal processing, and climate science. In this setting, one observes an ordered sequence of random vectors $X_1, X_2, \ldots, X_n \in \mathbb{R}^p$ that are independent and identically distributed, except for possible structural changes in mean vectors. In this paper, we consider the model
\begin{equation}
    \label{eq:model}
    X_j = \mu_j + \Sigma_p^{1/2} \bz_j, \qquad j = 1,\ldots,n,
\end{equation}
where $\{\mu_j\}$ is a sequence of mean vectors, $\{\bz_j\in \mathbb{R}^p\} $ are i.i.d.\ random vectors with independent, mean-zero, unit-variance entries, and $\Sigma_p$ is a general positive semi-definite covariance matrix.
Under the null hypothesis of no change point, the mean vectors are constant over time, that is,
\[
H_0:\ \mu_j \equiv \mu, \qquad j=1,\ldots,n.
\]
We are interested in testing $H_0$ against the alternative that the mean vectors undergo a finite number of abrupt changes:
\begin{equation}
\label{eq:H_a_MC}
H_a:\ 
\mu_1 = \cdots = \mu_{k_1} \neq \mu_{k_1+1} = \cdots = \mu_{k_s}
\neq \mu_{k_s+1} = \cdots = \mu_n,
\end{equation}
where $s \ge 1$ and $\{k_j\}_{j=1}^s \in [1, n)$ denote the (unknown) number and locations of change points. We henceforth refer to $H_a$ as the multiple change point (MC) alternative.

In the classical regime where the dimension $p$ is fixed and small, change point detection has been extensively studied; see \cite{csorgo1997limit} and \cite{aue2013structural} for comprehensive reviews of classical methods. To motivate the ideas, consider the case $s=1$, henceforth referred to as the single change-point (SC) alternative. That is, we are interested in testing $H_0$ against the alternative 
\[H_a^{\rm SC}: \mu_1 =\cdots = \mu_{k_1} \neq \mu_{k_1+1} =\cdots = \mu_n.\]
For the moment, assume that the covariance matrix $\Sigma_p$ is invertible and known. The classical CUSUM procedure compares the sample means of the two segments ${X_1,\ldots,X_k}$ and ${X_{k+1},\ldots,X_n}$ after normalization by $\Sigma_p^{-1/2}$ over all possible change-point locations $k$. Specifically, the test statistic is
\[
\max_{1 \le k \le n-1} \tilde{C}(k)^T \Sigma_p^{-1} \tilde{C}(k),
\]
where $\tilde{C}(k) =(n-k)^{-1} \sum_{j=k+1}^n X_j - k^{-1} \sum_{j=1}^k X_j$ denotes the difference between the sample means of the two segments separated at $k$. In the low-dimensional regime and under the null hypothesis $H_0$, $\Sigma_p$ can be consistently estimated by the pooled sample covariance matrix 
\begin{equation}\label{eq:sample_cov_mat}
S_n = \frac{1}{n} \sum_{j=1}^n (X_j - \bar{X})(X_j -\bar{X})^T, \qquad \text{with } ~\bar{X} = \frac{1}{n}\sum_{j=1}^n X_j.
\end{equation}
The CUSUM procedure can therefore be implemented by replacing $\Sigma_p$ with $S_n$.  
Importantly, under $H_0$, the distribution of the resulting test statistic, normalized by $S_n^{-1}$, is invariant with respect to $\Sigma_p$, which highlights the robustness of classical methods to the unknown covariance structure.   

High-dimensional data are now ubiquitous in many scientific fields. When the dimension $p$ is comparable to or larger than the sample size $n$, the performance of traditional methods deteriorates substantially. First, accurate estimation of $\Sigma_p^{-1}$ becomes challenging in this regime, particularly in the absence of structural assumptions, such as low-rank structure or sparsity. Second, the accumulation of noise across coordinates fundamentally alters the limiting distributions of the test statistics.

There is a substantial literature on change point testing and estimation for the mean of high-dimensional data. Broadly speaking, existing methods can be classified into three main categories. The first class focuses on detecting sparse mean changes, where only a small subset of components of the mean vector undergo change over time. These methods typically rely on the $\ell_\infty$-norm of the mean difference $\tilde{C}(k)$ and include, among others, \cite{jirak2015change} and \cite{yu2021high}. The second class targets dense alternatives, in which the signal may be distributed across many or all coordinates. For example, \cite{zhang2010detecting} and \cite{zhang2018unsupervised} consider tests based on the $\ell_2$-norm of $\tilde{C}(k)$ without normalization by $\Sigma_p^{-1}$, while \cite{wang2022inference} proposes a modified procedure motivated by U-statistic techniques. The third class comprises methods that do not presuppose specific alternative structures, such as projection-based approaches  \cite{wang2018high} and \cite{enikeeva2019high}. Other notable works on various forms of high-dimensional change point detection include \cite{shao2010testing,matteson2014nonparametric,kirch2015detection,cho2016change,liu2020unified,zhang2022adaptive,dornemann2026sequentialeigenvaluestatisticschangepoint,dornemann2024linear}.

While most existing methods largely ignore covariance normalization and instead rely directly on the norm of $\tilde{C}(k)$, we propose a new class of test statistics that achieves data-driven covariance normalization using ridge-regularized sample covariance estimators. Specifically, let $\Lambda=\{\lambda_1,\ldots,\lambda_M\}\subset\mathbb{R}_+$ be a finite collection of ridge regularization parameters. For each $\lambda_i\in\Lambda$, we transform the observations according to
\begin{equation}\label{eq:transformed_observations}
    Y_j(\lambda_i)
    =(S_n+\lambda_i I_p)^{-1/2}X_j,
    \qquad j=1,\ldots,n,\quad i=1,\ldots,M.
\end{equation}
Targeting dense alternatives, we construct CUSUM-type statistics based on the $\ell_2$-norm of the differences between segment means of the transformed observations and scan over a flexible collection of candidate change-point locations. We then aggregate the statistics across $\lambda_i\in\Lambda$ by taking their maximum, thereby improving robustness to the choice of the ridge parameter. The proposed methodology is presented in detail in Section~\ref{sec:method}.

Importantly, the resulting test statistics are \emph{rotation-invariant}, in the sense that their values are unchanged under arbitrary orthogonal transformations of the data, that is, when $X_j$ is replaced by $Q^TX_j$ for any orthogonal matrix $Q$. This invariance is a desirable property when not much prior knowledge about $\Sigma_p$ and the alternative is available.

We work in a high-dimensional regime where the dimension $p$ is comparable to the sample size $n$, in the sense that $p/n \to \gamma \in (0,\infty)$. Detecting dense alternatives in this regime is well motivated by a variety of applications. For example, KEGG pathway–level analyses of DNA copy number alteration patterns often involve a few hundred genes, and substantial empirical evidence suggests that such alterations in a cancer cell occur across a large fraction of genes rather than being confined to a few isolated components \citep{Graham2017CNApatterns}. In climate science, spatio-temporal fields are often represented using tens to a few hundred spatial locations or principal components, and regime changes typically induce coherent mean shifts across a large fraction of these components \citep{Defriez2016ClimateRegime}. In neuroimaging, brain activity is frequently summarized over a few hundred regions of interest, and transitions between cognitive or behavioral states are associated with changes across many regions, resulting in dense mean shifts in moderately high-dimensional signals \citep{Kringelbach2020BrainStates}. 

Technically, in the regime $p/n\to\gamma$, fundamental results from \emph{Random Matrix Theory} (RMT) indicate that, although the pooled sample covariance matrix $S_n$ is an inconsistent estimator of $\Sigma_p$ under $H_0$, it exhibits stable limiting spectral behavior under mild conditions, determined solely by the spectral distribution of $\Sigma_p$ and the limiting ratio $\gamma$. The introduction of a ridge term $\lambda I_p$ shifts the spectrum of $S_n$ away from singularity while preserving its essential spectral structure. As a consequence, the ridge-regularized matrix $(S_n+\lambda I_p)^{-1}$ retains meaningful structural information about the underlying precision matrix $\Sigma_p^{-1}$. We refer to \cite{bodnar2026reviving} for a systematic study of the asymptotic properties of ridge inverses in high-dimensional settings. The resulting $\ell_2$-norm–based procedures are therefore expected to achieve robustness with respect to the unknown covariance structure.  Ridge regularization has been successfully employed in high-dimensional mean testing \citep{chen2011regularized,li2020adaptable}, high-dimensional general linear hypothesis testing \citep{li2020high,li2025ridge}, and covariance estimation \citep{ledoit2004well}. In particular, the proposed methodology builds upon the ridge-regularized Hotelling's $T^2$ (RHT) test for comparing two high-dimensional mean vectors developed by \cite{li2020adaptable}. 

Our work focuses on the case where the number of change points $s$ is fixed but arbitrary, while the simpler case $s=1$ is discussed explicitly for the reader’s convenience. Moreover, we assume that the change-point locations $k_j$ are bounded away from the endpoints (1 and $n$) of the observation sequence and scale proportionally with the sample size, in the sense that $k_j/n \to \tilde{B}_j \in (0,1)$ for all $j=1,\ldots,s$. When $s$ is unknown, we adopt flexible scanning techniques in which differences between sample means are compared across multi-scale adjacent segments of the observations.

The rest of the paper is organized as follows. In Section~\ref{sec:method}, we present the proposed test statistics for both single and multiple change-point scenarios. In Section~\ref{sec:asymptotics}, after introducing the necessary preliminaries from RMT, we derive the asymptotic null distributions of the proposed test statistics and study their asymptotic power under two frameworks: a class of deterministic local alternatives and a class of probabilistic local alternatives, in which the mean shifts are treated as realizations from prior distributions. Section~\ref{sec:simulation} reports simulation studies under various representative settings. Finally, Section~\ref{sec:discussion} concludes with a discussion and possible future extensions, particularly to time series settings. Technical details and additional simulation results are provided in the Supplementary Material.

%% file: 2_method.tex
\section{Method}\label{sec:method}
As in the Introduction, suppose that a finite deterministic set of ridge regularization parameters,
\[
\Lambda=\{\lambda_1,\ldots,\lambda_M\}\subset\mathbb{R}_+,
\]
and a small trimming parameter $\varepsilon\in(0,0.5)$ are given. We present the proposed ridge-regularized CUSUM procedures separately for two settings: the single change-point case and the multiple change-point case. The motivation for considering a grid of ridge regularization parameters, together with our recommended choice of the grid, is deferred to the end of this section.

Recall the definition of the transformed observations in \eqref{eq:transformed_observations}. For $r\in[0,1]$, define
\[
k(r)=\lfloor nr\rfloor+1.
\]
We say that a transformed observation $Y_j(\lambda)$ belongs to the segment $[t_1,t_2)$ if $k(t_1)\leq j<k(t_2)$. 
The average of the observations belonging to a segment is referred to as the corresponding segment mean.

Denote the set of all ordered triples in $[0,1]$ by
\[ \calT = \big\{(t_1,t_2,t_3) \in [0,1]^3\, :\, t_1<t_2<t_3 \big\}. \]
Then, any element of $t = (t_1,t_2,t_3)\in \calT$ defines two adjacent segments $[t_1,t_2)$ and $[t_2,t_3)$.  We consider the difference between the corresponding sample segment means of the transformed observations.  
Specifically, denote
\begin{equation}
    \label{eq:segment_mean_diff}
    C_\lambda(t)
    = \frac{1}{k(t_3)-k(t_2)}
    \sum_{j=k(t_2)}^{k(t_3)-1} Y_j{(\lambda)}
    -
    \frac{1}{k(t_2)-k(t_1)}
    \sum_{j=k(t_1)}^{k(t_2)-1} Y_j{(\lambda)}.
\end{equation}
We further define the associated effective sample size by
\begin{equation}
N(t)
=
\frac{\bigl(k(t_2)-k(t_1)\bigr)\bigl(k(t_3)-k(t_2)\bigr)}
     {k(t_3)-k(t_1)}.
\label{eq:def_effective_sample_size}    
\end{equation}
With these notations, we define the ridge-regularized statistic as
\begin{equation}
    \label{eq:def_stat}
    V_\lambda(t)
    =
    N(t) \|C_\lambda(t)\|_2^2 
\end{equation}
where $\|\cdot\|_2$ is the Euclidean ($\ell_2$-) norm of a vector or the spectral norm of a matrix.

The theoretical analysis in Section~\ref{sec:asymptotics} shows that the asymptotic mean and variance of $V_\lambda(t)$ depend on the ridge parameter $\lambda$ and the population covariance matrix $\Sigma_p$. To account for these dependencies, we introduce the following data-driven normalization quantities:
\begin{equation}
\label{eq:def_Theta_hat}
\begin{aligned}
&m_n(z) = p^{-1}\tr\bigl[\bigl( \frac{n}{n-1} S_n - z I_p\bigr)^{-1}\bigr], \qquad z<0;\\
&\hat{\Omega}(z)
= \big[1-\gamma_n - \gamma_n z m_n(z)\big]^{-1}, \quad \text{with }\gamma_n = p/(n-1),
\qquad z<0;\\
&\hat{\Theta}(z)
=\gamma_n^{-1}\bigl[1-\hat{\Omega}^{-1}(z)\bigr]
=1 + z m_n(z),
\qquad z<0;\\
&\hat{\Gamma}(z_1,z_2)
=
\begin{cases}
\displaystyle
2\gamma_n^{-1} 
\hat{\Omega}^{-1}(z_1)\hat{\Omega}^{-1}(z_2)
\left[
\frac{z_1\hat{\Omega}(z_1)
-z_2\hat{\Omega}(z_2)}
{z_1-z_2}
-1
\right],
& z_1,\, z_2<0,\, z_1\neq z_2,\\[12pt]
\displaystyle 2\gamma_n^{-1} 
\hat{\Omega}^{-2}(z)
\left[
\frac{d}{dz}{z\hat{\Omega}(z)}-1
\right],& z_1=z_2=z<0.
\end{cases}
\end{aligned}
\end{equation}
Notably, the diagonal definition of $\hat{\Gamma}(z,z)$ is obtained as the continuous extension of $\hat{\Gamma}(z_1,z_2)$ as $z_1\to z_2=z$. For notational simplicity, we henceforth write $\hat{\Gamma}(z,z)$ as $\hat{\Gamma}(z)$.  The derivative appearing in  $\hat{\Gamma}(z)$ can be evaluated using
\begin{equation}
m_n'(z) =\frac{d}{dz}m_n(z) = p^{-1}\tr\bigl[\bigl(\frac{n}{n-1} S_n -z I_p\bigr)^{-2}\bigr].    
\end{equation}
Importantly, these quantities depend only on the eigenvalues of $S_n$ and the aspect ratio $\gamma_n$.

We then define the standardized statistic by
\begin{equation}\label{eq:def_standardized_stat}
D_\lambda(t)
=
\sqrt{p}\,
\frac{p^{-1}V_\lambda(t)-\hat{\Theta}(-\lambda)}
     { \sqrt{\hat{\Gamma}(-\lambda)}}.
\end{equation}
In Section~\ref{sec:asymptotics}, we show that, under mild regularity conditions, for each fixed $(\lambda,t)$, $D_\lambda(t)$ has asymptotic mean zero and unit variance under $H_0$. Consequently, the standardized statistics are asymptotically comparable across candidate segmentations $t$ and ridge parameters $\lambda$.

\subsection*{Single change point}

As a simple yet informative special case, we first consider the single change-point alternative. Recall the trimming parameter $\varepsilon\in(0,0.5)$. We scan over all binary partitions of $[0,1]$ such that both segments have length at least $\varepsilon$. Specifically, define
\[
\calT_{\rm sc}(\varepsilon)
=
\big\{
(0,t_2,1): t_2\in[\varepsilon,1-\varepsilon]
\big\}.
\]
We propose to test $H_0$ against the SC alternative using
\begin{equation}
\label{eq:def_test_single_point}
\SC(\varepsilon,\Lambda)
=
\max_{\lambda\in\Lambda}
\sup_{t\in\calT_{\rm sc}(\varepsilon)}
D_\lambda(t).
\end{equation}
For each candidate change-point location and ridge parameter $\lambda$,
$D_\lambda(t)$ measures the standardized difference between the two
segment means of the transformed observations. The statistic
$\SC(\varepsilon,\Lambda)$ aggregates information across both candidate
change-point locations and ridge parameters by taking the maximum.
The asymptotic critical value for implementing the test at a prescribed
significance level is derived in Section~\ref{sec:asymptotics}.

\subsection*{Multiple change points}

When multiple change points are present under the alternative, the scanning strategy in \eqref{eq:def_test_single_point} may suffer from a loss of power, since at least one of the two segments $[0,t_2)$ or $[t_2,1)$ may contain multiple change points. In such cases, mean shifts in opposite directions can partially cancel, resulting in little or no difference between the corresponding segment means. This issue is particularly pronounced when the mean shifts follow a nonmonotonic pattern. It is therefore desirable to develop a procedure that remains powerful without requiring prior knowledge of the number of change points $s$ under the alternative.

Motivated by this consideration, we scan over multi-scale adjacent segments whose lengths are at least $\varepsilon$. Specifically, we define
\begin{equation}
\label{eq:def_calS}
\calT_{\rm mc}(\varepsilon)
=
\big\{
(t_1,t_2,t_3)\in[0,1]^3:
t_2-t_1\geq\varepsilon,\;
t_3-t_2\geq\varepsilon
\big\}.
\end{equation}
We then propose to test $H_0$ against the MC alternative $H_a$ using
\[
\MC(\varepsilon,\Lambda)
=
\max_{\lambda\in\Lambda}
\sup_{t\in\calT_{\rm mc}(\varepsilon)}
D_\lambda(t).
\]

The scanning set $\calT_{\rm mc}(\varepsilon)$ is more flexible than those considered by \citet{zhang2018unsupervised} and \citet{wang2022inference}. In their procedures, either the left endpoint is fixed at $t_1=0$ or the right endpoint is fixed at $t_3=1$, while the remaining indices vary subject to the trimming constraint. In contrast, our procedure allows all three endpoints to vary, so that the pair of adjacent segments can be located anywhere within the observation period rather than being anchored at a boundary. This additional flexibility may improve sensitivity to change-point configurations occurring in the interior of the observation period.

The increased flexibility, however, comes at a higher computational cost. Given the transformed data $\{Y_j(\lambda)\}$, evaluating a single $D_\lambda(t)$ requires $O(pn)$ operations. The full scan involves $O(n^3)$ candidate triples $t$ for each $\lambda\in\Lambda$, resulting in $O(n^3|\Lambda|)$ evaluations of $D_\lambda(t)$. Consequently, a direct implementation of $\MC(\varepsilon,\Lambda)$ has overall computational complexity $O(pn^4|\Lambda|)$.

To alleviate this computational burden, we further consider a discretized version of the scanning set. Specifically, define
\[
\calT_{\rm disc}(\varepsilon)
=
\Big\{
(t_1,t_2,t_3)\in\calT_{\rm mc}(\varepsilon):
t_2-t_1=a\varepsilon,\;
t_3-t_2=b\varepsilon
\text{ for some }a,b\in\mathbb N_+
\Big\}.
\]
Thus, while the two segments remain free to move over the observation period, their lengths are restricted to integer multiples of $\varepsilon$. The discretized scanning set contains $O(n\varepsilon^{-2})$ candidate triples. The corresponding test statistic is
\[
T_{\rm disc}(\varepsilon,\Lambda)
=
\max_{\lambda\in\Lambda}
\sup_{t\in\calT_{\rm disc}(\varepsilon)}
D_\lambda(t).
\]
Given the transformed data, evaluating $T_{\rm disc}(\varepsilon,\Lambda)$ has computational complexity $O(pn^2\varepsilon^{-2}|\Lambda|),$
representing a substantial reduction from the full scan.

\subsection*{Design of $\Lambda$, effect of $\varepsilon$, and further discussion}

Consider first a homogeneous procedure with a single ridge parameter $\lambda$. The power analysis in Section~\ref{subsec:asymptotic_power} shows that the power-optimal choice of $\lambda$ depends critically on the interaction between the unknown mean shifts and the population covariance matrix $\Sigma_p$. In particular, the numerical studies in Section \ref{subsubsec:effect_lambda} reveal that, depending on the unknown alternative, the optimal choice may lie in the small-, moderate-, or large-ridge regime. Thus, estimating the power-optimal ridge parameter is generally difficult without additional assumptions on the mean shifts.

To improve robustness to the choice of $\lambda$, we aggregate the test statistics over a finite collection $\Lambda$ by taking their maximum. We recommend that $\Lambda$ include small, moderate, and large ridge parameters. To adapt their scale to the data, we estimate the overall spectral scale of $S_n$ using its non-leading eigenvalues. Let $\alpha_j(A)$ denote the $j$th largest eigenvalue of a symmetric matrix $A$, and define
\[
\mathfrak{S}
=
\frac{1}{p-L}\sum_{j=L+1}^{p}\alpha_j(S_n),
\]
rounded to a fixed number of significant digits. Here, the leading $L$ eigenvalues are excluded because they may be inflated by the presence of change points. In practice, $L$ can be selected by inspecting the empirical spectral scree plot. We then set
\[
\Lambda
=
\big\{
0.1(\gamma_n\vee 1),\,1,\,10
\big\}\times\mathfrak{S}.
\]
This choice is motivated by the study in Section \ref{subsubsec:effect_lambda}. It covers a range of regularization levels while adapting their overall scale to the empirical spectrum of $S_n$. 

The trimming parameter $\varepsilon$ ensures that all scanned segments contain at least $\lfloor n\varepsilon\rfloor$ observations, thereby avoiding contrasts based on very small subsamples. Such trimming is standard in change-point analysis; see, for example, \cite{csorgo1997limit,zhang2018unsupervised,wang2022inference}. Larger values of $\varepsilon$ generally yield more stable inference but may reduce power for detecting changes near the boundaries or closely spaced change points. Conversely, smaller values allow changes to be detected over shorter time scales, at the cost of increased scanning multiplicity, larger critical values, and greater computational burden. Throughout the numerical studies, we use $\varepsilon=0.1$.

From a theoretical perspective, keeping $\varepsilon$ bounded away from zero ensures that all segment lengths are of order $n$, which facilitates the asymptotic normality and the asymptotic equicontinuity of the process $D_\lambda(t)$ over the trimmed scanning sets. Our current theory does not cover the moving-domain setting where $\varepsilon=\varepsilon_n\downarrow0$.


While $D_\lambda(t)$ is motivated by the ridge-regularized Hotelling's $T^2$ (RHT) statistic of \citet{li2020adaptable} for two-sample mean testing, it differs from the RHT statistic in the choice of the scaling matrix. The RHT statistic uses a ridge-regularized sample covariance matrix obtained by centering each observation $X_j$ by its corresponding group mean, whereas $D_\lambda(t)$ uses $S_n$, in which all observations are centered by the overall mean $\bar{X}$. This difference leads to different asymptotic means and variances. The choice of scaling matrix reflects the different problem settings. In two-sample mean testing, group membership is known, so within-group centering provides an unbiased covariance estimator under both $H_0$ and $H_a$. In change-point detection, however, the locations and number of potential change points are unknown, and the test scans over a large collection of candidate triples $t$. Applying the RHT scaling matrix would therefore require a different covariance estimator for each candidate triple, which may also be biased when the candidate segmentation does not match the true change-point configuration. Repeatedly constructing and inverting these matrices would additionally incur a substantial computational burden. Using the common scaling matrix $S_n$ avoids these difficulties and allows the same covariance normalization to be used throughout the scan.





%% file: 3_theory.tex
\section{Asymptotic Theory}\label{sec:asymptotics}

We begin with some necessary preliminaries from RMT and then develop the asymptotic theory of the proposed tests under the null hypothesis and various local alternatives. Lastly, we examine the effect of the parameter $\lambda$ through numerical studies in illustrative settings. Throughout the analysis, the ridge parameter set $\Lambda$ is assumed to be fixed and finite, and the trimming parameter $\varepsilon$ is also held fixed.

\input{3_1_settings}

\input{3_2_Preliminary}

\input{3_3_Null}

\input{3_4_Power}

\input{3_5_Power_Comparison}

\input{3_6_Effect_Lambda}

%% file: 3_1_settings.tex
\subsection{Settings}\label{subsec:settings}
We make the following assumptions on the model. 
Let $\tau_{1,p}\ge \cdots \ge \tau_{p,p}\ge 0$ denote the eigenvalues of $\Sigma_p$, and define
\begin{equation}\label{eq:def_PSD}
H_p(\tau) =\frac{1}{p}\sum_{j=1}^p \mathbbm{1}_{[\tau_{j,p},\,\infty)}(\tau)
\end{equation}
to be the spectral distribution function of $\Sigma_p$. We henceforth refer to $H_p$ as the \emph{population spectral distribution function} (PSD).  

\begin{enumerate}[label={\bf C\arabic*}]
\item\label{enum:moment_conditions} \emph{Moment conditions.} The entries of $\{\bz_j\}$ have finite moments of all orders.
\item\label{enum:high_dimensional_regime}
\emph{High-dimensional regime.}
As $p,n\to\infty$, we assume that $\gamma_n = p/(n-1) \to\gamma \in(0,\infty)$.
\item\label{enum:location_breaks}
\emph{Location of breaks.} We assume that $s$ is fixed. If $s>0$, the locations $k_s$ is such that for some $\tilde{B}_1< \tilde{B}_2 < \cdots  < \tilde{B}_s \in (0,1)$, $k_j/n \to \tilde{B}_j$, $j=1,\ldots, s$, as $p, n\to\infty$.
\item\label{enum:spectrum_bound} \emph{Boundedness of the spectral norm.} The covariance matrix $\Sigma_p$ is nonnegative definite and satisfies $\limsup_{p\to\infty}\tau_{1,p}<\infty$.
\item\label{enum:stability_sigma} \emph{Asymptotic stability of the PSD.} The PSD $H_p$ converges weakly to a probability distribution function $H$ as $p\to\infty$, where $H$ is not the Dirac measure at $0$. 
We henceforth refer to $H$ as the limiting PSD.
\item\label{enum:convergence_rate} \emph{Convergence rate.} We require the convergence of $\gamma_n$ and $H_p$ as in \ref{enum:high_dimensional_regime} and \ref{enum:stability_sigma} is such that $\sqrt{n}|\gamma_n - \gamma | \longrightarrow 0$ and  $\sqrt{n} \|H_p-H\|_\infty \longrightarrow0$. 
\end{enumerate}
Condition~\ref{enum:moment_conditions} specifies the moment assumptions
used in deriving the asymptotic results. Condition~\ref{enum:high_dimensional_regime}
places the problem in the high-dimensional regime where $p$ and $n$
grow proportionally. Condition~\ref{enum:location_breaks} assumes that
the number of change points is fixed and their locations are
asymptotically bounded away from the boundaries. Conditions \ref{enum:spectrum_bound} and \ref{enum:stability_sigma} impose,
respectively, a uniform bound on the spectral norm of $\Sigma_p$ and
the convergence of its PSD to a nondegenerate limiting distribution.

The $\sqrt{n}$-convergence rates in
Condition~\ref{enum:convergence_rate} are technical assumptions that
allow us to express the limiting distribution of the standardized
process $\{D_\lambda(t)\}$ in terms of a Gaussian process
whose covariance kernel depends only on the limiting PSD $H$ and the
limiting aspect ratio $\gamma$. Without these rates, a Gaussian
approximation can still be obtained, but its covariance kernel
generally depends on the finite-dimensional quantities $H_p$ and
$\gamma_n$. We therefore impose Condition~\ref{enum:convergence_rate} throughout
to simplify the presentation.

%% file: 3_2_Preliminary.tex
\subsection{Preliminaries on Random Matrix Theory}\label{subsec:preliminary}
In this section, we present some fundamental results from RMT concerning the asymptotic spectral behavior of $S_n$ under the null hypothesis $H_0$. These results are well established in the RMT literature; see, for example, \cite{bai2004clt,paul2014random,knowles2017anisotropic,li2020high,bodnar2026reviving}.

We consider the spectral distribution function of $S_n$, defined as 
\[ F_p(\tau) = \frac{1}{p} \sum_{j=1}^p \mathbbm{1}_{[\alpha_j,\, \infty) }(\tau), \]
where $\alpha_j = \alpha_j(S_n)$. We henceforth refer to $F_p$ as the \emph{empirical spectral distribution} (ESD). 

Recall that the Stieltjes transform $\varphi_G(\cdot)$ of any function $G$ of bounded variation on $\mathbb{R}$ is defined by 
\[\varphi_G(z) = \int_{-\infty}^{\infty} \frac{dG(\tau)}{\tau - z}, \qquad z \in \mathbb{C}^+ \coloneqq \{u + iv:\, v>0\}.\]
Consider any compactly supported distribution function $F$, excluding the Dirac measure at $0$. Then, for any $c>0$, the famous Mar\v{c}enko-Pastur (M-P) equation in RMT is defined for any $z\in \mathbb{C}^+$ as 
\begin{equation}
    \label{eq:MP_eq}
    \varphi = \int \frac{dF(\tau)}{\tau (1-c -c z \varphi) -z}.  
\end{equation}
It is well known that there exists a unique solution $\varphi = \varphi(z; c, F) \in \mathbb{C}^+$ to \eqref{eq:MP_eq}. The function $\varphi(z)$ is the Stieltjes transform of a probability measure with bounded support in $[0,\infty)$. Unfortunately, except in extreme cases where $F$ has a simple structure, there is no explicit closed-form for $\varphi(z)$ or the associated probability measure. Moreover, the function $\varphi(z)$ is analytic on $\mathbb{C}^+$ and admits a smooth extension to the negative real axis $\mathbb{R}_-$. With a slight abuse of notation, we continue to denote this extension as 
\[ \varphi(x; c, F)\coloneqq \lim_{z\in\mathbb{C}^+\to x} \varphi(z; c, F), \qquad  \text{ for} \quad  x\in\mathbb{R}_-.\] 
The subsequent analysis will primarily focus on this domain.  For further details of M-P equation, see Chapter 3 of \cite{bai2010spectral} and \cite{paul2014random}. We are particularly interested in the M-P equation evaluated as $F = H$ and $c = \gamma$. We shall write $\varphi(z) = \varphi(z;\gamma, H)$, $z\in \mathbb{R}_-$, in the subsequent analysis.

The asymptotic behavior of the sample covariance matrix $S_n$ under the null hypothesis $H_0$ is closely related to $\varphi(x)$ in the following sense. For $z<0$, define the limiting counterpart of $\hat{\Omega}(z)$, $\hat{\Theta}(z)$, and $\hat{\Gamma}(z)$ as 
\begin{equation}
\label{eq:def_Theta}
\begin{aligned}
&{\Omega}(z)
= \big[1-\gamma - \gamma z \varphi(z)\big]^{-1},
\qquad z<0,\\
&{\Theta}(z)
=\gamma^{-1}\bigl[1-{\Omega}^{-1}(z)\bigr]
=1 + z \varphi(z),
\qquad z<0,\\
&{\Gamma}(z_1,z_2)
=
\begin{cases}
\displaystyle
2\gamma^{-1} 
{\Omega}^{-1}(z_1){\Omega}^{-1}(z_2)
\left[
\frac{z_1{\Omega}(z_1)
-z_2{\Omega}(z_2)}
{z_1-z_2}
-1
\right],
& z_1,\, z_2<0,\, z_1\neq z_2,\\[12pt]
\displaystyle 2\gamma^{-1} 
{\Omega}^{-2}(z)
\left[
\frac{d}{dz}{z{\Omega}(z)}-1
\right],& z_1=z_2=z<0.
\end{cases}
\end{aligned}
\end{equation}
Again, we shall write $\Gamma(z,z)$ as $\Gamma(z)$. Further, define the \emph{deterministic equivalent} matrix 
\begin{equation}\label{eq:deterministic_equivalent}
\calD_p(z) = \Big[ \Omega^{-1}(z) \Sigma_p - z I_p \Big]^{-1}.
\end{equation}

\begin{lemma}
\label{lemma:estimate_Stieltjes}
Suppose that $H_0$ holds and $\Lambda$ is any finite set in $\mathbb{R}_+$. Under Conditions~\ref{enum:moment_conditions}--\ref{enum:convergence_rate}, the following statements hold.
\begin{itemize}
    \item[(i)] Following the main theorem of \cite{bai2004clt},
    \[
    \sup_{\lambda \in \Lambda}
    \sqrt{n}\, |m_n(-\lambda) - \varphi(-\lambda)|
    \stackrel{P}{\longrightarrow} 0, 
    \qquad
    \sup_{\lambda \in \Lambda}
    \sqrt{n}\, |m'_n(-\lambda) - \varphi'(-\lambda)|
    \stackrel{P}{\longrightarrow} 0 .
    \]
    Moreover, it is shown in Section 5 of \cite{bai2004clt} that we can find a constant $\calK>0$ such that 
    \[ \inf_{\lambda \in \Lambda} \Big|\Omega^{-1}(-\lambda)\Big| =  \inf_{\lambda\in \Lambda} |1 -\gamma +\gamma \lambda \varphi(-\lambda)| > \calK. \]
    Consequently, 
    \begin{align*}
    \sup_{\lambda \in \Lambda}\sqrt{n}\, |\hat{\Omega}(-\lambda) &- \Omega(-\lambda)| \stackrel{P}{\longrightarrow}0,  \qquad
    \sup_{\lambda \in \Lambda} \sqrt{n}\, |\hat{\Theta}(-\lambda) - \Theta(-\lambda)| \stackrel{P}{\longrightarrow} 0,\\  
    \sup_{\lambda_1,\lambda_2 \in \Lambda} &\sqrt{n}\, |\hat{\Gamma}(-\lambda_1,-\lambda_2) - \Gamma(-\lambda_1,-\lambda_2)|
    \stackrel{P}{\longrightarrow} 0 .    
    \end{align*} 
    \item[(ii)]  
    For any sequence of symmetric matrices $A$ with $\|A\|_2<\infty$,
    \[
    \sup_{\lambda\in\Lambda}\sqrt{n} \big|p^{-1}\tr\!\left[(S_n + \lambda I_p)^{-1}A\right] - p^{-1}\tr\!\left[\calD_p(-\lambda)A\right]\big|\stackrel{P}{\longrightarrow} 0.
    \]
    For any sequence of deterministic vectors $e$ such that $\|e\|_2 =1$,
    \[ \sup_{\lambda\in \Lambda} \sqrt{n} \big| e^T(S_n + \lambda I_p)^{-1} e - e^T\calD_p(-\lambda)e \big|\stackrel{P}{\longrightarrow} 0  \]
    These results follow from Theorem~3.6 of \cite{knowles2017anisotropic}.
    \item[(iii)] The results can be extended to the alternative hypothesis $H_a$. See Section \ref{subsubsec:shift_sample_cov} for details.
\end{itemize}
\end{lemma}

The results in (ii) of Lemma \ref{lemma:estimate_Stieltjes} are particularly insightful for the proposed method. It demonstrates that  $\calD_p(-\lambda)$ can be viewed as the \emph{deterministic equivalent} of the ridge-regularized inverse $(S_n+\lambda I_p)^{-1}$, when $p/n\to \gamma$. Importantly, $\calD_p(-\lambda)$ shares the same eigenvectors as $\Sigma_p^{-1}$, while its eigenvalues are transformed through a nonlinear mapping induced by the ridge regularization.  In this sense, $(S_n+\lambda I_p)^{-1}$ preserves the essential spectral structure of the underlying precision matrix $\Sigma_p^{-1}$.

%% file: 3_3_Null.tex
\subsection{Asymptotic Null Distribution}
\label{subsec:asymptotic_null}

In this section, we derive the asymptotic distributions of the proposed statistics under the null hypothesis $H_0$. Unless otherwise specified, weak convergence of stochastic processes, denoted by $\rightsquigarrow$, is understood as weak convergence in $\ell^\infty(\Lambda\times\calA)$, where $\calA$ is taken to be $\calT_{\rm sc}(\varepsilon)$, $\calT_{\rm mc}(\varepsilon)$, or $\calT_{\rm disc}(\varepsilon)$, depending on the process under consideration. In contrast, $\stackrel{D}{\longrightarrow}$ denotes convergence in distribution of random variables.

We first introduce several auxiliary functions. For an ordered triple $t = (t_1,t_2,t_3)\in\calT$, define
\begin{equation}
\label{eq:def_u}
u(x\,;\,t)
=
-\frac{1}{t_2-t_1}\mathbbm{1}_{[t_1,t_2)}(x)
+
\frac{1}{t_3-t_2}\mathbbm{1}_{[t_2,t_3)}(x),
\qquad x\in[0,1].
\end{equation}
Define the inner-product kernel
\begin{equation}
\label{eq:def_kappa}
\kappa(t,t')
=
\int_0^1 u(x\,;\,t)u(x\,;\,t')\,dx,
\qquad t,t'\in\calT,
\end{equation}
and, for notational simplicity, write $\kappa(t)=\kappa(t,t)$. The normalized kernel on $\calT$ is defined by
\begin{equation}
\label{eq:kernel_A}
K_{\calT}(t,t')
=
\frac{\kappa(t,t')^2}
{\kappa(t)\kappa(t')},
\qquad t,t'\in\calT.
\end{equation}

Similarly, define the normalized kernel on $\Lambda$ by
\begin{equation}
\label{eq:kernel_Lambda}
K_{\Lambda}(\lambda,\lambda')
=
\frac{\Gamma(-\lambda,-\lambda')}
{\Gamma^{1/2}(-\lambda)\Gamma^{1/2}(-\lambda')},
\qquad \lambda,\lambda'\in\Lambda.
\end{equation}
The positive definiteness of $K_{\Lambda}$ follows from
\citet{li2020adaptable,li2020high}.

We now define a centered Gaussian process on $\Lambda\times\calA$,
\[
\big\{G_{\lambda}(t): \lambda\in\Lambda,\ t\in\calA\big\},
\]
with covariance kernel
\[
\mE\big[G_{\lambda}(t)G_{\lambda'}(t')\big]
=
K_{\calT}(t,t')K_{\Lambda}(\lambda,\lambda').
\]
It is straightforward to verify that the covariance kernel is
nonnegative definite, and hence such a centered Gaussian process exists.
Moreover, for each $\lambda\in\Lambda$, the increment variance $\mE\left|G_{\lambda}(t)-G_{\lambda}(t')\right|^2$
is Hölder continuous in $t$ and $t'$. Since Gaussian increments have
moments of all orders, Kolmogorov's continuity theorem implies that
$G_{\lambda}$ admits a modification with almost surely continuous sample
paths on $\calA$.
\begin{theorem}
    \label{thm:weak_convergence_process}
Suppose that Conditions \ref{enum:moment_conditions}--\ref{enum:convergence_rate} hold. Fix a finite subset $\Lambda\subset\mathbb{R}_+$ and $\varepsilon>0$. Let $\calA$ denote either $\calT_{\rm sc}(\varepsilon)$, $\calT_{\rm mc}(\varepsilon)$, and $\calT_{\rm disc}(\varepsilon)$. Under the null hypothesis $H_0$, we have 
\[\big\{ D_{\lambda}(t), \, t \in \calA, \, \lambda \in \Lambda \big\} \rightsquigarrow \big\{ G_{\lambda} (t),\, t\in \calA, \, \lambda \in \Lambda  \big\}.\]
\end{theorem}

\begin{corollary}
    \label{corollary:convergence_test_statistic}
    Under the conditions of Theorem~\ref{thm:weak_convergence_process}, the continuous mapping theorem and Slutsky's theorem yield
    \[\SC(\varepsilon, \Lambda)\, \stackrel{D}{\longrightarrow} \,\max_{\lambda \in \Lambda} \sup_{t\in \calT_{\rm sc}(\varepsilon) } G_\lambda(t), \quad \MC(\varepsilon, \Lambda)\, \stackrel{D}{\longrightarrow}\, \max_{\lambda \in \Lambda} \sup_{t\in \calT_{\rm mc}(\varepsilon) } G_\lambda(t), \]  
    \[T_{\rm disc}(\varepsilon, \Lambda) \stackrel{D}{\longrightarrow} \max_{\lambda\in\Lambda}\sup_{t\in \calT_{\rm disc}(\varepsilon)}G_\lambda(t).\]
\end{corollary}


In practice, we approximate the critical value of the test at any prescribed significance level as follows. First, construct the $|\Lambda|\times|\Lambda|$ matrix
\[\widetilde{K}_{\Lambda} = \Bigl[\frac{\hat{\Gamma}(-\lambda_i,-\lambda_j)}{\hat{\Gamma}^{1/2}(-\lambda_i) \hat{\Gamma}^{1/2}(-\lambda_j) }\Bigr]_{i,j=1}^{|\Lambda|}.\]
Although the limiting kernel $K_{\Lambda}(\lambda_1,\lambda_2)$ is nonnegative definite, the finite-sample estimator $\widetilde{K}_{\Lambda}$ need not be nonnegative definite. In this case, we replace $\widetilde{K}_{\Lambda}$ by its spectral projection onto the cone of positive semidefinite matrices, obtained by setting all negative eigenvalues to zero while leaving the corresponding eigenvectors unchanged. The resulting matrix is still denoted by $\widetilde{K}_{\Lambda}$.

Second, for the scan index set $\calA = \mathcal T_{\rm sc}(\varepsilon)$, $\mathcal T_{\rm mc}(\varepsilon)$ or $\calT_{\rm disc}(\varepsilon)$, construct a sufficiently fine grid
$\{\tau_1,\tau_2,\ldots,\tau_Q\}\subset \calA.$ 
Next, construct the $Q\times Q$ matrix
\[ \widetilde{K}_{\calT} =\bigl[K_{\calT}(\tau_i,\tau_j)\bigr]_{i,j=1}^{Q}.\] 
By construction, $\widetilde{K}_{\mathcal T}$ is positive semidefinite.

Third, we approximate the limiting Gaussian process $\{G_{\lambda}(t)\}$ by its values on the selected grid. Specifically, let
\[W\sim\mathcal{MN}\!\left(0,~\widetilde{K}_{\Lambda},~\widetilde{K}_{\calT}\right),\]
that is,
$W =\widetilde{K}_{\Lambda}^{1/2} \mathbf{E} \widetilde{K}_{\calT}^{1/2}$,
where $\mathbf{E}$ is a $|\Lambda|\times Q$ matrix whose entries are independent standard normal random variables. The asymptotic critical value at significance level $\alpha\in(0,1)$ is then approximated by the empirical $(1-\alpha)$-quantile of the simulated maxima of $W$, denoted by $\xi_{\rm sc}(1-\alpha)$, $\xi_{\rm mc}(1-\alpha)$ or $\xi_{\rm disc}(1-\alpha)$ for the three scan index sets, respectively.

The corresponding tests reject the null hypothesis at asymptotic significance level $\alpha$ whenever
\[
\SC(\varepsilon,\Lambda)>\xi_{\rm sc}(1-\alpha),\qquad
\MC(\varepsilon,\Lambda)>\xi_{\rm mc}(1-\alpha), \qquad T_{\rm disc}(\varepsilon, \Lambda) > \xi_{\rm disc}(1-\alpha).
\]




%% file: 3_4_Power.tex
\subsection{Asymptotic Power Under Local Alternatives}\label{subsec:asymptotic_power}

In this section, we investigate the asymptotic power of the proposed procedures under local alternatives. The analysis is carried out first for the instructive special case of a single change point and then extended to the general multiple change-point setting. Throughout this section, in addition to the ridge parameters $\Lambda$ and $\varepsilon$, we assume that the significance level $\alpha\in(0,1)$ is fixed.

\subsubsection{Shift of the sample covariance matrix under $H_a$} 
\label{subsubsec:shift_sample_cov}
We define the oracle sample covariance matrix constructed from the correctly centered observations by
\[ \tilde{S}_n = \frac{1}{n} \sum_{j=1}^n \Sigma_p^{1/2} \bigl(\bz_j-\bar{\bz}\bigr)\bigl(\bz_j-\bar{\bz}\bigr)^{T}\Sigma_p^{1/2},\]
where $\bar{\bz} = n^{-1}\sum_{j=1}^n \bz_j$. Let $R_n = S_n - \tilde{S}_n$. Then, it is immediate that $R_n = 0$ under $H_0$. Thus, $R_n$ represents the shift of $S_n$ under $H_a$. While the Frobenius norm of $R_n$ depends on the mean vectors $\{\mu_j\}$, the rank of $R_n$ is at most $3(s+1)$ when there are $s$ change points under $H_a$.  Consequently, we can obtain a deterministic rank bound on the shift of $m_n(-\lambda)$ as stated in Lemma \ref{lemma:Stieltjes_under_alternative}.

\begin{lemma}
\label{lemma:Stieltjes_under_alternative}
Fix any $\lambda,\, \lambda_1,\, \lambda_2 >0$. Let $\tilde{m}_n(-\lambda)= p^{-1} \tr\bigl[(n/(n-1) \tilde{S}_n+\lambda I_p)^{-1}\bigr]$.
Then, 
\[\bigl| m_n(-\lambda) - \tilde{m}_n(-\lambda) \bigr|
\le {3(s+1)}/{(p\lambda)},
\quad
\bigl| m_n'(-\lambda) - \tilde{m}_n'(-\lambda) \bigr|
\le {3(s+1)}/{(p\lambda^2)},
\]
\[ \Big|p^{-1}\tr\big[(S_n + \lambda I_p)^{-1}A \big] - p^{-1} \tr\big[(\tilde{S}_n + \lambda I_p)^{-1} A\big]\Big| \leq 6 \|A\|_2(s+1)/(p\lambda), \]
where $A$ is a general sequence of symmetric matrices.

Moreover, let $\tilde{\Theta}(-\lambda)$ and $\tilde{\Gamma}(-\lambda_1, -\lambda_2)$ be the counterpart of $\hat\Theta(-\lambda)$ and $\hat\Gamma(-\lambda_1, -\lambda_2)$ when $m_n(-\lambda)$, $m_n(-\lambda_1)$ and $m_n(-\lambda_2)$ are replaced by $\tilde{m}_n(-\lambda)$, $\tilde{m}_n(-\lambda_1)$, and $\tilde{m}_n(-\lambda_2)$ in \eqref{eq:def_Theta_hat}.  Then,
\[ |\hat{\Theta}(-\lambda)- \tilde{\Theta}(-\lambda)| \leq \frac{3(s+1)}{p}, \quad  |\hat{\Gamma}(-\lambda_1,-\lambda_2) - \tilde{\Gamma}(-\lambda_1,-\lambda_2)| \leq \calK \frac{3(s+1)}{p}, \]
where $\calK$ is a constant depending only on $\gamma_n$. Consequently, the convergence results in Lemma~\ref{lemma:estimate_Stieltjes} continue to hold under $H_a$, given that $s$ is fixed.
\end{lemma}

\subsubsection{Power analysis for the single change-point case}
\label{subsubsec:power_single_change_point}

We first consider the case $s=1$ as a benchmark for understanding the power behavior of the proposed test. Under $H_a^{\rm SC}$, the power is determined by the mean shift
$\delta_p=\mu_n-\mu_1$. Denote the power of the proposed test at asymptotic size $\alpha$ by
\[
\Upsilon_{\rm sc}(\delta_p,\varepsilon,\Lambda)
=
\mP\Big(
\SC(\varepsilon,\Lambda)
>
\tilde{\xi}_{\rm sc}(1-\alpha;\varepsilon,\Lambda)
\mid \delta_p
\Big).
\]
Here, $\tilde{\xi}_{\rm sc}(1-\alpha;\varepsilon,\Lambda)$ is the $(1-\alpha)$-quantile of $\max_{\lambda\in\Lambda} \sup_{t\in\calT_{\rm sc}(\varepsilon)} G_\lambda(t)$, and $\mP(\cdot\mid\delta_p)$ denotes the distribution of the observations conditional on $\delta_p$.

We next define the drift process governing the asymptotic power. Recall the deterministic equivalent $\calD_p(-\lambda)$ of the ridge inverse defined in \eqref{eq:deterministic_equivalent} and Lemma~\ref{lemma:estimate_Stieltjes}. Let
\begin{equation}
\label{eq:def_q_p}
q_p(\lambda,\delta_p)
=
\sqrt{p}\,
\delta_p^T\calD_p(-\lambda)\delta_p,
\qquad \lambda\in\Lambda.
\end{equation}
For $t=(0,t_2,1)\in\calT_{\rm sc}(\varepsilon)$, define
\[
\eta_{\rm sc}(t)
=
\begin{cases}
\displaystyle
\frac{t_2}{1-t_2}(1-\tilde{B}_1)^2,
& 0\leq t_2<\tilde{B}_1,\\[8pt]
\displaystyle
\frac{1-t_2}{t_2}\tilde{B}_1^{\,2},
& \tilde{B}_1\leq t_2\leq1.
\end{cases}
\]
Here, $\tilde{B}_1$ is the limiting scaled location of the change point specified in Condition~\ref{enum:location_breaks}. We then define the drift process by
\begin{equation}
\label{eq:def_drift_process}
\mathcal{S}_{\lambda}(t;\delta_p)
=
\eta_{\rm sc}(t)
\frac{q_p(\lambda,\delta_p)}
{\gamma\Gamma^{1/2}(-\lambda)}
[1-\gamma\Theta(-\lambda)]^2,
\qquad
t\in\calT_{\rm sc}(\varepsilon),\quad \lambda\in\Lambda.
\end{equation}
By Lemma~\ref{lemma:estimate_Stieltjes}, the factor
$[1-\gamma\Theta(-\lambda)]^2$ is bounded away from zero. 

\begin{theorem}
\label{thm:power_single_deterministic}
Suppose that Conditions \ref{enum:moment_conditions}--\ref{enum:convergence_rate} hold. Under the single change point alternative $H_a^{\rm SC}$, we assume the mean shift $\delta_p$ is local in the sense that 
there exist constants $0< \underline{\calK}\leq \overline{\calK}$ such that $\underline{\calK}\leq q_p(\lambda, \delta_p) \leq \overline{\calK}$ for all sufficiently large $p$ and $\lambda\in\Lambda$. Then, 
\begin{equation*}
\Upsilon_{\rm sc}(\delta_p, \varepsilon, \Lambda) - \mathbb{P}\,\Big( \max_{\lambda\in\Lambda}\sup_{t\in\calT_{\rm sc}(\varepsilon)} \Big\{ G_{\lambda}(t) +  \mathcal{S}_{\lambda}\big(t;\delta_p \big)\Big\}  > \tilde{\xi}_{\rm sc}(1-\alpha;\varepsilon, \Lambda) \Big) \longrightarrow 0. 
\end{equation*}
\end{theorem}
Theorem~\ref{thm:power_single_deterministic} shows that the asymptotic local power is governed by the drift process. In particular, for each $\lambda\in\Lambda$, the function $\eta_{\rm sc}(t)$ characterizes the attenuation of the signal resulting from using a candidate split point $t_2$ that differs from the true limiting change-point location $\tilde{B}_1$. It is straightforward to verify that $\eta_{\rm sc}(t)$ is uniquely maximized at $t_2=\tilde{B}_1$, yielding the familiar population drift shape of the classical CUSUM scan statistic; see, for example, \cite{csorgo1997limit}. On the other hand, the quantity 
\[\SNR(\lambda, \delta_p) \coloneqq \frac{q_p(\lambda,\delta_p) [1-\gamma \Theta(-\lambda)]^2 }{\Gamma^{1/2}(-\lambda)}\]
can be interpreted as the (asymptotic) signal-to-noise ratio (SNR) associated with the parameter $\lambda$. 

It is therefore natural to investigate how the SNR varies with $\lambda$ and whether a power-optimal ridge parameter can be identified. We examine this question through numerical studies under illustrative settings in Section~\ref{subsubsec:effect_lambda}. The results show that the SNR-maximizing choice of $\lambda$ generally depends on the mean shift $\delta_p$, the covariance structure $\Sigma_p$, and the aspect ratio $\gamma$. Depending on their interaction, the optimal $\lambda$ may lie in the small-, moderate-, or large-ridge regime. Consequently, consistently estimating the power-optimal ridge parameter is generally intractable without additional structural assumptions on $\delta_p$. This motivates the proposed composite procedure, which aggregates the test statistics over a collection of ridge parameters to improve robustness with respect to the choice of $\lambda$.

\begin{remark}
\label{remark:regime_trivial_power}
Theorem \ref{thm:power_single_deterministic} shows that the proposed test has non-trivial power in the regime $\sqrt{n}\|\delta_p\|_2^2 \asymp 1$. If instead $\sqrt{n}\|\delta_p\|_2^2 \to 0$, the power degenerates to the significance level $\alpha$. On the other hand, if $\sqrt{n}\|\delta_p\|_2^2 \to \infty$, the power converges to $1$. \cite{aston2018high} introduced the concept of \emph{high-dimensional efficiency} (HDE), which has been widely used in the power analysis of change point detection methods. Under their framework, the HDE of the proposed test is $\|\delta_p\|_2^2$.
\end{remark}

\subsubsection{Power analysis for the multiple change-points case}\label{subsubsec:power_multiple}
We next consider the general alternative $H_a$ under which the number of change points $s$ is unknown. For simplicity, we present the analysis only for the scanning set $\calT_{\rm mc}(\varepsilon)$. All results in this section apply directly to the computationally efficient scanning set $\calT_{\rm disc}(\varepsilon)$ after replacing the scanning set. 

We collect the segment mean vectors across the $(s+1)$ segments into the matrix
\[\mathcal{U}=\bigl(\mu_1,\ \mu_{k_1+1},\ \mu_{k_2+1},\ \ldots,\ \mu_{k_s+1}\bigr) \in \mathbb{R}^{p \times (s+1)}.\]
Notably, when $s=1$, the mean shift $\delta_p = \calU c$,  where $c = (1,-1)^T$ is the contrast vector. Denote the power of the proposed decision rule given $\calU$ as 
\[\Upsilon_{\rm mc}(\calU, \varepsilon, \Lambda) = \mP\Big( \MC(\varepsilon,\Lambda) > \tilde{\xi}_{\rm mc}(1-\alpha;\varepsilon, \Lambda)  \mid \calU \Big).\]
Here, $\tilde{\xi}_{\rm mc}(1-\alpha; \varepsilon,\Lambda)$ is the $(1-\alpha)$-quantile of $\max_{\lambda \in\Lambda} \sup_{t\in\calT_{\rm mc}(\varepsilon)} G_\lambda(t)$, and the probability measure $\mP(\cdot\mid\calU)$ is the distribution of the observations given $\calU$.

Similar to the SC case, we define the drift process for the MC setting in the following way. Let
\[v_j(x) = \mathbbm{1}_{[\tilde{B}_{j-1},\, \tilde{B}_{j})}(x), \quad x \in [0,1],\quad j=1,\dots, s+1,\] 
where we set $\tilde{B}_0 = 0$ and $\tilde{B}_{s+1} = 1$. Recall the definition of $u(x\, ;\, t)$ in \eqref{eq:def_u} and $\kappa(t)$ in \eqref{eq:def_kappa}. We define a $(s+1)$-vector $\psi(t)$ for $t\in\calT$, whose $j$th element $\psi_j(t)$ is 
\[ \psi_j(t) = \frac{1}{ \sqrt{\kappa(t)}} \int_0^1 u(x; t) v_j(x) dx, \quad j=1,\dots, s+1.\]
Analogously to \eqref{eq:def_drift_process}, we define the drift process
\begin{equation}
    \label{eq:def_drift_multiple}
    \mathcal{S}_{\lambda}\big(t \, ; \, \mathcal{U}\big) =  \frac{ \sqrt{p}  \big(\psi(t) \big)^T \calU^T \calD_p(-\lambda)  \calU  \psi(t)  }{\gamma\Gamma^{1/2}(-\lambda)} [1-\gamma \Theta(-\lambda)]^2, ~~ t\in \calT_{\rm mc}(\varepsilon), ~~ \lambda\in\Lambda.
\end{equation}
This is the natural extension of \eqref{eq:def_drift_process} to the MC setting. Accordingly, we retain the notation $\mathcal{S}_{\lambda}$.

\begin{theorem}
    \label{thm:power_multiple_deterministic}
    Suppose that Conditions \ref{enum:moment_conditions}--\ref{enum:convergence_rate} hold. Under the general alternative $H_a$, we assume that the shifts are local in the sense that there exist constants $0< \underline{\calK}\leq \overline{\calK}$ such that for all sufficiently large $p$ and $\lambda \in \Lambda$,
    \[ \underline{\calK}  \leq \sqrt{p} \|\,\calU^T \calD_p(-\lambda) \calU \|_2 \leq \overline{\calK}. \]
    Then, as $n,p\to\infty$, 
    \begin{equation*}
    \Upsilon_{\rm mc}(\calU, \varepsilon, \Lambda) -  \mathbb{P}\, \Big( \max_{\lambda\in\Lambda}
\sup_{t\in\calT_{\rm mc}(\varepsilon)}
\Big\{ G_\lambda(t)
+ \mathcal{S}_\lambda\big(t; \mathcal{U}\big)
 \Big\}
>
\tilde{\xi}_{\rm mc}(1-\alpha;\varepsilon,\Lambda)
\Big)\longrightarrow 0.     
    \end{equation*}
    \end{theorem}

As in the SC case, Theorem~\ref{thm:power_multiple_deterministic} shows that the asymptotic local power is governed by the drift process. Unlike the SC case, however, the effects of the ridge parameter $\lambda$ and the candidate segmentation $t$ do not decouple in the drift process \eqref{eq:def_drift_multiple}.

To facilitate a more refined analysis of the effect of $\lambda$, we next introduce a structural assumption on $\mathcal U$. Specifically, rather than treating $\mathcal U$ as a deterministic sequence of mean vectors, we regard it as a realization of a prior distribution and adopt a Bayesian framework, referred to as the \emph{probability alternatives} (\textbf{PA}) framework. The primary advantage of the \textbf{PA} framework is that it naturally incorporates structural information about the mean shifts through the choice of the prior distribution. This framework was first proposed by \citet{li2020adaptable} in the context of two-sample mean testing and was subsequently extended by \citet{li2020high}. We focus on the following class of prior distributions with separable covariance structure. 

\begin{block}{PA}
    Assume that, under $H_a$, 
    \[ \calU = p^{-3/4} \mathcal{G}W\Xi^T,\]
    where $W$ is a $p\times m$ random matrix with independent entries $w_{ij}$ satisfying $\mE w_{ij} =0$, $\mE w_{ij}^2 =1$, and having finite moments of all orders; $\mathcal{G}$ is a $p\times p$ deterministic matrix and $\Xi$ is a fixed  $(s+1)\times m$ matrix.   Here, $m \geq 1$ is any fixed integer.  
    Assume that  $\|{\mathcal{G}}\|_2 <\infty$ and  $p^{-1} \tr(\mathcal{G}\mathcal{G}^T) \geq \calK$ for some constant $\calK>0$, as $n,p\to\infty$,
\end{block}
To better understand \textbf{PA}, first observe that $\calU$ has mean zero, row-wise covariance matrix $p^{-3/2} \mathcal{G}\mathcal{G}^T$, and column-wise covariance matrix $\Xi\Xi^T$. The factor $p^{-3/4}$ provides the scaling under which  
$$\sqrt{p}\| \calU^T\calU \|_2 = O_p(1),$$
so that the proposed test has non-trivial local power.

We now define the drift process under the \textbf{PA} framework. Let
\[ \calQ_p(\lambda, \mathcal{G}) = p^{-1}\tr (\calD_p(-\lambda) \mathcal{G}\mathcal{G}^T), \qquad \lambda \in \Lambda.\]
which generalizes the quadratic term $q_p(\lambda,\delta_p)$ in \eqref{eq:def_q_p}. Next, define
\[ \eta_{\rm mc} (t; \Xi) = \big\|\Xi^T \psi(t)\big\|_2^2,  \qquad t\in\calT_{\rm mc}(\varepsilon),\]
which is the MC analogue of $\eta_{\rm sc}(t)$. The corresponding drift process is defined by
\[ \calS_\lambda^{\rm PA}\big(t; \mathcal{G}, \Xi\big) =\eta_{\rm mc}(t; \Xi) \frac{\calQ_p(\lambda, \mathcal{G})}{\gamma \Gamma^{1/2}(-\lambda)}[1-\gamma \Theta(-\lambda)]^2, \qquad t\in\calT_{\rm mc}(\varepsilon) ,  \quad \lambda\in\Lambda. \]

\begin{theorem}
    \label{thm:power_multiple_PA}
    Suppose that Conditions \ref{enum:moment_conditions}--\ref{enum:convergence_rate} hold. Assume that, under the alternative $H_a$, the matrix $\calU$ has a prior given by \textbf{PA}. 
    Then, 
    \begin{equation*}
    \begin{split}
    \Upsilon_{\rm mc}(\calU, \varepsilon, \Lambda)-
    \mP\Big(\max_{\lambda\in\Lambda}\sup_{t\in\calT_{\rm mc}(\varepsilon)} \Big\{G_{\lambda}(t) + \calS_\lambda^{\rm PA}\big(t; \mathcal{G}, \Xi\big)\Big\} > \tilde{\xi}_{\rm mc}(1-\alpha; \varepsilon,\Lambda)\Big) 
    \stackrel{\mP_{\mathcal{U}}}{\longrightarrow} 0, 
    \end{split}
    \end{equation*}
    where the convergence  $\xrightarrow{\mathbb{P}_{\mathcal{U}}}$ is in probability with respect to the prior distribution of $\calU$. 
\end{theorem}

Theorem~\ref{thm:power_multiple_PA} shows that under the separable covariance structure, the row-wise covariance
$\mathbf{G} \coloneqq p^{-3/2}\mathcal{G}\mathcal{G}^T$ affects the asymptotic power through $\mathcal Q_p(\lambda,\mathcal{G})$, thereby interacting with both the data covariance matrix $\Sigma_p$ and the ridge parameter $\lambda$. In contrast, the column-wise covariance $\Xi\Xi^T$ influences the asymptotic power solely through the quantity $\eta_{\rm mc}$, which captures its interaction with the true change-point configuration represented by $\psi$. Moreover, no distributional features of $W$ beyond $\mathbf{G}$ and the imposed moment conditions affect the asymptotic power. 

As in the single change-point setting, the ridge parameter affects the asymptotic power through the quantity 
\begin{equation}\label{eq:SNR_PA}
\SNR_{\rm PA}(\lambda, \calG) \coloneqq \frac{\mathcal Q_p(\lambda,\mathcal{G})[1-\gamma \Theta(-\lambda)]^2}{\Gamma^{1/2}(-\lambda)},
\end{equation}
which we refer to as the SNR under \textbf{PA}. Like the SC case, the maximizer of the SNR generally depends on $\mathcal{G}$, the covariance matrix $\Sigma_p$, and the aspect ratio $\gamma$. Consequently, consistently estimating the power-optimal ridge parameter is generally intractable unless additional structural assumptions are imposed on $\calG$. It again motivates the composite test procedure that aggregates over a collection of ridge parameters.

%% file: 3_5_Power_Comparison.tex
\subsubsection{Power Comparison}
\label{subsubsec:power_comparison}

We compare the proposed procedure with several representative approaches under the SC setting. For $t=(0,t_2,1)\in\calT_{\rm sc}(\varepsilon)$, let
\[
C_\infty(t)
=
\frac{1}{n+1-k(t_2)}
\sum_{j=k(t_2)}^{n}X_j
-
\frac{1}{k(t_2)-1}
\sum_{j=1}^{k(t_2)-1}X_j
\]
denote the difference between the two segment means. Existing methods differ primarily in how this mean contrast is normalized and in the choice of norm used to aggregate the signal. 

At one extreme, traditional covariance-normalized procedures consider the $\ell_2$-norm of the mean contrast after normalization by the full inverse sample covariance matrix:
\[
T_{\rm cov}
=
\sup_{t\in\calT_{\rm sc}(\varepsilon)}
N(t)\|C_0(t)\|_2^2,
\qquad
C_0(t)=S_n^{-1/2}C_\infty(t).
\]
Such normalization accounts for the covariance structure but becomes unstable or infeasible when $p$ is comparable to or larger than $n$.

At the other extreme, a class of existing $\ell_2$-based procedures directly uses the unnormalized mean contrast:
\[
T_{\ell_2}
=
\sup_{t\in\calT_{\rm sc}(\varepsilon)}
N(t)\|C_\infty(t)\|_2^2.
\]
Representative examples include \cite{zhang2010detecting,horvath2012change}. The U-statistic procedure of \cite{wang2022inference} is closely related to this approach: it removes the squared terms and employs the self-normalization technique of \citet{shao2010self} to obtain a pivotal limiting distribution.

The proposed ridge-regularization scheme interpolates between these two extremes. In particular,
\[
V_\lambda(t)
=
N(t)\|C_\lambda(t)\|_2^2,
\qquad
C_\lambda(t)
=
(S_n+\lambda I_p)^{-1/2}C_\infty(t).
\]
As $\lambda\to0$, the ridge transformation approaches full covariance normalization whenever $S_n$ is invertible, recovering the principle underlying $T_{\rm cov}$. Conversely, for any fixed observation samples,
\[
\sqrt{\lambda}(S_n+\lambda I_p)^{-1/2}C_\infty(t)
\longrightarrow C_\infty(t),
\qquad \lambda\to\infty,
\]
so that the large-ridge regime approaches the unnormalized $\ell_2$ procedure. Notably, the additional factor $\sqrt{\lambda}$ does not affect the standardized statistic $D_\lambda(t)$, since the corresponding scale is simultaneously absorbed by the normalization through $\Theta(-\lambda)$ and $\Gamma^{1/2}(-\lambda)$. Thus, $\lambda$ controls the degree of covariance normalization, ranging from full normalization to essentially no normalization.

Finally, for sparse alternatives, $\ell_\infty$-based procedures such as those of \citet{jirak2012change,yu2021high} detect changes through large coordinatewise contrasts and can be represented by
\[
T_{\ell_\infty}
=
\sup_{t\in\calT_{\rm sc}(\varepsilon)}
N(t)\|C_\infty(t)\|_\infty^2.
\]

Thus, the covariance-normalized, ridge-regularized, and unnormalized $\ell_2$ procedures primarily differ in the degree of covariance normalization, whereas the $\ell_\infty$ procedures employ a different signal aggregation strategy tailored to sparse alternatives.

Given these relationships, the local power analysis in Section~\ref{subsubsec:power_single_change_point} provides a unified framework for comparing $T_{\rm cov}$, $T_{\ell_2}$, and the proposed ridge-regularized procedure. Note that under the local alternative regime considered therein, the effect of the ridge parameter on local power is characterized by
\[
\SNR(\lambda,\delta_p)
=
\frac{q_p(\lambda,\delta_p)[1-\gamma\Theta(-\lambda)]^2}
{\Gamma^{1/2}(-\lambda)}.
\]
The limiting regimes $\lambda\to0$ and $\lambda\to\infty$ correspond to the covariance-normalized and unnormalized $\ell_2$ procedures, respectively. If the SNR is maximized at a finite positive value of $\lambda$, a ridge-regularized procedure with $\lambda$ near the optimum can achieve higher local power than both limiting procedures. If instead the maximum occurs as $\lambda\to0$ or $\lambda\to\infty$, then $T_{\rm cov}$ (when applicable) or $T_{\ell_2}$ is asymptotically optimal within this continuum of covariance normalization. The numerical results in Section~\ref{subsubsec:effect_lambda} demonstrate that the optimal ridge parameter can lie in the small, moderate, or large regime, depending on the interaction among $\delta_p$, $\Sigma_p$, and $\gamma$.

This observation further motivates the proposed composite procedure. Since sufficiently small and large ridge parameters approximate $T_{\rm cov}$ and $T_{\ell_2}$, respectively, choosing $\Lambda$ to contain small, moderate, and large values of $\lambda$ allows the procedure to adapt to different degrees of covariance normalization. Aggregating across these values therefore provides robustness to the unknown power-optimal ridge parameter and maintains strong performance across a broad range of alternatives.

The comparison with the $\ell_\infty$-based procedures is less direct, as their local power generally does not admit an explicit form comparable to the SNR derived above. Nevertheless, the difference between the two approaches can be illustrated in the simplified setting where $\Sigma_p=I_p$, the observations are Gaussian, and $p\asymp n$. In this setting, $\|C_\infty(t)\|_\infty \asymp \sqrt{\log (p)}/\sqrt{p}$ under $H_0$. Therefore, the detection boundary under $H_a^{\rm SC}$ for $\ell_\infty$-based procedures is of the order
\[\|\delta_p\|_\infty \asymp\sqrt{\log(p)}/\sqrt{p}.\]
That is, such procedures have asymptotic power tending to one when $\|\delta_p\|_\infty \gg \sqrt{\log(p)}/\sqrt{p}$, whereas their power approaches the nominal level when $\|\delta_p\|_\infty \ll \sqrt{\log(p)}/\sqrt{p}$.

In contrast, the proposed $\ell_2$-based procedure has nontrivial local power when
$\|\delta_p\|_2\asymp n^{-1/4}$, with asymptotically trivial and overwhelming power when
$\|\delta_p\|_2\ll n^{-1/4}$ and $\|\delta_p\|_2\gg n^{-1/4}$, respectively.

To compare the two detection regimes, suppose
$\|\delta_p\|_2\asymp n^{-1/4}$ and let $\|\delta_p\|_0$ denote the number of nonzero coordinates of
$\delta_p$. If these nonzero coordinates are of comparable magnitude, then
$
\|\delta_p\|_\infty
\asymp
{n^{-1/4}}/{\sqrt{\|\delta_p\|_0}}$. Consequently,
\[
\|\delta_p\|_\infty
\gg
\sqrt{\frac{\log n}{n}}
\quad\Longleftrightarrow\quad
\|\delta_p\|_0\ll \frac{\sqrt n}{\log n}
\quad\text{and}\quad 
\|\delta_p\|_\infty
\ll
\sqrt{\frac{\log n}{n}}
\quad\Longleftrightarrow\quad
\|\delta_p\|_0\gg \frac{\sqrt n}{\log n}.
\]
Hence, sparse signals, characterized by
$\|\delta_p\|_0\ll \sqrt n/\log n$, lie above the $\ell_\infty$ detection boundary even at the local $\ell_2$ scale, favoring $\ell_\infty$-based procedures. Conversely, for dense signals with
$\|\delta_p\|_0\gg \sqrt n/\log n$, the coordinatewise signal falls below the $\ell_\infty$ detection boundary while the aggregate $\ell_2$ signal remains nontrivial, favoring the proposed $\ell_2$-based procedure. This comparison highlights the complementary nature of the two approaches: $\ell_\infty$-based procedures are tailored to sparse changes, whereas the proposed procedure is designed for dense alternatives.

%% file: 3_6_Effect_Lambda.tex
\subsubsection{Numerical Studies of the Effect of the Ridge Parameter}
\label{subsubsec:effect_lambda}
In this section, we numerically investigate the effect of $\lambda$ on the asymptotic local power under illustrative settings. We focus on the \textbf{PA} framework and study the SNR defined in \eqref{eq:SNR_PA} as a function of $\lambda$ under different choices of the PSD $H_p$ (see \eqref{eq:def_PSD}), the signal structure $\calG$, and the aspect ratio $\gamma$.

We consider a simple covariance model in which $\Sigma_p$ has two distinct eigenvalues, denoted by $\alpha_{(1)}$ and $\alpha_{(2)}$, with proportions $H_p(\{\alpha_{(1)}\})$ and $1-H_p(\{\alpha_{(1)}\})$, respectively. We normalize $\Sigma_p$ so that $p^{-1}\tr(\Sigma_p)=1$. 
We consider three choices of  $\mathbf{G}=\calG\calG^T$: $\mathbf{G}\propto I_p$, $\mathbf{G}\propto\Sigma_p$, $\mathbf{G}\propto\Sigma_p^2$, where in each case $\mathbf{G}$ is normalized so that $p^{-1}\tr(\mathbf{G})=1$. These settings represent progressively stronger alignment of the signal with the high-variance eigendirections of $\Sigma_p$, ranging from an isotropic signal under $\mathbf{G}\propto I_p$ to a strongly covariance-aligned signal under $\mathbf{G}\propto\Sigma_p^2$.

\begin{figure}[htbp]
\centering
\includegraphics[width=0.8\linewidth]{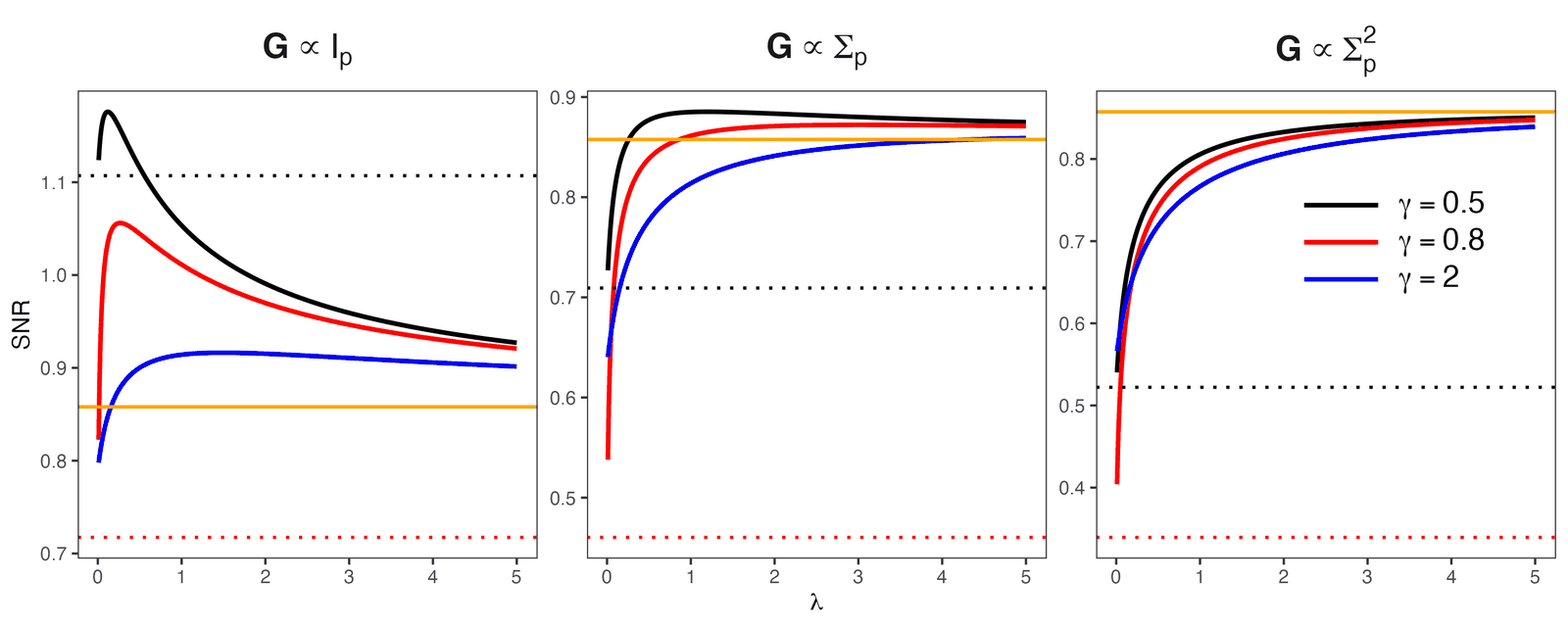}
\caption{$\SNR_{\rm PA}$ as a function of $\lambda$ for $\alpha_{(1)}=0.4$ and $H_p({\alpha_{(1)}})=0.5$. The solid black, red, and blue curves show the SNR of the proposed ridge-regularized procedure for $\gamma=0.5$, $0.8$, and $2$, respectively. The dotted horizontal lines indicate the limits as $\lambda\to0$ (black for $\gamma=0.5$ and red for $\gamma=0.8$), corresponding to $T_{\rm cov}$, which is applicable only when $\gamma<1$. The dashed horizontal lines indicate the common limit as $\lambda\to\infty$ for all three values of $\gamma$, corresponding to $T_{\ell_2}$.}
\label{fig:snr_a_5_x_4}
\end{figure}

Under each setting, $\SNR_{\rm PA}$ is evaluated numerically. Figure \ref{fig:snr_a_5_x_4} displays the SNR curves for $\alpha_{(1)}=0.4$ and $H_p({\alpha_{(1)}})=0.5$, while three additional settings are reported in Section 1 of the Supplementary Material. The main findings can be summarized as follows. First, when $\mathbf{G}$ is not strongly aligned with the covariance structure, a small or moderate value of $\lambda$ is generally preferred, demonstrating the potential advantage of ridge regularization over the two limiting procedures. In contrast, when $\mathbf{G}\propto\Sigma_p^2$, the SNR is typically maximized as $\lambda\to\infty$, favoring the unnormalized procedure $T_{\ell_2}$. Second, when $\gamma$ is relatively small, smaller values of $\lambda$ tend to be favored, whereas for larger $\gamma$, the SNR is generally less sensitive to $\lambda$ over a reasonably broad range. Notably, the asymptotic SNR of $T_{\ell_2}$ is independent of $\gamma$, suggesting a potential advantage of the unnormalized approach in the regime $p\gg n$. Third, when the spectrum of $\Sigma_p$ is highly dispersed, particularly when a small proportion of eigenvalues are substantially larger than the others, the advantage of a moderate $\lambda$ over the unnormalized limit $\lambda\to\infty$ becomes more pronounced. This advantage diminishes as the spectrum becomes more homogeneous. In the extreme case $\Sigma_p\propto I_p$, the unnormalized procedure $T_{\ell_2}$ effectively uses the true covariance structure up to a scalar factor, leaving little room for improvement through additional covariance normalization.

%% file: 5_simulation.tex
\section{Simulation Studies}\label{sec:simulation}
In this section, we evaluate the finite-sample performance of the proposed procedures and compare them with several representative competing methods through Monte Carlo simulations. We consider three distributions for the noise vectors $\bz_j$: standard normal, Poisson$(10)$, and $t(6)$, with the latter two standardized to have mean zero and unit variance. Note that the $t(6)$ setting violates the moment condition in \ref{enum:moment_conditions}. We report empirical sizes under all three distributions, but restrict the power comparisons to the Gaussian setting, as the empirical power differs only slightly across the three distributions.

We set $n=400$ and consider $p$ ranging from $200$ to $800$. We consider four models for $\Sigma_p$ in \eqref{eq:model}:
(i) \emph{Identity}: $\Sigma_p=I_p$;
(ii) \emph{Toeplitz}: $\Sigma_p$ is a Toeplitz matrix with $(i,j)$th entry $0.3^{|i-j|}$;
(iii) \emph{Poly Decay}: the eigenvalues $\tau_j$ of $\Sigma_p$ decay polynomially according to $\tau_j=0.01+(p-j+0.1)^2$;
and
(iv) \emph{Exp Decay}: the eigenvalues $\tau_j$ of $\Sigma_p$ decay exponentially according to $\tau_j=\exp(-3j/p)$.
For all four models, $\Sigma_p$ is further rescaled so that $\tr(\Sigma_p)=p$. The Exp Decay model represents a challenging setting with a relatively dispersed spectrum and is included to evaluate the performance of the proposed procedures under substantial spectral heterogeneity.

For computational efficiency, we implement the single change-point scanning statistic $T_{\rm sc}$ and the discretized multiple change-point scanning statistic $T_{\rm disc}$. The ridge parameter set $\Lambda$ and the trimming parameter $\varepsilon$ are chosen according to the recommendations in Section~\ref{sec:method}. We compare the proposed methods with the following representative competitors.
\begin{itemize}
\item \textbf{Cov-Norm.} The classical covariance-normalized CUSUM procedure based on $T_{\rm cov}$ introduced in Section~\ref{subsubsec:power_comparison}. This procedure corresponds to the limiting case of the proposed ridge-regularized method as $\lambda\to0$ and is applicable only when $p<n-1$. We use the same scanning sets as the proposed procedures, namely $\calT_{\rm sc}(\varepsilon)$ under the SC setting and $\calT_{\rm disc}(\varepsilon)$ under the MC setting.

\item \textbf{U-Stat.} The unnormalized U-statistic-based procedure of \citet{wang2022inference}. As discussed in Section~\ref{subsubsec:power_comparison}, this procedure is closely related to the unnormalized $\ell_2$-norm approach, which corresponds to the limiting regime $\lambda\to\infty$ of the proposed method. We implement statistic (2.3) of \citet{wang2022inference} under the SC setting and the multiple change-point procedure in (2.4)--(2.5) of their work under the MC setting.

\item \textbf{$\ell_\infty$-Sparse.} The $\ell_\infty$-norm procedure based on $T_{\ell_\infty}$ introduced in Section~\ref{subsubsec:power_comparison}. This procedure is designed for sparse alternatives and is included as a benchmark for evaluating performance under sparse signal models. We again use the same scanning sets as the proposed procedures, namely $\calT_{\rm sc}(\varepsilon)$ under the SC setting and $\calT_{\rm disc}(\varepsilon)$ under the MC setting.
\end{itemize}

\subsection{Null Validation}\label{sec:null}
We assess the empirical sizes of the proposed test statistics at the nominal level $5\%$.  The critical values are approximated using the procedure introduced in Section \ref{subsec:asymptotic_null}. The results are reported in Table \ref{tab:size}.

\begin{table}[ht]
\centering
\caption{Empirical sizes $\times 100\%$ of the proposed test statistics
at the nominal level $\alpha=0.05$ for $n=400$.}
\label{tab:size}
\resizebox{\linewidth}{!}{
\begin{tabular}{lccccccccccccccccc}
\toprule
& & \multicolumn{4}{c}{Identity}
  & \multicolumn{4}{c}{Toeplitz}
  & \multicolumn{4}{c}{Poly Decay}
  & \multicolumn{4}{c}{Exp Decay} \\
\cmidrule(r){3-6}
\cmidrule(r){7-10}
\cmidrule(r){11-14}
\cmidrule(l){15-18}
 $n=400$ & $p=$
& $300$ & $500$ & $600$ & $800$
& $300$ & $500$ & $600$ & $800$
& $300$ & $500$ & $600$ & $800$
& $300$ & $500$ & $600$ & $800$ \\
\midrule

\multirow{3}{*}{$T_{\rm sc}$}
& Normal
& 5.21 & 4.75 & 5.07 & 5.41
& 5.91 & 4.91 & 5.10 & 5.52
& 6.88 & 5.82 & 5.31 & 4.93
& 7.24 & 6.14 & 5.66 & 5.43 \\
& $t(6)$
& 5.59 & 4.69 & 4.89 & 4.97
& 5.78 & 4.95 & 5.24 & 5.50
& 7.00 & 5.90 & 5.36 & 4.99
& 6.81 & 5.73 & 5.40 & 5.62 \\
& Poi$(10)$
& 5.50 & 4.90 & 4.85 & 5.09
& 6.18 & 5.23 & 5.06 & 5.28
& 6.82 & 5.48 & 5.57 & 4.91
& 6.90 & 5.77 & 5.43 & 5.23 \\

\midrule

\multirow{3}{*}{$T_{\rm disc}$}
& Normal
& 6.01 & 4.70 & 4.89 & 5.27
& 6.54 & 5.13 & 5.08 & 5.37
& 8.27 & 6.27 & 5.72 & 5.35
& 8.22 & 6.04 & 6.27 & 5.81 \\
& $t(6)$
& 6.50 & 4.64 & 4.99 & 5.06
& 7.07 & 5.21 & 5.06 & 5.62
& 8.48 & 5.98 & 6.27 & 5.54
& 8.60 & 6.17 & 6.19 & 5.78 \\
& Poi$(10)$
& 5.98 & 4.75 & 4.99 & 5.48
& 7.05 & 5.18 & 5.38 & 5.31
& 8.86 & 5.57 & 5.88 & 5.46
& 7.96 & 6.15 & 5.79 & 5.72 \\

\bottomrule
\end{tabular}
}
\end{table}

The results support the accuracy of the proposed Gaussian approximation across a broad range of covariance and distributional settings. Overall, the empirical sizes are reasonably well controlled at the nominal level, with some mild inflation when $p$ is relatively small. This inflation can be attributed in part to approximating the finite-sample distribution of $D_{\lambda}(t)$, which is expected to be right-skewed and $\chi^2$-like, by a symmetric Gaussian distribution. The results under the $t(6)$ distribution are comparable to those under the Gaussian and Poisson distributions, suggesting that the proposed procedures remain robust to moderately heavy-tailed noise even when the moment condition in \ref{enum:moment_conditions} is violated.

Compared with the single change-point procedure, the multiple change-point procedure generally exhibits greater variability and somewhat more pronounced size inflation. This is partly due to the shorter segments included in the multiple change-point scan, which lead to greater finite-sample variability and a less accurate Gaussian approximation. Moreover, the maximization over a substantially larger collection of candidate segmentations can further amplify these finite-sample approximation errors, resulting in more noticeable size distortion.

\subsection{Power Comparison}\label{sec:power_comparison}

We compare the empirical power of the proposed procedures with that of the competing methods. The study is divided into two scenarios: (i) the SC setting, where $s$ is known to be at most $1$, and (ii) the MC setting, where $s=2$ and is treated as unknown. Let $\delta$ denote the mean shift, generated according to one of the following three models:
\begin{itemize}
\item \textbf{i.i.d.} $\delta\sim N(0,cI_p)$, corresponding to a dense alternative with homoscedastic signal across coordinates.
\item \textbf{$\Sigma_p$-correlated.} $\delta\sim N(0,c\Sigma_p)$, so that the signal is aligned with the covariance structure of $\Sigma_p$.
\item \textbf{Sparse.} Three coordinates are selected uniformly at random and assigned the values $\pm5c$ with independent random signs, while the remaining $p-3$ coordinates are set to zero.
\end{itemize}
Here, $c>0$ controls the signal strength. Under the SC setting, the mean vectors are generated as
\begin{equation}
\tag{SC}
\mu_j
=
\delta\,\mathbbm{1}(j>\lfloor n/2\rfloor),
\qquad
j=1,\ldots,n.
\end{equation}
Under the MC setting, the mean vectors are generated as
\begin{equation}
\tag{MC}
\mu_j
=
\delta\,\mathbbm{1}\!\left(
\lfloor0.35n\rfloor
\le
j
\le
\lfloor0.65n\rfloor
\right),
\qquad
j=1,\ldots,n.
\end{equation}
This configuration is commonly referred to as the \emph{epidemic alternative}. It represents a temporary departure from the baseline mean followed by a return to the original state and has been widely adopted as a benchmark model in the change-point literature. 

Since the asymptotic critical value for the $\ell_\infty$-Sparse procedure is difficult to obtain in the high-dimensional regime, we calibrate the critical values of all competing procedures by Monte Carlo simulation under each $(\Sigma_p,p,n)$ configuration to achieve the nominal significance level. This calibration ensures comparable Type~I error control across methods, so that differences in empirical power primarily reflect the ability of the competing test statistics to distinguish the alternative from the null.
\begin{figure}[ht]
\centering
\setlength{\tabcolsep}{1pt}
\renewcommand{\arraystretch}{0}
\begin{tabular}{cccc}
\includegraphics[width=0.24\textwidth]{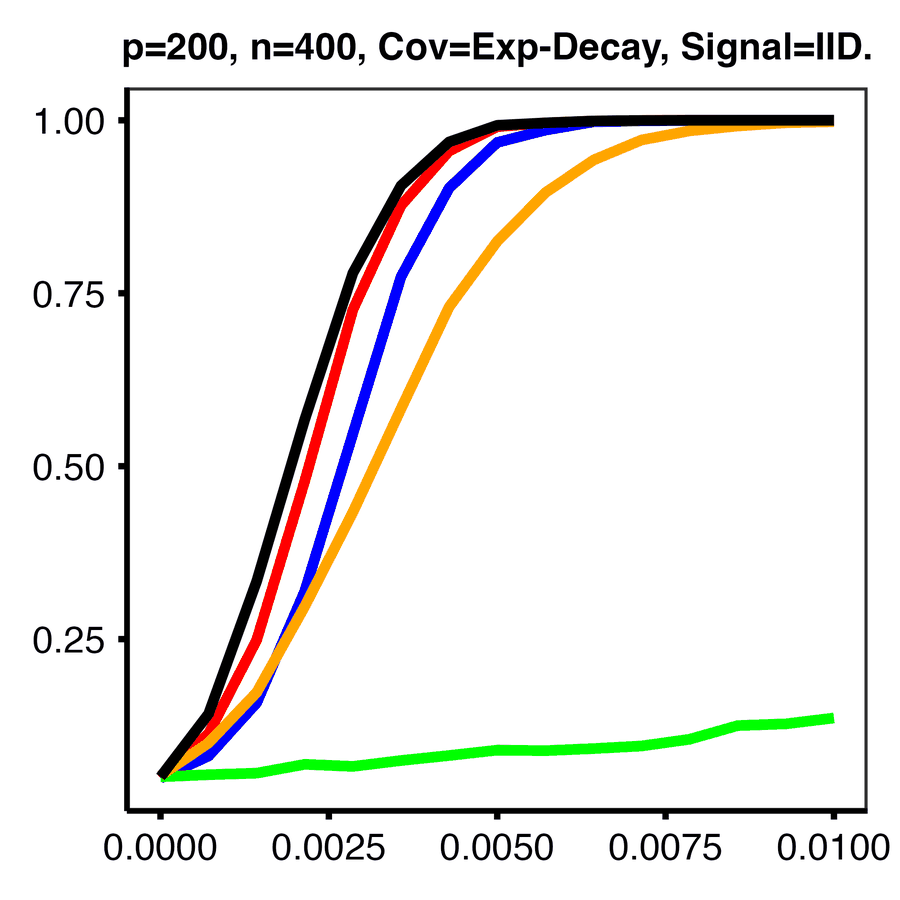} &
\includegraphics[width=0.24\textwidth]{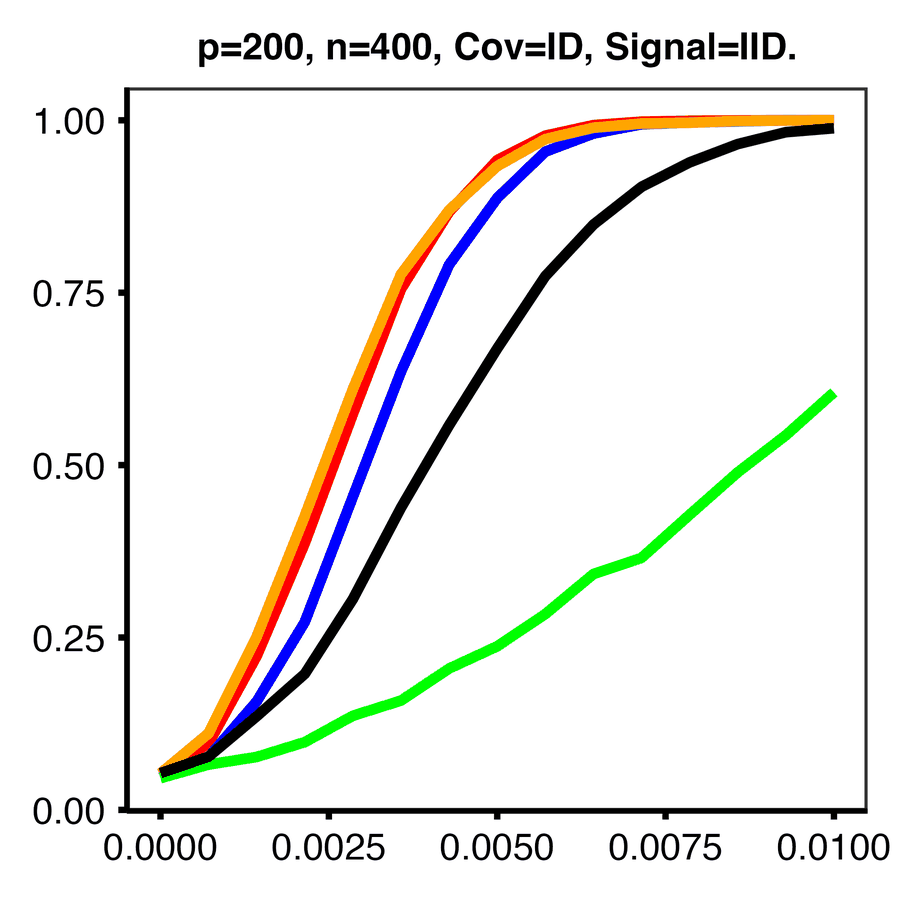} &
\includegraphics[width=0.24\textwidth]{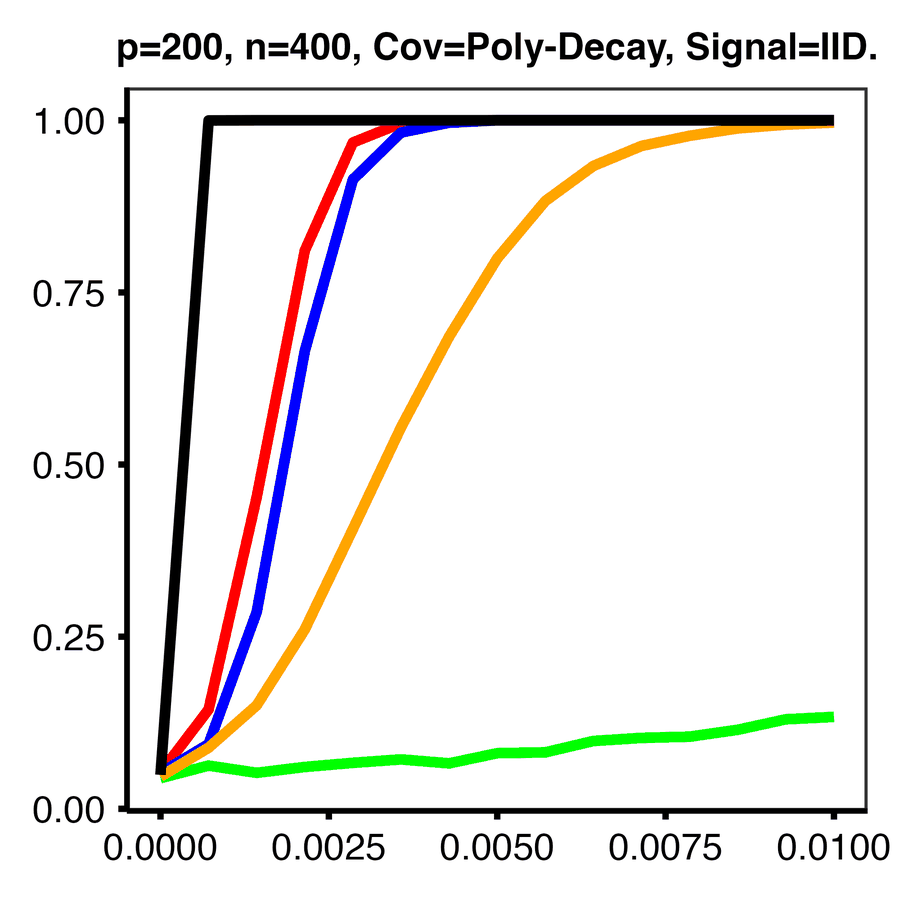} &
\includegraphics[width=0.24\textwidth]{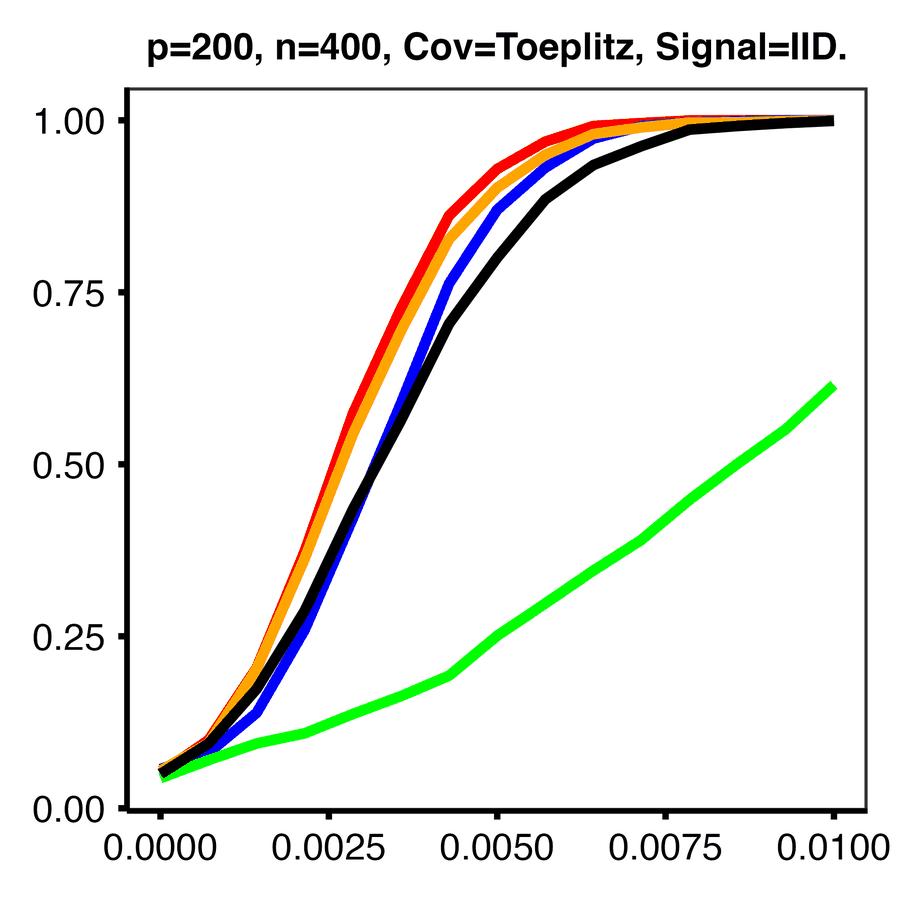}
\\
\includegraphics[width=0.24\textwidth]{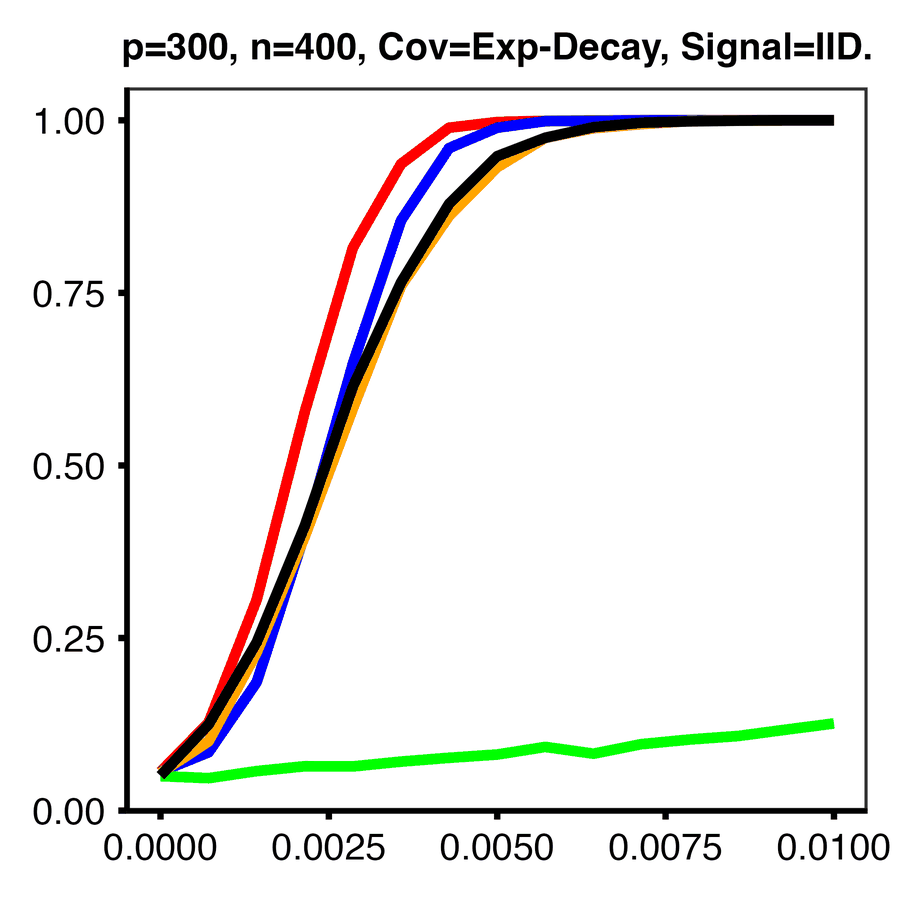} &
\includegraphics[width=0.24\textwidth]{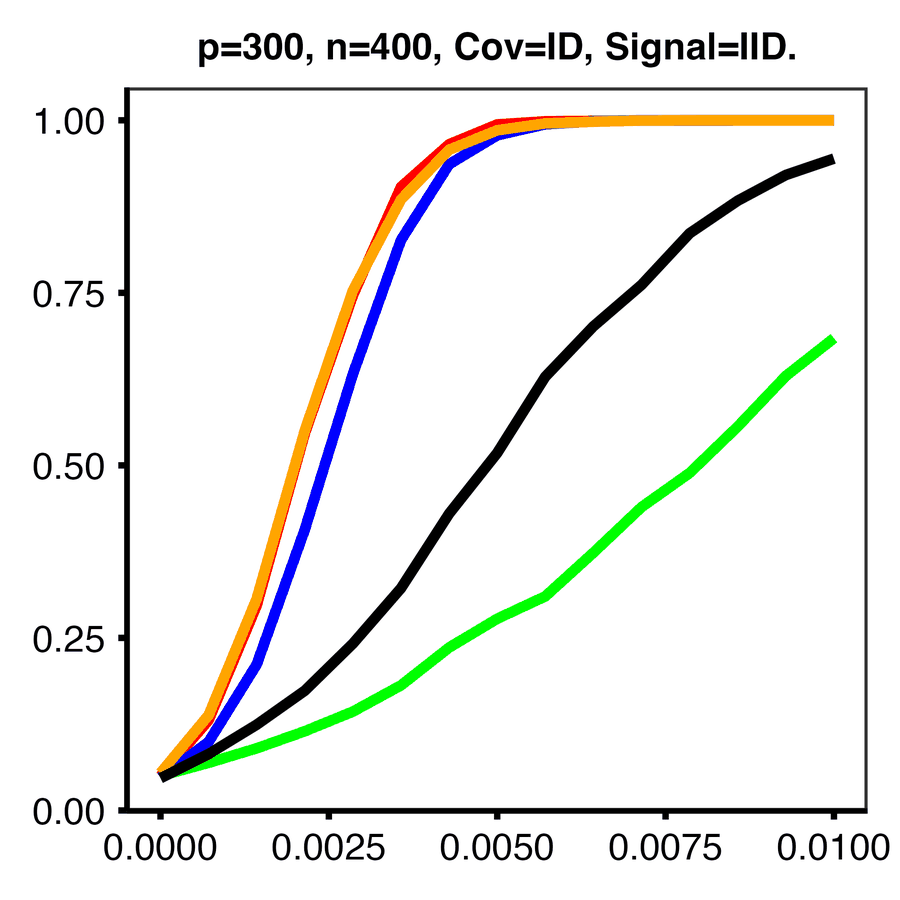} &
\includegraphics[width=0.24\textwidth]{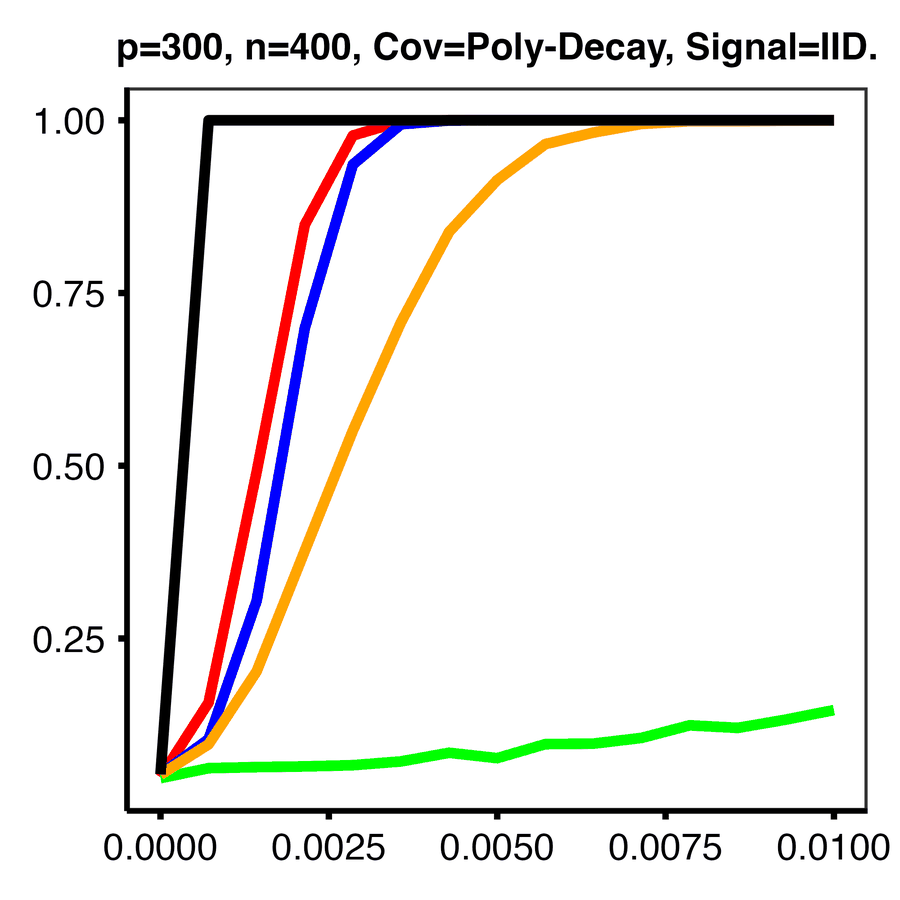} &
\includegraphics[width=0.24\textwidth]{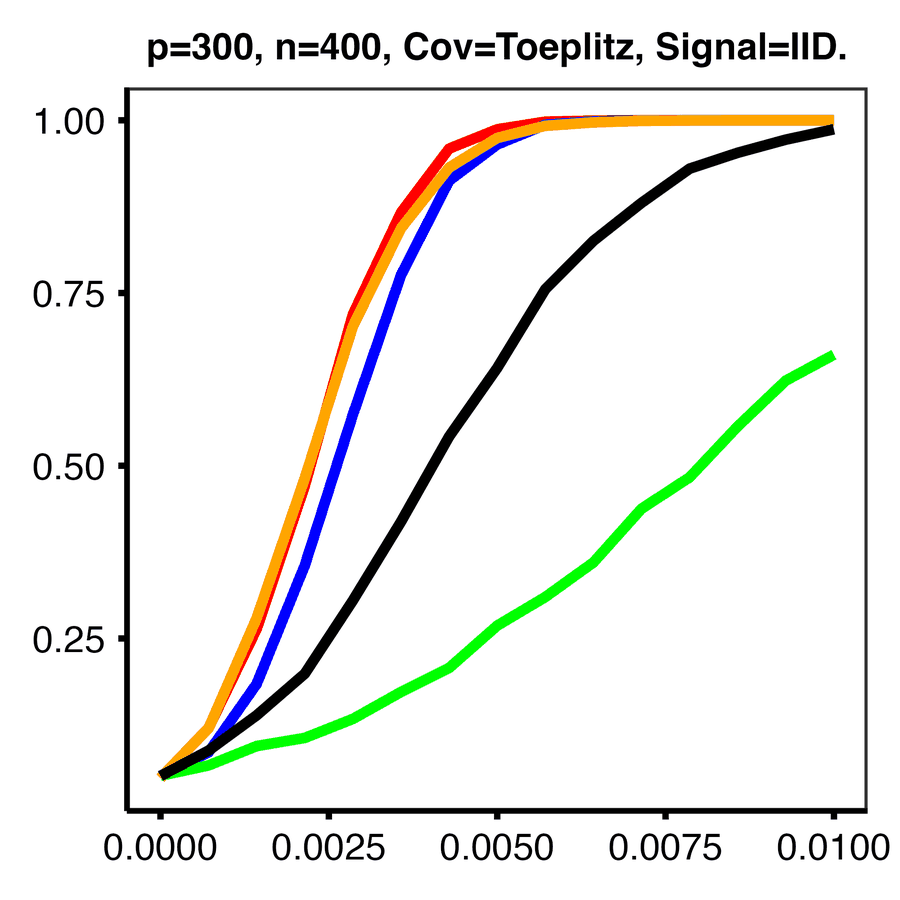}
\\
\includegraphics[width=0.24\textwidth]{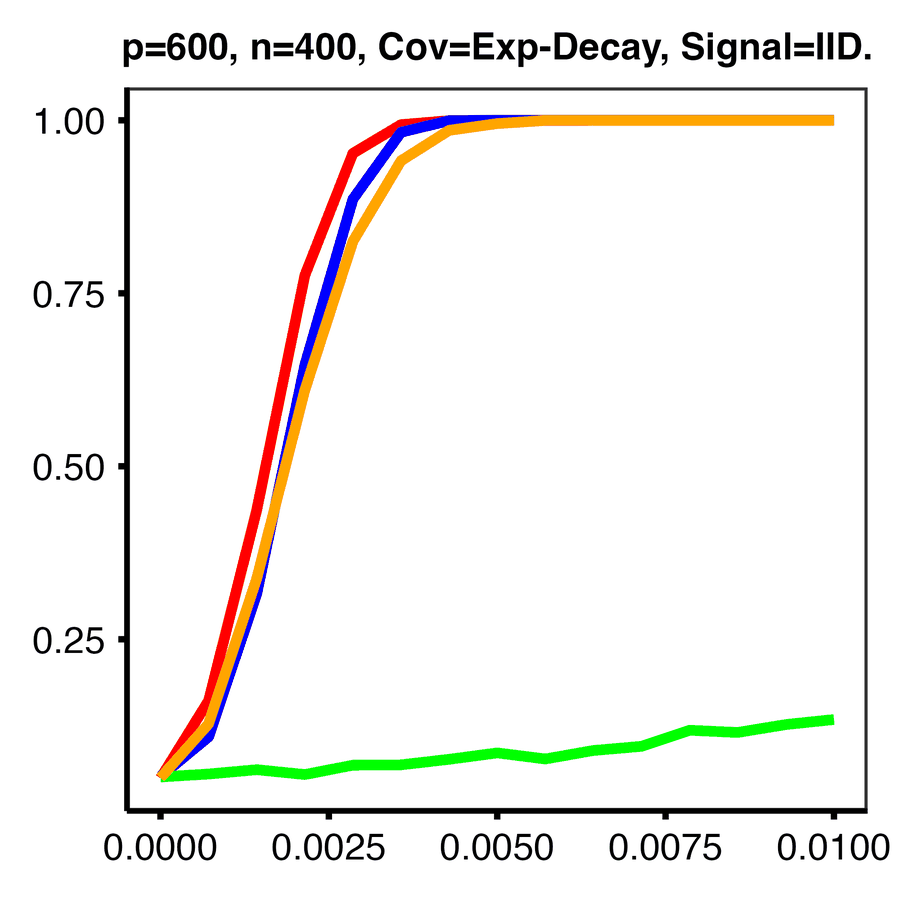} &
\includegraphics[width=0.24\textwidth]{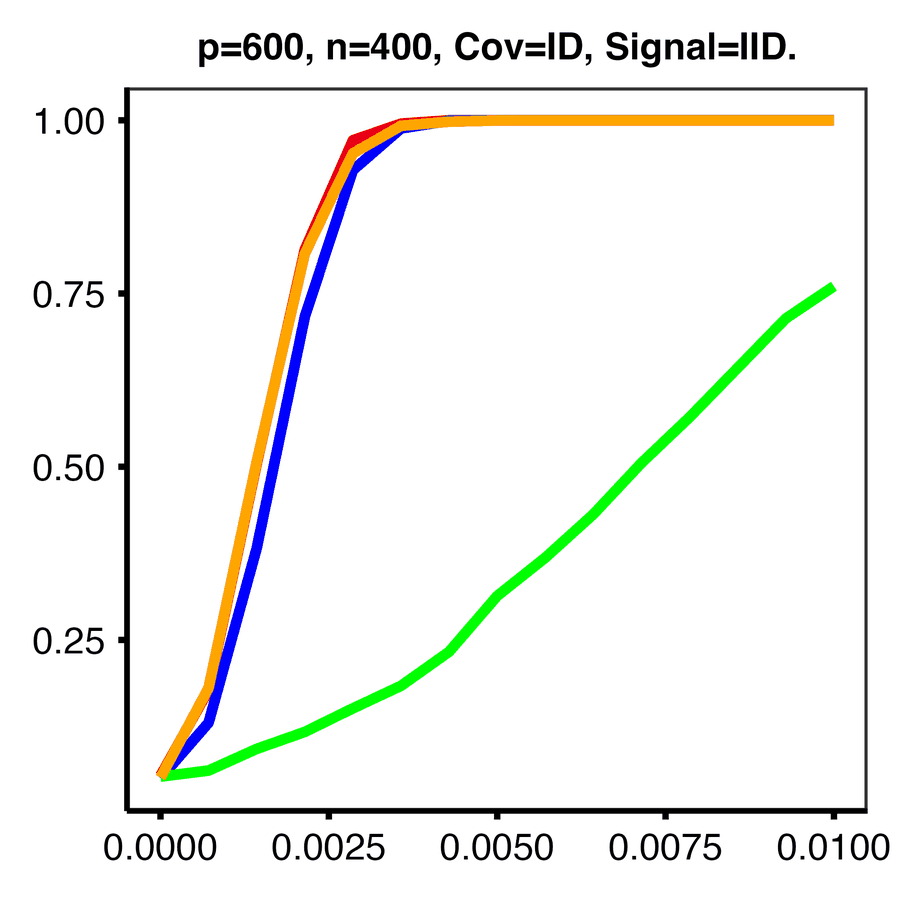} &
\includegraphics[width=0.24\textwidth]{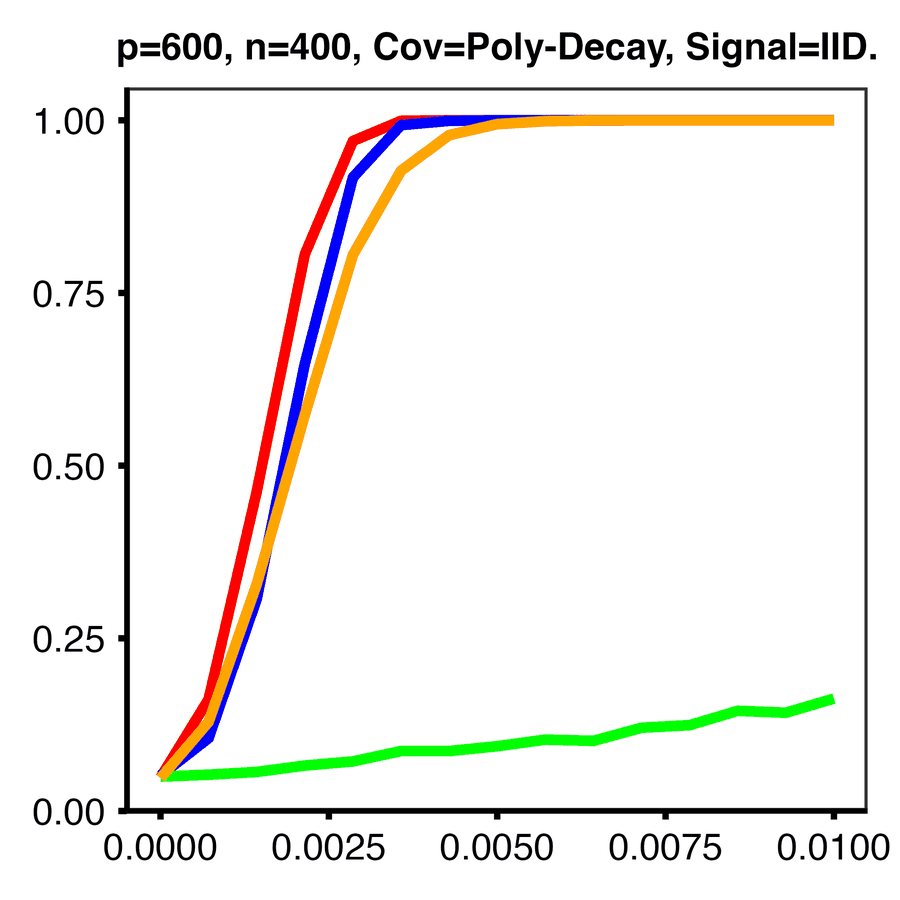} &
\includegraphics[width=0.24\textwidth]{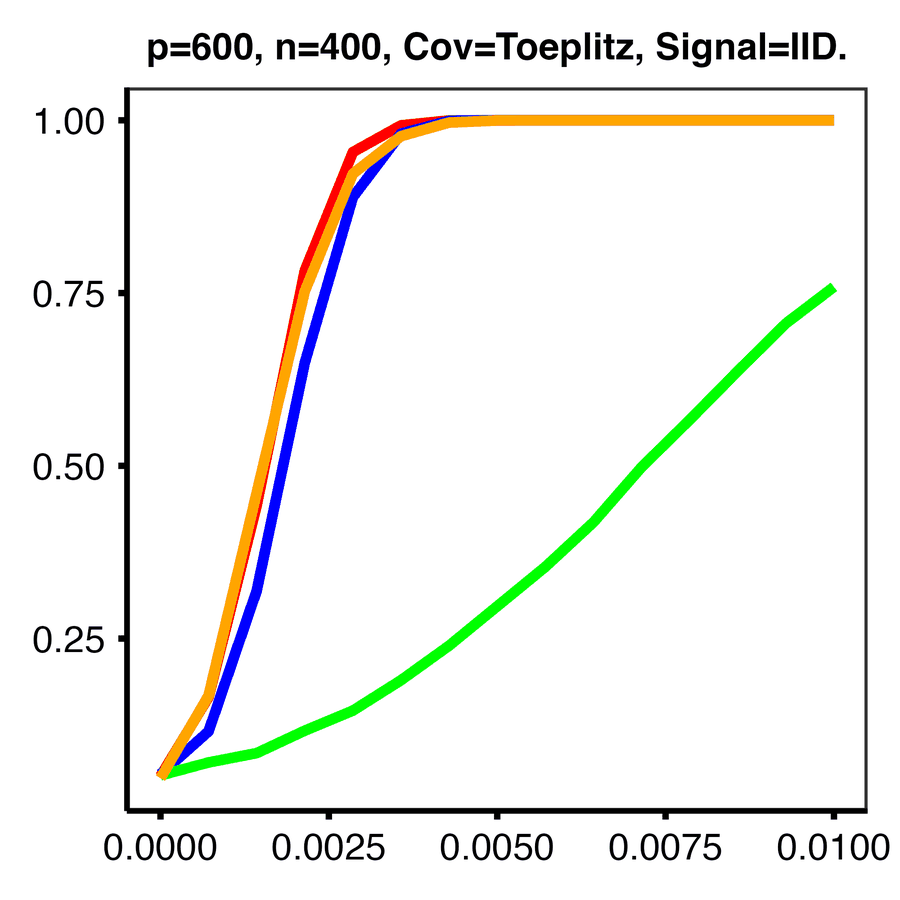}
\\
\includegraphics[width=0.24\textwidth]{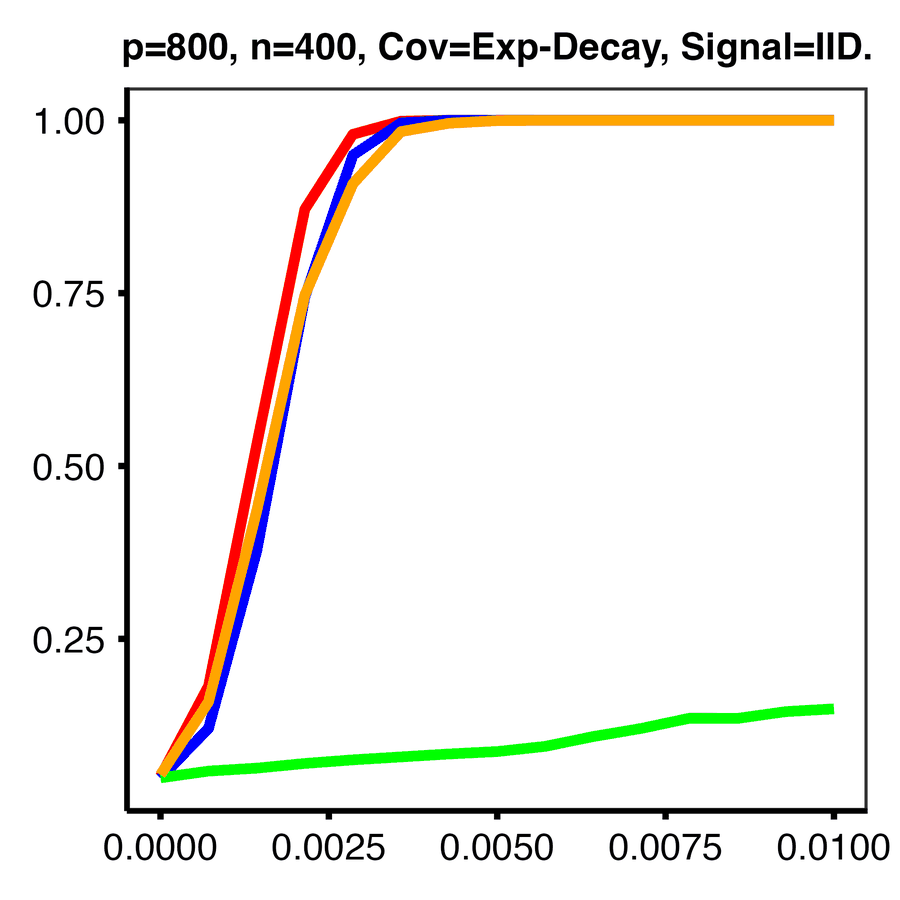} &
\includegraphics[width=0.24\textwidth]{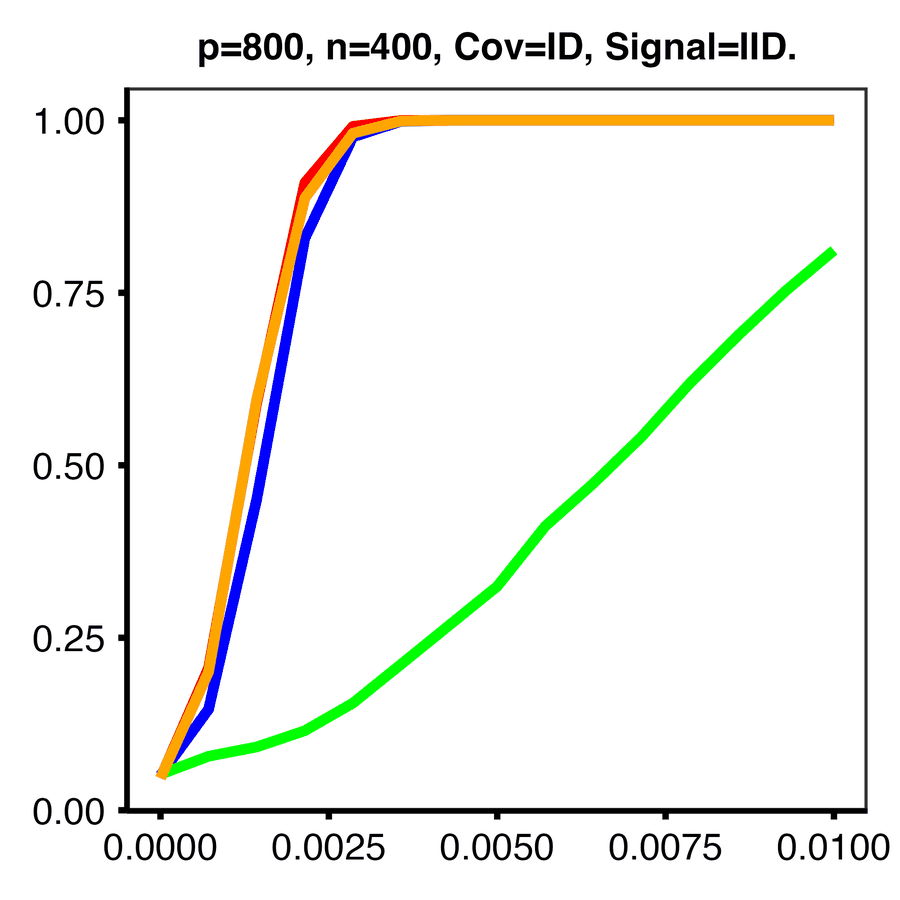} &
\includegraphics[width=0.24\textwidth]{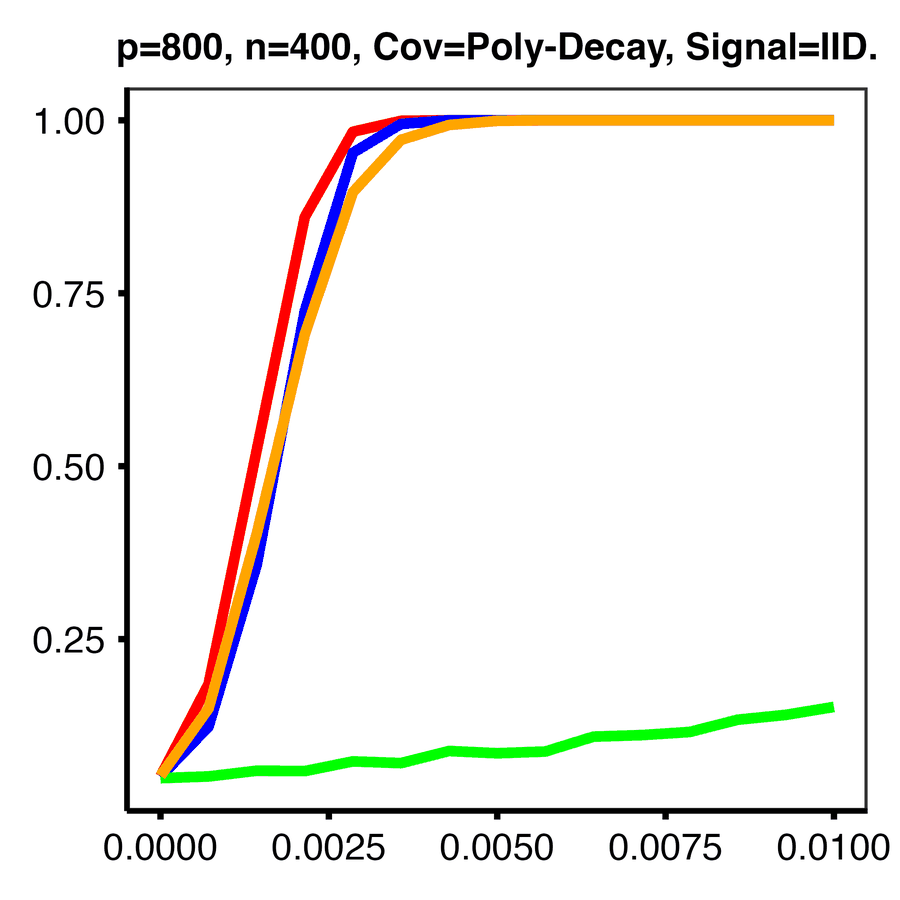} &
\includegraphics[width=0.24\textwidth]{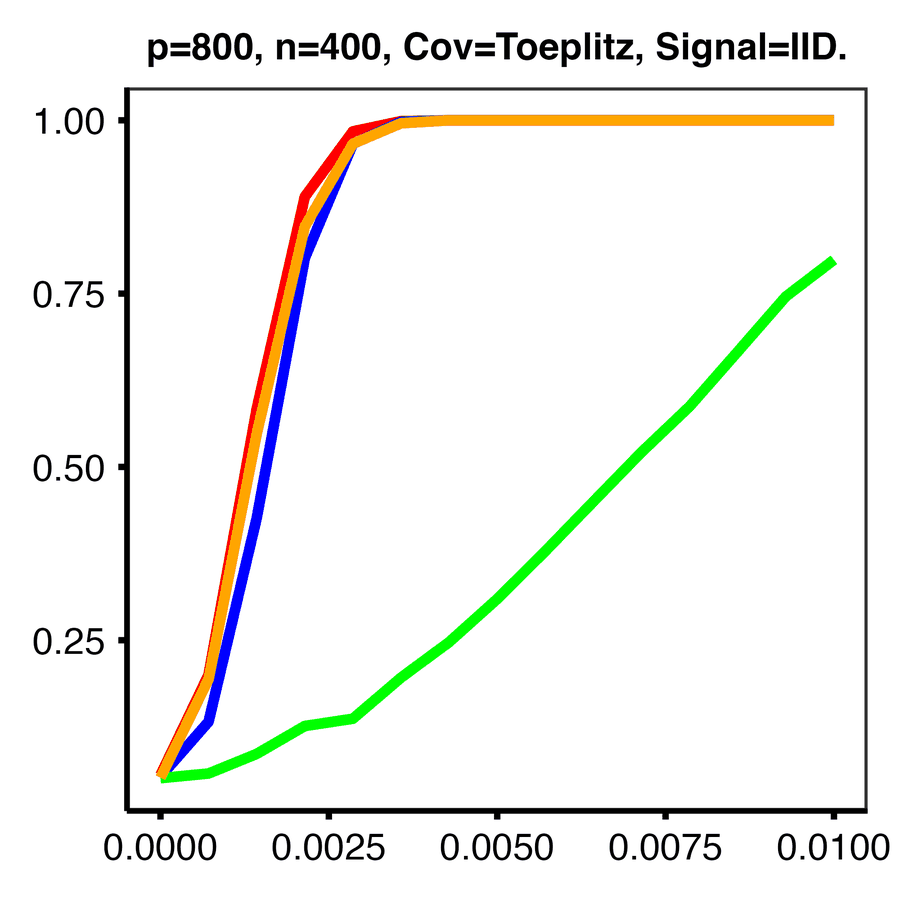}
\end{tabular}
\caption{Power curves under the SC setting with i.i.d.\ mean shifts $\delta$. Rows correspond to p= 200, 300, 600, and 800. Columns correspond to the Exp Decay, Identity, Poly Decay, and Toeplitz covariance models. Blue: $T_{\rm disc}$; Red: $T_{\rm sc}$; Black (when p<n): Cov-Norm; Orange: U-Stat; Green: $\ell_\infty$-Sparse.}
\label{fig:power_SC_iid}
\end{figure}
\begin{figure}[ht]
\centering
\setlength{\tabcolsep}{1pt}
\renewcommand{\arraystretch}{0}
\begin{tabular}{cccc}
\includegraphics[width=0.24\textwidth]{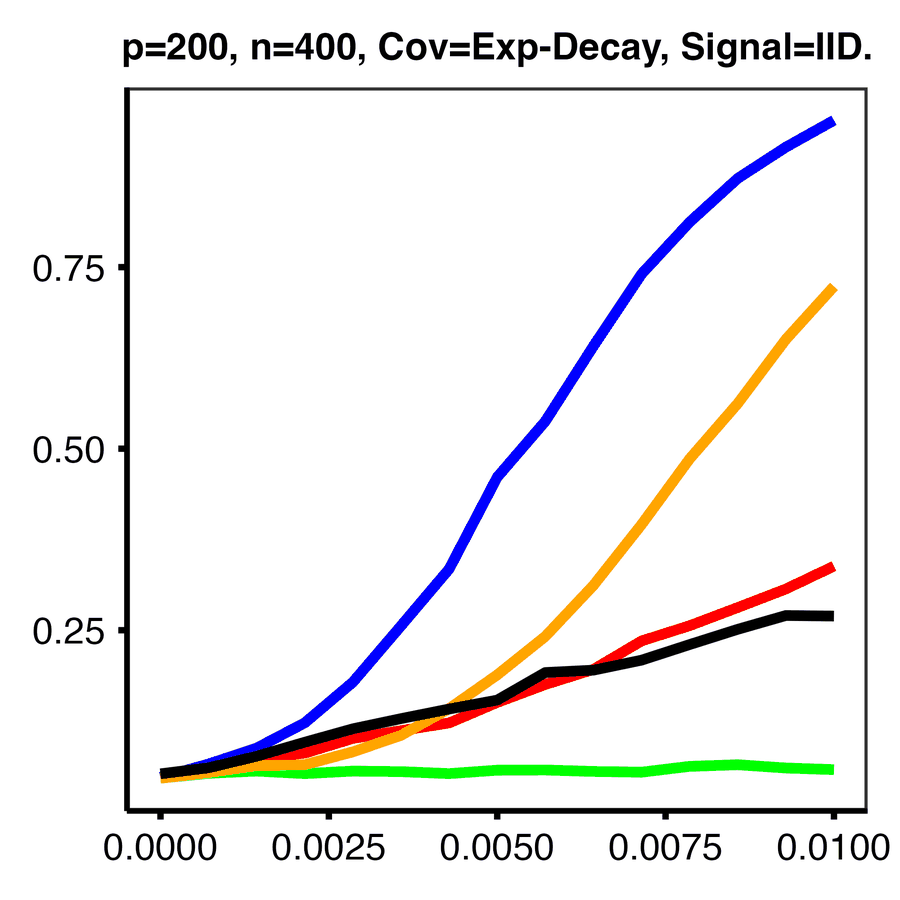} &
\includegraphics[width=0.24\textwidth]{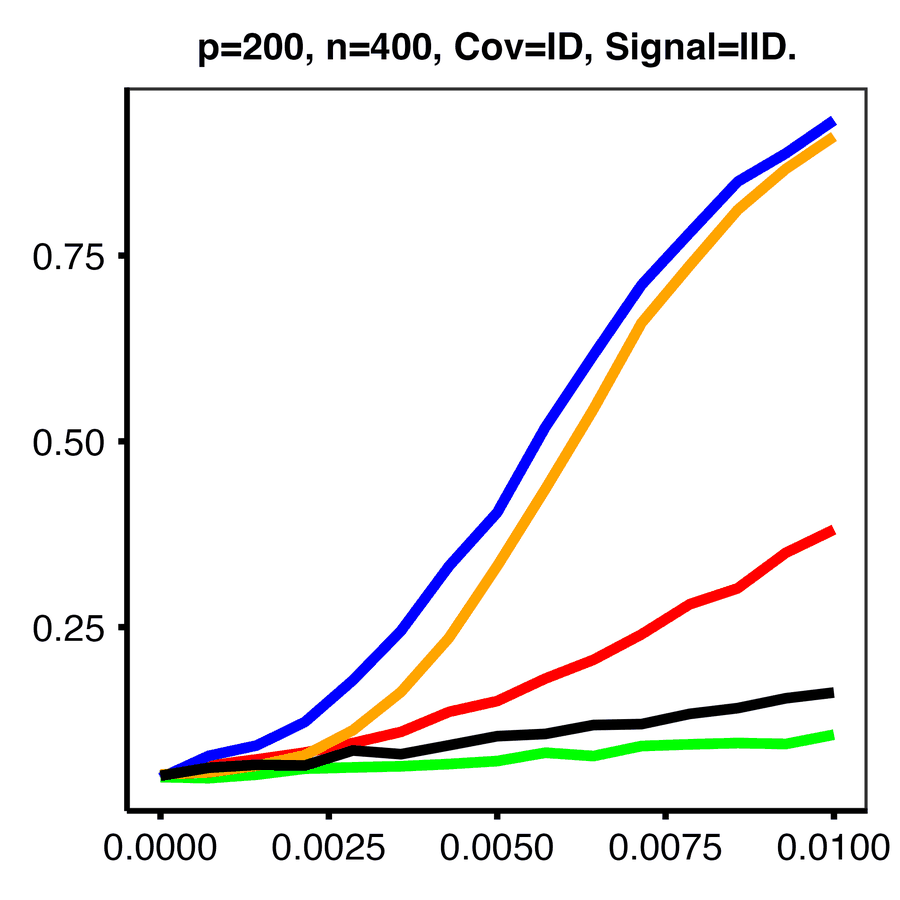} &
\includegraphics[width=0.24\textwidth]{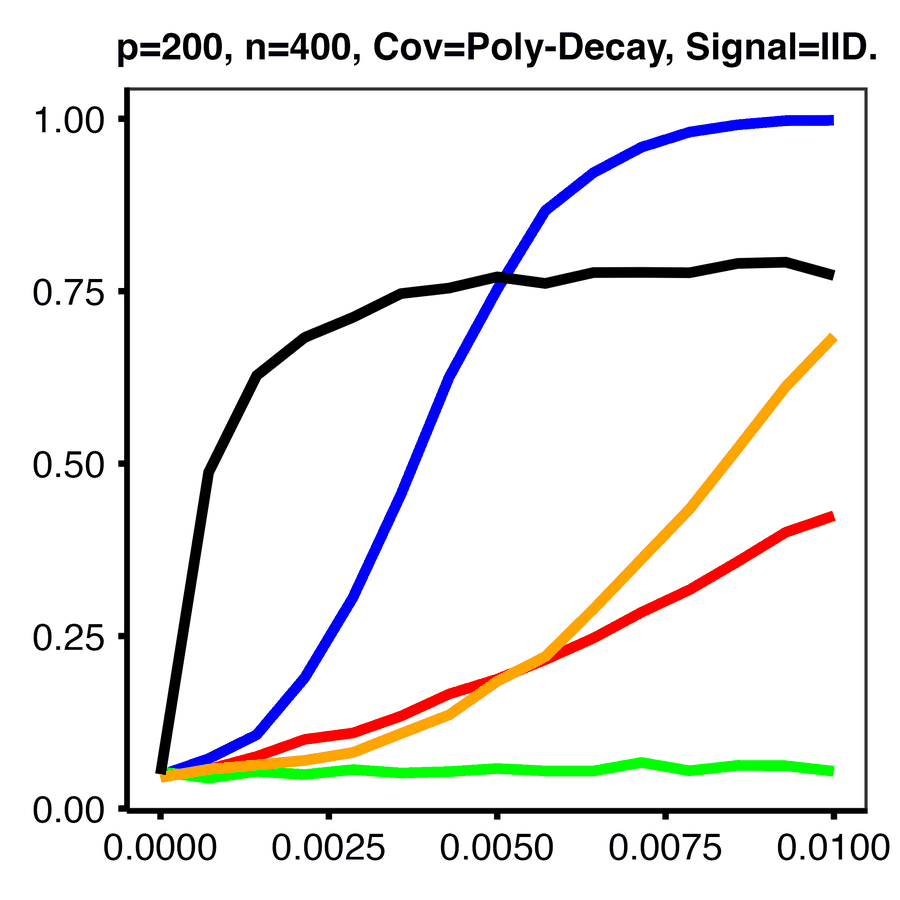} &
\includegraphics[width=0.24\textwidth]{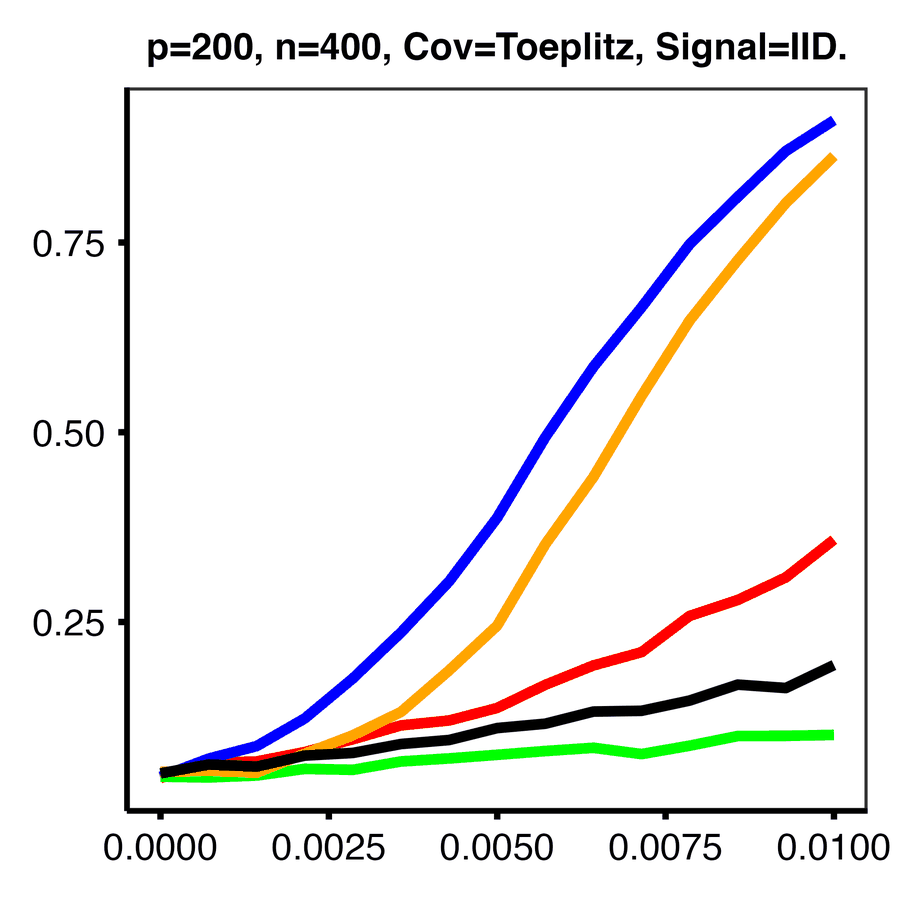}
\\
\includegraphics[width=0.24\textwidth]{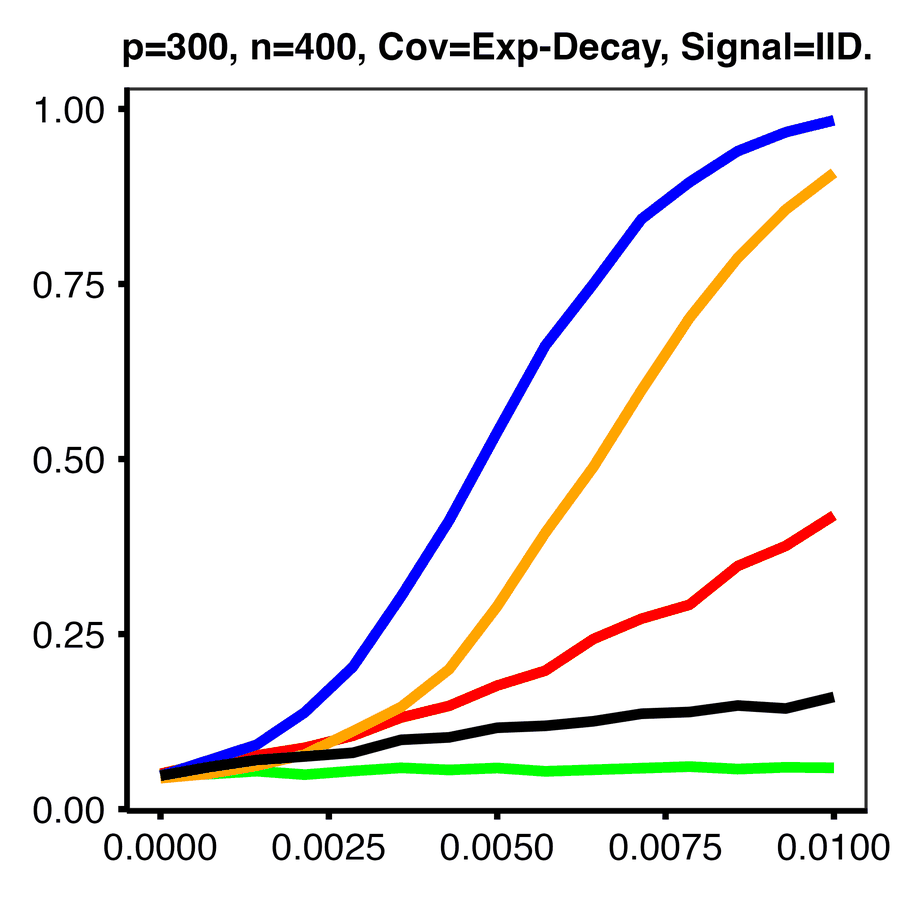} &
\includegraphics[width=0.24\textwidth]{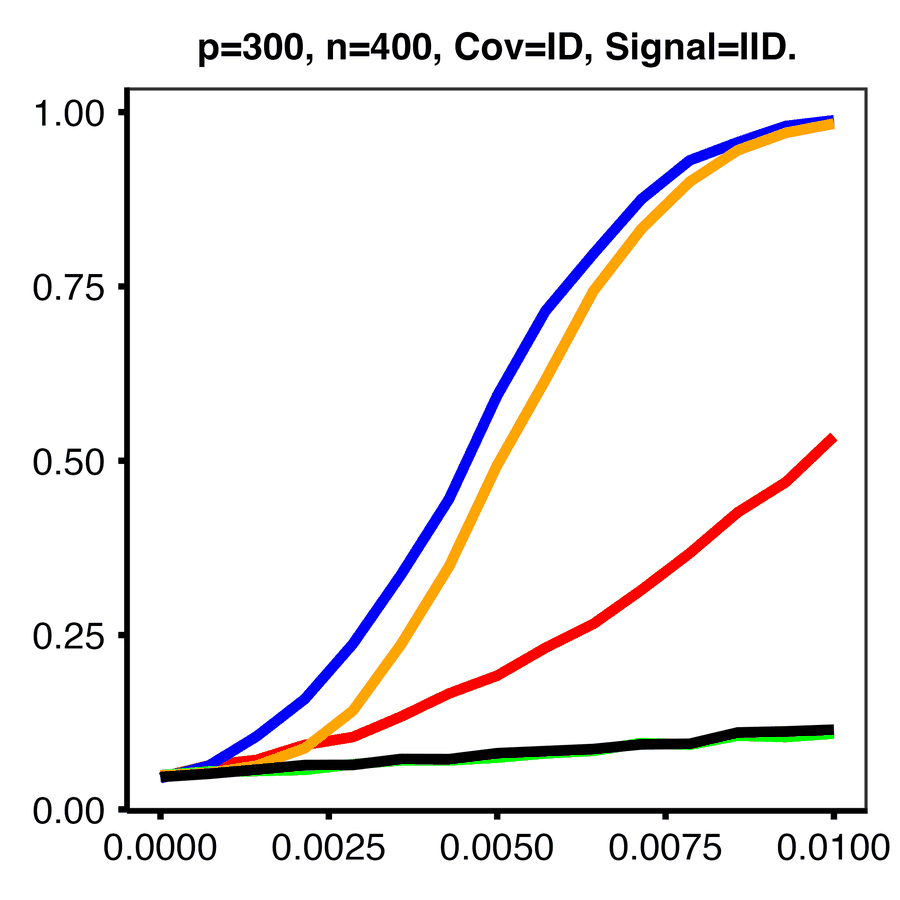} &
\includegraphics[width=0.24\textwidth]{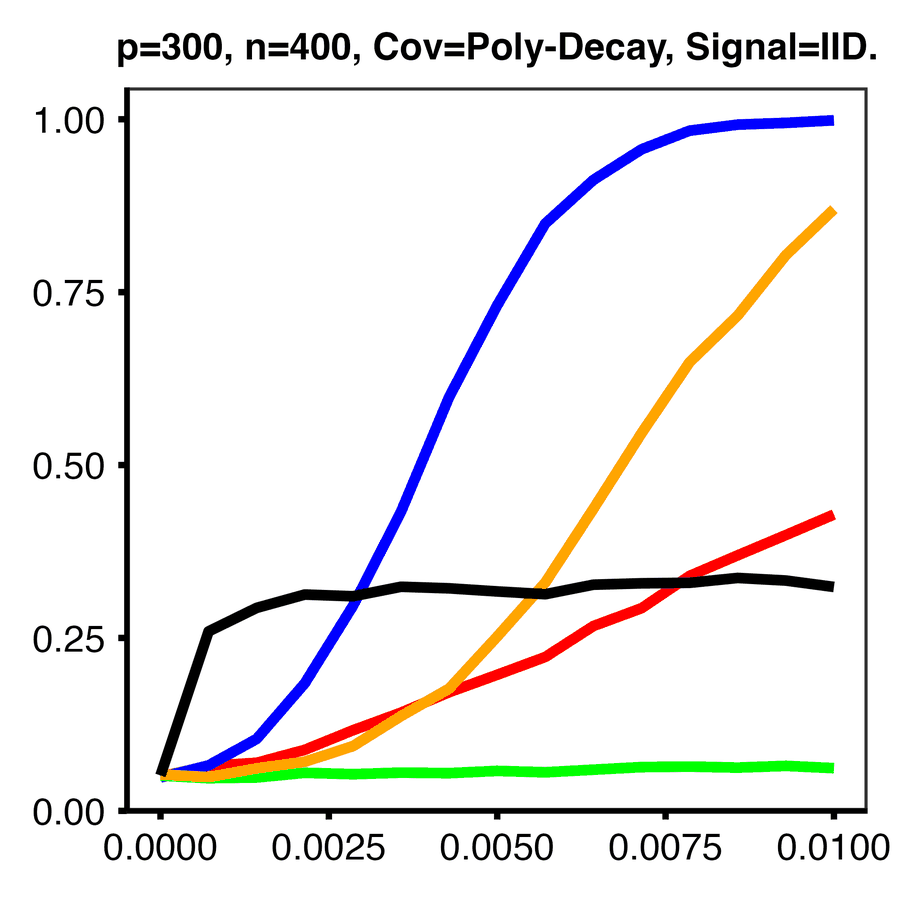} &
\includegraphics[width=0.24\textwidth]{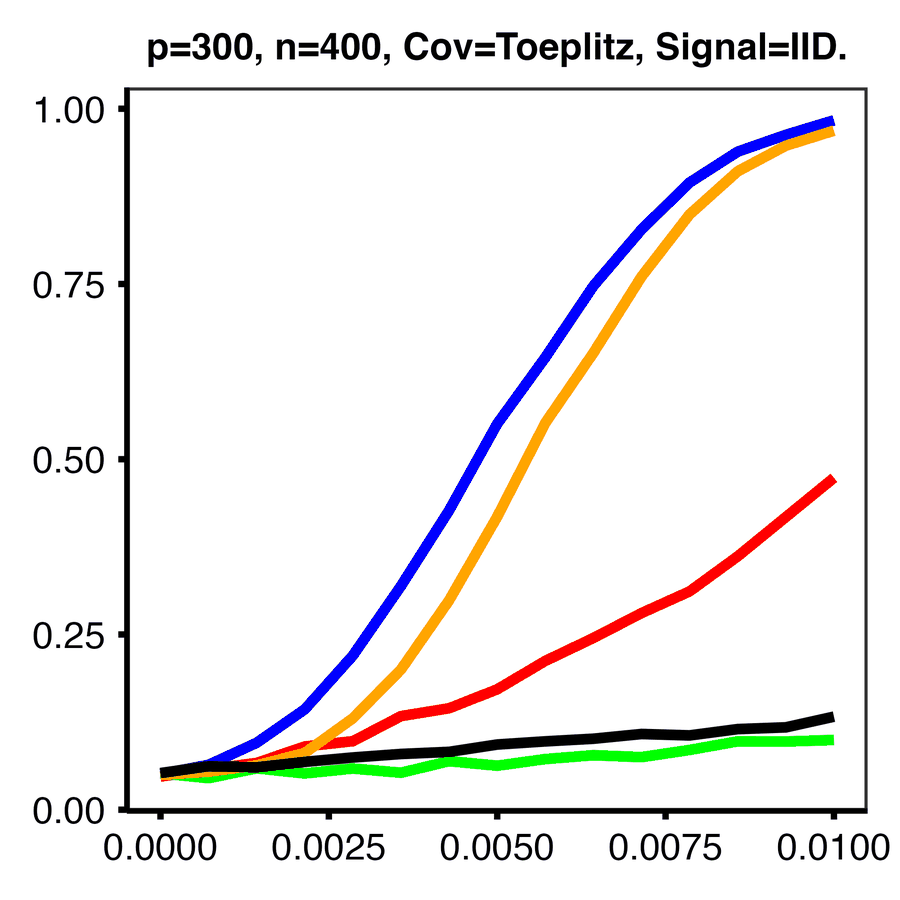}
\\
\includegraphics[width=0.24\textwidth]{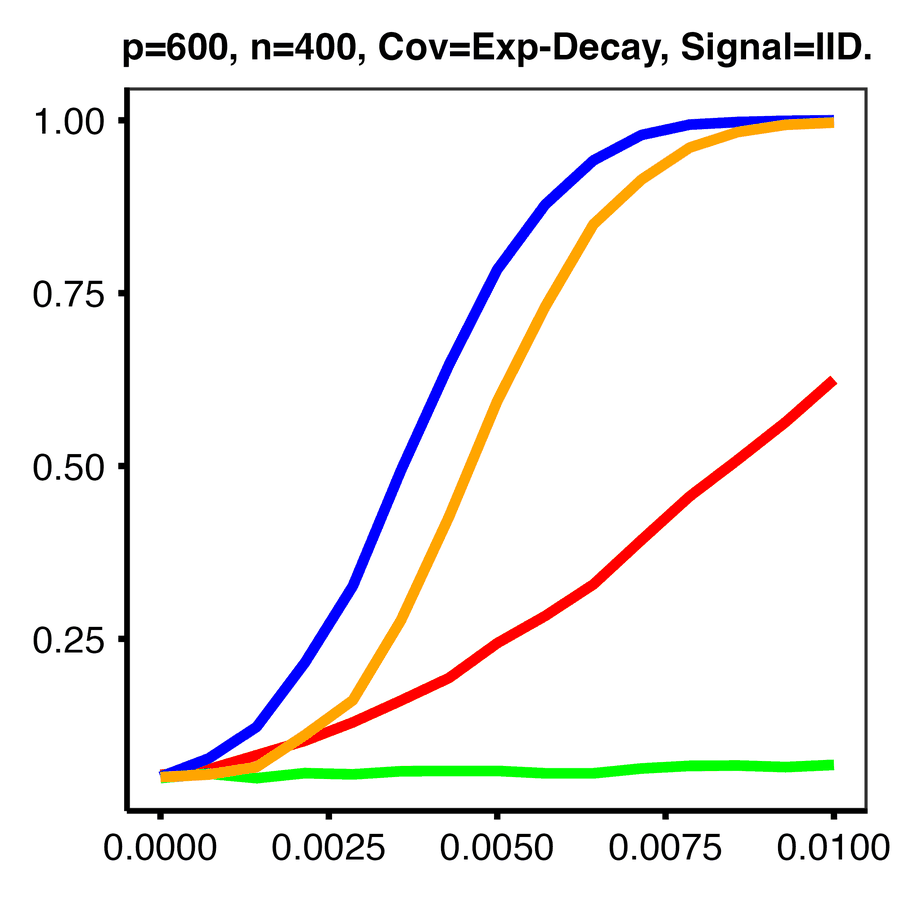} &
\includegraphics[width=0.24\textwidth]{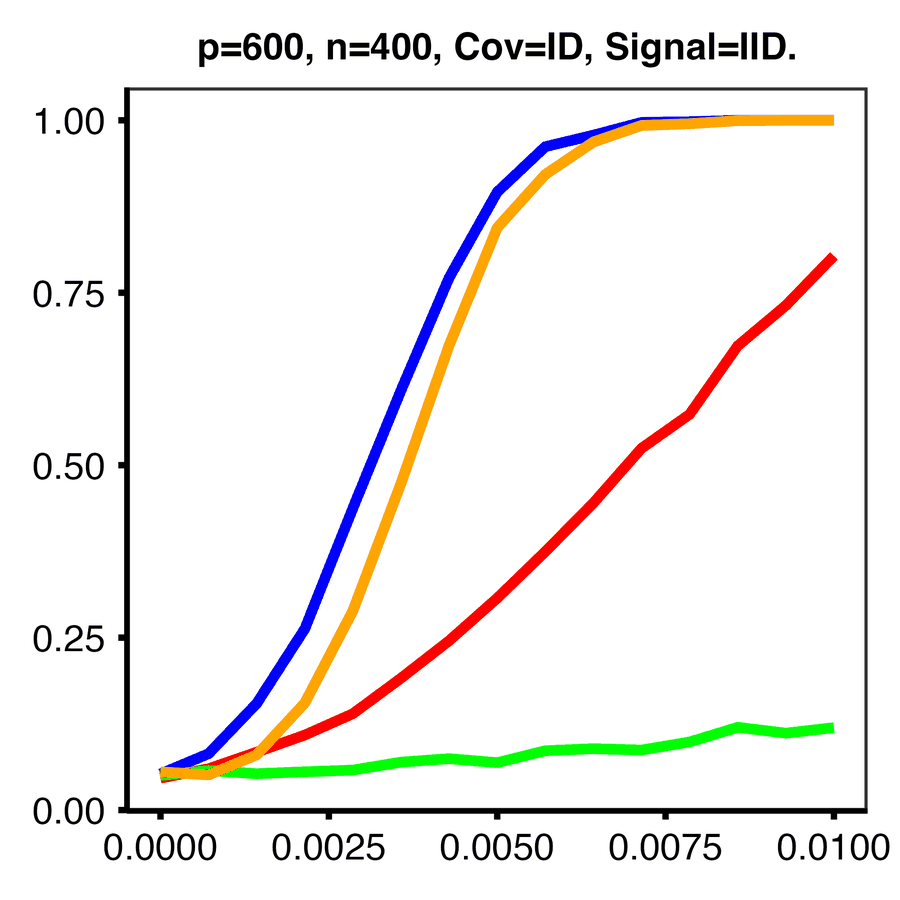} &
\includegraphics[width=0.24\textwidth]{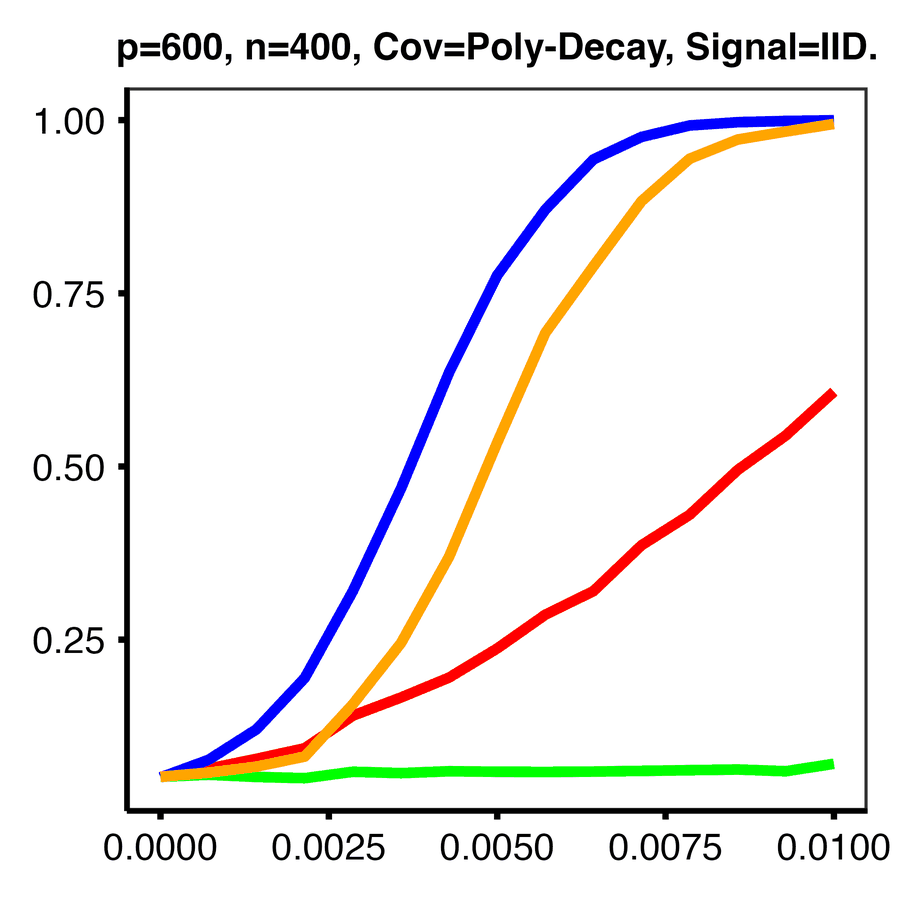} &
\includegraphics[width=0.24\textwidth]{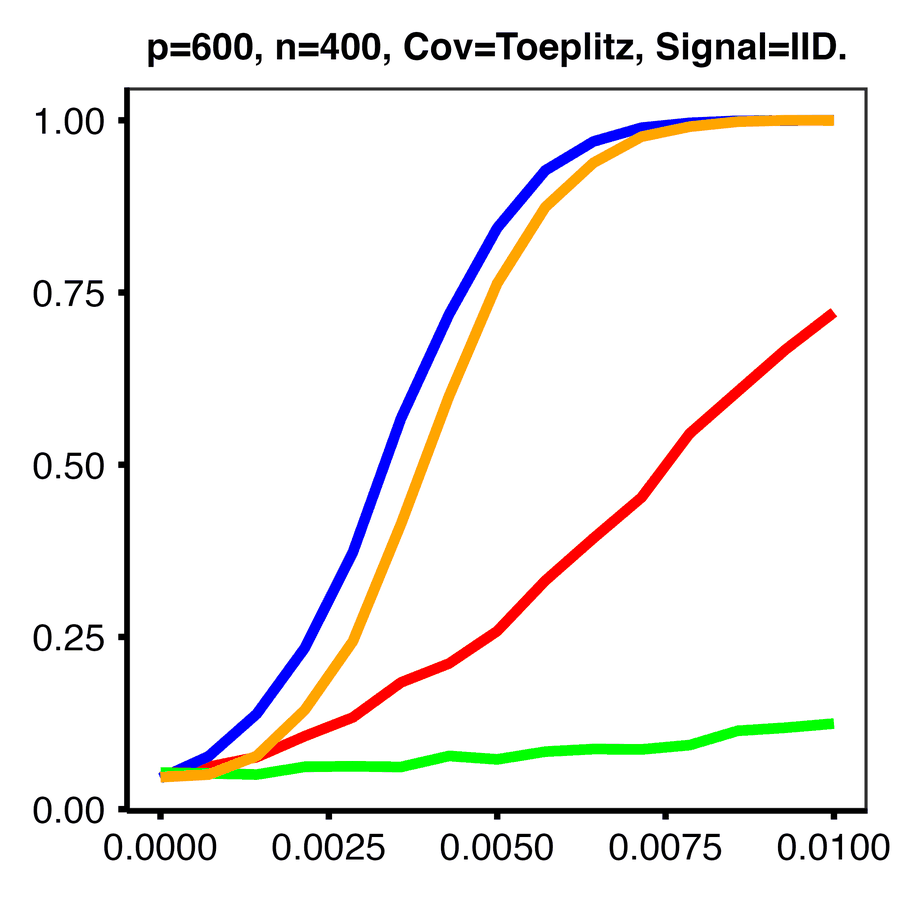}
\\
\includegraphics[width=0.24\textwidth]{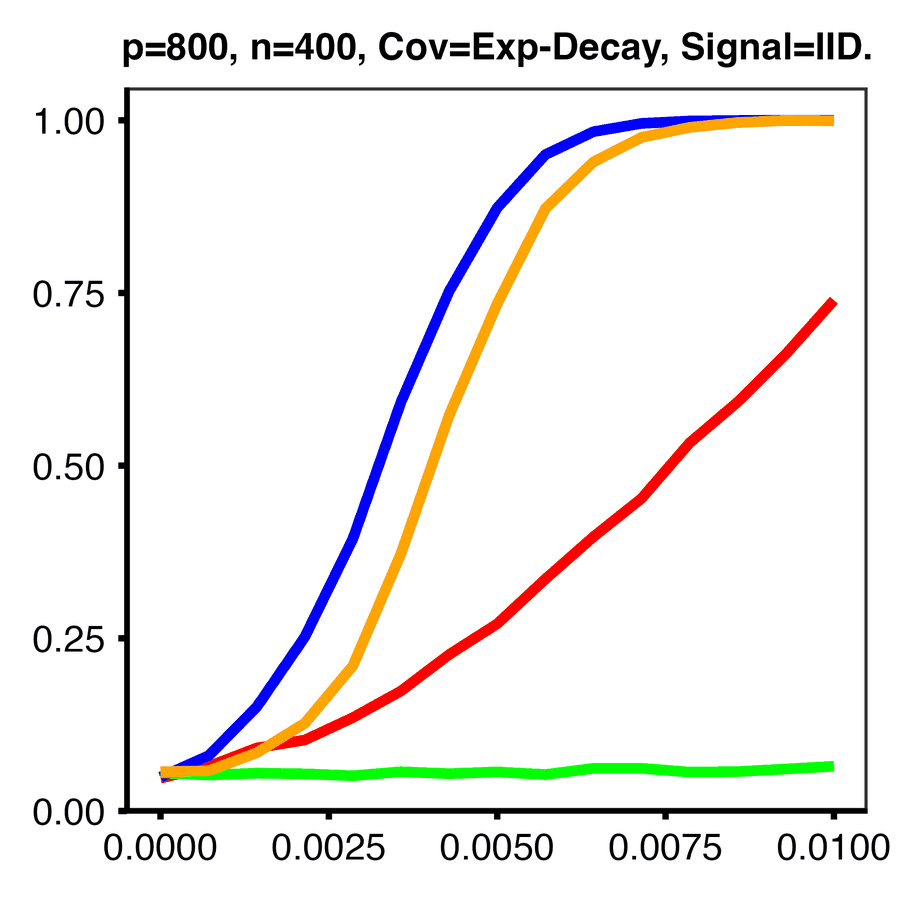} &
\includegraphics[width=0.24\textwidth]{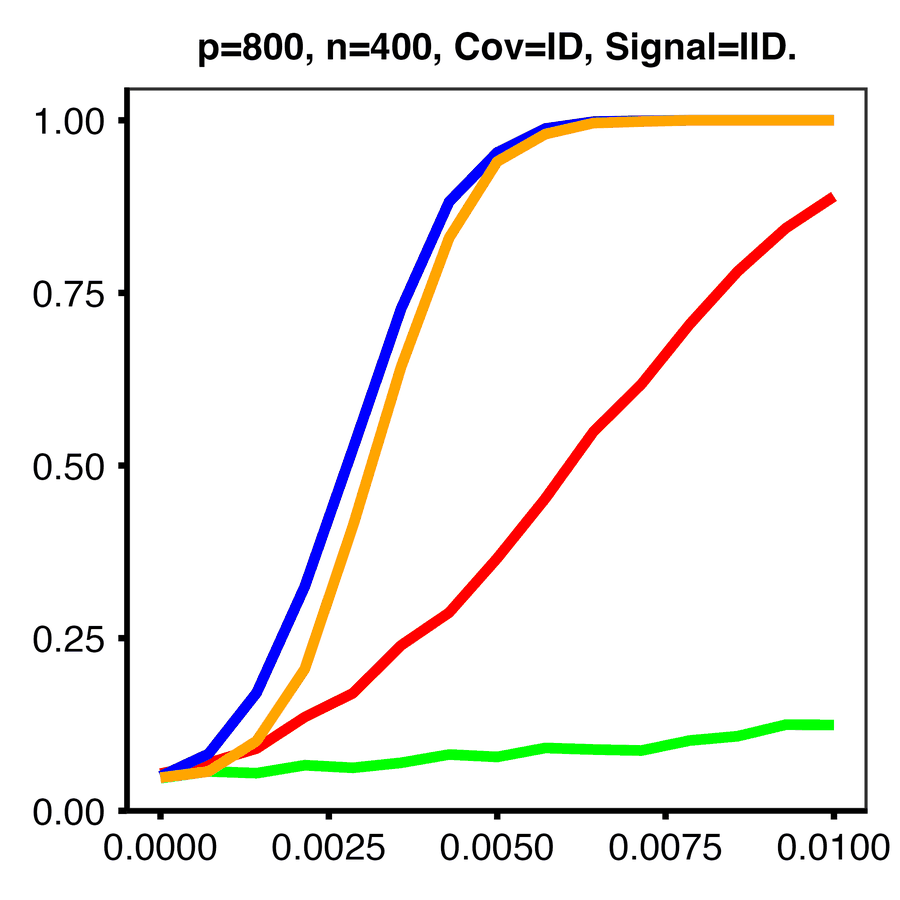} &
\includegraphics[width=0.24\textwidth]{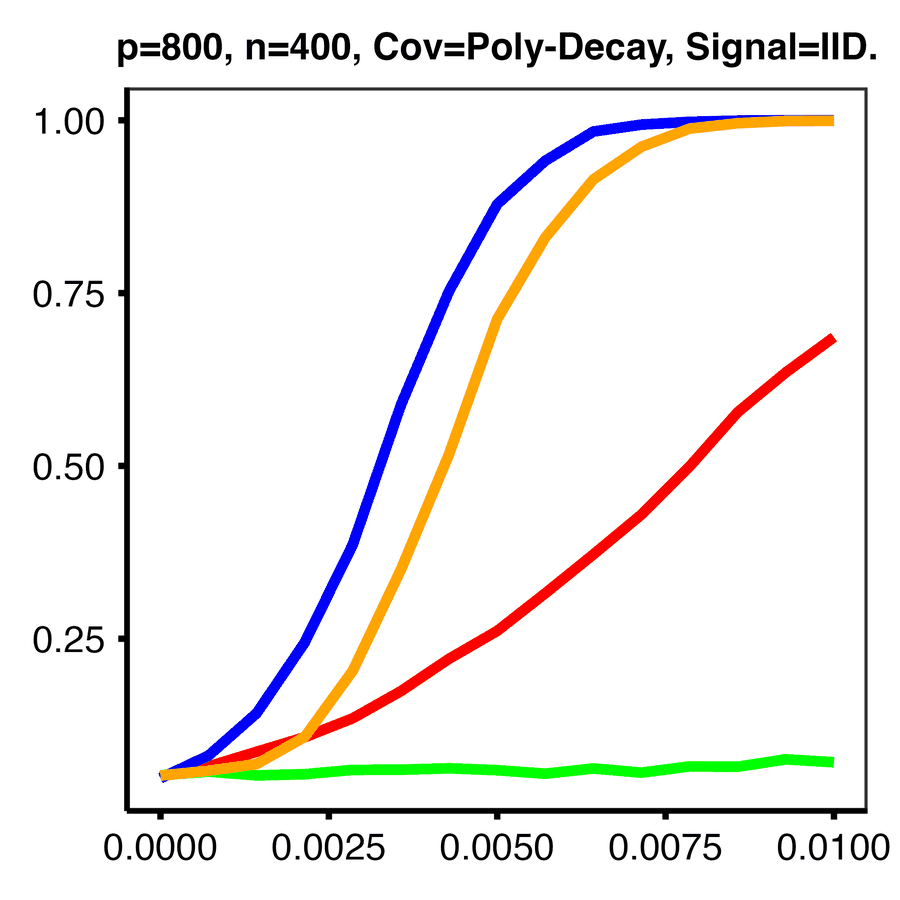} &
\includegraphics[width=0.24\textwidth]{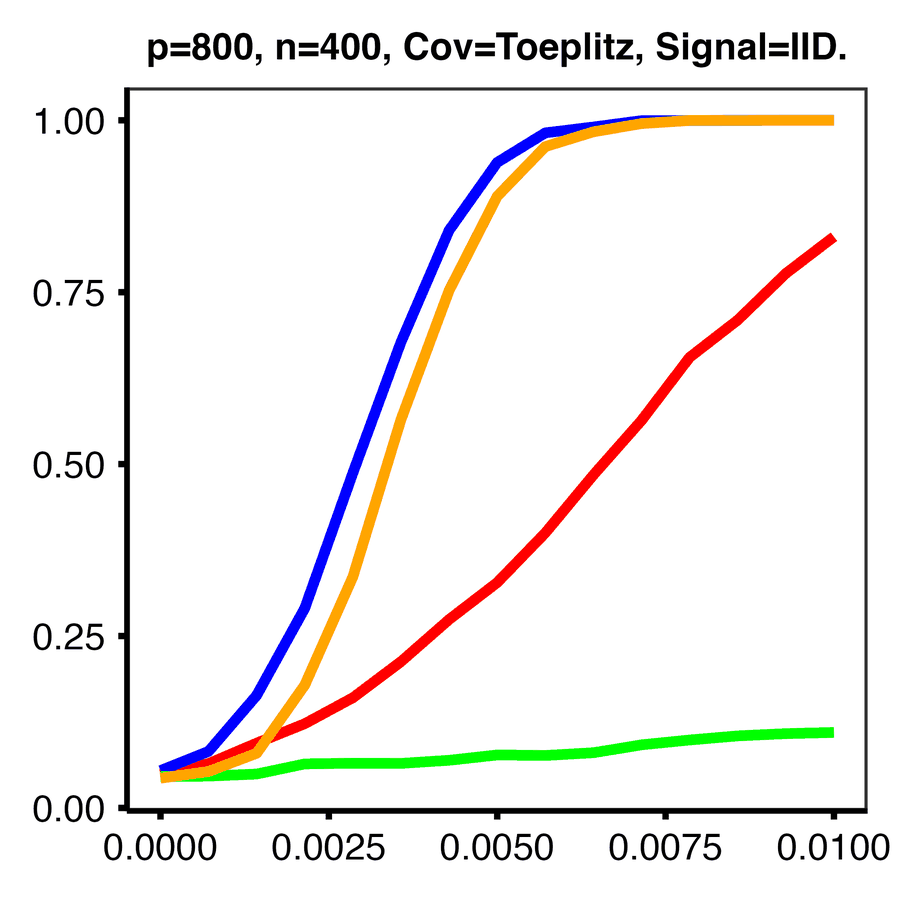}
\end{tabular}
\caption{Power curves under the MC setting with i.i.d.\ mean shifts $\delta$. Rows correspond to p= 200, 300, 600, and 800. Columns correspond to the Exp Decay, Identity, Poly Decay, and Toeplitz covariance models. Blue: $T_{\rm disc}$; Red: $T_{\rm sc}$; Black (when p<n): Cov-Norm; Orange: U-Stat; Green: $\ell_\infty$-Sparse.}
\label{fig:power_MC_iid}
\end{figure}

At the nominal significance level of $0.05$, we investigate the empirical power of all methods as a function of the signal strength parameter $c$. Figures \ref{fig:power_SC_iid} and \ref{fig:power_MC_iid} display the results under the i.i.d.\ mean shifts for the SC and MC settings separately. Results for the remaining settings are reported in Section S.2 of the Supplementary Material.

We first compare the two proposed scanning procedures. Under the SC setting, $T_{\rm sc}$ is slightly more powerful than $T_{\rm disc}$, as expected from its scanning strategy tailored to a single change point. In contrast, under the MC setting, $T_{\rm disc}$ exhibits a substantial power advantage because its more flexible scanning set can better capture the underlying multiple change-point configuration, whereas $T_{\rm sc}$ may suffer from signal cancellation across multiple changes.

The comparison with U-Stat \citep{wang2022inference} depends strongly on the spectral dispersion of $\Sigma_p$ and the alignment of the mean shift $\delta_p$ with its eigendirections. When $\Sigma_p=I_p$, the unnormalized U-Stat procedure effectively uses the correct covariance scaling and therefore provides a strong benchmark. Nevertheless, the proposed procedure performs comparably or slightly better, partly because its large-$\lambda$ component closely approximates the unnormalized regime. The small advantage may also be attributed to the direct data-driven standardization employed by the proposed method. When the spectrum of $\Sigma_p$ is highly dispersed, as in the Exp Decay and Poly Decay models, the proposed procedure exhibits a substantial power advantage over U-Stat. This advantage is particularly pronounced for i.i.d.\ and sparse mean shifts, but becomes smaller when the mean shift is strongly aligned with the leading eigendirections of $\Sigma_p$. These observations are consistent with the SNR analysis in Section~\ref{subsubsec:effect_lambda}.

Compared with the $\ell_\infty$-Sparse procedure, the proposed method is substantially more powerful under dense alternatives, in agreement with the detection-boundary comparison in Section~\ref{subsubsec:power_comparison}. Interestingly, even under sparse alternatives, the proposed procedures remain competitive and can outperform the $\ell_\infty$-based method when the spectrum of $\Sigma_p$ is highly dispersed, highlighting the benefit of covariance normalization in such settings.

Finally, Cov-Norm exhibits competitive power when $p$ is relatively small compared with $n$. Its performance deteriorates as $p$ approaches $n$, reflecting the increasing instability of the inverse sample covariance matrix in this regime.




%% file: 8_discussion.tex
\section{Discussion}\label{sec:discussion}

In this paper, we developed a computationally tractable procedure for detecting multiple change points in the mean vectors of independent high-dimensional observations. The proposed method combines ridge-regularized covariance normalization with CUSUM-type scanning and aggregates information across multiple ridge parameters. Using tools from random matrix theory, we established its asymptotic behavior in the regime where the dimension is comparable to the sample size and developed an efficient data-driven procedure for estimating the asymptotic critical values. We further conducted a detailed local power analysis, which characterizes the effects of the ridge parameter, covariance structure, and mean shifts on the performance of the proposed tests. Extensive simulation studies demonstrate strong power across a broad range of alternatives and covariance structures.

An important direction for future research is to extend the proposed framework to temporally dependent observations. In this setting, the segment mean contrasts are naturally normalized by an estimator of the long-run covariance matrix. The spectral behavior of high-dimensional sample autocovariance matrices has received considerable attention; see, for example, \cite{liu2015marcenko}, although existing results often require restrictive assumptions on the dependence structure. More recently, \cite{deitmar2026marchenko} adopt a frequency-domain approach and establish asymptotic Mar\v{c}enko--Pastur laws for Daniell-smoothed periodograms under mild conditions, while \cite{deitmar2026spectral} study finite-lag linear processes and derive deterministic equivalents for long-run sample covariance matrices. These developments provide promising theoretical foundations for extending the proposed ridge-regularized CUSUM framework to time series. A key challenge is to develop consistent estimators of the normalization quantities associated with these deterministic equivalents, analogous to those derived for independent observations in the present work. Addressing this problem requires substantial additional theoretical development and is left for future research.

%% file: Supp_Material.tex
\setcounter{page}{1}
\setcounter{section}{0}
\renewcommand{\thesection}{S.\arabic{section}}
\setcounter{equation}{0}
\renewcommand{\theequation}{S.\arabic{section}.\arabic{equation}}
\setcounter{subsection}{0}
\renewcommand{\thesubsection}{S.\arabic{section}.\arabic{subsection}}
\setcounter{table}{0}
\renewcommand{\thetable}{S.\arabic{table}}
\setcounter{figure}{0}
\renewcommand{\thefigure}{S.\arabic{figure}}

\setcounter{theorem}{0}
\renewcommand\thetheorem{S.\arabic{section}.\arabic{theorem}}
\renewcommand\thelemma{S.\arabic{section}.\arabic{lemma}}

\setcounter{proposition}{0}
\renewcommand\theproposition{S.\arabic{proposition}}

\newcommand{\bZ}{\mathbf{Z}}
\newcommand{\bOmega}{\widetilde{\Omega}_n}
\newcommand{\bA}{\mathbf{A}}
\newcommand{\calC}{\mathcal{C}}
\newcommand{\calJ}{\mathcal{J}}
\newcommand{\bD}{\mathbf{D}}
\newcommand{\bY}{\mathbf{Y}}
\newcommand{\bB}{\mathbf{B}}
\newcommand{\bG}{\mathbf{G}}
\newcommand{\bJ}{\mathbf{J}}

\begin{center}
{\Large \bf \centering
Supplementary Material to \\
``Adaptable High-Dimensional Change Point Detection via Ridge Regularization''
}
\end{center}


\begin{center}
{
Haoran Li\footnote{Corresponding author.}~~ and ~~Haotian Xu\\[0.5em]
}    
\end{center}

\begin{center}
{Department of Mathematics and Statistics, Auburn University}    
\end{center}

\vspace{1cm}

This supplementary material is organized as follows.
\begin{itemize}
\item[1.] Section \ref{sec:additional_details_effect_lambda} provides additional numerical results complementing the analysis in Section \ref{subsubsec:effect_lambda} of the manuscript.
\item[2.] Section \ref{sec:additional_simulation} provides additional simulation results.
\item[3.] Section \ref{sec:proof_weak_convergence_process} proves Theorem \ref{thm:weak_convergence_process}.
\item[4.] Section \ref{sec:proof_stieltjes_under_alternative} proves Lemma \ref{lemma:Stieltjes_under_alternative}.
\item[5.] Section \ref{sec:proof_power_single_deterministic} proves Theorems \ref{thm:power_single_deterministic} and \ref{thm:power_multiple_deterministic}.
\item[6.] Section \ref{sec:proof_thm_power_PA} proves Theorem \ref{thm:power_multiple_PA}.
\item[7.] Section \ref{s_sec:technical_lemma} collects the technical lemmas used throughout the proofs.
\end{itemize}

\section{Additional Details on the Effect of the Ridge Parameter}
\label{sec:additional_details_effect_lambda}

This section provides additional numerical results complementing the analysis in Section \ref{subsubsec:effect_lambda} of the manuscript.

\begin{figure}[ht]
\centering
\includegraphics[width=0.8\linewidth]{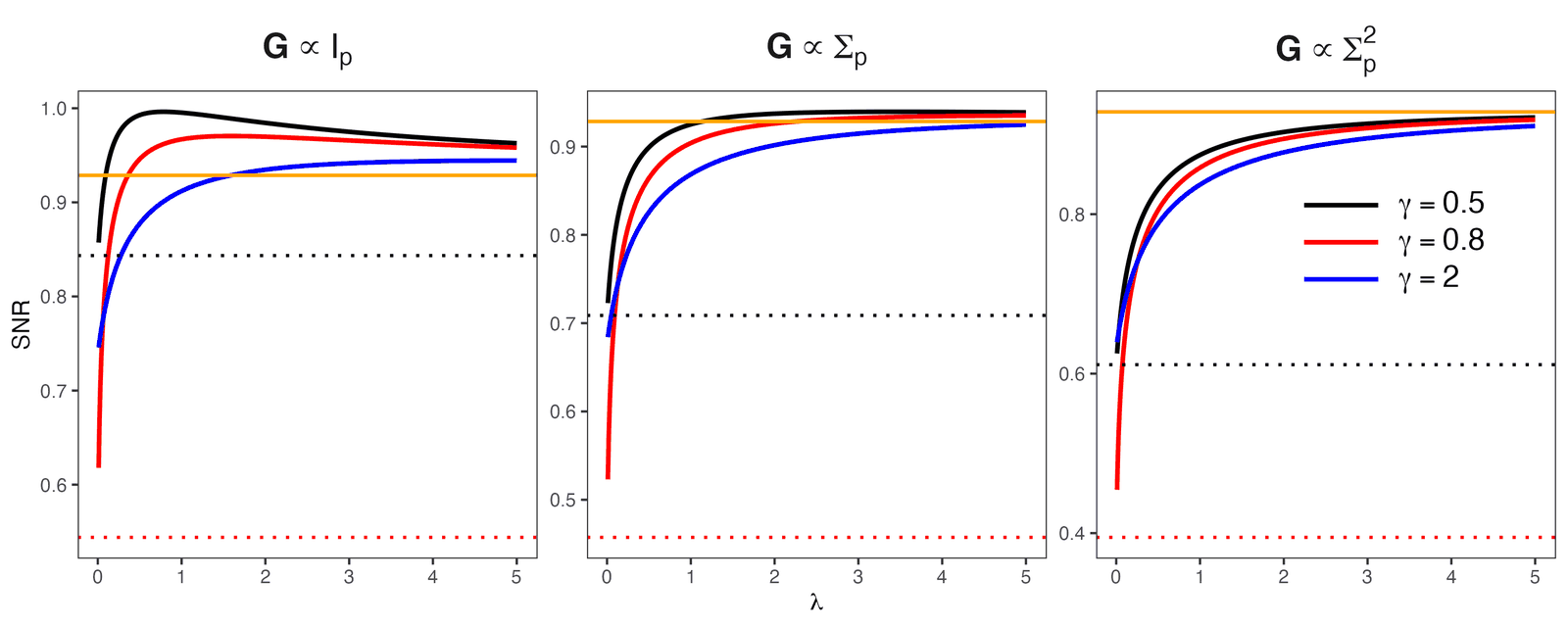}
\caption{$\SNR_{\rm PA}$ as a function of $\lambda$ for $\alpha_{(1)}=0.6$ and $H_p({\alpha_{(1)}})=0.5$. The solid black, red, and blue curves show the SNR of the proposed ridge-regularized procedure for $\gamma=0.5$, $0.8$, and $2$, respectively. The dotted horizontal lines indicate the limits as $\lambda\to0$ (black for $\gamma=0.5$ and red for $\gamma=0.8$), corresponding to $T_{\rm cov}$, which is applicable only when $\gamma<1$. The dashed horizontal lines indicate the common limit as $\lambda\to\infty$ for all three values of $\gamma$, corresponding to $T_{\ell_2}$.}
\label{fig:snr_a_5_x_6}
\end{figure}

\begin{figure}[htbp]
\centering
\includegraphics[width=0.8\linewidth]{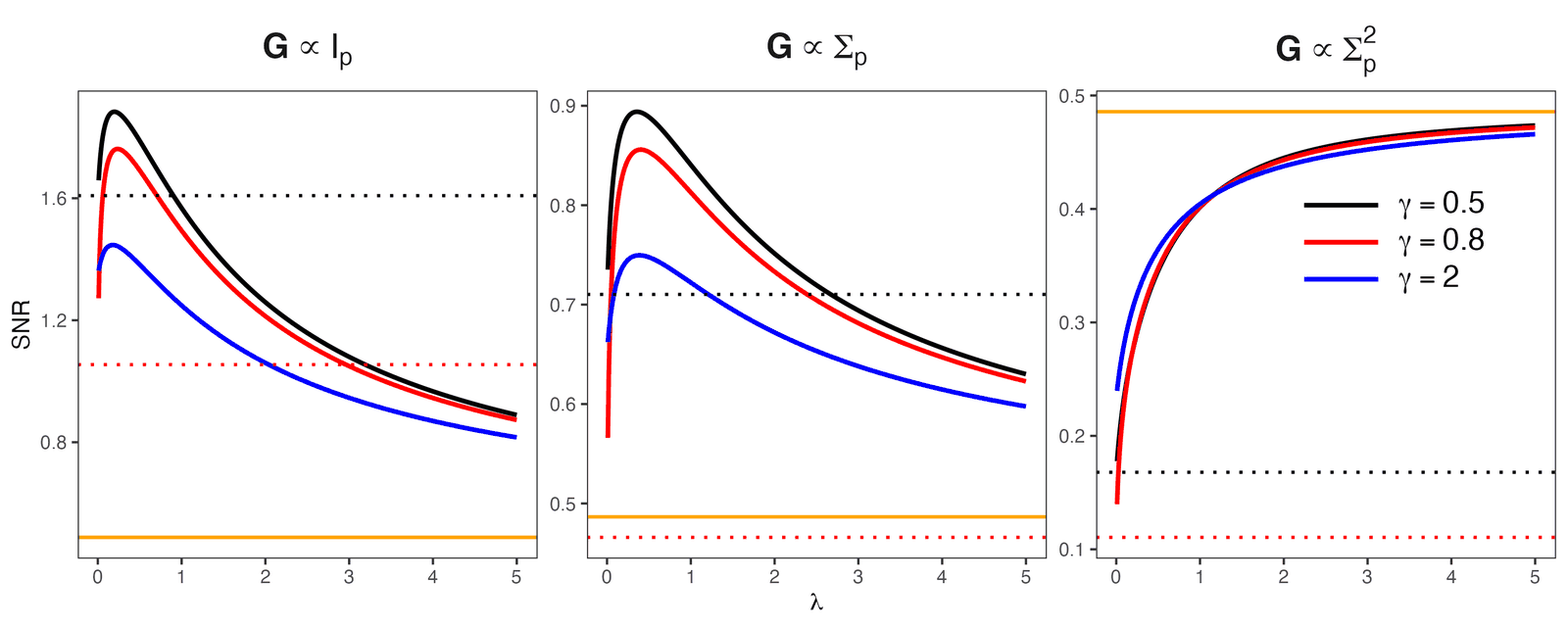}
\caption{Same as Figure \ref{fig:snr_a_5_x_6} but for $\alpha_{(1)}=0.4$ and $H_p({\alpha_{(1)}})=0.9$.  
}
\label{fig:snr_a_9_x_4}
\end{figure}

\begin{figure}[htbp]
\centering
\includegraphics[width=0.8\linewidth]{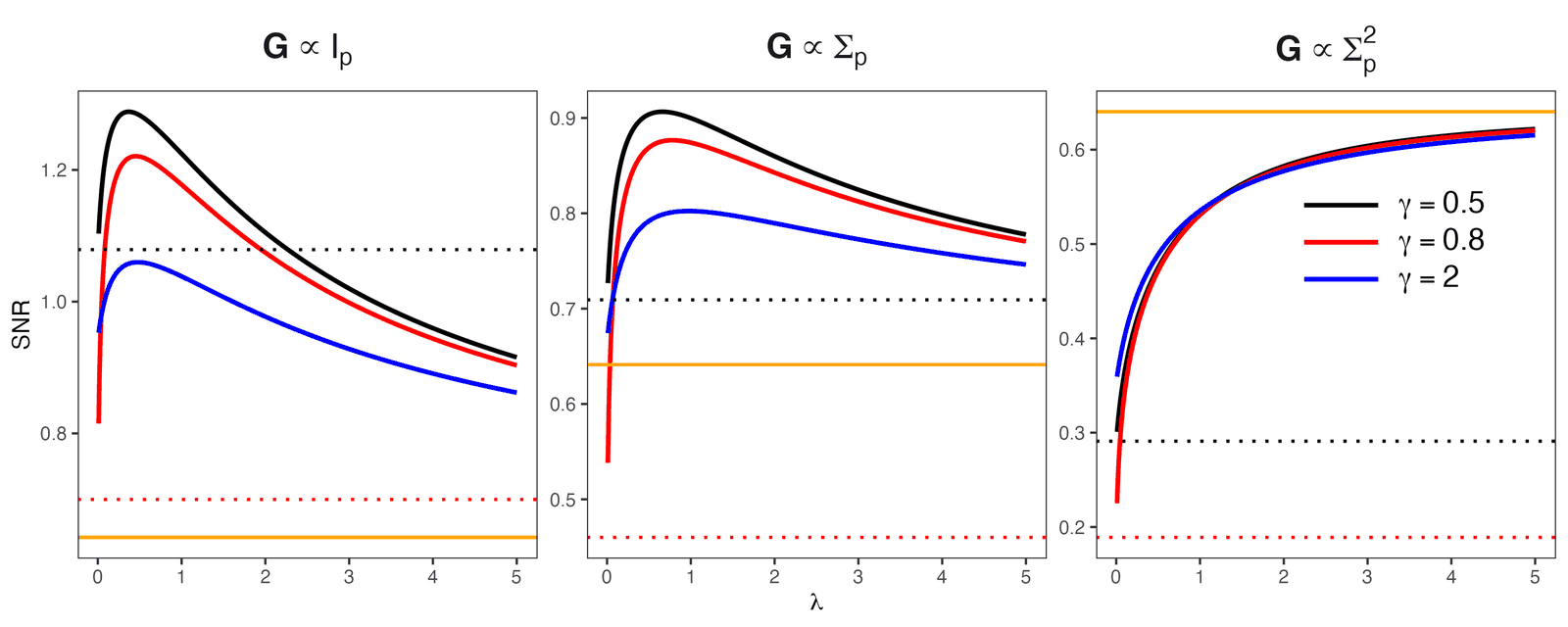}
\caption{Same as Figure \ref{fig:snr_a_5_x_6} but for $\alpha_{(1)}=0.8$ and $H_p({\alpha_{(1)}})=0.9$.} 
\label{fig:snr_a_9_x_6}
\end{figure}

\clearpage

\section{Additional Simulation Results} \label{sec:additional_simulation}

This section provides additional power curves complementing those displayed in Section \ref{sec:power_comparison}.

\begin{figure}[ht]
\centering
\setlength{\tabcolsep}{1pt}
\renewcommand{\arraystretch}{0}
\begin{tabular}{cccc}
\includegraphics[width=0.24\textwidth]{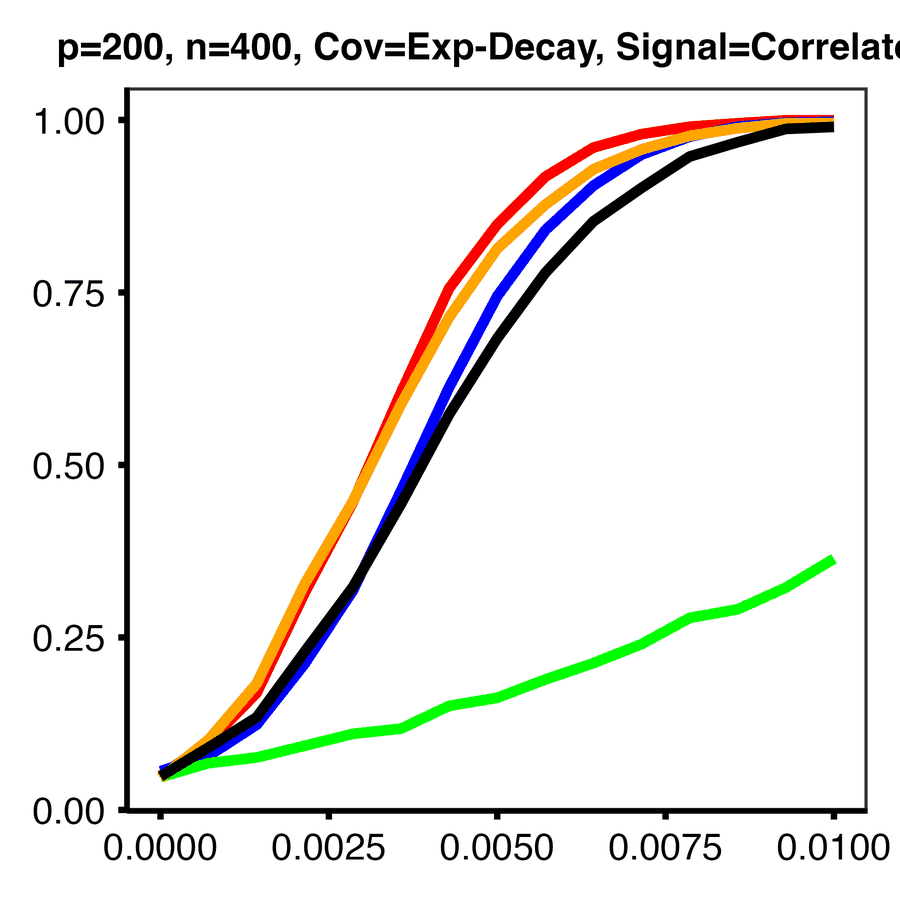} &
\includegraphics[width=0.24\textwidth]{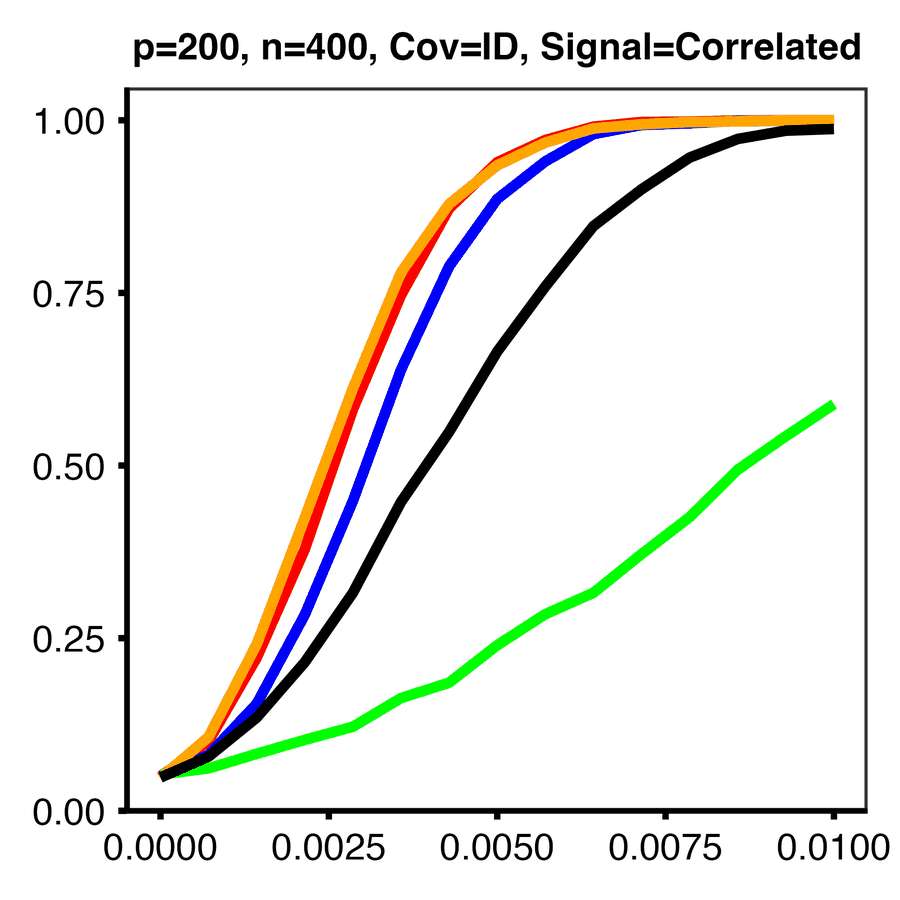} &
\includegraphics[width=0.24\textwidth]{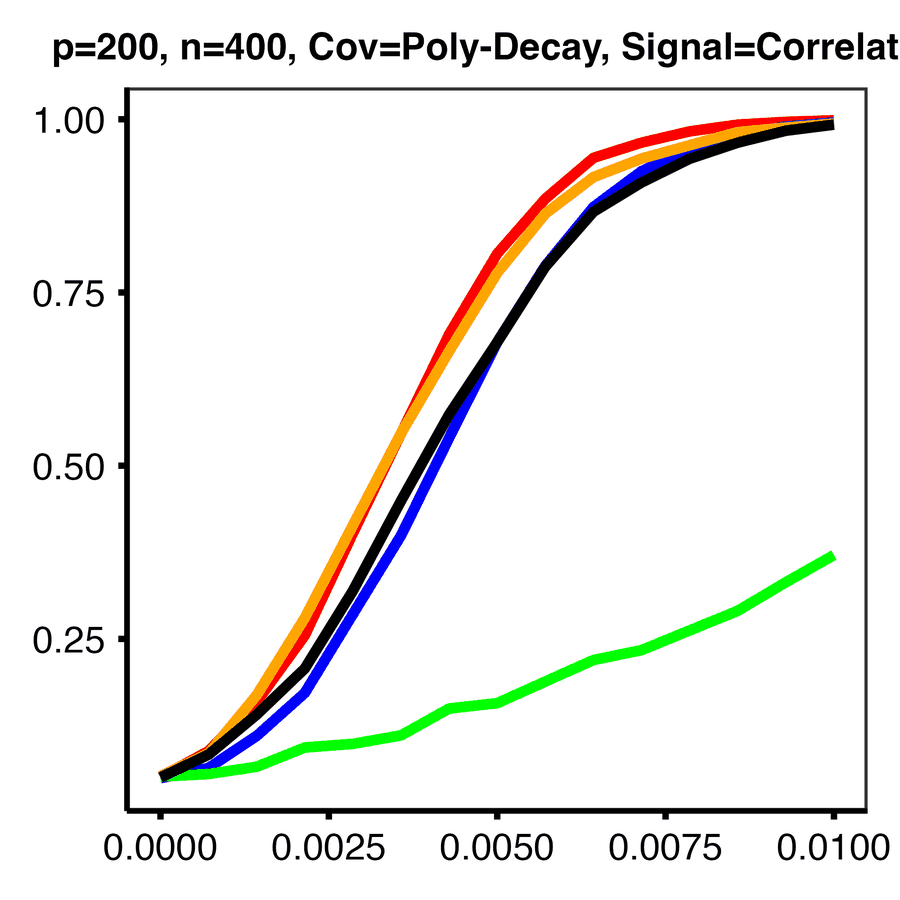} &
\includegraphics[width=0.24\textwidth]{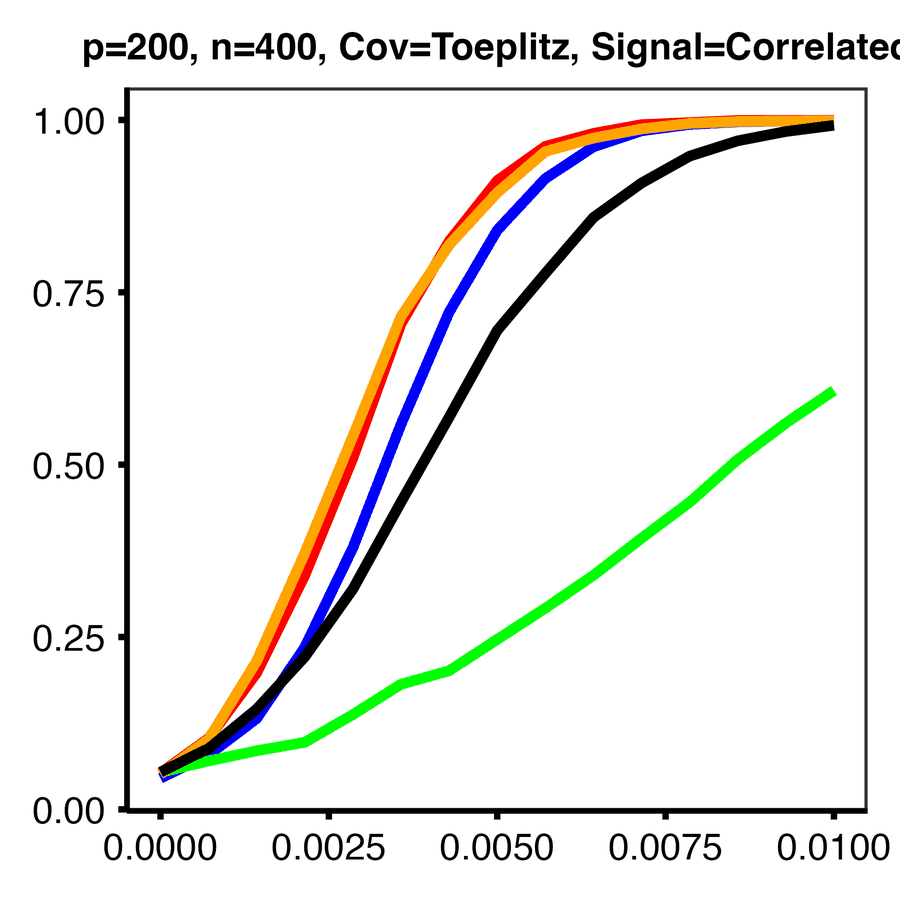}
\\
\includegraphics[width=0.24\textwidth]{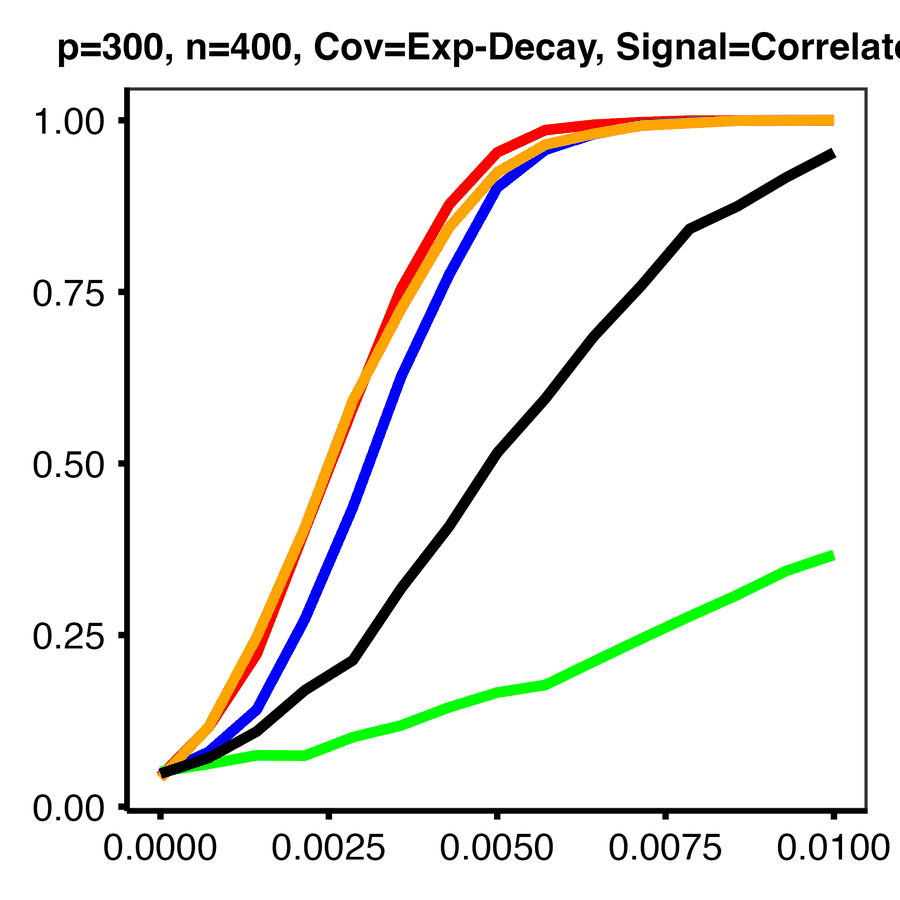} &
\includegraphics[width=0.24\textwidth]{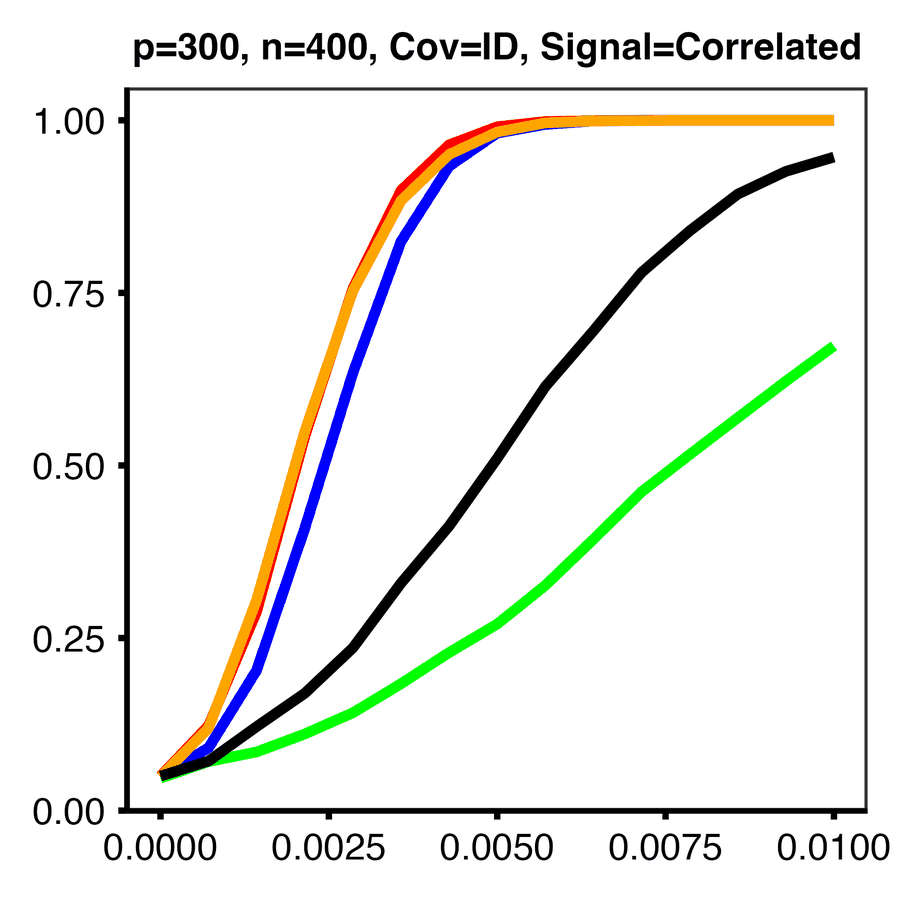} &
\includegraphics[width=0.24\textwidth]{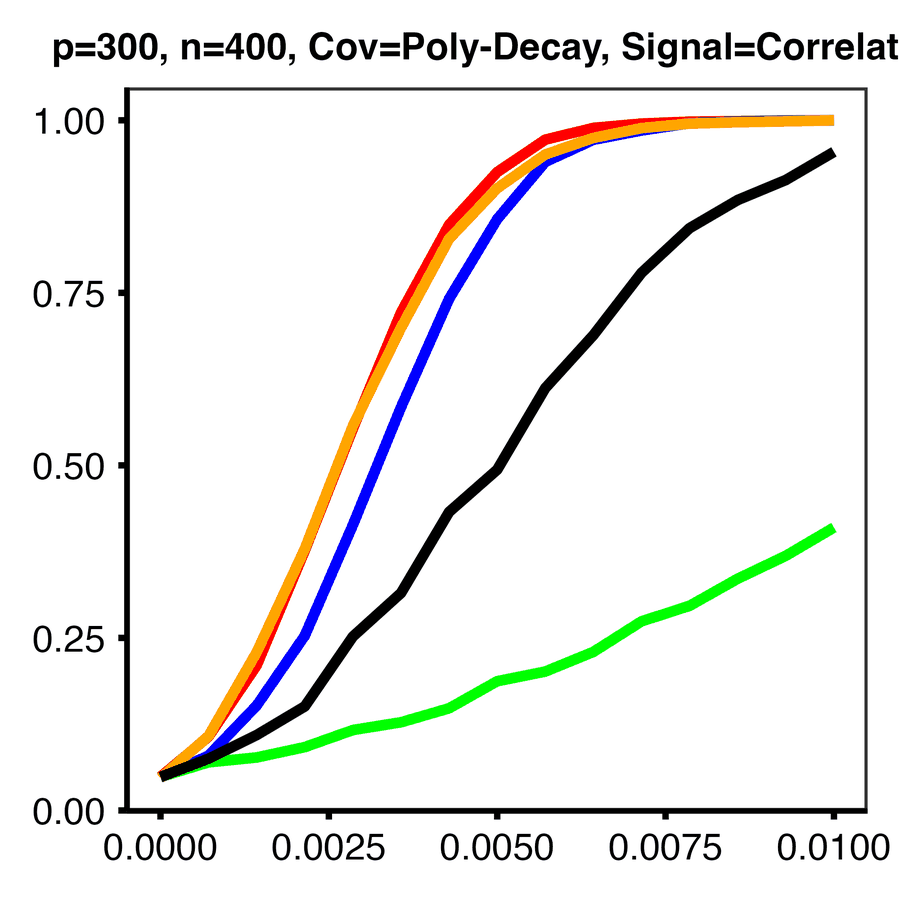} &
\includegraphics[width=0.24\textwidth]{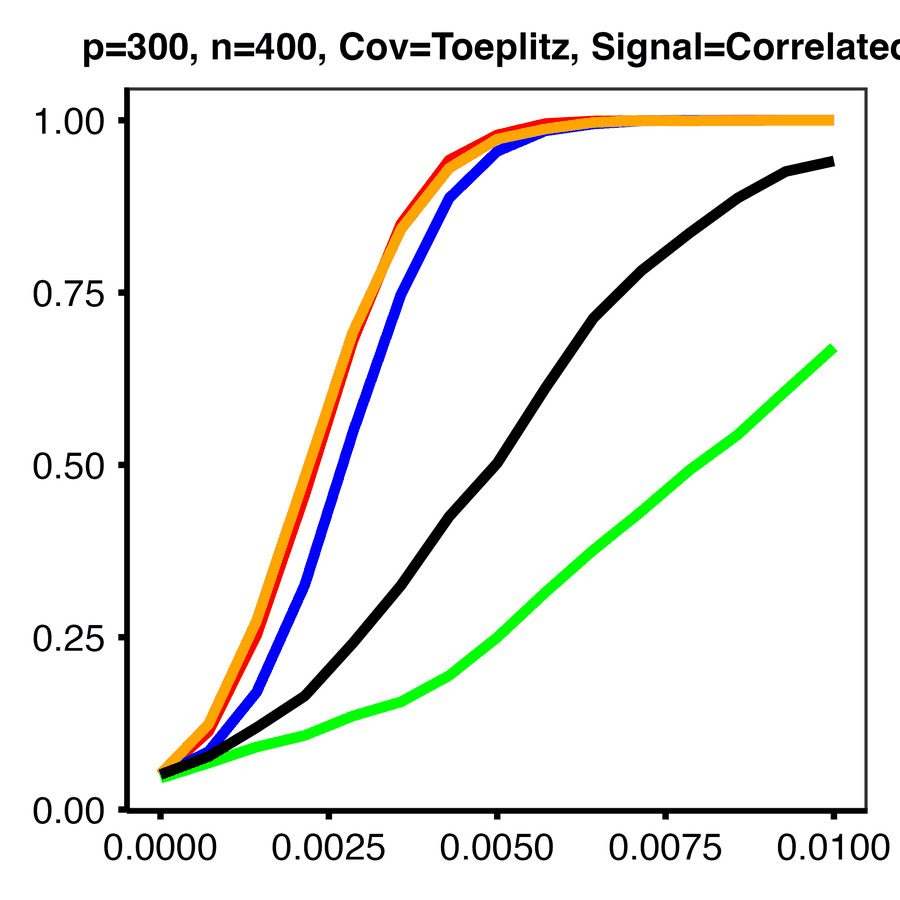}
\\
\includegraphics[width=0.24\textwidth]{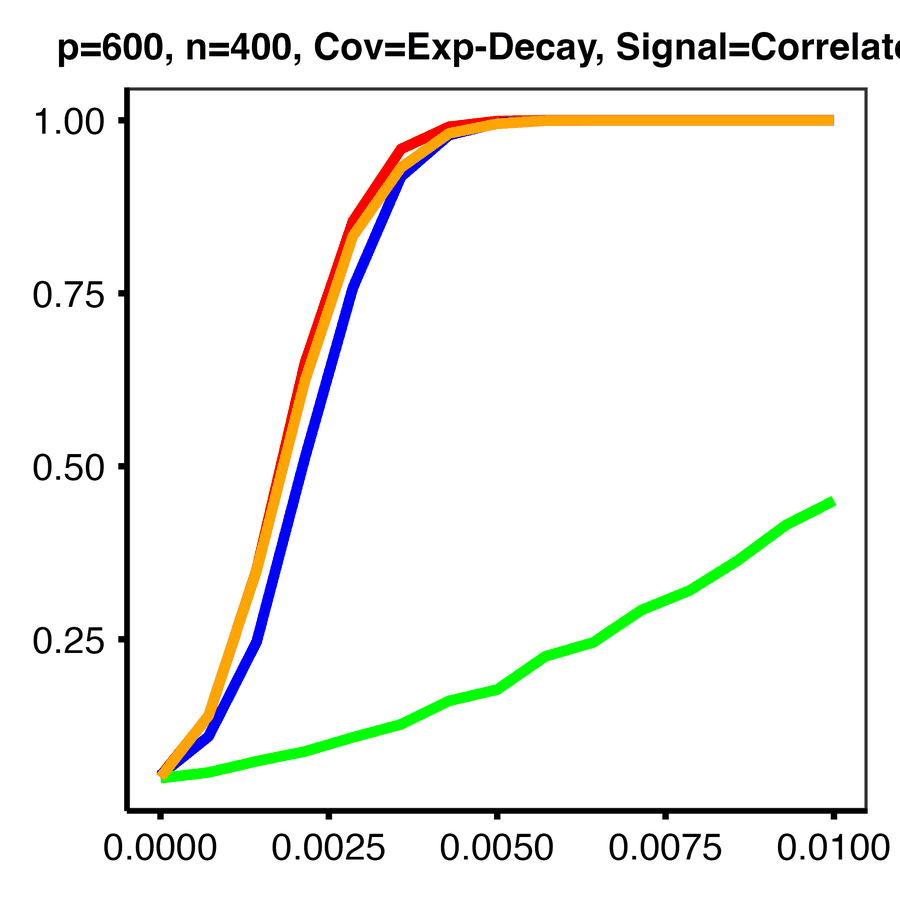} &
\includegraphics[width=0.24\textwidth]{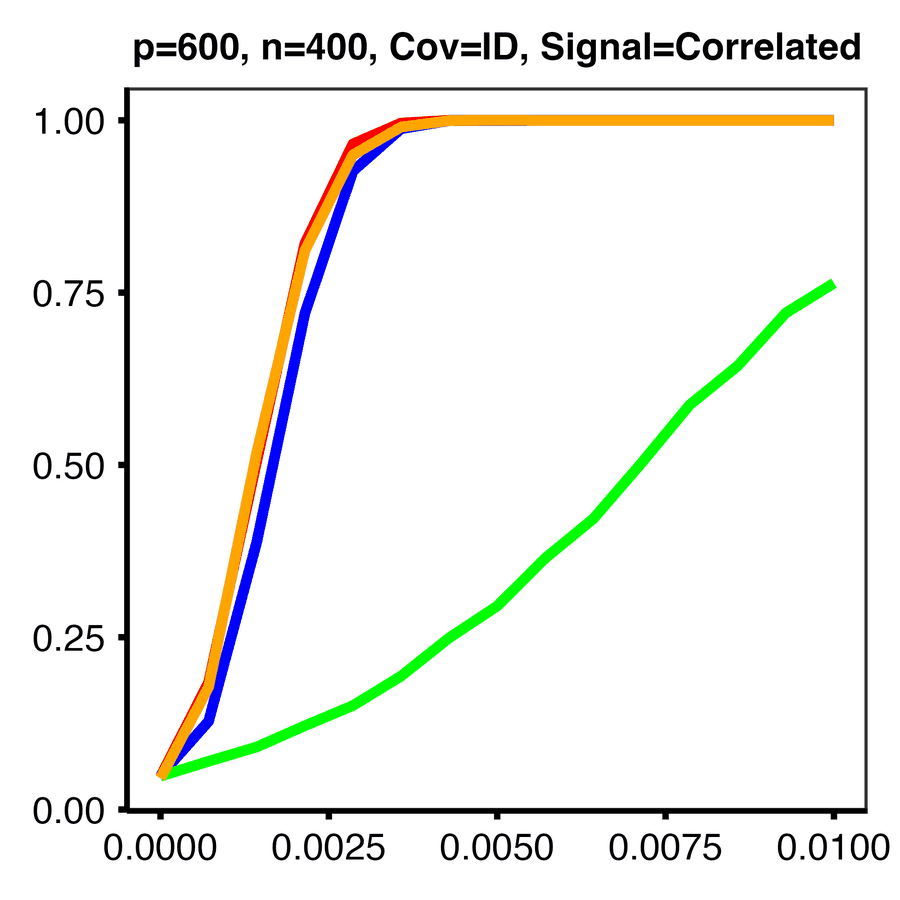} &
\includegraphics[width=0.24\textwidth]{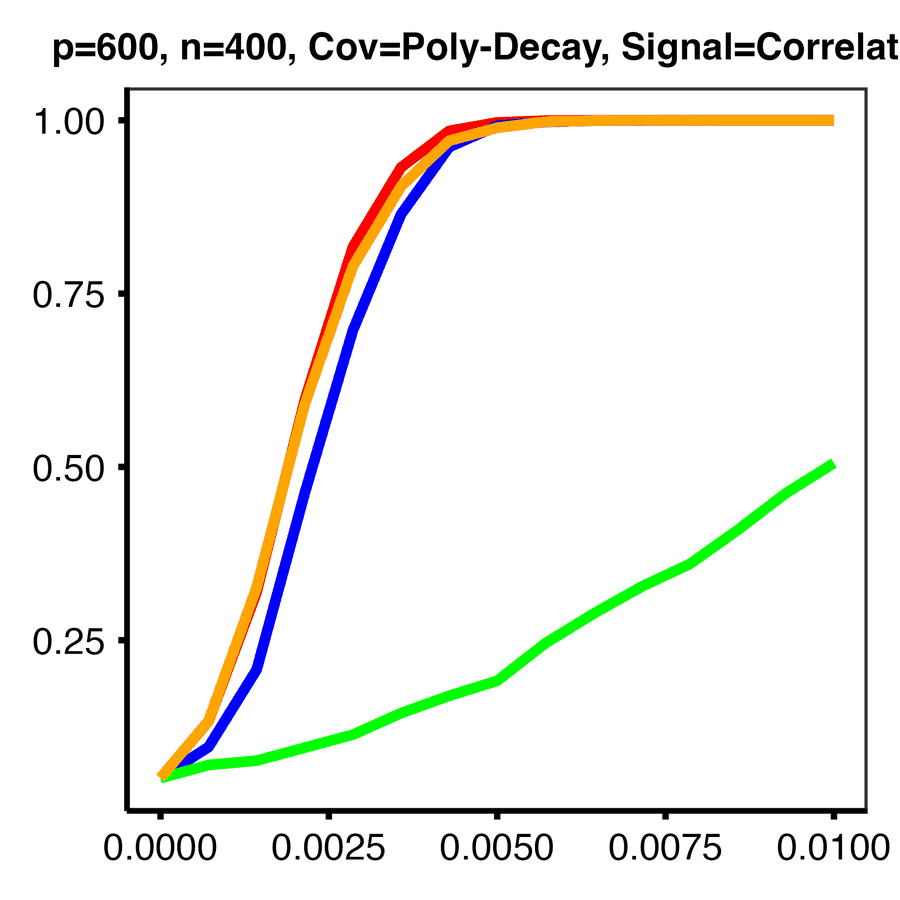} &
\includegraphics[width=0.24\textwidth]{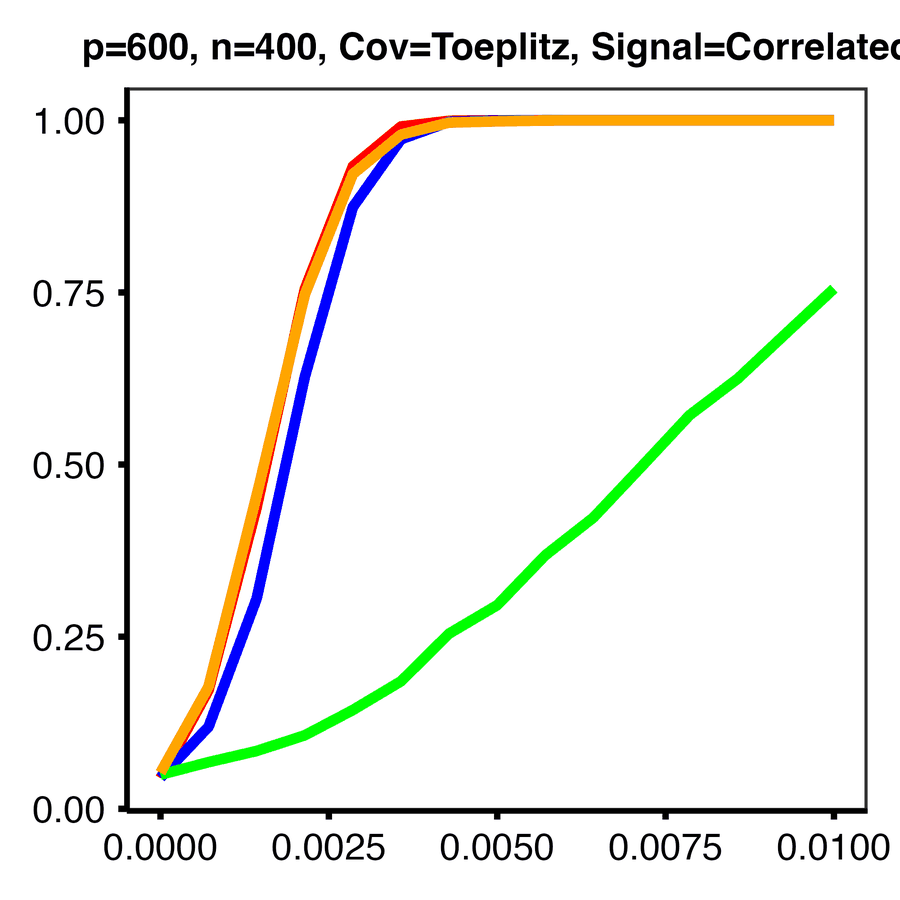}
\\
\includegraphics[width=0.24\textwidth]{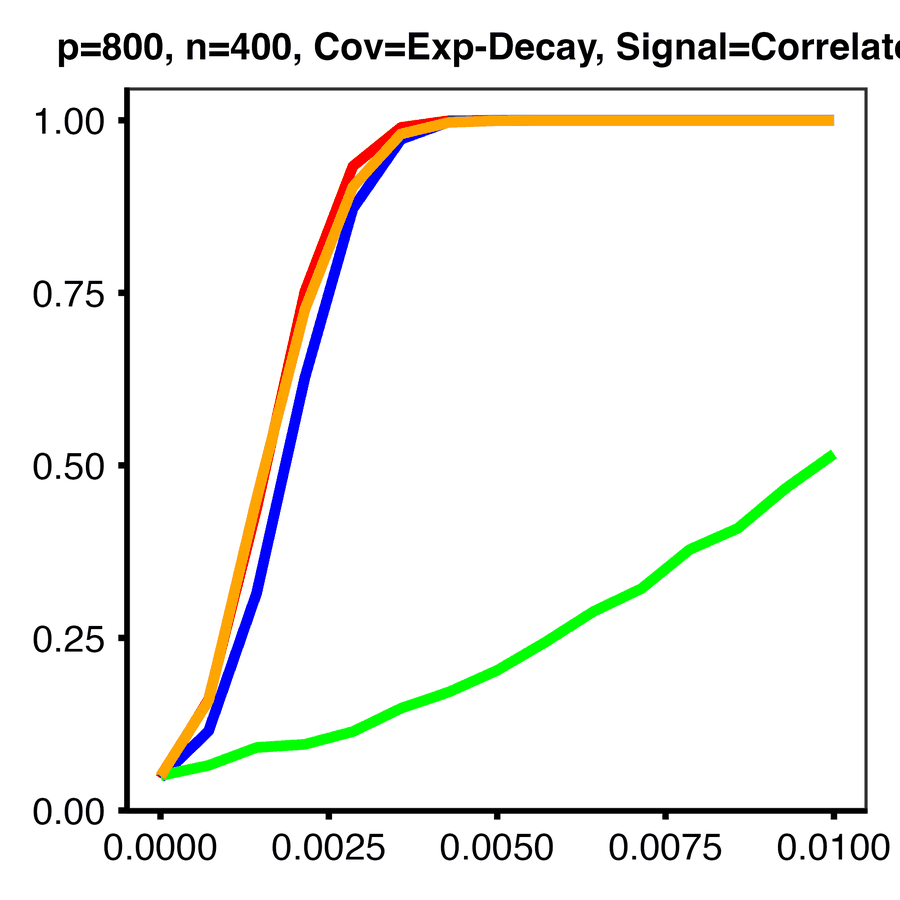} &
\includegraphics[width=0.24\textwidth]{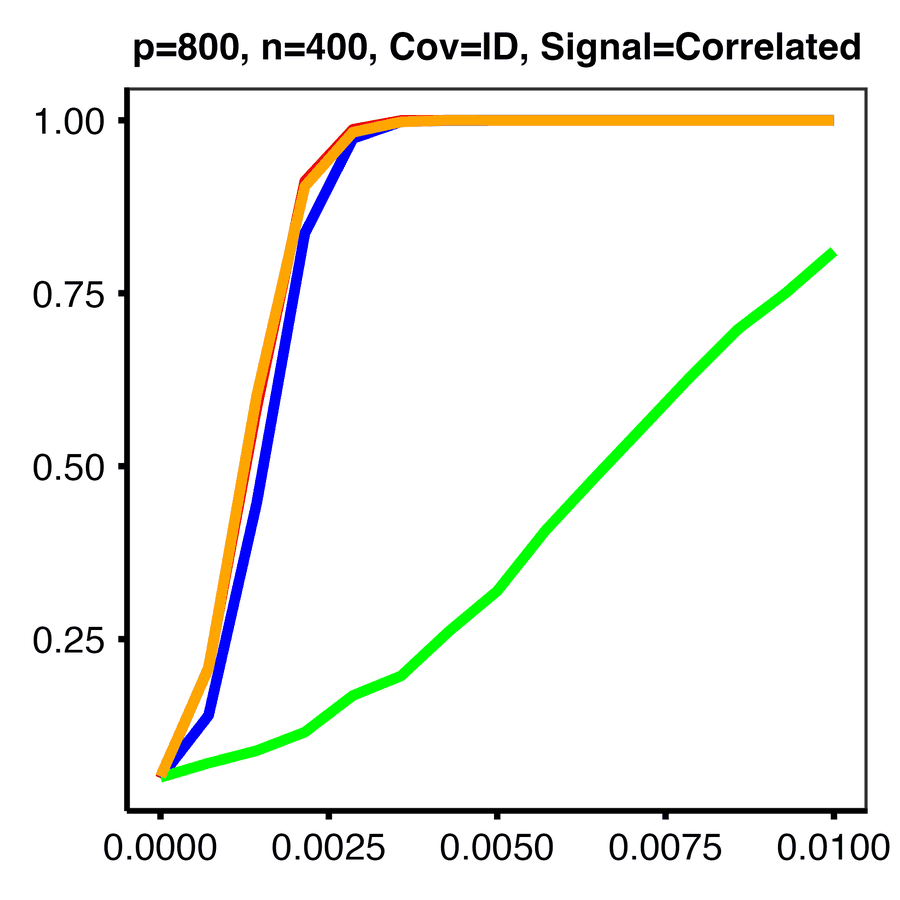} &
\includegraphics[width=0.24\textwidth]{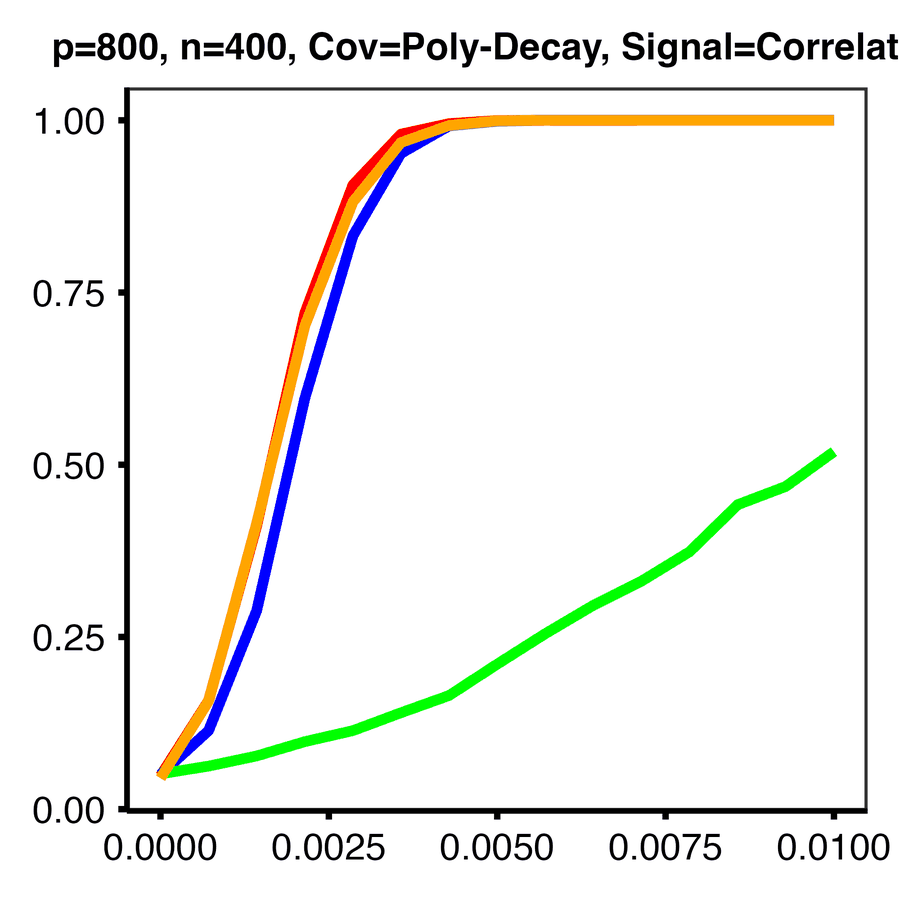} &
\includegraphics[width=0.24\textwidth]{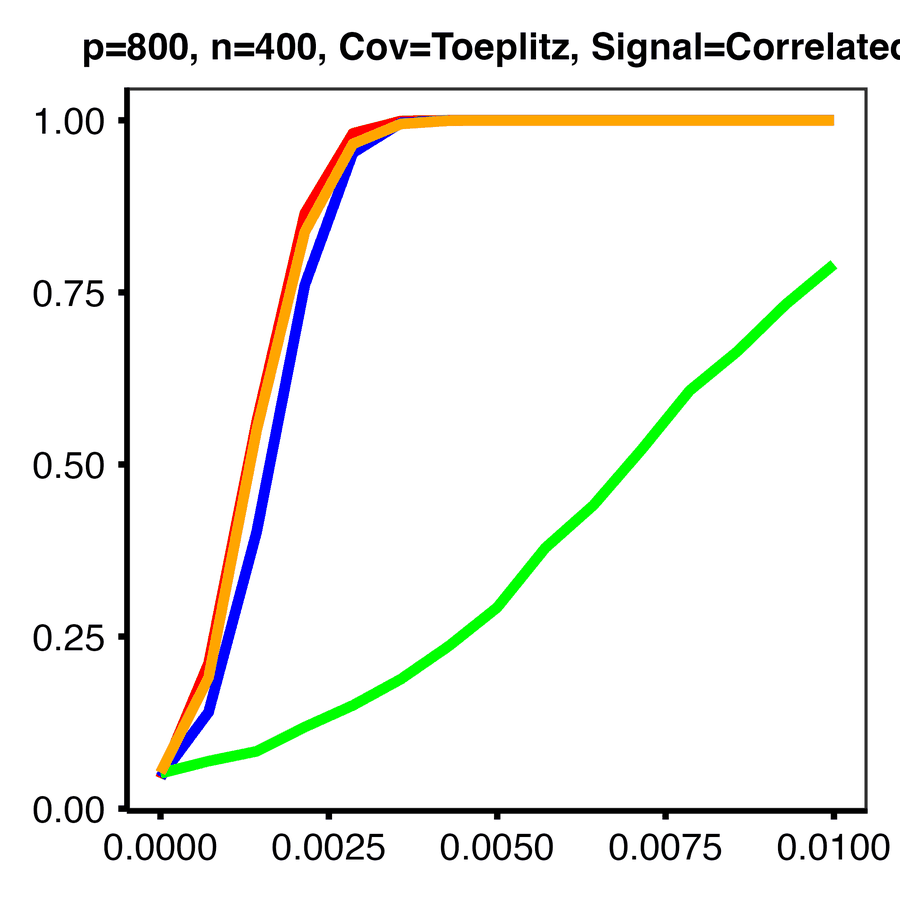}
\end{tabular}
\caption{Power curves under the SC setting with $\Sigma_p$-correlated mean shifts $\delta$. Rows correspond to p= 200, 300, 600, and 800. Columns correspond to the Exp Decay, Identity, Poly Decay, and Toeplitz covariance models. Blue: $T_{\rm disc}$; Red: $T_{\rm sc}$; Black (when p<n): Cov-Norm; Orange: U-Stat; Green: $\ell_\infty$-Sparse.}
\label{fig:power_SC_correlated}
\end{figure}

\begin{figure}[ht]
\centering
\setlength{\tabcolsep}{1pt}
\renewcommand{\arraystretch}{0}
\begin{tabular}{cccc}
\includegraphics[width=0.24\textwidth]{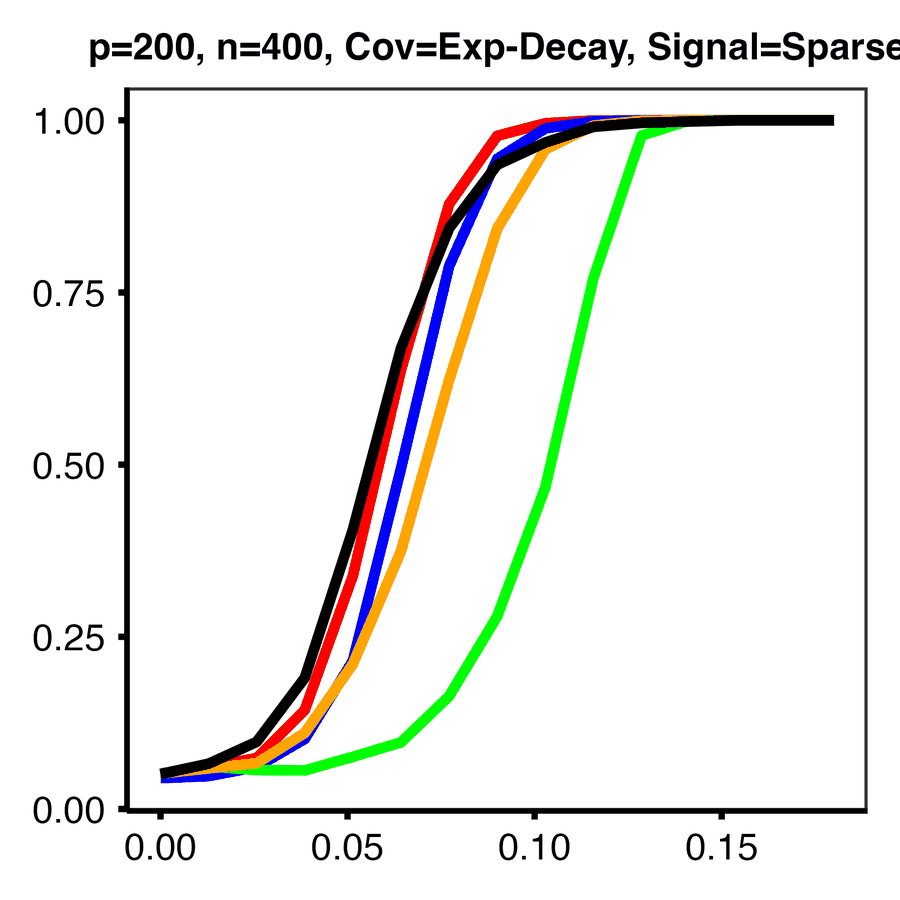} &
\includegraphics[width=0.24\textwidth]{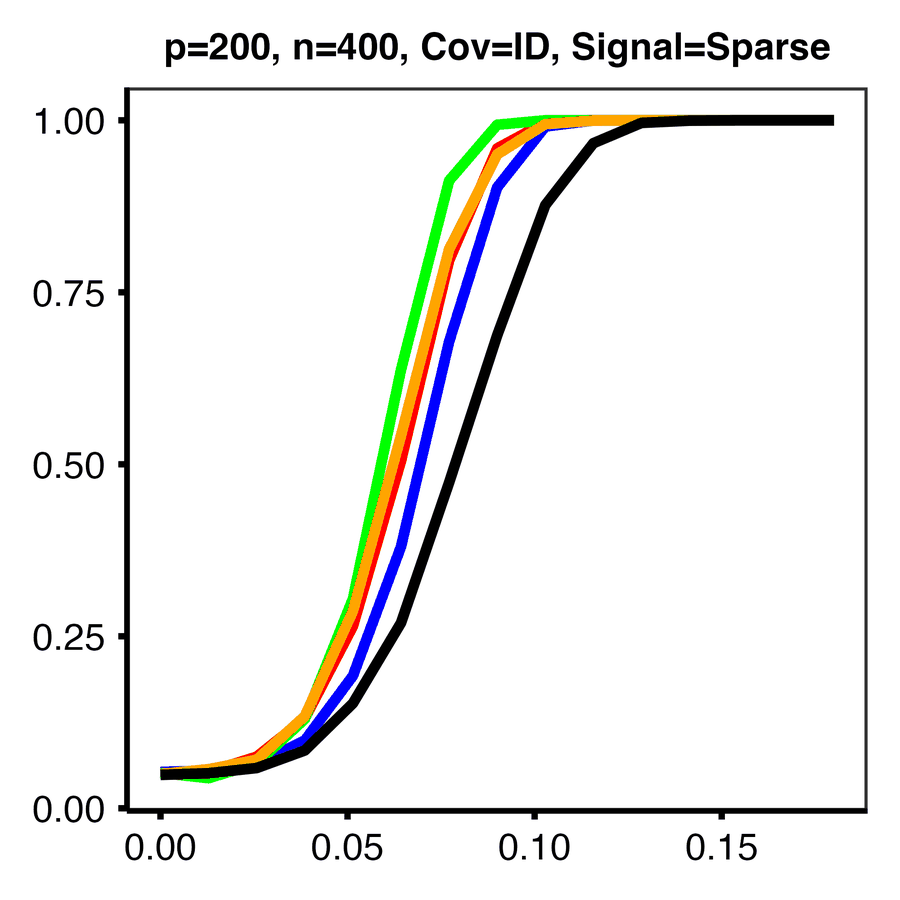} &
\includegraphics[width=0.24\textwidth]{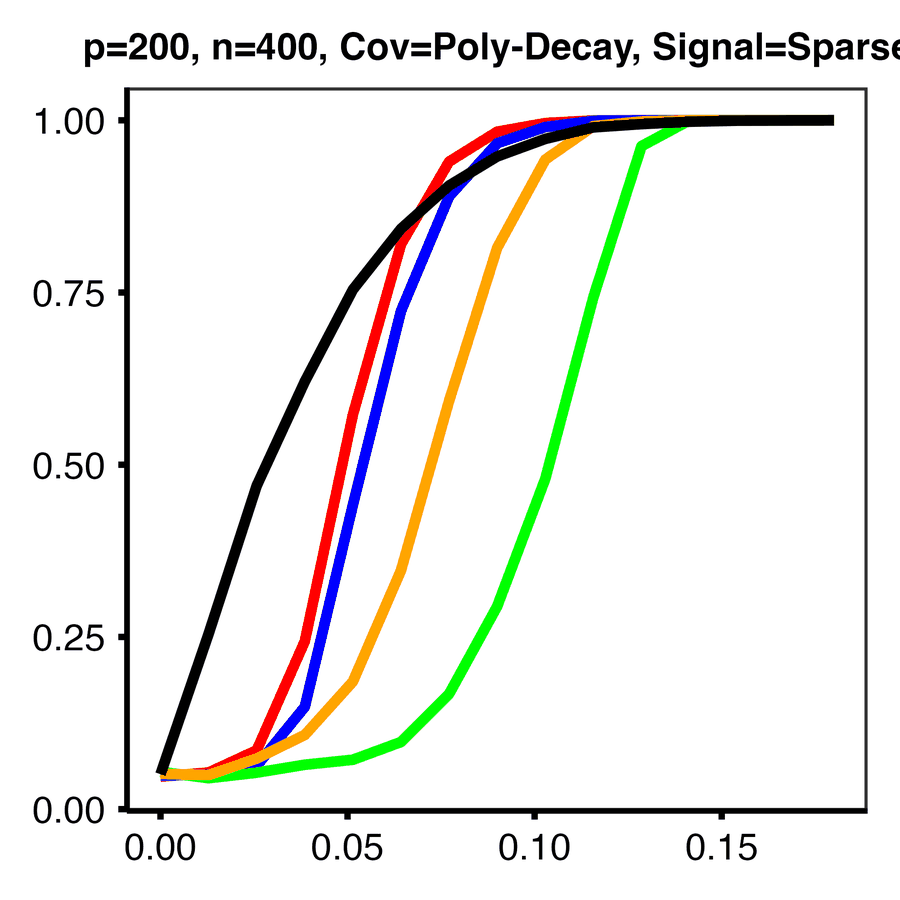} &
\includegraphics[width=0.24\textwidth]{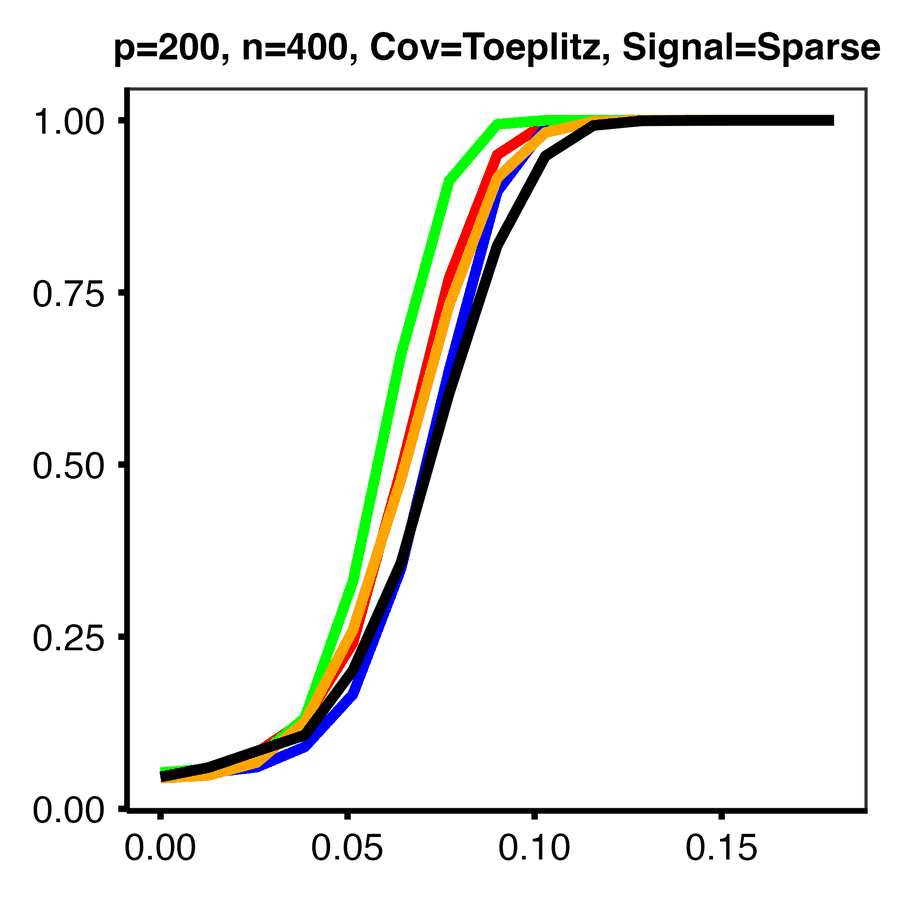}
\\
\includegraphics[width=0.24\textwidth]{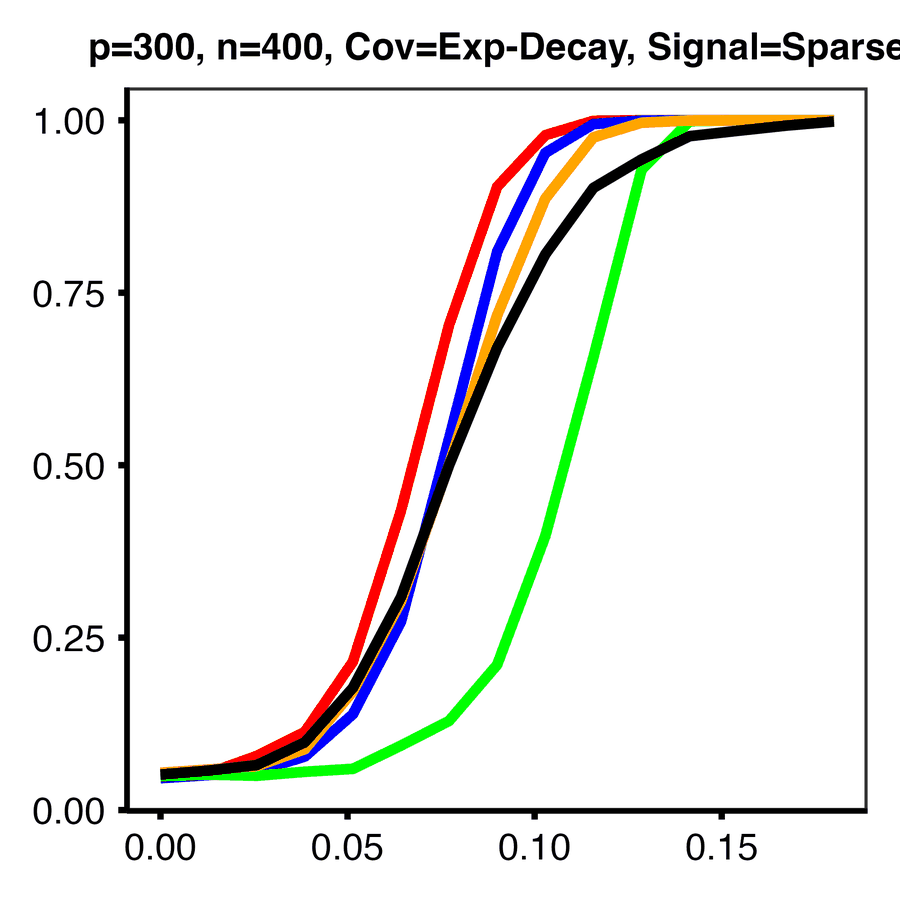} &
\includegraphics[width=0.24\textwidth]{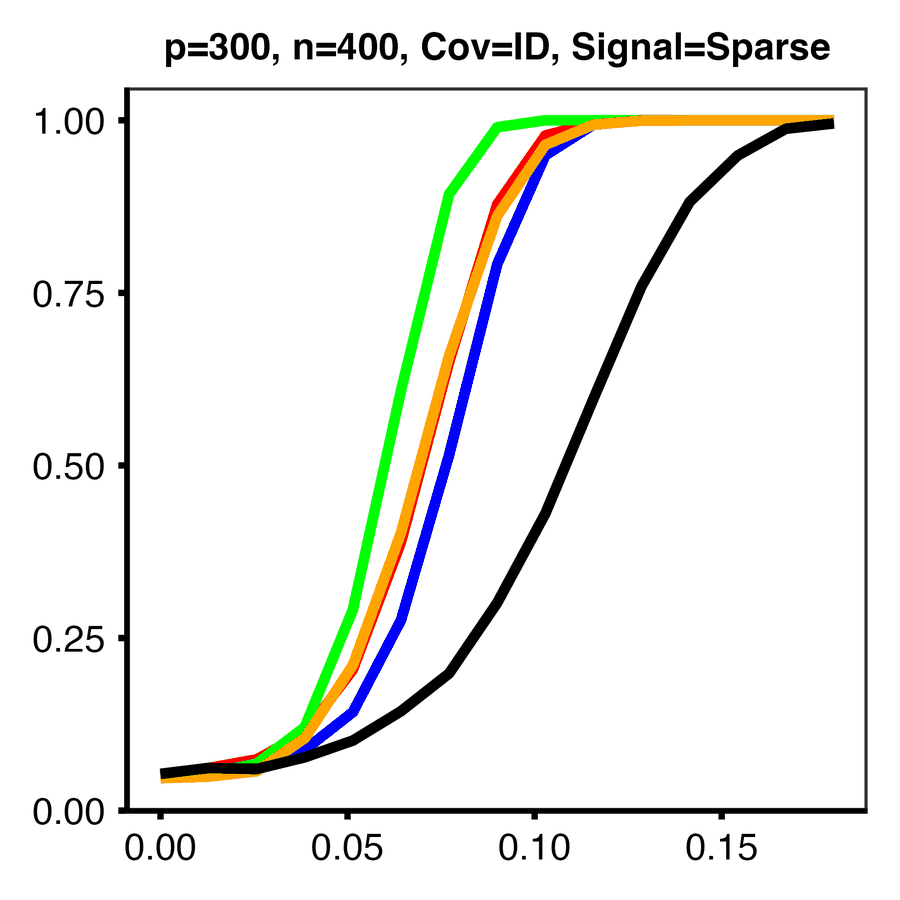} &
\includegraphics[width=0.24\textwidth]{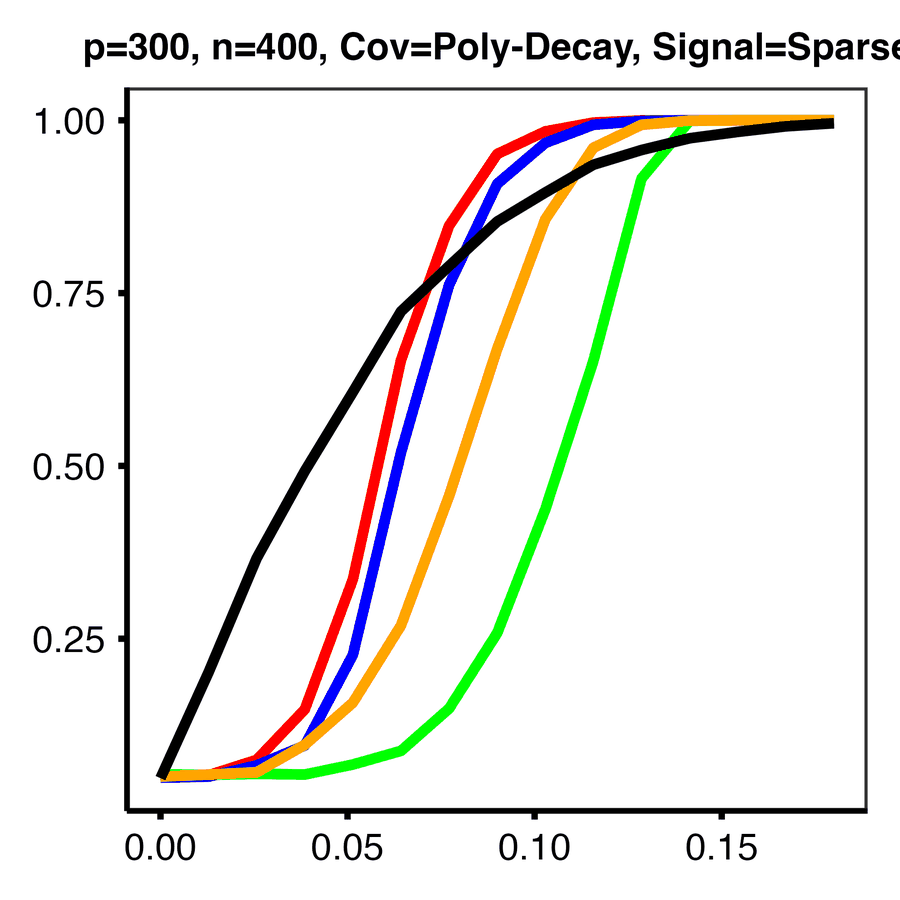} &
\includegraphics[width=0.24\textwidth]{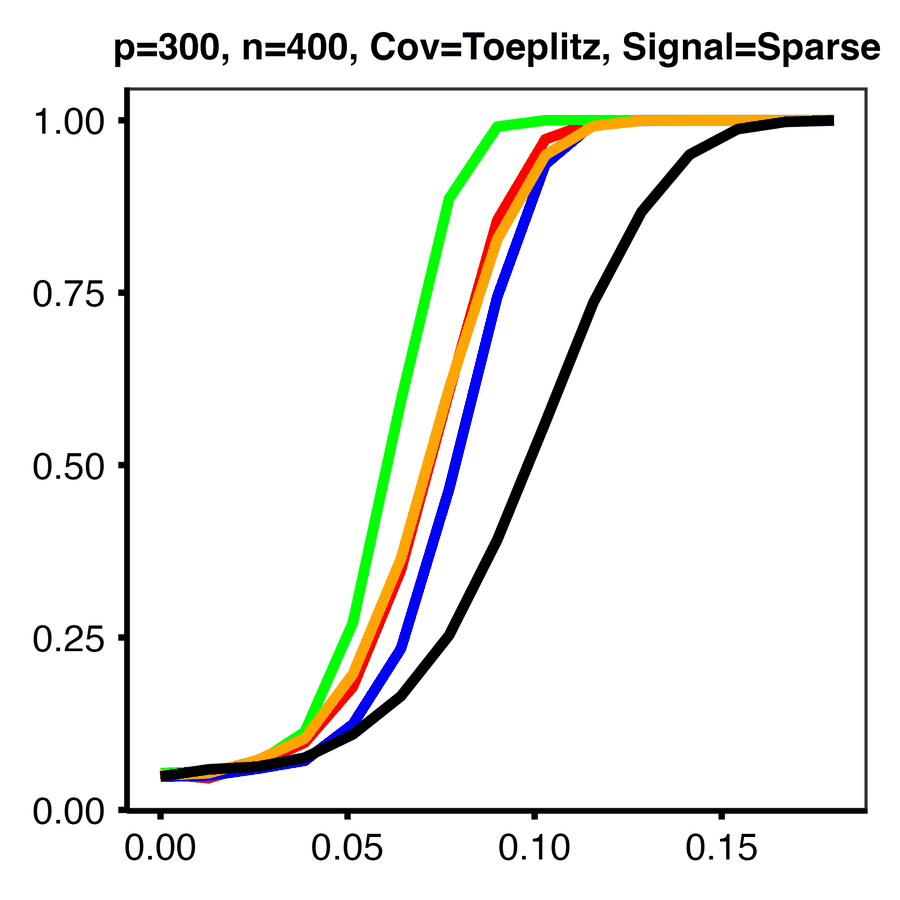}
\\
\includegraphics[width=0.24\textwidth]{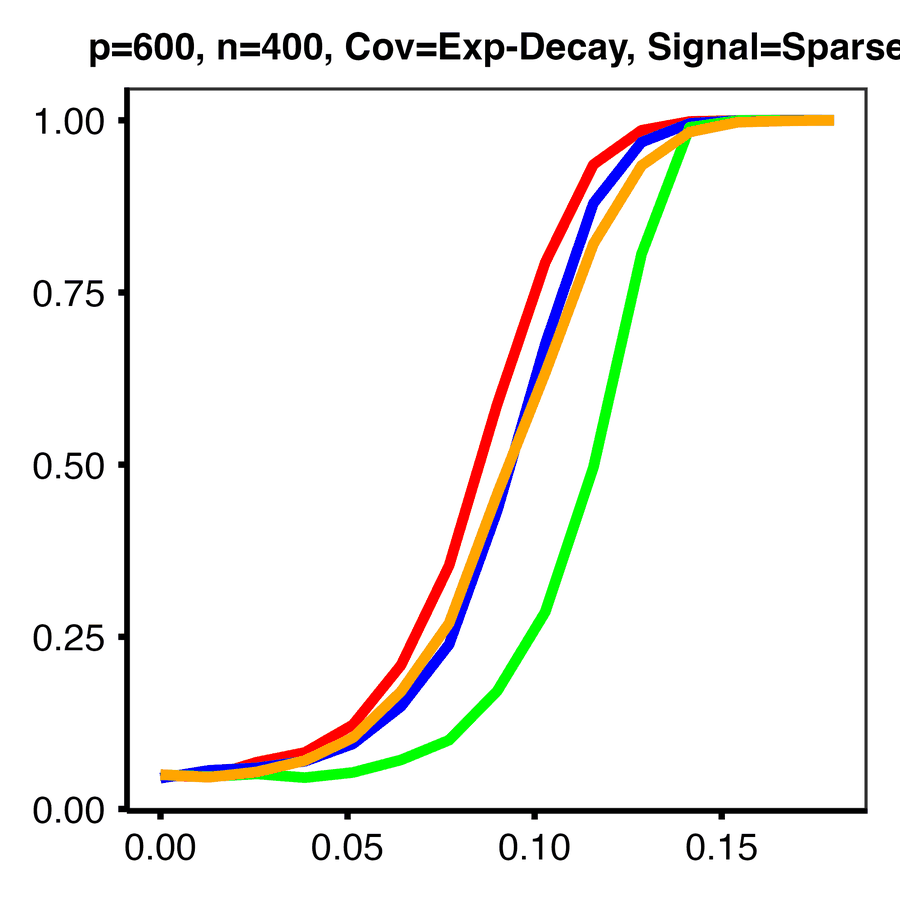} &
\includegraphics[width=0.24\textwidth]{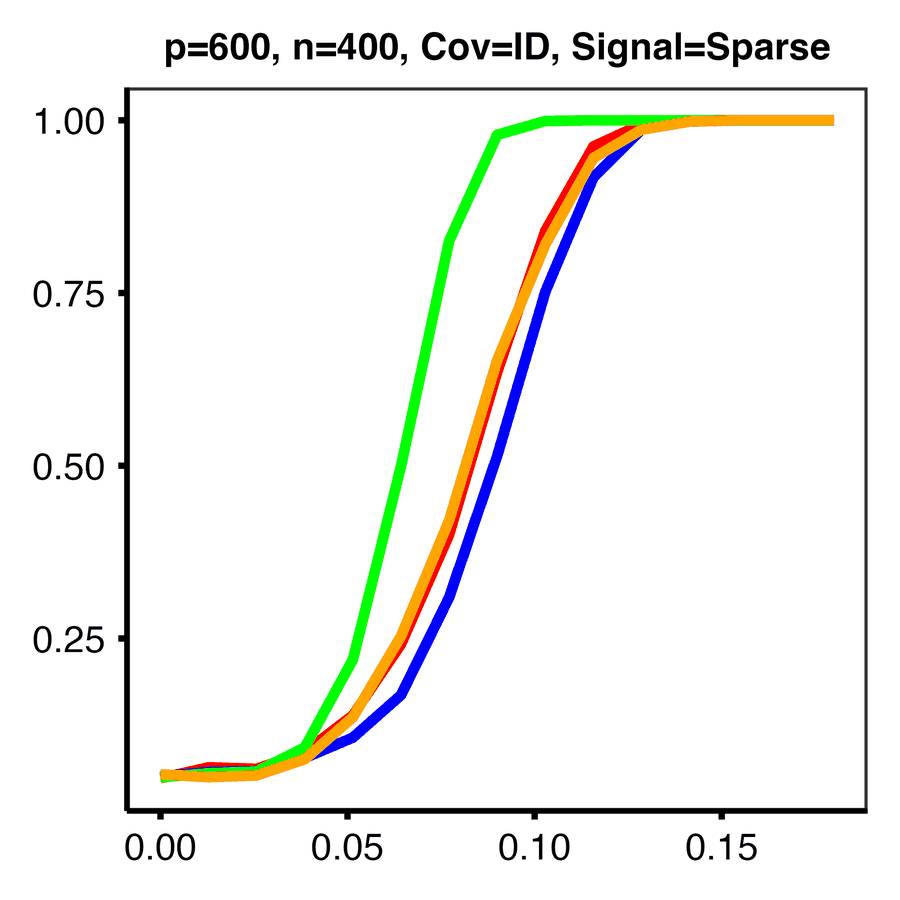} &
\includegraphics[width=0.24\textwidth]{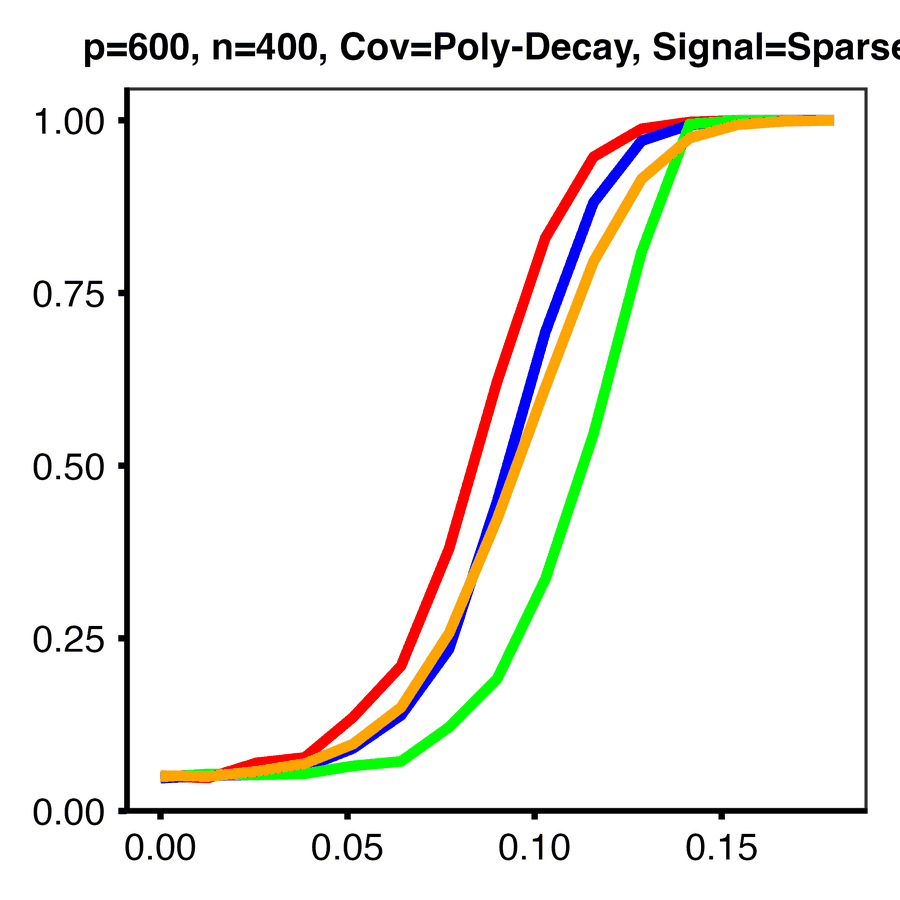} &
\includegraphics[width=0.24\textwidth]{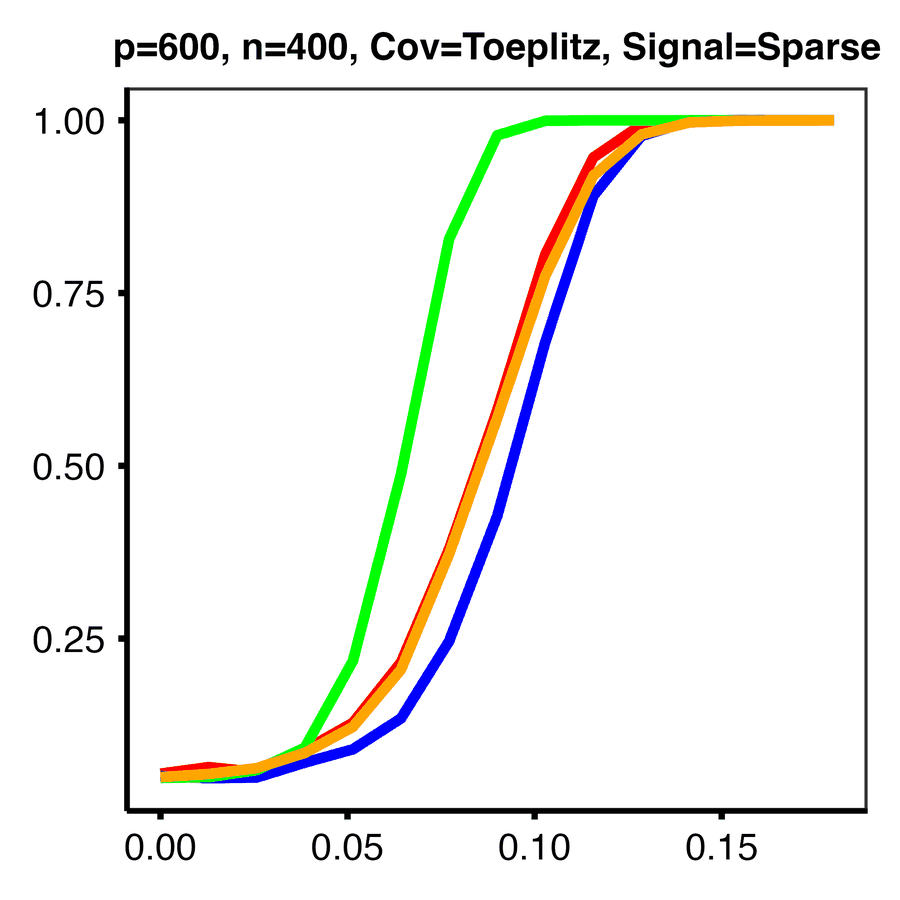}
\\
\includegraphics[width=0.24\textwidth]{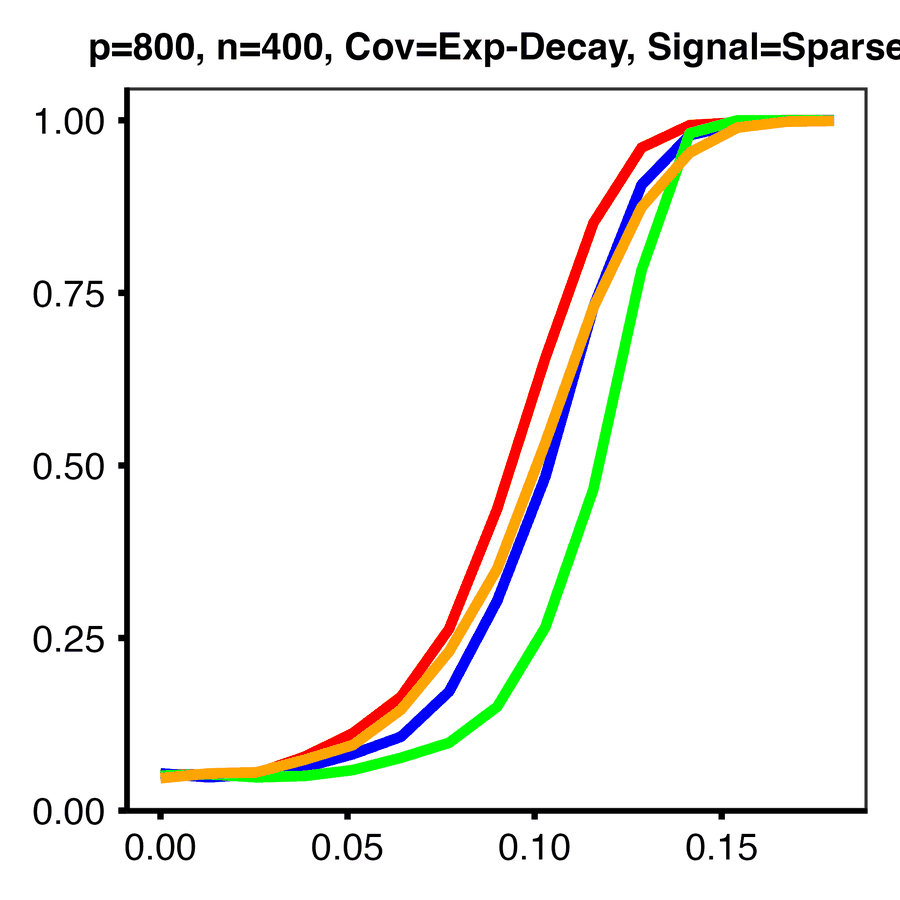} &
\includegraphics[width=0.24\textwidth]{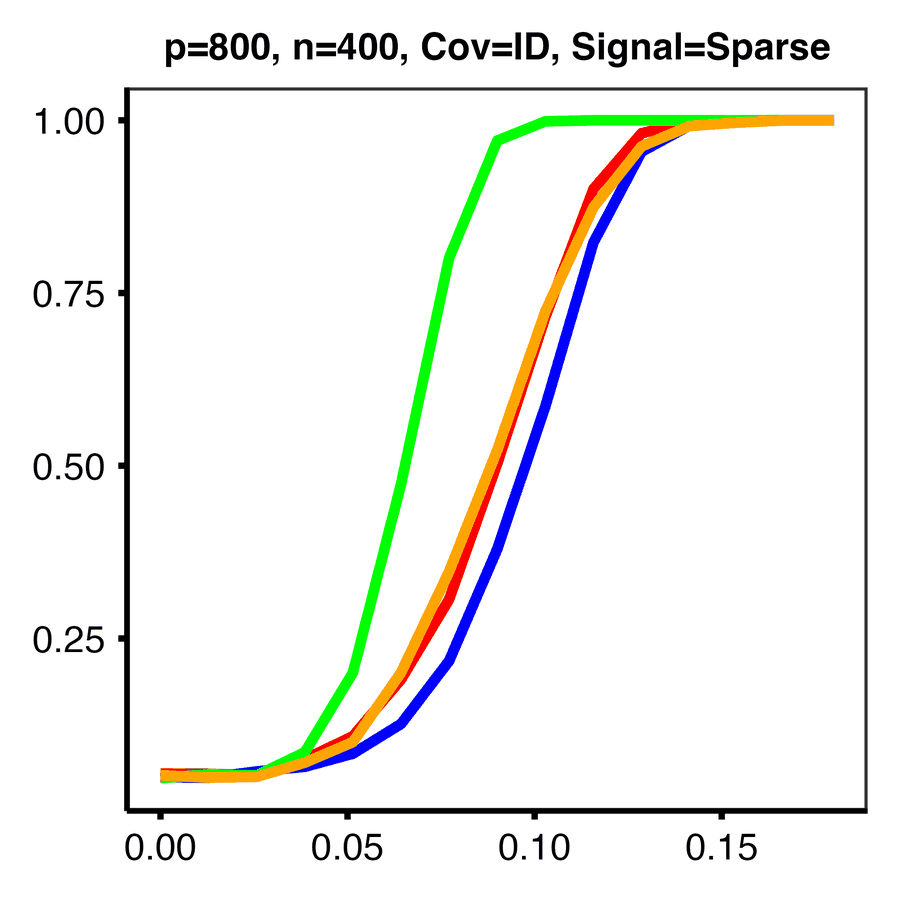} &
\includegraphics[width=0.24\textwidth]{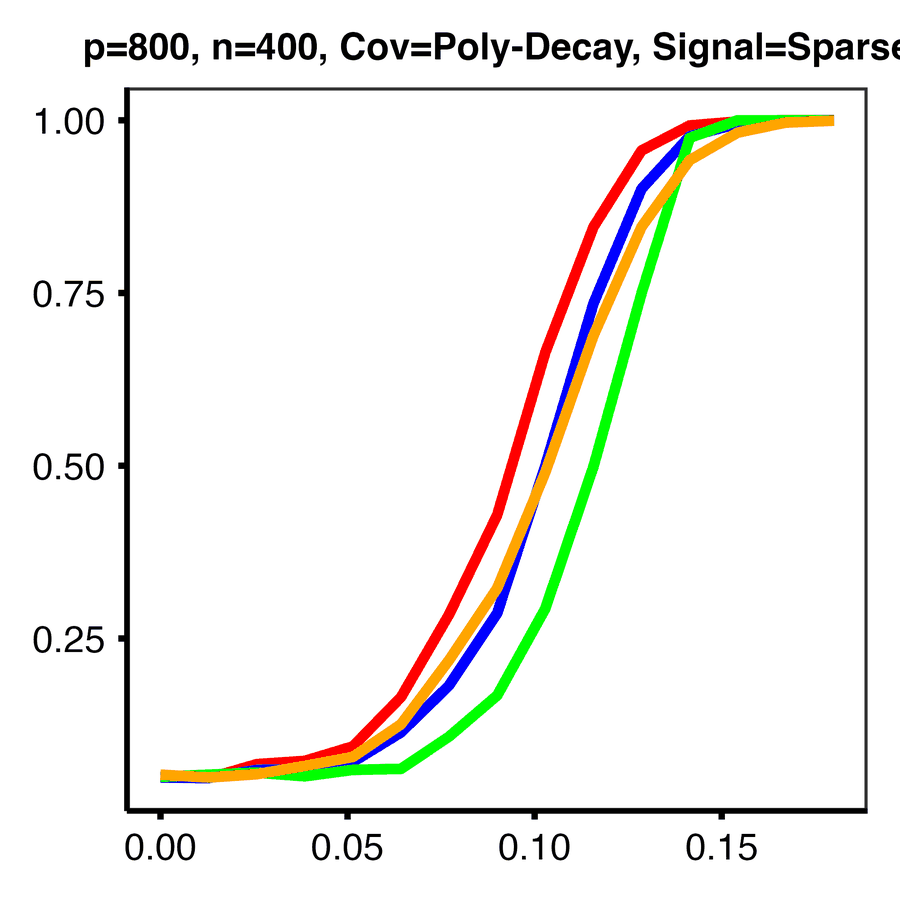} &
\includegraphics[width=0.24\textwidth]{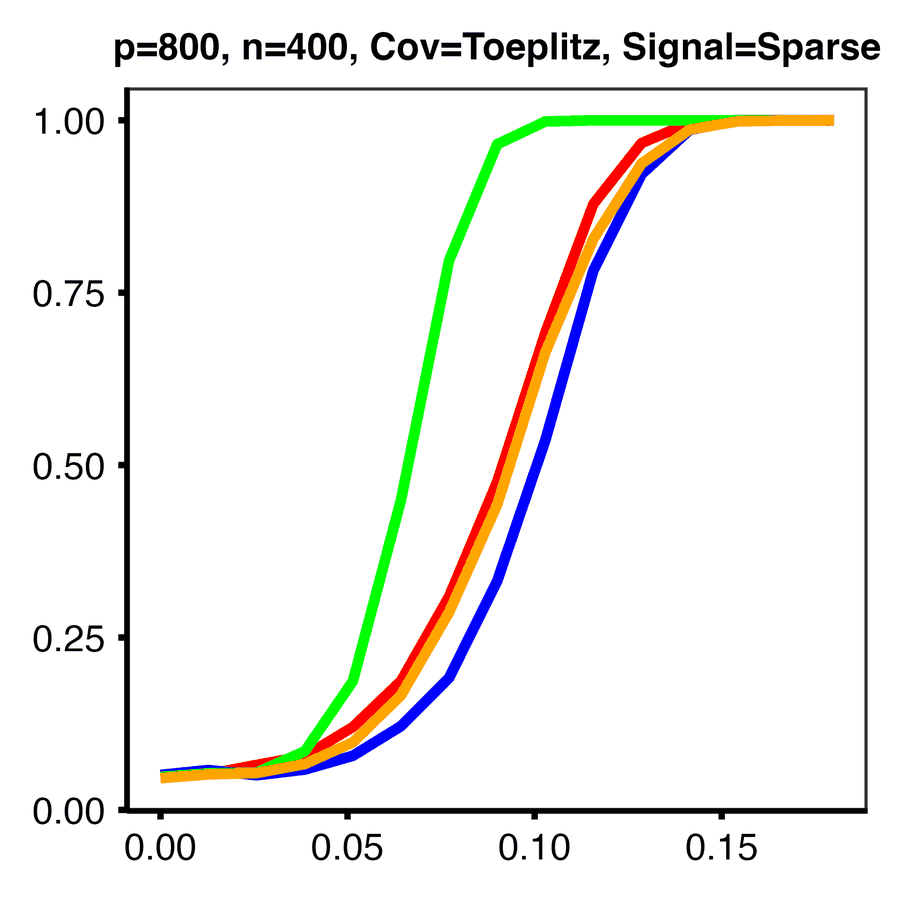}
\end{tabular}
\caption{Power curves under the SC setting with sparse mean shifts $\delta$. Rows correspond to p= 200, 300, 600, and 800. Columns correspond to the Exp Decay, Identity, Poly Decay, and Toeplitz covariance models. Blue: $T_{\rm disc}$; Red: $T_{\rm sc}$; Black (when p<n): Cov-Norm; Orange: U-Stat; Green: $\ell_\infty$-Sparse.}
\label{fig:power_SC_sparse}
\end{figure}

\begin{figure}[ht]
\centering
\setlength{\tabcolsep}{1pt}
\renewcommand{\arraystretch}{0}
\begin{tabular}{cccc}
\includegraphics[width=0.24\textwidth]{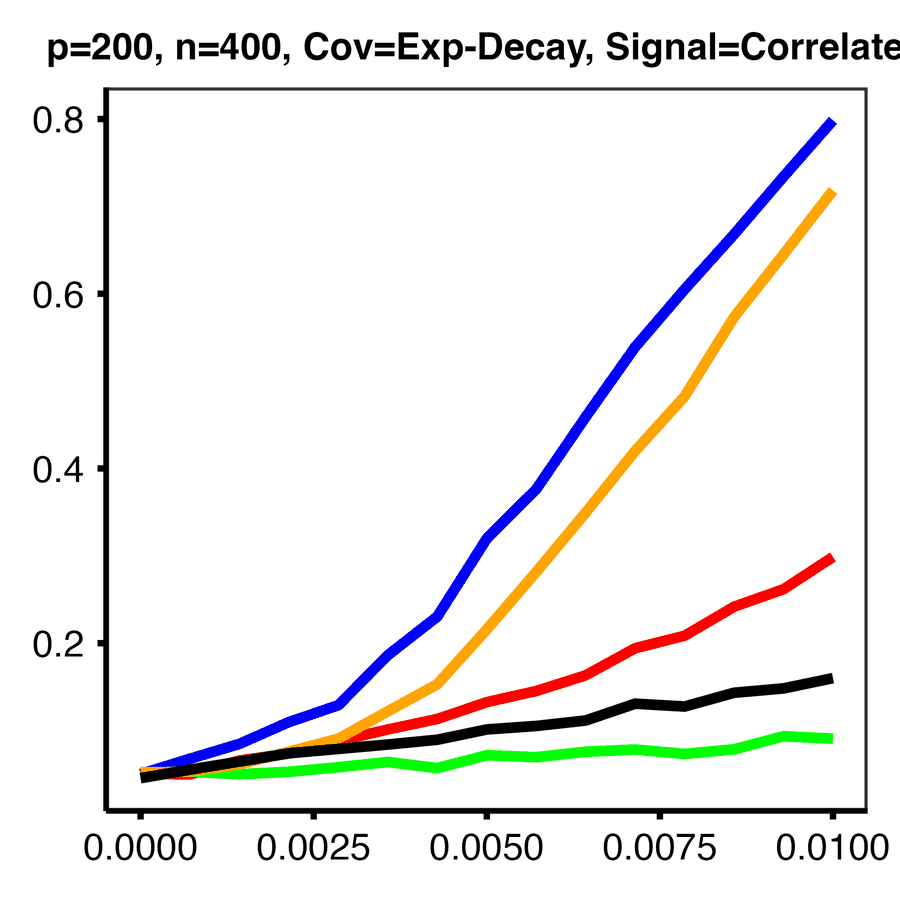} &
\includegraphics[width=0.24\textwidth]{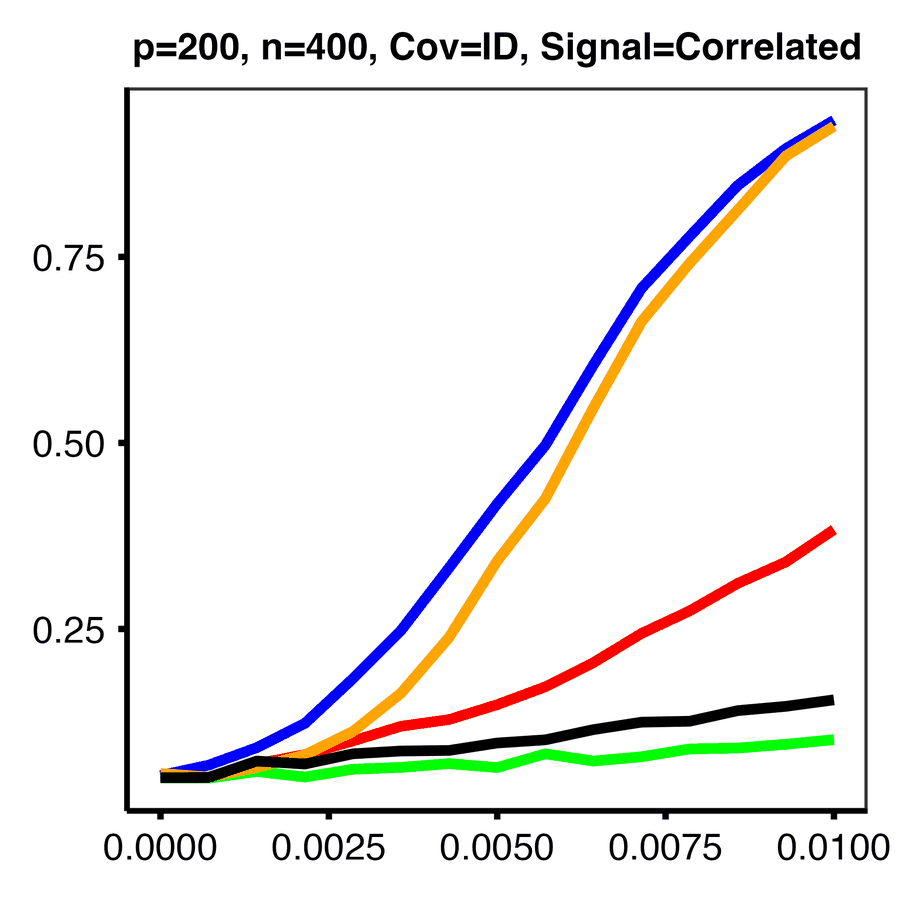} &
\includegraphics[width=0.24\textwidth]{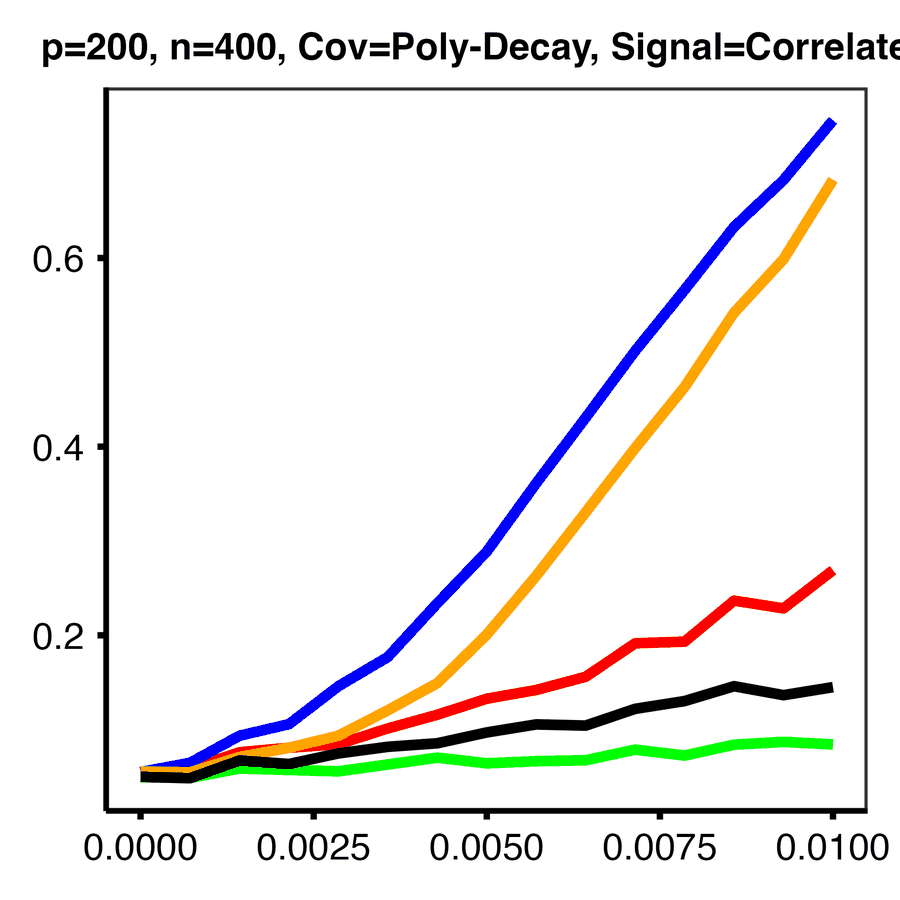} &
\includegraphics[width=0.24\textwidth]{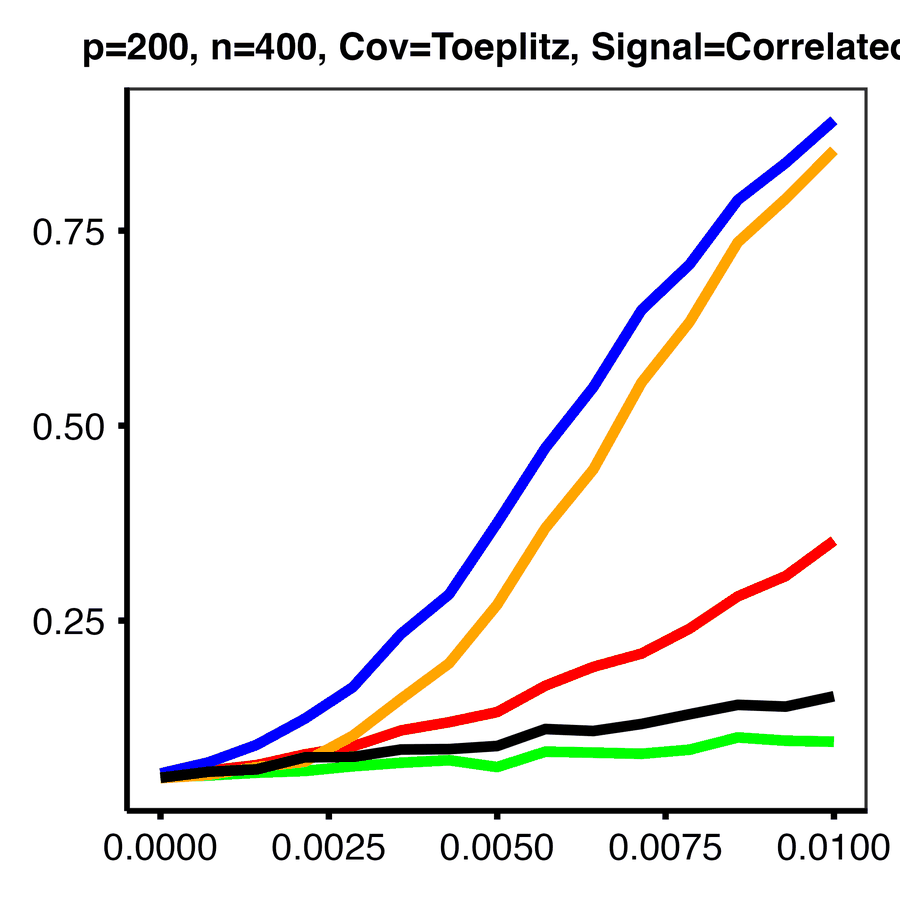}
\\
\includegraphics[width=0.24\textwidth]{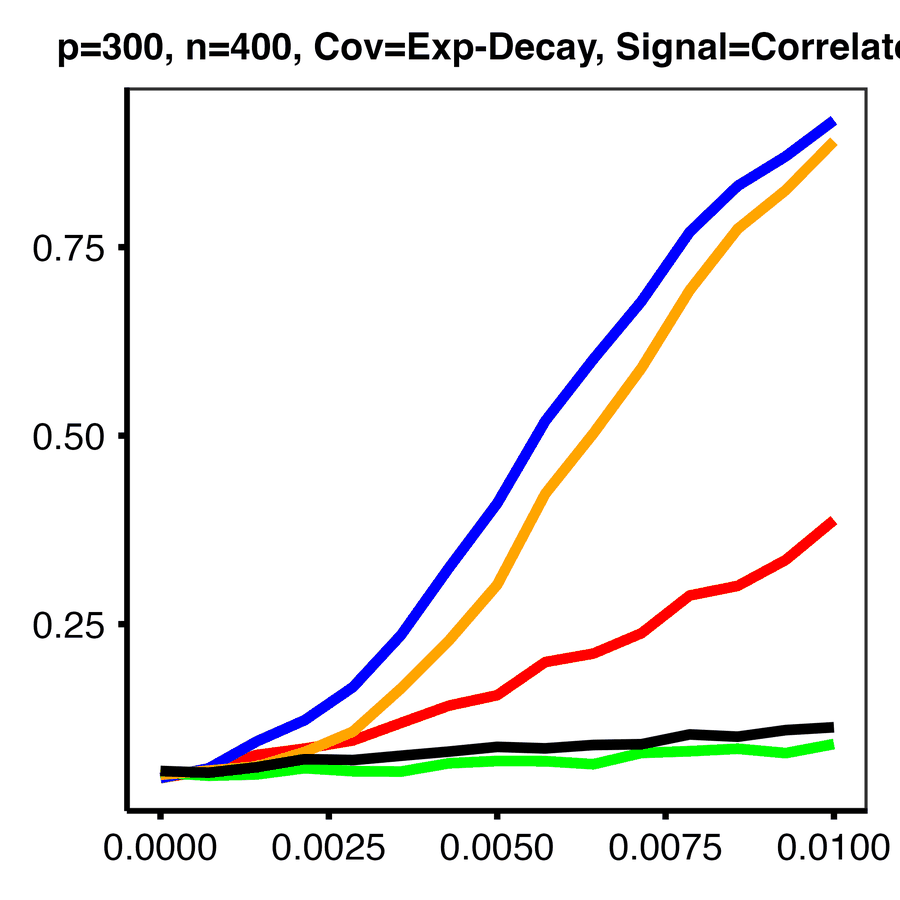} &
\includegraphics[width=0.24\textwidth]{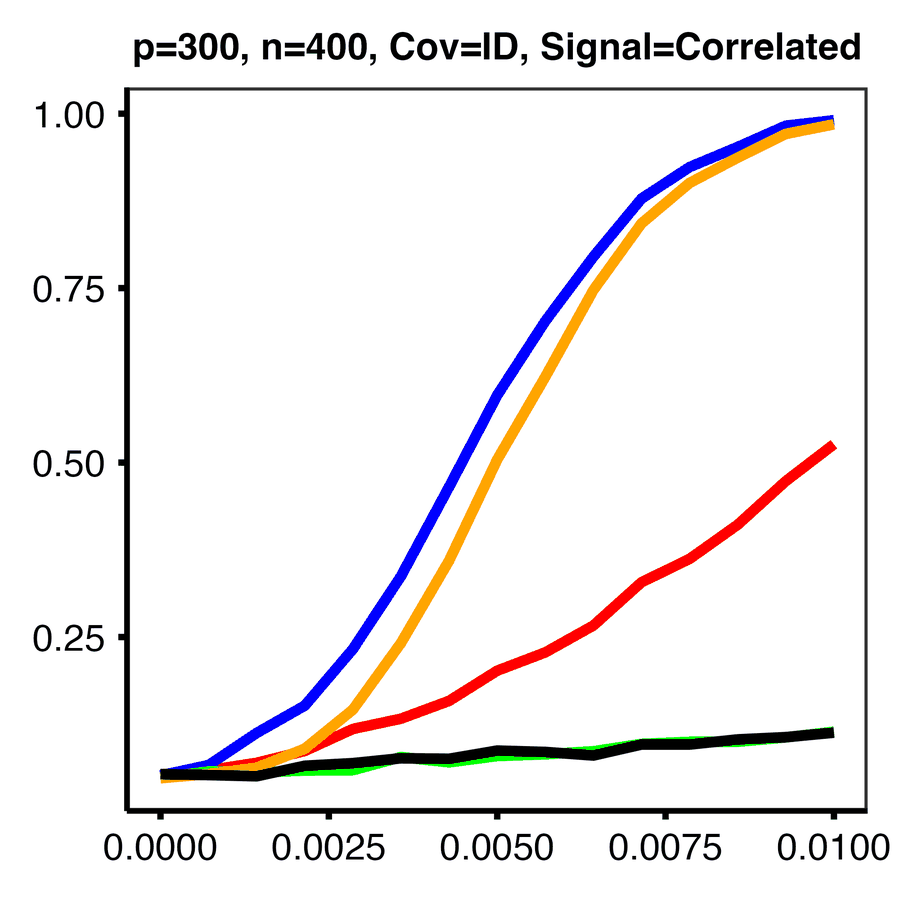} &
\includegraphics[width=0.24\textwidth]{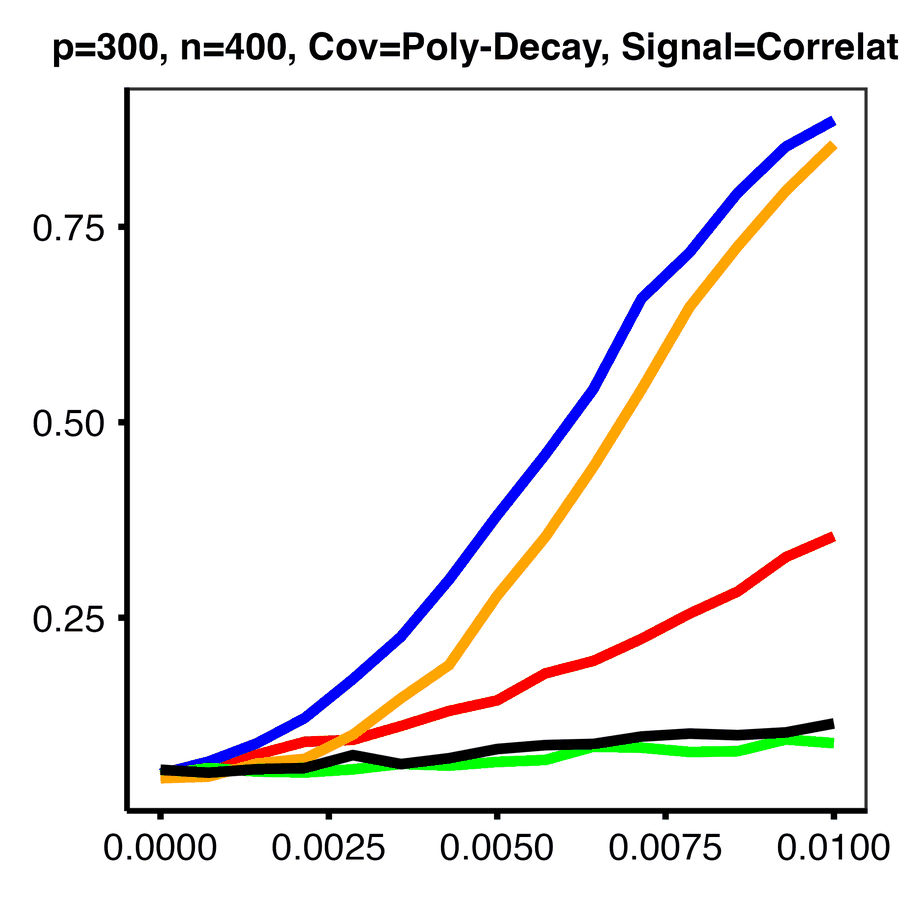} &
\includegraphics[width=0.24\textwidth]{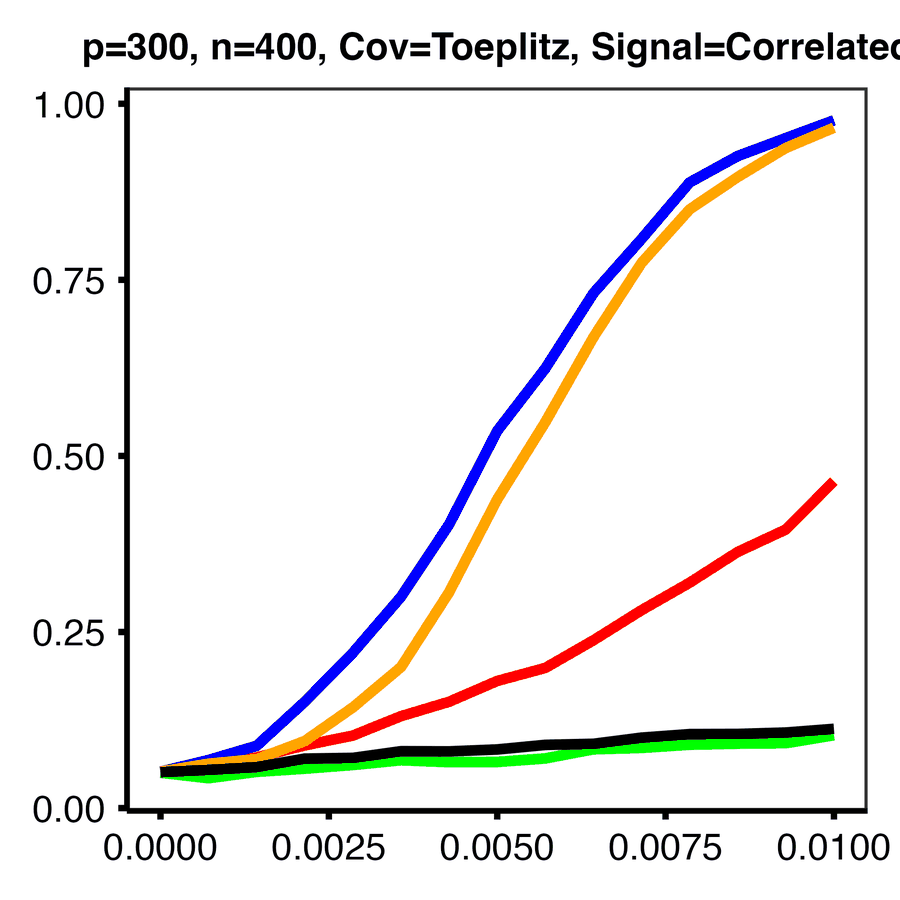}
\\
\includegraphics[width=0.24\textwidth]{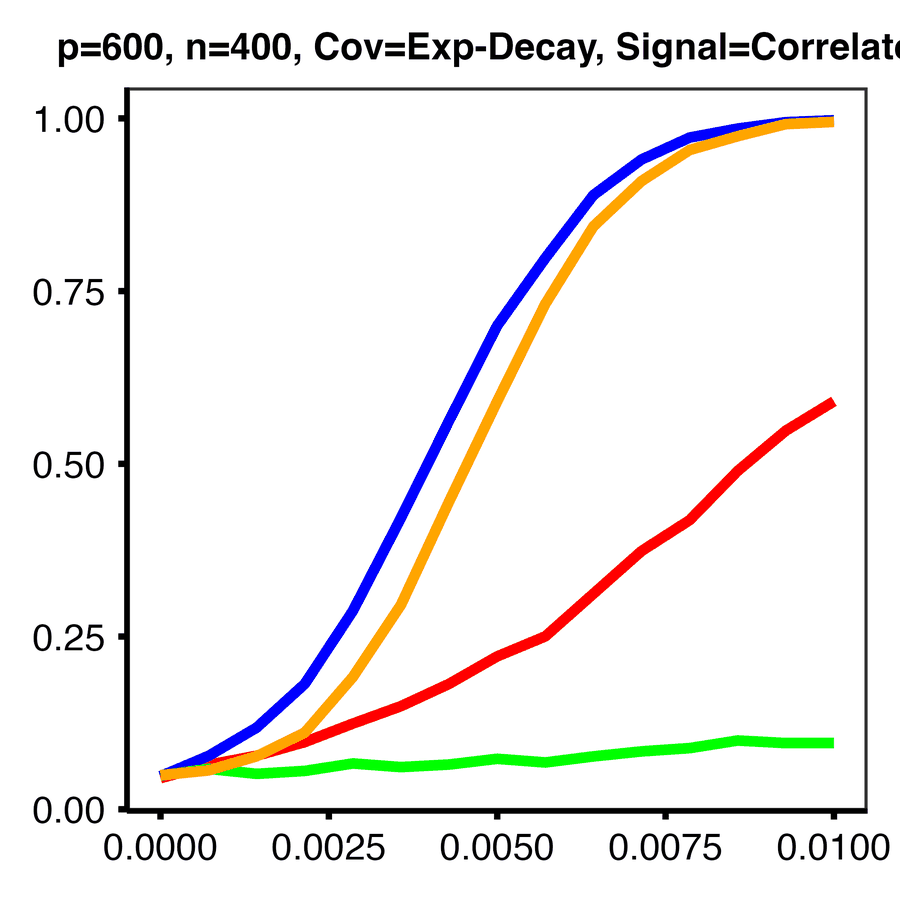} &
\includegraphics[width=0.24\textwidth]{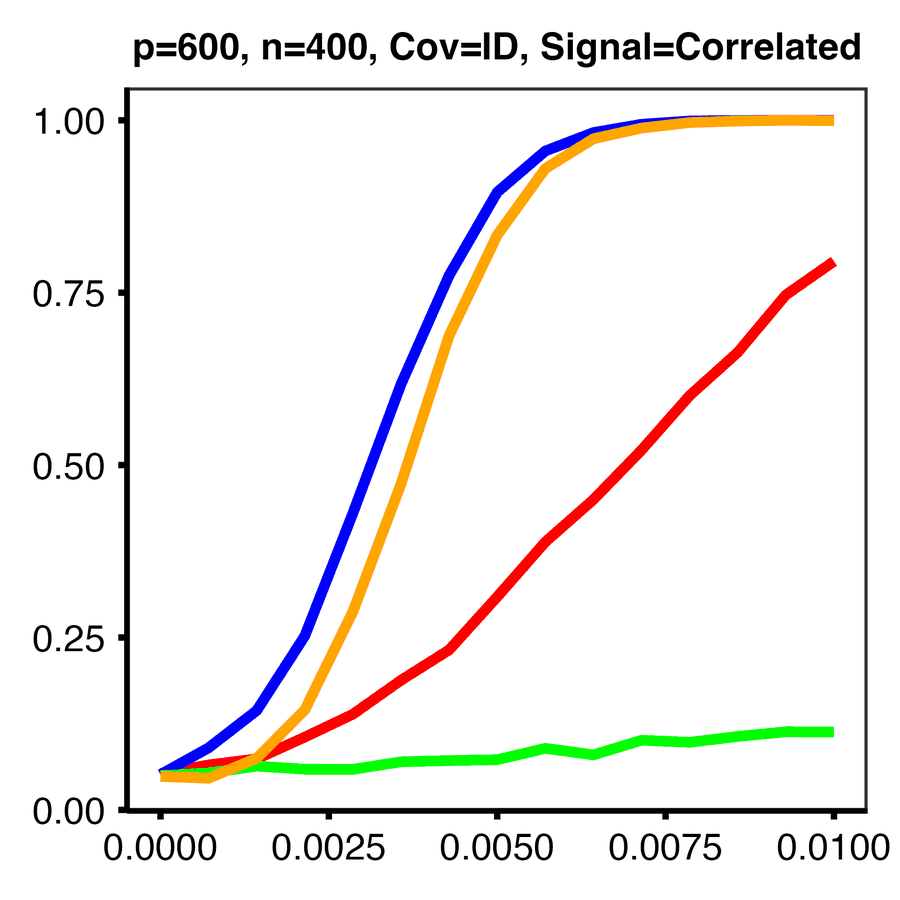} &
\includegraphics[width=0.24\textwidth]{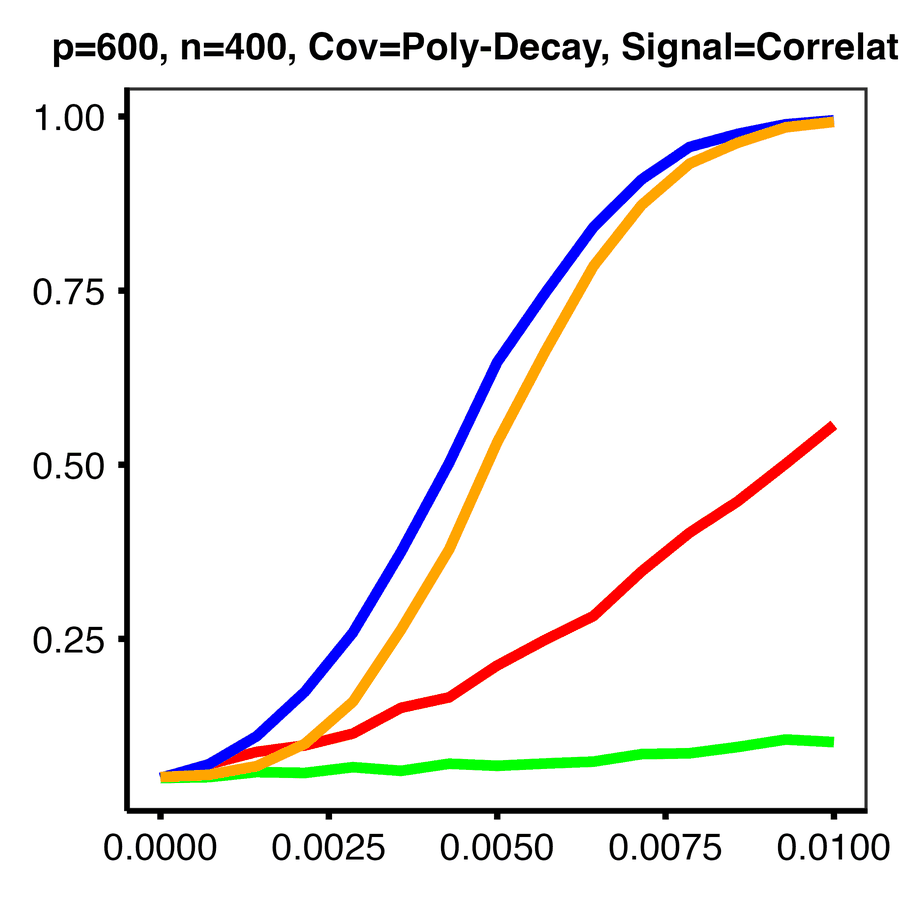} &
\includegraphics[width=0.24\textwidth]{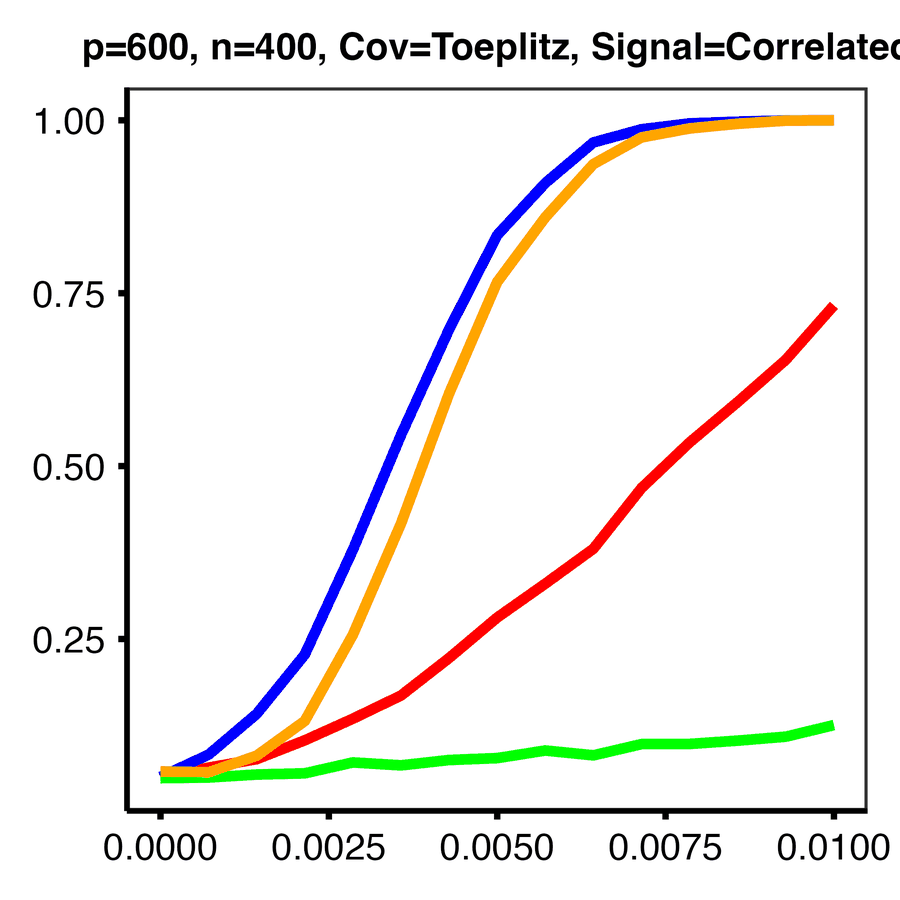}
\\
\includegraphics[width=0.24\textwidth]{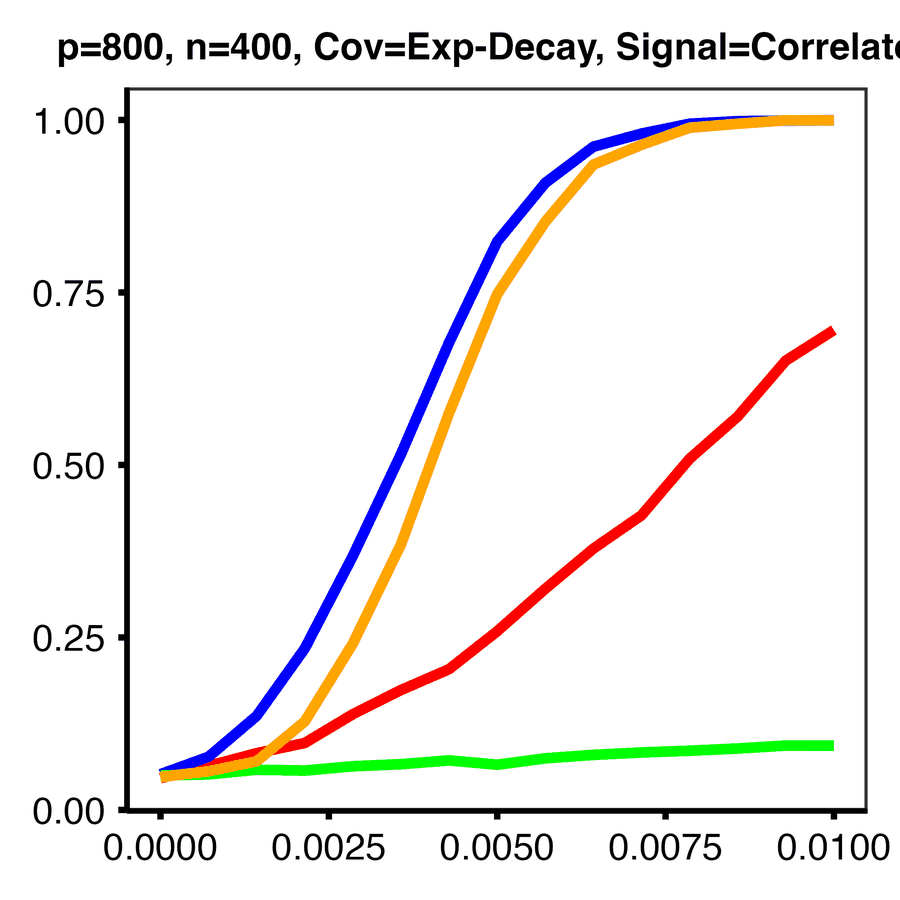} &
\includegraphics[width=0.24\textwidth]{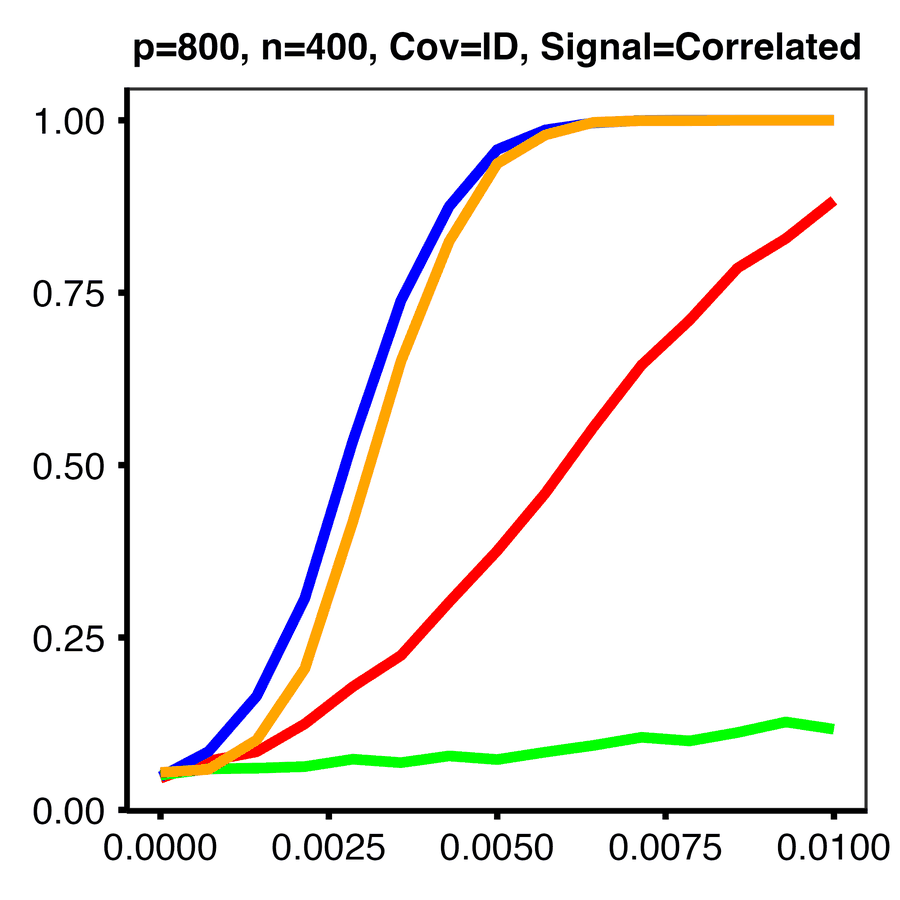} &
\includegraphics[width=0.24\textwidth]{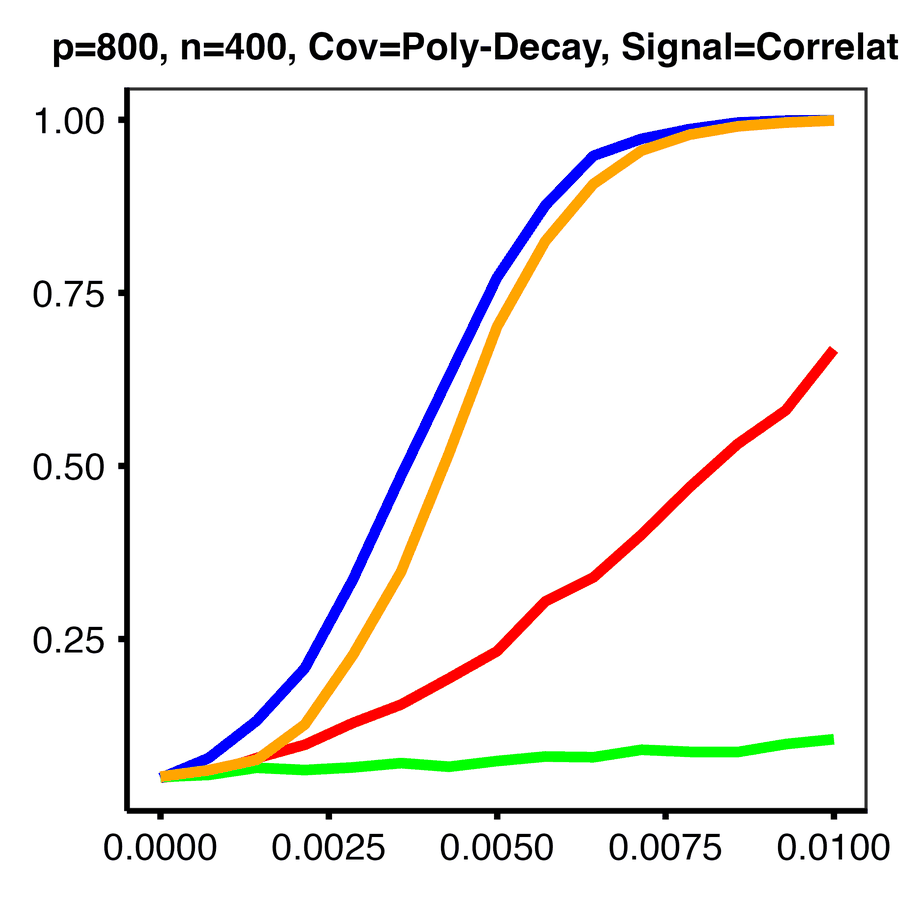} &
\includegraphics[width=0.24\textwidth]{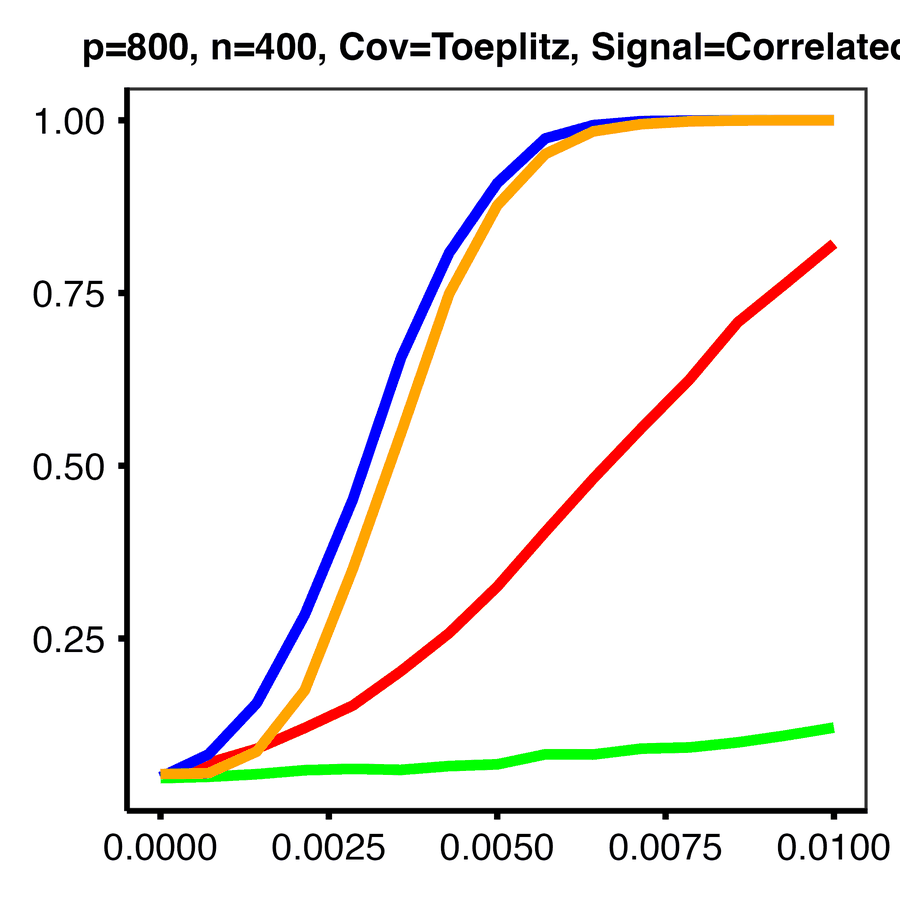}
\end{tabular}
\caption{Power curves under the MC setting with $\Sigma_p$-correlated mean shifts $\delta$. Rows correspond to p= 200, 300, 600, and 800. Columns correspond to the Exp Decay, Identity, Poly Decay, and Toeplitz covariance models. Blue: $T_{\rm disc}$; Red: $T_{\rm sc}$; Black (when p<n): Cov-Norm; Orange: U-Stat; Green: $\ell_\infty$-Sparse.}
\label{fig:power_MC_correlated}
\end{figure}

\begin{figure}[ht]
\centering
\setlength{\tabcolsep}{1pt}
\renewcommand{\arraystretch}{0}
\begin{tabular}{cccc}
\includegraphics[width=0.24\textwidth]{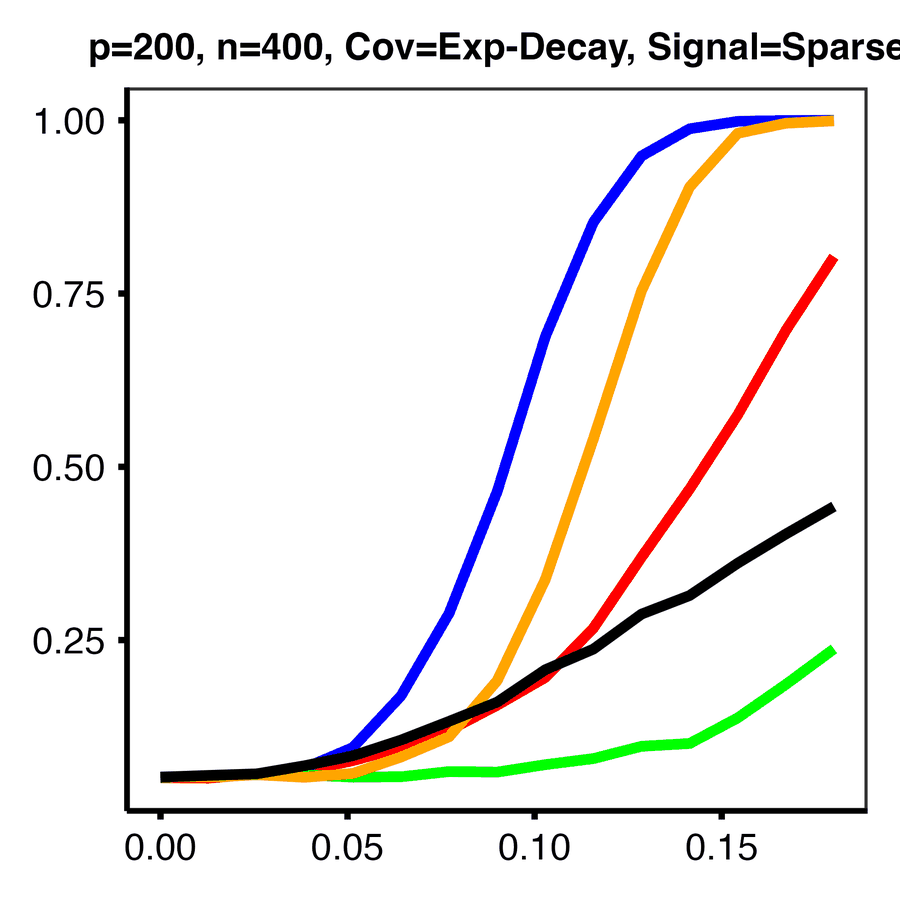} &
\includegraphics[width=0.24\textwidth]{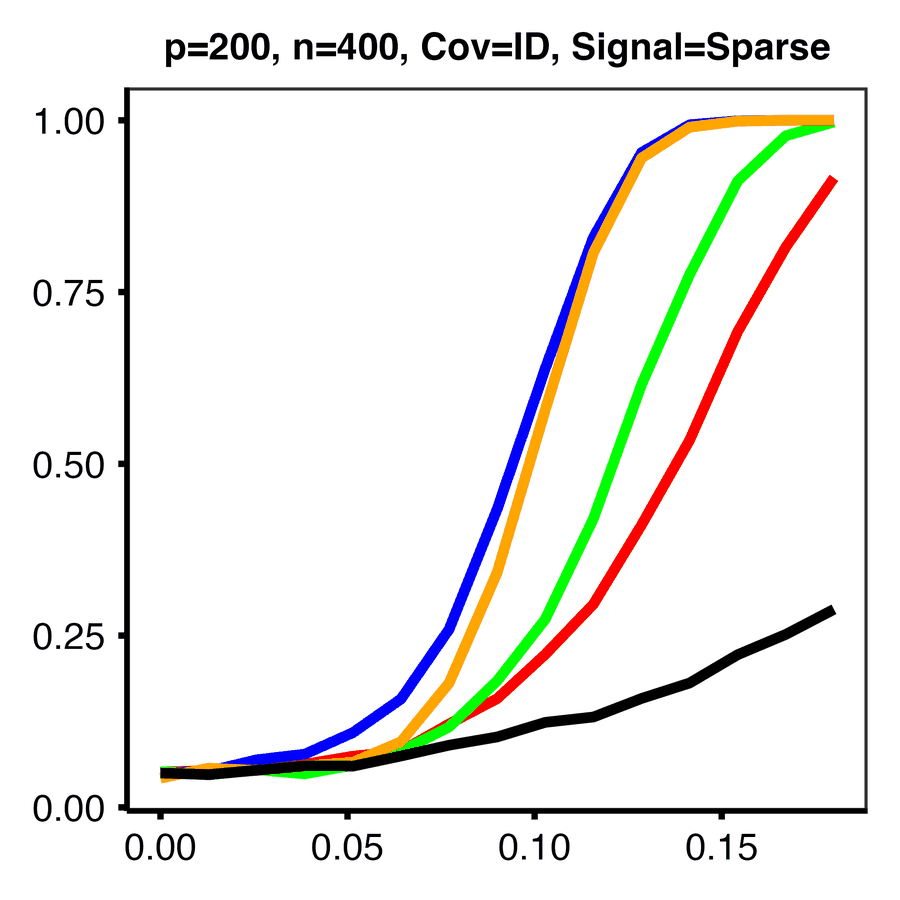} &
\includegraphics[width=0.24\textwidth]{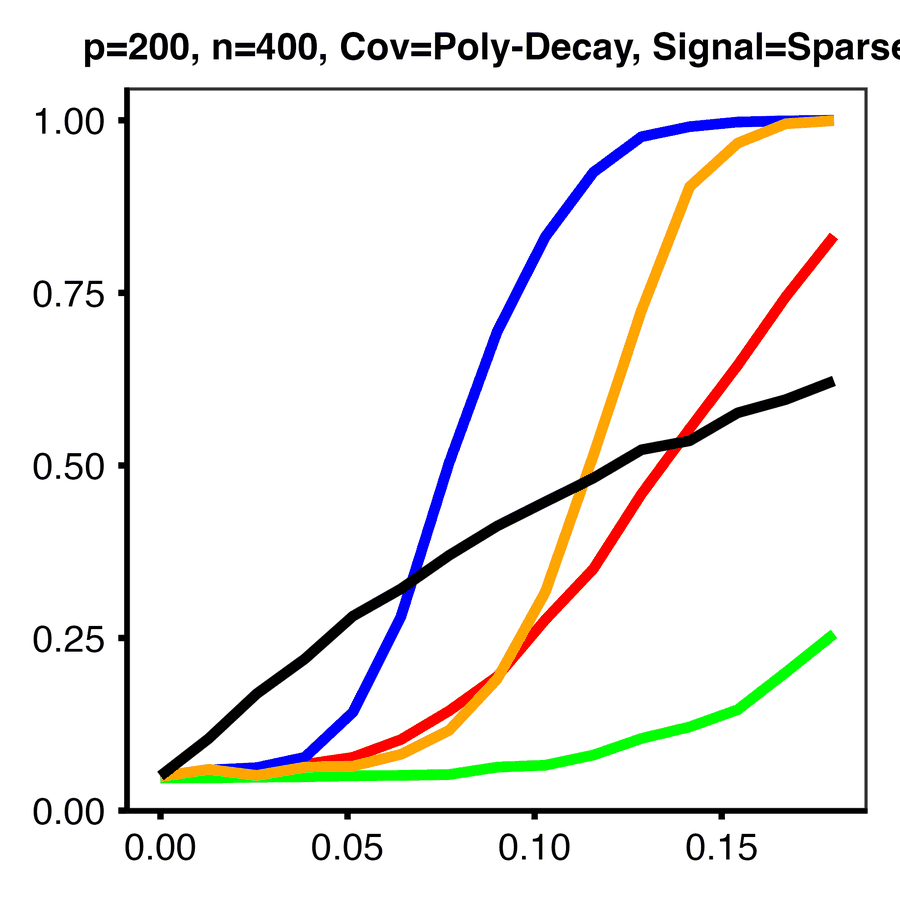} &
\includegraphics[width=0.24\textwidth]{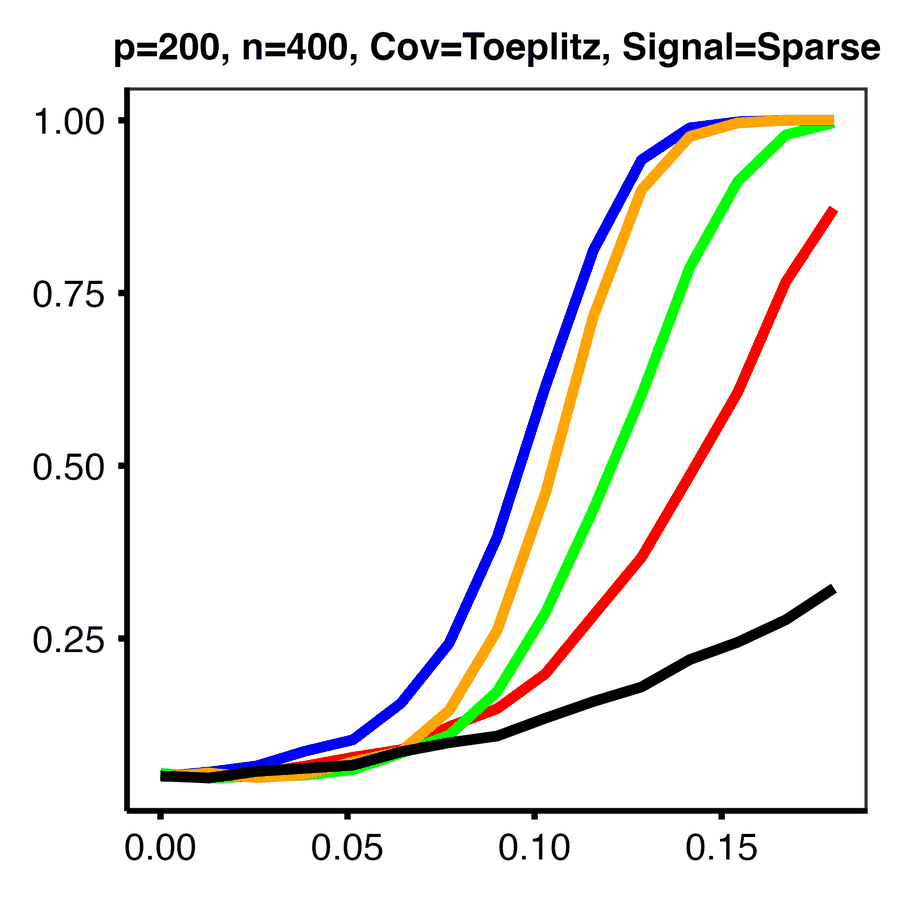}
\\
\includegraphics[width=0.24\textwidth]{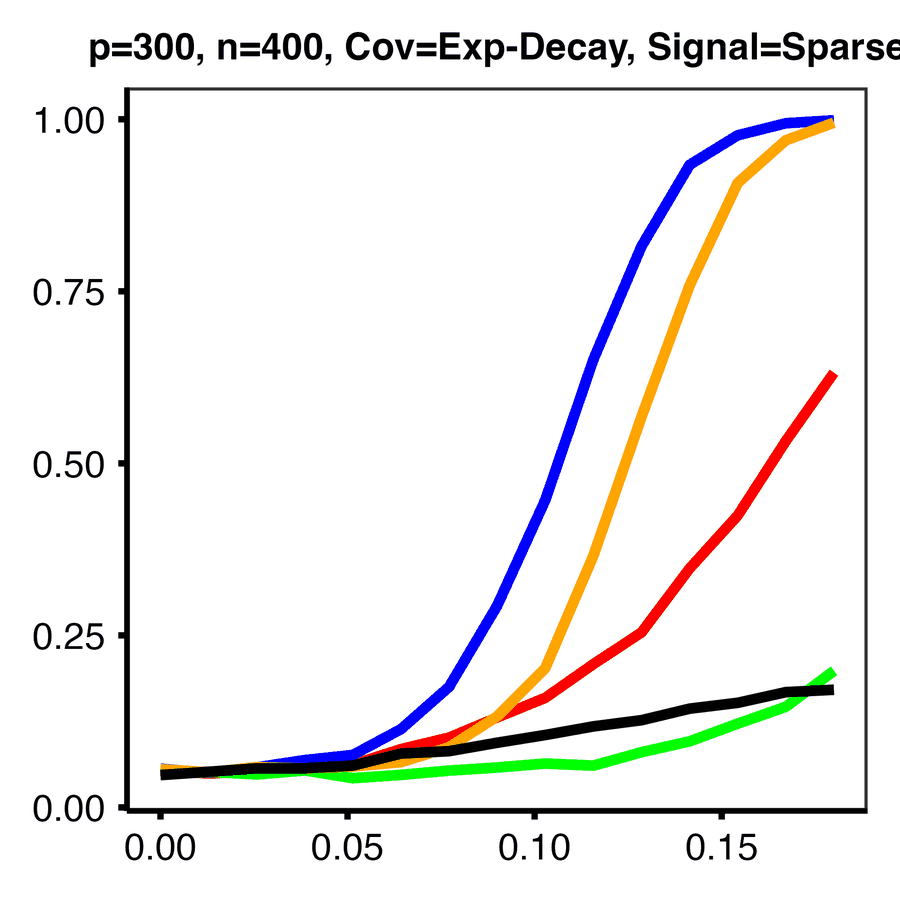} &
\includegraphics[width=0.24\textwidth]{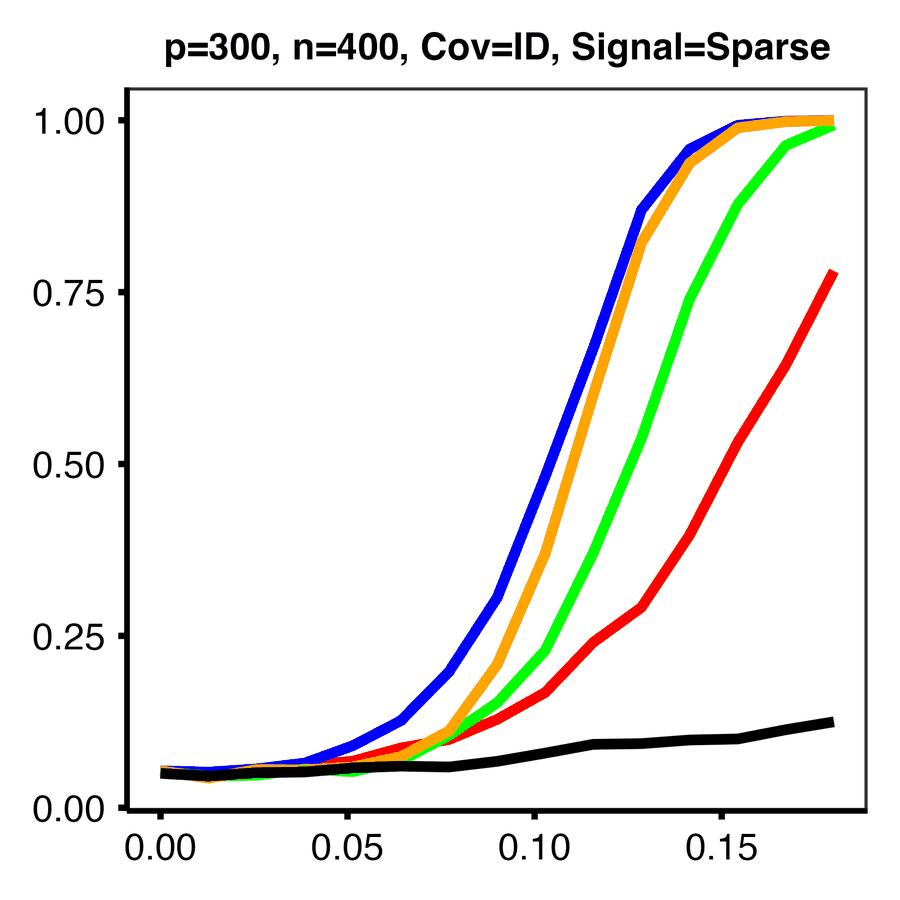} &
\includegraphics[width=0.24\textwidth]{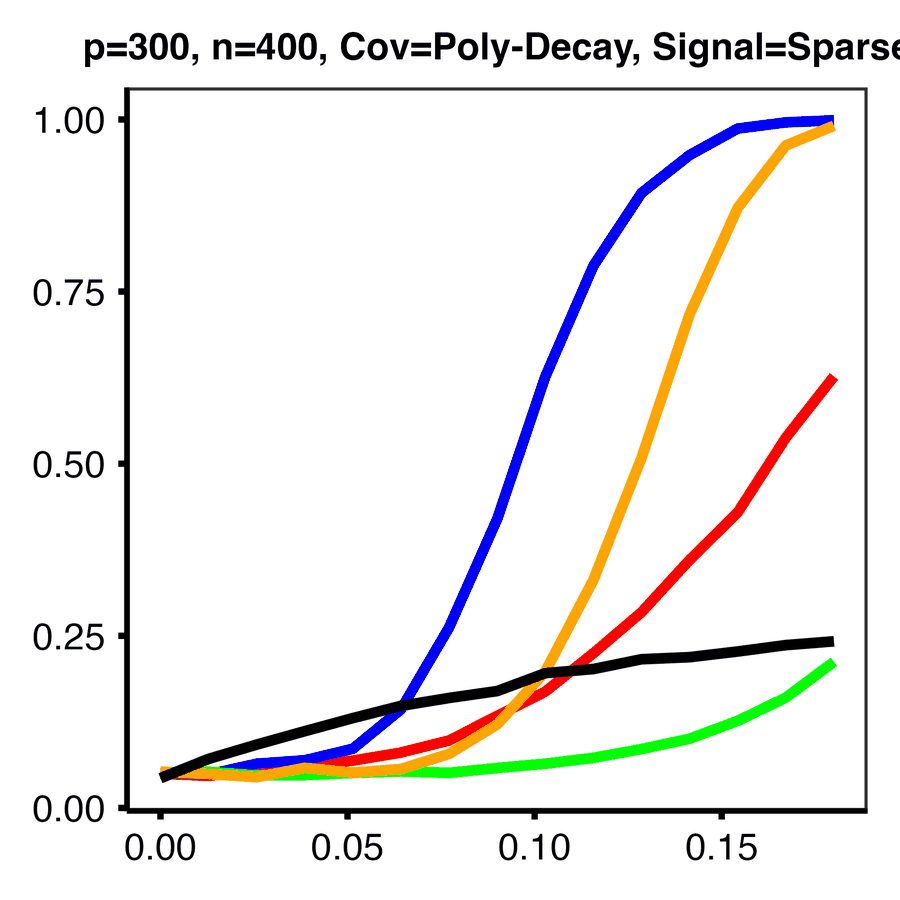} &
\includegraphics[width=0.24\textwidth]{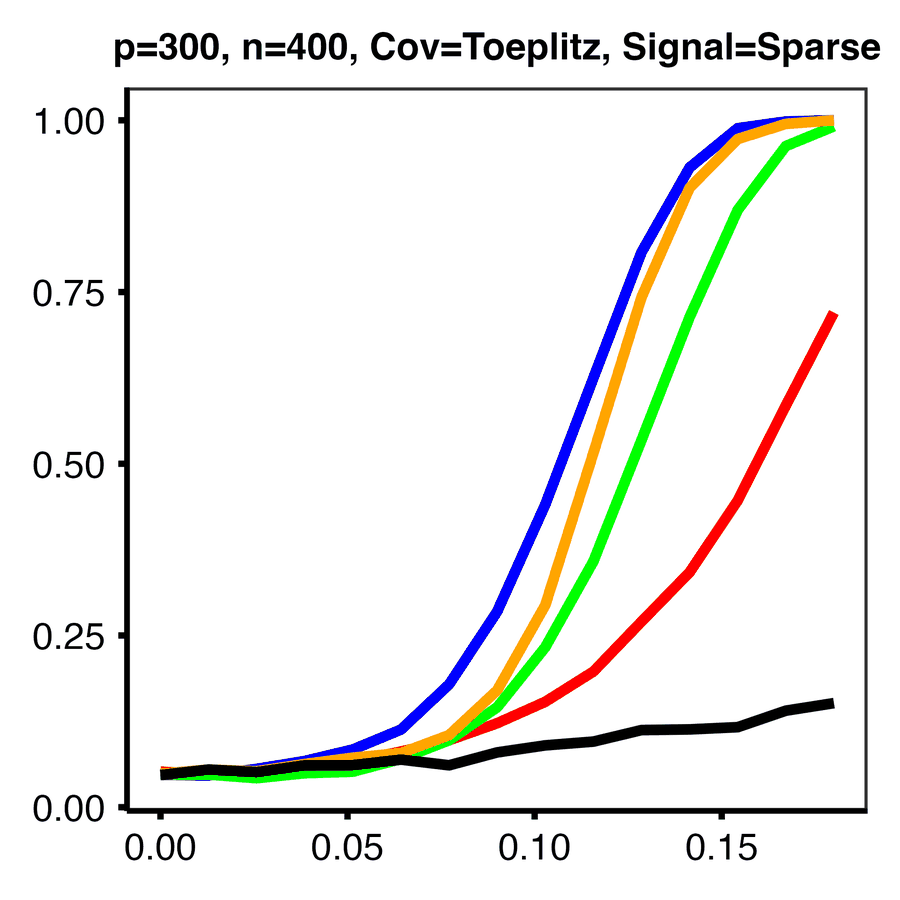}
\\
\includegraphics[width=0.24\textwidth]{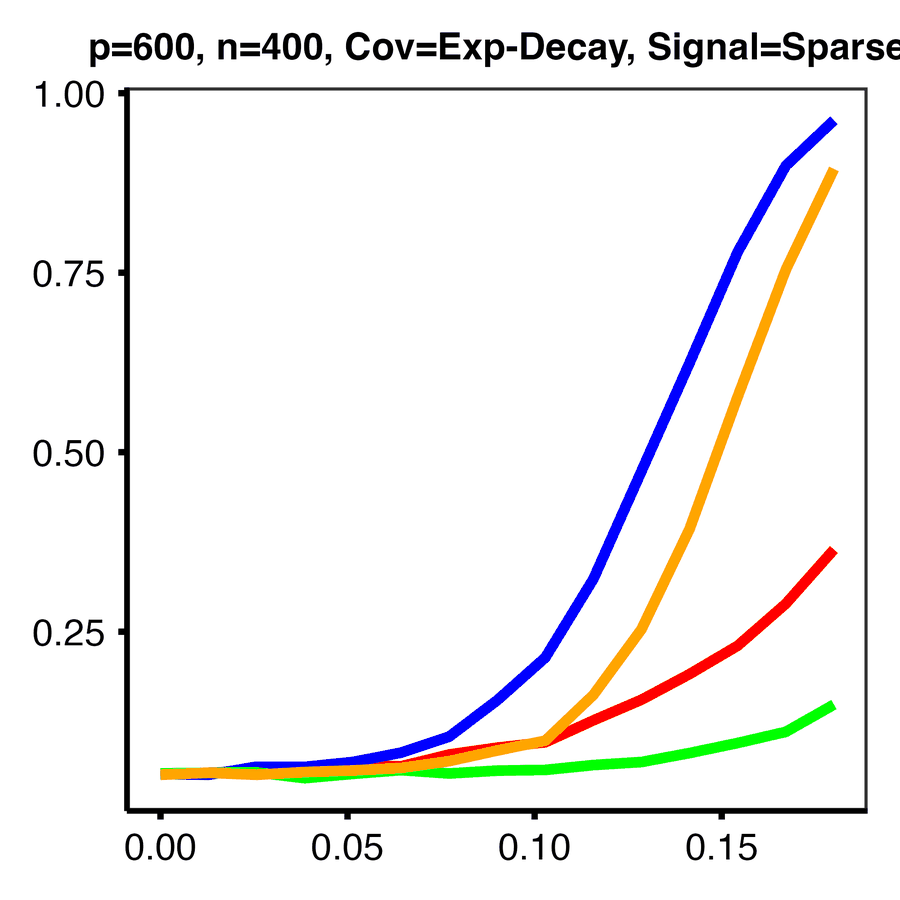} &
\includegraphics[width=0.24\textwidth]{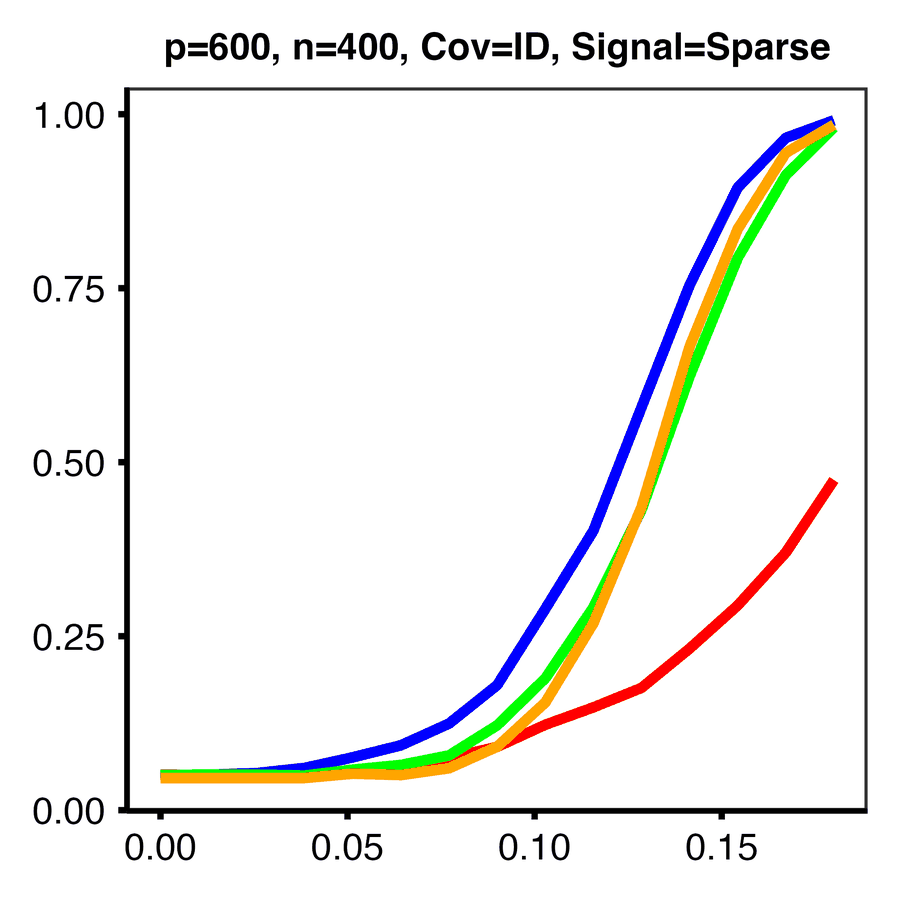} &
\includegraphics[width=0.24\textwidth]{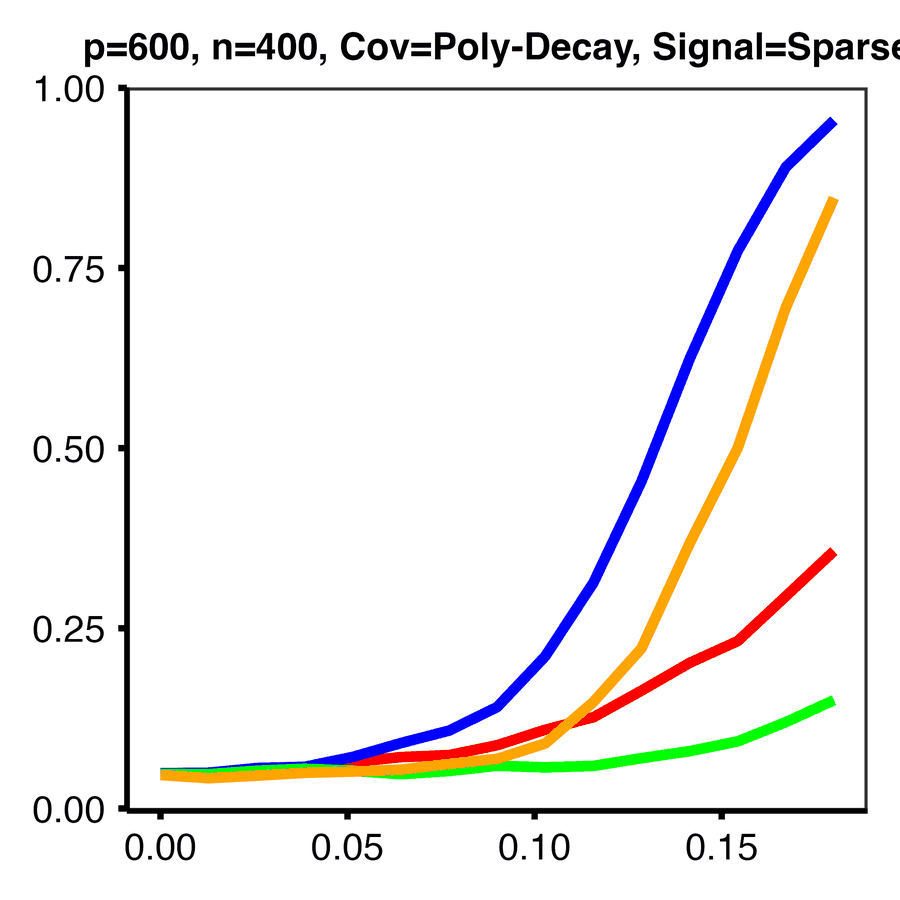} &
\includegraphics[width=0.24\textwidth]{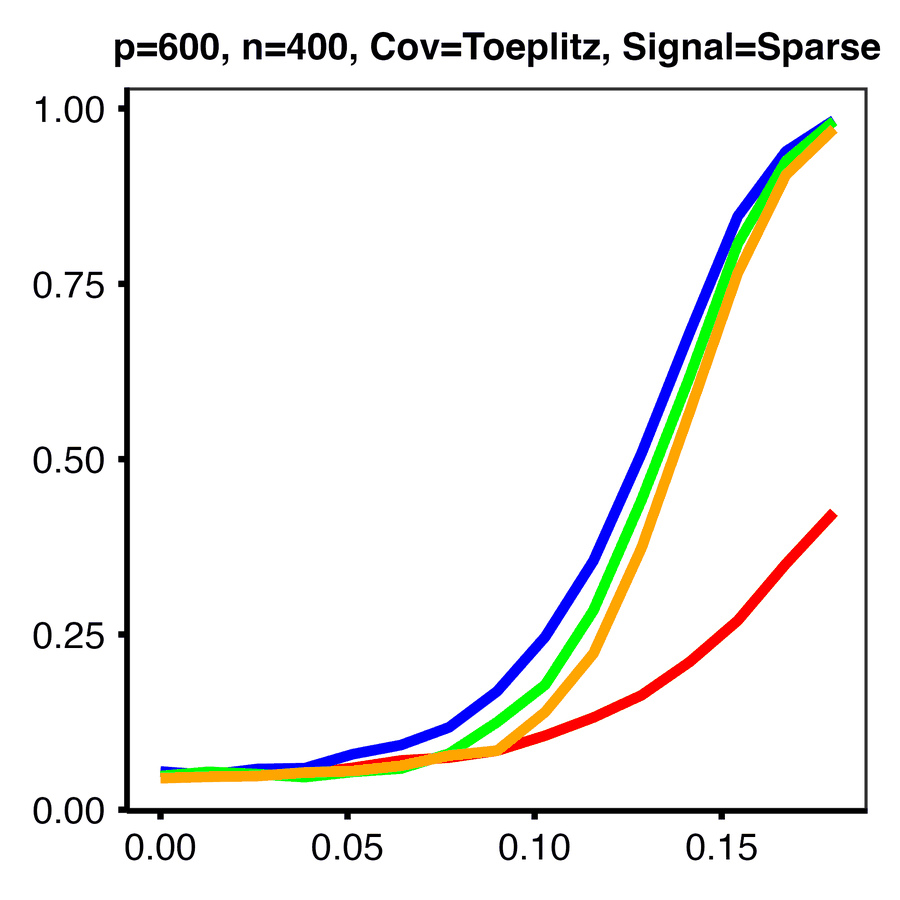}
\\
\includegraphics[width=0.24\textwidth]{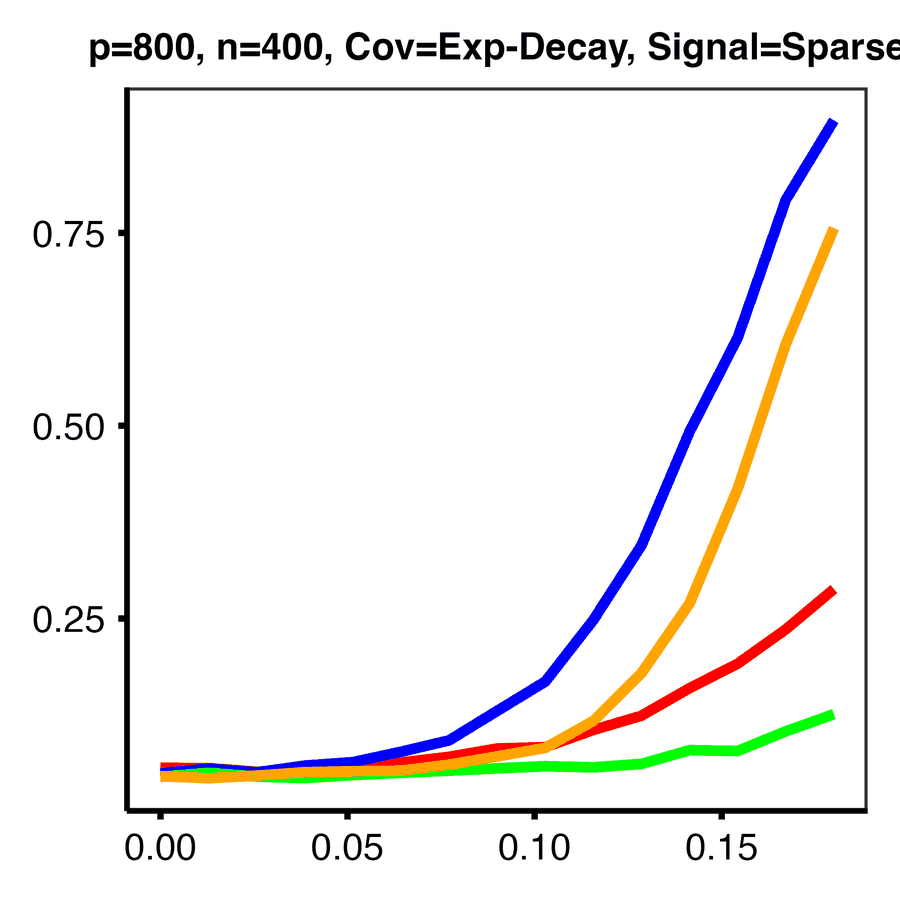} &
\includegraphics[width=0.24\textwidth]{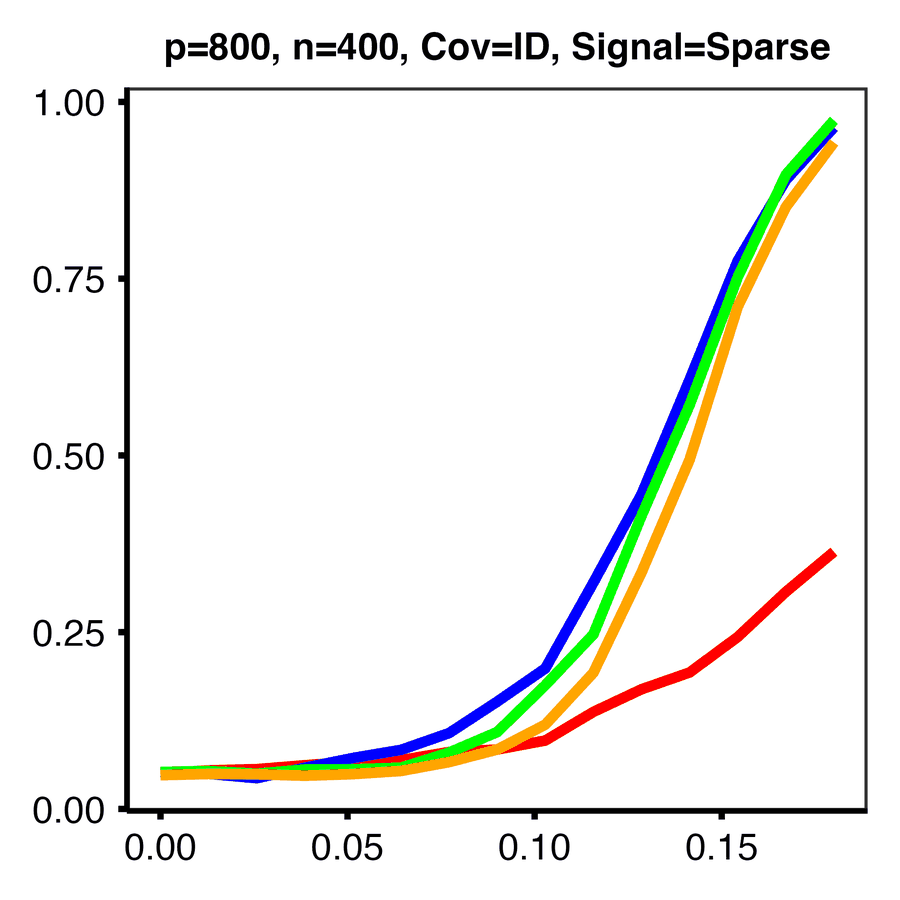} &
\includegraphics[width=0.24\textwidth]{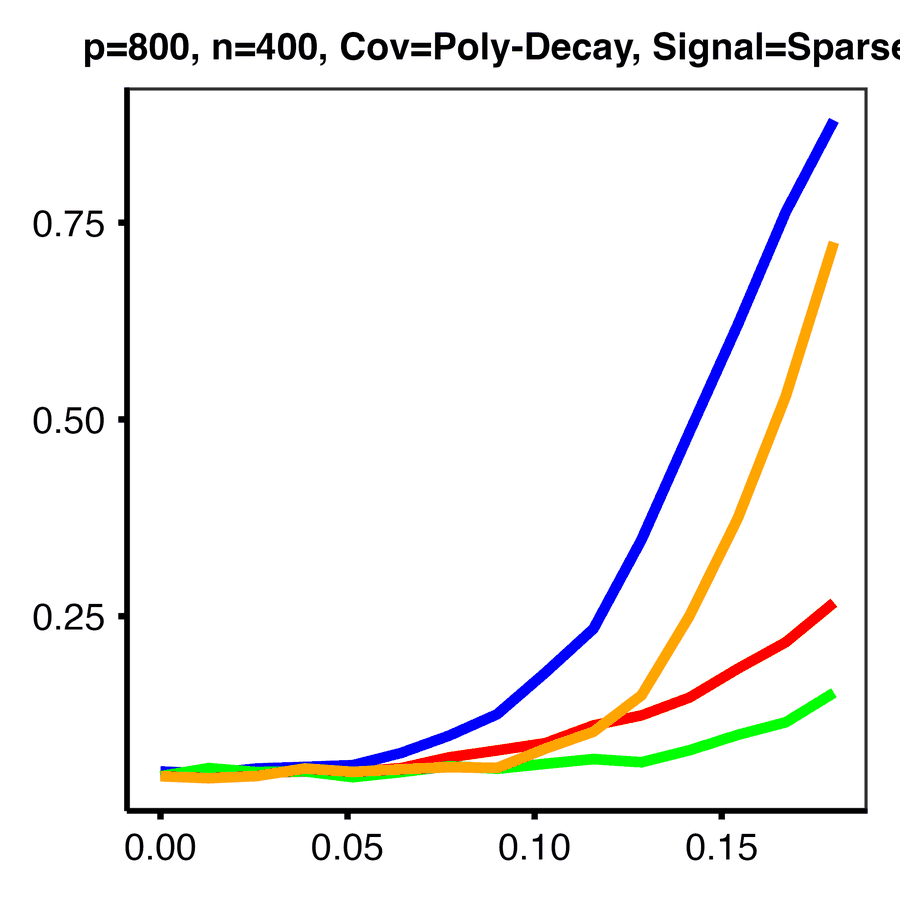} &
\includegraphics[width=0.24\textwidth]{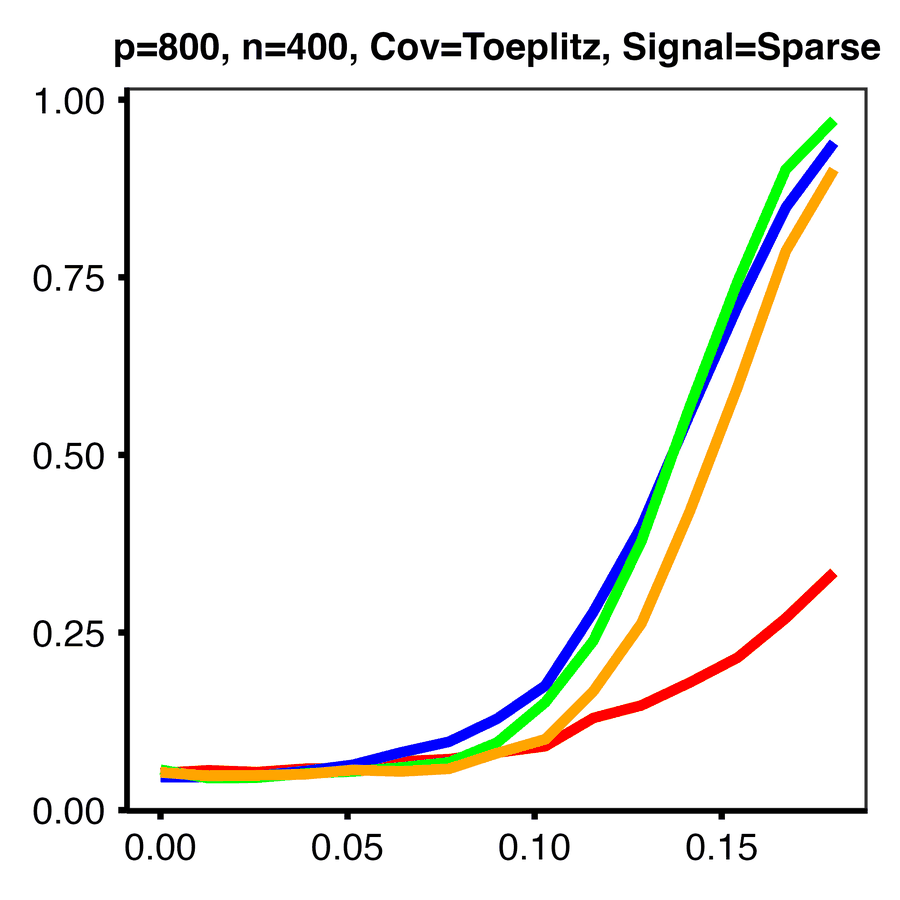}
\end{tabular}
\caption{Power curves under the MC setting with sparse mean shifts $\delta$. Rows correspond to p= 200, 300, 600, and 800. Columns correspond to the Exp Decay, Identity, Poly Decay, and Toeplitz covariance models. Blue: $T_{\rm disc}$; Red: $T_{\rm sc}$; Black (when p<n): Cov-Norm; Orange: U-Stat; Green: $\ell_\infty$-Sparse.}
\label{fig:power_MC_sparse}
\end{figure}

\clearpage

{\large The remainder of this Supplementary Material is devoted to the proofs of the technical results in the main manuscript.}

\section*{Notation List}\label{sec:notation_list}
For the benefit of readers, we collect the definitions of commonly used notation in the Supplementary Material in the following list. 

\begin{enumerate}
\itemsep2ex
\item Without further specification, we use $\calK$ to denote a generic positive constant, whose value may differ from line to line. We reserve $\calK_0$ for a universal positive constant whose value remains fixed throughout the paper. 

\item For a deterministic sequence $a_n$, we write $X_n=O_{L_1}(a_n)$ if  $\mE |X_n|  =  O(|a_n|)$. Namely, there exists a constant $\calK>0$, independent of $n$, such that $\mE|X_n|\leq \calK |a_n|$ for all sufficiently large $n$. Similarly, we write $X_n=o_{L_1}(a_n)$ if $\mE |X_n| = o(|a_n|)$. 
 Similarly, we write $X_n=O_{L_2}(a_n)$ and $X_n=o_{L_2}(a_n)$ if $\bigl(\mE|X_n|^2\bigr)^{1/2}=O(|a_n|)$ and $\bigl(\mE|X_n|^2\bigr)^{1/2}=o(|a_n|)$,  respectively.

\item $ e_i \mbox{ is the canonical vector with 1 on the }i\mbox{th coordinate}$.

\item For a  deterministic vector $a$, we denote $a_j$ to be the $j$th element of $a$. Similarly, for a collect of vectors $\{u_q\}$, we denote $u_{qj}$ to be the $j$-th entry of the $q$-th vector in the collection.

\item $\|\cdot\|_2$ is the $L_2$-norm of a vector or the operator norm of a matrix. 

\item $\|\cdot\|_\infty$ is the $L_\infty$-norm of a vector or a matrix. 

\item We write $\bZ=(\bz_1,\bz_2,\ldots,\bz_n)$. For any $j$, let $\bZ_j$ denote the matrix obtained from $\bZ$ by replacing $\bz_j$ with the zero vector. Likewise, $\bZ_{ij}$ denotes the matrix obtained from $\bZ$ by replacing both $\bz_i$ and $\bz_j$ with the zero vector.


\item $\bOmega = n^{-1} \Sigma_p^{1/2} \bZ \bZ^T \Sigma_p^{1/2}$. 

\item $ \sigma(\bz_1,\dots, \bz_j) \mbox{ is the }\sigma\mbox{-algebra generated by }\bz_1,\dots, \bz_j$. 

\item $\mE_j[\cdot]  = \mE[\cdot\mid \sigma(\bz_1,\dots,\bz_j)]$, $j = 0,\dots,n $, with the convention $ \mE_0[\cdot]= \mE[\cdot]$.

\item $\bA(\lambda)   = \Sigma_p^{1/2} (\bOmega + \lambda I)^{-1} \Sigma_p^{1/2} =\Sigma_p^{1/2} ( \frac{1}{n} \Sigma_p^{1/2} \bZ\bZ^T\Sigma_p^{1/2}+ \lambda I)^{-1} \Sigma_p^{1/2}$. 

\item $\bA_j(\lambda)  =\Sigma_p^{1/2} ( \frac{1}{n} \Sigma_p^{1/2} \bZ_j\bZ_j^T\Sigma_p^{1/2}+ \lambda I)^{-1} \Sigma_p^{1/2}$. 

\item $\bA_{ij}(\lambda) = \Sigma_p^{1/2} ( \frac{1}{n} \Sigma_p^{1/2} \bZ_{ij}\bZ_{ij}^T\Sigma_p^{1/2} + \lambda I)^{-1} \Sigma_p^{1/2}$. 

\item $\beta_j(\lambda) = \Big(1+n^{-1}\bz_j^T \bA_j(\lambda)\bz_j\Big)^{-1}$. 

\item $\beta_j^{\tr}(\lambda) = \Big(1+n^{-1} \tr\bA_j(\lambda)\Big)^{-1}$. 

\item $\beta^{\mE}(\lambda) = \Big(1+\mE n^{-1} \tr\bA_1(\lambda)\Big)^{-1}$. 

\item $\theta_j(\lambda) = \frac{1}{n}\bz_j^T \bA_j(\lambda)\bz_j - \frac{1}{n}\tr\bA_j(\lambda)$. 

\item $\varrho_{iqj} = \frac{1}{n} \bz_j^T \bA_j(\lambda_i) \bZ_ju_q v_q^T \bZ_j^T \bA_j(\lambda_i) \bz_j - \frac{1}{n} u_q^T\bZ_j^T \bA_j^{2}(\lambda_i)\bZ_jv_q$.

\item For any $t = (t_1,t_2,t_3) \in \calT_{\rm sc}(\varepsilon) ~\text{or}~ \calT_{\rm mc}(\varepsilon)$, we define  
{\small
\begin{equation*}
u(t) = \sqrt{n} \Big( \underbrace{0, \ldots, 0}_{k(t_1)-1},  \underbrace{\dfrac{-1}{k(t_2)- k(t_1)}, \ldots,\dfrac{-1}{k(t_2)- k(t_1)}}_{k(t_2) -k(t_1)},
 \underbrace{\dfrac{1}{k(t_3)- k(t_2)}, \ldots,\dfrac{1}{k(t_3)- k(t_2)}}_{k(t_3) -k(t_2)}, \underbrace{0,\ldots, 0}_{n-k(t_3)+1}
\Big)^T.
\end{equation*}
}

\item For vectors $u =(u_1,\dots, u_n)$ and $v = (v_1,\dots, v_n)$ in $\mathbb{R}^n$, we call
\[ H_{\lambda}(u,v) =   p^{-1} u^T \bZ^T \bA(\lambda) \bZ v = p^{-1} \sum_{i,j=1} u_i v_j \bz_i^T \bA(\lambda) \bz_j \] 
If $u = v$, we shall simply write $H_{\lambda}(u,u) = H_{\lambda}(u)$. 
\end{enumerate}

\section{Proof of Theorem \ref{thm:weak_convergence_process}}
\label{sec:proof_weak_convergence_process}

Under the null hypothesis $H_0$, we may assume without loss of generality that $\mu_j\equiv0$ for $j=1,\ldots,n$. We consider the un-normalized process $\{p^{-1}V_{\lambda}(t)\}$. 
Since $\calT_{\rm sc}(\varepsilon)$ and $\calT_{\rm disc}(\varepsilon)$ are subsets of $\calT_{\rm mc}(\varepsilon)$, it suffices to show the weak convergence in $\ell^\infty(\Lambda\times \calT_{\rm mc}(\varepsilon))$.

Using the notation introduced in \emph{Notation List}, we can express
\begin{align*}
p^{-1}V_\lambda(t)
&=
\frac{1}{p\|u(t)\|_2^2}
u^T(t)\bZ^T\Sigma_p^{1/2}
(S_n+\lambda I_p)^{-1}
\Sigma_p^{1/2}\bZ u(t)\\
&=
\frac{1}{p\|u(t)\|_2^2}
\sum_{i,j=1}^n
u_i(t)u_j(t)\,
\bz_i^T\Sigma_p^{1/2}
(S_n+\lambda I_p)^{-1}
\Sigma_p^{1/2}\bz_j.
\end{align*}

To analyze the asymptotic behavior of $V_\lambda(t)$, it is desirable to remove the dependence of the sample covariance matrix $S_n$ on the observations $\bz_i$ and $\bz_j$. However, this dependence is complicated by the centering term in
\[
S_n
=
\frac{1}{n}
\sum_{j=1}^n
\Sigma_p^{1/2}\bz_j\bz_j^T\Sigma_p^{1/2}
-
\Sigma_p^{1/2}\bar{\bz}\bar{\bz}^T\Sigma_p^{1/2},
\]
where the rank-one correction $\bar{\bz}\bar{\bz}^T$ couples all observations.

To overcome this difficulty, we first transform the problem into one involving the uncentered sample covariance matrix and subsequently translate the resulting asymptotic analysis back to $V_\lambda(t)$. Specifically, we define the uncentered sample covariance matrix
\[
\bOmega
=
\frac{1}{n}
\sum_{j=1}^n
\Sigma_p^{1/2}\bz_j\bz_j^T\Sigma_p^{1/2}.
\]
Then, by the Woodbury matrix identity (Lemma~\ref{lemma:woodbury}),
\[
(S_n+\lambda I_p)^{-1} = (\bOmega+\lambda I_p)^{-1} + \frac{ (\bOmega+\lambda I_p)^{-1} \Sigma_p^{1/2}\bar{\bz}\bar{\bz}^T\Sigma_p^{1/2}(\bOmega+\lambda I_p)^{-1}}{ 1-\bar{\bz}^T \Sigma^{1/2}_p(\bOmega+\lambda I_p)^{-1}\Sigma_p^{1/2}\bar{\bz}}.\]

Define
\[
h
=
\left(
\frac{1}{\sqrt{n}},
\frac{1}{\sqrt{n} },
\ldots,
\frac{1}{\sqrt{n}}
\right)^T = \frac{1}{\sqrt{n}} \mathbbm{1}_n^T.
\]
Consequently,
\begin{equation}\label{eq:relation_V_H}
p^{-1}V_\lambda(t)
=
\frac{1}{\|u(t)\|_2^2}
H_\lambda(u(t))
+
\frac{p/n}{\|u(t)\|_2^2}
\frac{
\bigl(H_\lambda(u(t),h)\bigr)^2
}{
1-(p/n)H_\lambda(h)
}.
\end{equation}
See \emph{Notation List} for the definition of $H_\lambda$.

Our analysis therefore proceeds by studying the two stochastic processes
\[
\left\{
H_\lambda(u(t)):
\lambda\in\Lambda,\;
t\in \calT_{\rm mc}(\varepsilon)
\right\},
\qquad
\left\{
H_\lambda(u(t),h):
\lambda\in\Lambda,\;
t\in\calT_{\rm mc}(\varepsilon) 
\right\},
\]
together with the collection 
\[
\left\{
H_\lambda(h):
\lambda\in\Lambda
\right\}.
\]
Indeed, we shall show that 
\[ \sqrt{p} \Big\| H_{\lambda}(u(t), h) \Big\|^2_\infty = o_p(1) \qquad \text{and} \qquad \mP\Big( \max_{\lambda\in\Lambda}\big|1-  (p/n) H_{\lambda}(h) \big|^{-1} \leq \calK_0 \Big)\to 1.\]
Here, $\|\cdot\|_\infty$-norm is taken over $t$ and $\lambda \in \Lambda$. 
 It implies that after the $\sqrt{p}$-scaling and standardization, $p^{-1}V_{\lambda}(t)$ has the same weak limit as that of $\|u(t)\|_2^{-2} H_{\lambda}(u(t))$.

\vspace{\baselineskip}

\begin{definition}
A vector $u\in\mathbb{R}^n$ is called \emph{regular} if $\sqrt{n}\,\|u\|_\infty\le\mathcal K_0$.  
\end{definition}
It is straightforward to verify that $u(t)$ is regular for every $t \in \calT_{\rm mc}(\varepsilon)$ and the vector $h$ is regular.

The key step in our proofs is the analysis of the following superposition of quadratic forms. Let
$\{u_q\}_{q=1}^m$ and $\{v_q\}_{q=1}^m$ be two finite collections of regular vectors, and $\{a_{iq}\}$ be a finite collection of coefficients. Define
\begin{equation}\label{eq:def_calH}
\calH = \calH(\{u_q\}, \{v_q\})
\coloneqq
\sum_{i=1}^M  \sum_{q=1}^m a_{iq} H_{\lambda_i}(u_q,v_q)
=
p^{-1}
\sum_{i=1}^M
\sum_{q=1}^m
a_{iq}
u_q^T
\bZ^T
\bA(\lambda_i)
\bZ
v_q.
\end{equation}

The reason to study $\calH$ is that every linear combination of finite-dimensional evaluations of the processes
$\{H_\lambda(u(t))\}$ and $\{H_\lambda(u(t),h)\}$ can be expressed in the form \eqref{eq:def_calH} for suitable choices of
$\{a_{iq}\}$, $\{u_q\}$, and $\{v_q\}$.

The remainder of this section is organized as follows. In Subsection~\ref{subsec:martingale_construct}, we express $\calH-\mE\calH$ as a sum of martingale differences, simplify the resulting decomposition, and derive moment bounds for the remainder terms. In Subsection \ref{subsec:convergence_mean}, we show the convergence of the mean $\mE \calH$.  In Subsection~\ref{subsec:variance_martingale}, we characterize the condition variance of the simplified martingale difference sequence.  In Subsection~\ref{subsec:increment_bound}, we derive an increment bound for $\calH$ under perturbations of the vectors $\{u_q\}$ and $\{v_q\}$. Subsection~\ref{subsec:asymptotic_normality_calH} establishes the asymptotic normality of $\calH$ after appropriate standardization. Subsection~\ref{subsec:weak_convergence_process} applies these results to establish the weak convergence of the processes $\{H_{\lambda}(u(t))\}$ and $\{H_{\lambda}(u(t),h)\}$. 
 
\subsection{Construction and simplification of the martingale decomposition}
\label{subsec:martingale_construct}

The goal of this subsection is to derive a simplified martingale decomposition of
$\sqrt{p}(\calH-\mE\calH)$.
Specifically, we show that
\[
\sqrt{p}\bigl(\calH-\mE\calH\bigr)
=
\sum_{j=1}^n
\sum_{i=1}^M
\sum_{q=1}^m
a_{iq}\beta^{\mE}(\lambda_i)
\Bigl[
\calH_{iqj}^{(1)}
+
\calH_{iqj}^{(2)}
+
\calH_{iqj}^{(3)}
\Bigr]
+ O_{L_2}\Big( \sum_{q=1}^m  r_{q} \Big)
\]
where the martingale terms are
\begin{align*}
   \calH_{iqj}^{(1)}  &= \mE_j [p^{-1/2} u_{qj} \bz_j^T \bA_j(\lambda_i) \bZ_j v_q], \\
   \calH_{iqj}^{(2)}  &= \mE_j [ p^{-1/2} v_{qj} \bz_j^T \bA_j(\lambda_i) \bZ_j u_q],\\
   \calH_{iqj}^{(3)}  &= - \mE_j [ p^{-1/2} \varrho_{iqj}] = - p^{-1/2}\mE_j\Big[\frac{1}{n} \bz_j^T \bA_j(\lambda_i) \bZ_ju_q v_q^T \bZ_j^T \bA_j(\lambda_i) \bz_j - \frac{1}{n} u_q^T\bZ_j^T \bA_j^{2}(\lambda_i)\bZ_jv_q\Big],
\end{align*} 
and the residual term is such that
\[ r_{q} = \max\left\{ n^{-1/4} \|u_q\|_2\|v_q\|_2,~~~  n^{1/4} \sqrt{\sum_{j=1}^n |u_{qj}|^2|v_{qj}|^2}\right\}.\]
Notably, for any fixed regular vector $u_q$ and $v_q$, $r_q = O(n^{-1/4})$.


For each $i$ and $q$, we write 
\begin{align*}
&\sqrt{p} \Big( H_{\lambda_i}(u_q, v_q) - \mE  H_{\lambda_i}(u_q, v_q)\Big) \\
& = \sqrt{p} \sum\limits_{j=1}^n \Big\{\mE_j[H_{\lambda_i}(u_q, v_q)] -\mE_{j-1}[H_{\lambda_i}(u_q,v_q)]\Big\}\\
& = \sqrt{p} \sum\limits_{j=1}^n  p^{-1}\Big\{\mE_j [  u_q^T \bZ^T \bA(\lambda_i)\bZ v_q   -  u_q^T \bZ_j^T \bA_j(\lambda_i) \bZ_jv_q ]\\
& \qquad \qquad\quad-\mE_{j-1}[u_q^T \bZ^T \bA(\lambda_i) \bZ v_q - u_q^T \bZ_j^T \bA_j(\lambda_i) \bZ_j v_q ]\Big\}\\
& = \sqrt{p} \sum\limits_{j=1}^n (\mE_j-\mE_{j-1}) [d_{iqj1} + d_{iqj2} + d_{iqj3}], 
\end{align*}
where 
\begin{align*}
d_{iqj1}&= p^{-1} u_q^T(\bZ - \bZ_j)^T\bA(\lambda_i)\bZ v_q, \\
d_{iqj2} &=p^{-1} u_q^T\bZ_j^T(\bA(\lambda_i)-\bA_j(\lambda_i))\bZ v_q,\\
d_{iqj3}&=p^{-1} u_q^T\bZ_j^T\bA_j(\lambda_i)(\bZ-\bZ_j) v_q = p^{-1} v_{qj}\bz_j^T \bA_j(\lambda_i) \bZ_j u_q. 
\end{align*}

We further decompose $d_{iqj1}$ and $d_{iqj2}$ as follows. Using the Woodbury matrix identity,
\[ \bA(\lambda) = \bA_j(\lambda)- \frac{n^{-1} \bA_j (\lambda )\bz_j \bz_j^T \bA_j(\lambda)}{1+ n^{-1} \bz_j^T \bA_j(\lambda) \bz_j} = \bA_j(\lambda) -\frac{1}{n} \beta_j(\lambda) \bA_j(\lambda) \bz_j \bz_j^T \bA_j(\lambda).\]
Here, we recall the definition of $\beta_j (\lambda)= (1 + n^{-1} \bz_j^T \bA_j(\lambda) \bz_j)^{-1}$ in the \emph{Notation List}.

Then,
\begin{align*}
    d_{iqj1} &= \frac{1}{p} u_{qj} \bz_j^T \bA_j(\lambda_i) \bZ_j v_q + \frac{1}{p} u_{qj} v_{qj} \bz_j^T \bA_j(\lambda_i) \bz_j \\
    &\quad - \frac{1}{pn} u_{qj} \Big(\bz_j^T \bA_j(\lambda_i) \bz_j\Big) \Big(\bz_j^T \bA_j(\lambda_i) \bZ_j v_{q}\Big) \beta_j(\lambda_i) - \frac{1}{pn} u_{qj} v_{qj} \Big(\bz_j^T \bA_j(\lambda_i)\bz_j\Big)^2 \beta_j(\lambda_i).\\
    & = d_{iqj1}^{(1)} + d_{iqj1}^{(2)} + d_{iqj1}^{(3)} + d_{iqj1}^{(4)}, \quad \text{say}.\\
    d_{iqj2} & = - \frac{1}{pn} v_{qj} \Big(\bz_j^T \bA_j(\lambda_i) \bz_j\Big) \Big(\bz_j^T \bA_j(\lambda_i) \bZ_j u_{q}\Big) \beta_j(\lambda_i) \\
            &\quad - \frac{1}{pn} \Big(\bz_j^T \bA_j(\lambda_i) \bZ_j v_q\Big) \Big(\bz_j^T \bA_j(\lambda_i)\bZ_j u_q\Big) \beta_j(\lambda_i)\\
            & = \frac{1}{p}v_{qj} [\beta_j(\lambda_i)-1] \bz_j^T \bA_j(\lambda_i) \bZ_j u_q - \frac{1}{p} \varrho_{iqj}\beta_j(\lambda_i) - \frac{1}{pn} u_q^T \bZ_j^T \bA_j^2(\lambda_i) \bZ_jv_q \beta_j(\lambda_i)\\
            & = d_{iqj2}^{(1)} + d_{iqj2}^{(2)} + d_{iqj2}^{(3)}, \quad \text{say}. 
\end{align*}

We claim that 
\begin{align}
   p^{1/2} \sum\limits_{j=1}^n &(\mE_j - \mE_{j-1}) d_{iqj1} = \beta^{\mE}(\lambda_i) \sum_{j=1}^n \calH_{iqj}^{(1)} + O_{L_2}(r_q). \label{eq:simplification_diqj1} \\
   p^{1/2} \sum\limits_{j=1}^n &(\mE_j - \mE_{j-1}) d_{iqj2} = (\beta^{\mE}(\lambda_i) -1) \sum_{j=1}^n \calH_{iqj}^{(2)} + \beta^{\mE}(\lambda_i) \sum_{j=1}^n \calH_{iqj}^{(3)} + O_{L_2}(r_q).\label{eq:simplification_diqj2}
\end{align}
Combining with the form of $d_{iqj3}$, the claimed martingale decomposition of $\sqrt{p}(\calH- \mE\calH)$ will follow. 


We first show the proof of \eqref{eq:simplification_diqj1} by checking the four terms in the decomposition of $d_{iqj1}$ separately. As for $d_{iqj1}^{(1)}$,
\begin{align*}
     &p^{1/2} \sum_{j=1}^n (\mE_j - \mE_{j-1}) d_{iqj1}^{(1)} =  p^{1/2} \sum_{j=1}^n (\mE_j - \mE_{j-1}) \frac{1}{p} u_{qj} \bz_j^T \bA_j(\lambda_i) \bZ_j v_q \\
     &= p^{1/2} \sum_{j=1}^n \mE_j \Big[\frac{1}{p} u_{qj} \bz_j^T \bA_j(\lambda_i) \bZ_j v_q\Big]= \sum_{j=1}^n \calH_{iqj}^{(1)}.
\end{align*}

As for $d_{iqj1}^{(2)}$, due to Lemma \ref{lemma:Burkholder_martingale_moments}, 
\begin{align*}
   & \mE \Big|p^{1/2} \sum_{j=1}^n (\mE_j -\mE_{j-1}) d_{iqj1}^{(2)}\Big|^2 \\
  & = \mE \Big|p^{1/2} \sum_{j=1}^n (\mE_j -\mE_{j-1})  \frac{1}{p} u_{qj} v_{qj} \bz_j^T \bA_j(\lambda_i) \bz_j \Big|^2\\
  &\leq \calK p^{-1} \sum_{j=1}^n |u_{qj} v_{qj}|^2  \mE \Big| (\mE_j -\mE_{j-1}) \bz_j^T \bA_j(\lambda_i) \bz_j  \Big|^2\\
  &  \leq\calK p^{-1} \sum_{j=1}^n |u_{qj}|^2|v_{qj}|^2 \mE\Big|\bz_j^T (\mE_j\bA_{j}(\lambda_i))\bz_j - \tr(\mE_j\bA_j(\lambda_i))  \Big|^2 \\
  & \leq \calK p^{-1} \sum_{j=1}^n |u_{qj}|^2|v_{qj}|^2 p \mE \Big\|\mE_{j} \bA_j(\lambda)\Big\|^2_2 \\
  & \leq \calK \sum_{j=1}^n |u_{qj}|^2 |v_{qj}|^2.
\end{align*}
Here, we are using the fact that 
\[\mE\Big|\bz_j^T (\mE_j\bA_{j}(\lambda_i))\bz_j - \tr(\mE_j\bA_j(\lambda_i))  \Big|^2 \leq \calK  \mE \tr \Big[\Big(\mE_{j} \bA_j(\lambda_i)\Big)^2\Big] \leq \calK p \mE \Big\|\mE_{j} \bA_j(\lambda_i)\Big\|_2^2  \leq \calK p \frac{\|\Sigma_p\|_2^2}{\lambda_i^2}, \]
obtained by applying Lemma \ref{lemma:Burkholder} and Lemma \ref{lemma:bound_beta}.

Next, as for $d_{iqj1}^{(3)}$, we aim to replace $\beta_j(\lambda)$ and $n^{-1} \bz_j^T \bA_j(\lambda_i)\bz_j$  with $\beta_j^{\tr}(\lambda_i)$ and $n^{-1}\tr(\bA_j(\lambda_i))$ separately and show the induced difference is negligible. Recall the definition of $\beta_j^{\tr}(\lambda)$ and $\theta_j(\lambda)$ in the \emph{Notation List}. We have 
\[ \beta_j(\lambda) - \beta_j^{\tr}(\lambda) = - \beta_j(\lambda) \beta_j^{\tr}(\lambda)\theta_j(\lambda). \]
\begin{align*}
    &\frac{1}{pn} u_{qj} \Big(\bz_j^T \bA_j(\lambda_i) \bz_j\Big) \Big(\bz_j^T \bA_j(\lambda_i) \bZ_j v_{q}\Big) \beta_j(\lambda_i) \\
    &=  \frac{1}{pn} u_{qj} \Big(\bz_j^T \bA_j(\lambda_i) \bz_j\Big) \Big(\bz_j^T \bA_j(\lambda_i) \bZ_j v_{q}\Big) \beta^{\tr}_j(\lambda_i) \\
    &\quad -\frac{1}{pn} u_{qj} \Big(\bz_j^T \bA_j(\lambda_i) \bz_j\Big) \Big(\bz_j^T \bA_j(\lambda_i) \bZ_j v_q \Big) \beta_j(\lambda_i) \beta_j^{\tr}(\lambda_i) \theta_j(\lambda_i) \\ 
    & = \frac{1}{p} u_{qj} \theta_j(\lambda_i) \Big(\bz_j^T \bA_j(\lambda_i) \bZ_j v_{q}\Big) \beta^{\tr}_j(\lambda_i) + \frac{1}{pn} u_{qj} \tr(\bA_j(\lambda_i)) \Big(\bz_j^T \bA_j(\lambda_i) \bZ_j v_{q}\Big) \beta^{\tr}_j(\lambda_i)\\
    & \quad-\frac{1}{pn} u_{qj} \Big(\bz_j^T \bA_j(\lambda_i) \bz_j\Big) \Big(\bz_j^T \bA_j(\lambda_i) \bZ_j v_q \Big) \beta_j(\lambda_i) \beta_j^{\tr}(\lambda_i) \theta_j(\lambda_i)\\
    & = \frac{1}{p} u_{qj} \theta_j(\lambda_i) \Big(\bz_j^T \bA_j(\lambda_i) \bZ_j v_{q}\Big) \beta^{\tr}_j(\lambda_i) + \frac{1}{p} u_{qj} \Big(\bz_j^T \bA_j(\lambda_i)\bZ_j v_{q} \Big)(1-\beta_j^{\tr}(\lambda_i))\\
    &\quad-\frac{1}{pn} u_{qj} \Big(\bz_j^T \bA_j(\lambda_i) \bz_j\Big) \Big(\bz_j^T \bA_j(\lambda_i) \bZ_j v_q \Big) \beta_j(\lambda_i) \beta_j^{\tr}(\lambda_i) \theta_j(\lambda_i).
\end{align*}

By Lemma \ref{lemma:Burkholder_martingale_moments}, Lemma \ref{lemma:bound_beta}, and Lemma \ref{lemma:comprehensive}, we conclude that the first term and the third term are such that

\begin{align*}
 \mE \Big| p^{1/2} \sum_{j=1}^n (\mE_j -\mE_{j-1}) \frac{1}{p} u_{qj} \theta_j(\lambda_i) \Big(\bz_j^T \bA_j(\lambda_i) \bZ_j v_{q}\Big) \beta^{\tr}_j(\lambda_i) \Big|^2\\
 \leq \calK  \sum_{j=1}^n |u_{qj}|^2 \mE \Big| \theta_j(\lambda_i) \Big(\frac{1}{\sqrt{p}} \bz_j^T \bA_j(\lambda_i) \bZ_j v_q \Big) \Big|^2 \leq \calK n^{-1/2}\|u_q\|_2^2 \|v_q\|_2^2.  
\end{align*}
\begin{align*}
 \mE \Big|p^{1/2}\sum_{j=1}^n (\mE_j -\mE_{j-1}) & \frac{1}{\sqrt{p}} u_{qj} \Big(\frac{1}{n} \bz_j^T \bA_j(\lambda_i) \bz_j\Big) \Big(\frac{1}{\sqrt{p}} \bz_j^T \bA_j(\lambda_i) \bZ_j v_q \Big) \beta_j(\lambda_i) \beta_j^{\tr}(\lambda_i) \theta_j(\lambda_i)\Big|^2\\
 \leq& \calK n^{-1/2}\|u_q\|_2^2 \|v_q\|_2^2. 
\end{align*}
Here, we are using the fact that 
\[\theta_j(\lambda_i)\Big(\frac{1}{n} \bz_j^T \bA_j(\lambda_i) \bz_j\Big) \Big(\frac{1}{\sqrt{p}} \bz_j^T \bA_j(\lambda_i) \bZ_j v_q \Big)\quad \text{and}\quad \theta_j(\lambda_i) \Big(\frac{1}{\sqrt{p}} \bz_j^T \bA_j(\lambda_i) \bZ_j v_q \Big)\]  are both of the form considered in Lemma \ref{lemma:comprehensive}. 

It follows then 
\begin{align*}
&p^{1/2} \sum_{j=1}^n (\mE_j -\mE_{j-1}) d_{iqj1}^{(3)} = p^{1/2} \sum_{j=1}^n \mE_j \frac{1}{p}u_{qj} \Big(\bz_j^T \bA_j(\lambda_i) \bZ_j v_q\Big)(\beta^{\tr}_j(\lambda_i)-1)\\
& \qquad + O_{L_2}(n^{-1/4} \|u_q\|_2 \|v_q\|_2) \\
& =  - \sum_{j=1}^{n} \calH_{iqj}^{(1)} + p^{-1/2} \sum_{j=1}^n \mE_j u_{qj} \Big(\bz_j^T \bA_j(\lambda_i) \bZ_j v_q\Big)\beta^{\tr}_j(\lambda_i) + O_{L_2}(r_q) 
\end{align*}

As for $d_{iqj1}^{(4)}$,
\begin{align*}
   &-  p^{1/2} \sum_{j=1}^n (\mE_j -\mE_{j-1}) d_{iqj1}^{(4)} \\
   & = p^{1/2} \sum_{j=1}^n (\mE_j -\mE_{j-1}) \frac{1}{pn} u_{qj}v_{qj} \Big(\bz_j^T \bA_j(\lambda_i) \bz_j\Big)^2 \beta_j(\lambda_i)\\
   & = p^{1/2} \sum_{j=1}^n (\mE_j -\mE_{j-1}) \frac{1}{pn} u_{qj}v_{qj} \Big(\bz_j^T \bA_j(\lambda_i) \bz_j\Big)^2 \beta^{\tr}_j(\lambda_i)\\
   &\quad -  p^{1/2} \sum_{j=1}^n (\mE_j -\mE_{j-1}) \frac{1}{pn} u_{qj}v_{qj} \Big(\bz_j^T \bA_j(\lambda_i) \bz_j\Big)^2 \beta^{\tr}_j(\lambda_i)\beta_j(\lambda_i)\theta_j(\lambda_i)\\
   & = p^{1/2} \sum_{j=1}^n (\mE_j - \mE_{j-1}) \frac{n}{p} u_{qj}v_{qj} \theta^2_j(\lambda_i) \beta_j^{\tr}(\lambda_i)\\
   &\qquad + p^{1/2} \sum_{j=1}^n (\mE_j -\mE_{j-1})2 \frac{n}{p} u_{qj}v_{qj} \theta_j(\lambda_i) \frac{1}{n}\tr(\bA_j(\lambda_i))  \beta_j^{\tr}(\lambda_i)\\
   &\qquad -  p^{1/2} \sum_{j=1}^n (\mE_j -\mE_{j-1}) \frac{n}{p} u_{qj}v_{qj} \theta_j(\lambda_i)  \Big(n^{-1}\bz_j^T \bA_j(\lambda_i) \bz_j\Big)^2 \beta^{\tr}_j(\lambda_i)\beta_j(\lambda_i)
\end{align*}

By Lemma \ref{lemma:concentration_quadrat} and Lemma \ref{lemma:comprehensive}, the terms on the right-hand side are such that 
\begin{align*}
    &\mE \Big|p^{1/2} \sum_{j=1}^n (\mE_j - \mE_{j-1}) u_{qj} v_{qj} \theta_j^2(\lambda_i) \beta_j^{\tr}(\lambda_i)  \Big|^2 \leq \calK p \sum_{j=1}^n |u_{qj}|^2 |v_{qj}|^2 \mE |\theta_j(\lambda_i)|^4\\
    & \qquad \leq \calK \sum_{j=1}^n |u_{qj}|^2 |v_{qj}|^2 n^{-1}.  \\ 
    & \mE \Big| p^{1/2} \sum_{j=1}^n (\mE_j -\mE_{j-1})2 \frac{n}{p} u_{qj}v_{qj} \theta_j(\lambda_i) \frac{1}{n}\tr(\bA_j(\lambda_i))  \beta_j^{\tr}(\lambda_i)\Big|^2 \\
    &\qquad \leq\calK p \sum_{j=1}^n |u_{qj}|^2 |v_{qj}|^2 \mE |\theta_j(\lambda_i)|^2  \leq \calK \sum_{j=1}^n |u_{qj}|^2|v_{qj}|^2.\\ 
    & \mE \Big|  p^{1/2} \sum_{j=1}^n (\mE_j -\mE_{j-1}) \frac{n}{p} u_{qj}v_{qj} \theta_j(\lambda_i)  \Big(n^{-1}\bz_j^T \bA_j(\lambda_i) \bz_j\Big)^2 \beta^{\tr}_j(\lambda_i)\beta_j(\lambda_i) \Big|^2\\
    &\qquad \leq \calK p \sum_{j=1}^n |u_{qj}|^2 |v_{qj}|^2 \mE \Big[ |\theta_j(\lambda_i)|^2 \Big|n^{-1} \bz_j^T \bA_j(\lambda_i)\bz_j\Big|^4  \Big]\\
    &\qquad \leq \calK \sqrt{n} \sum_{j=1}^n |u_{qj}|^2|v_{qj}|^2. 
\end{align*}
We conclude that
\[\sqrt{p} \sum_{j=1}^n (\mE_j -\mE_{j-1}) d_{iqj1}^{(4)} = O_{L_2}(r_q). \]

Combining the analysis of $d_{iqj1}^{(1)}$,  $d_{iqj1}^{(2)}$,  $d_{iqj1}^{(3)}$, and  $d_{iqj1}^{(4)}$, up to now, we showed that 
\[p^{1/2} \sum\limits_{j=1}^n (\mE_j - \mE_{j-1}) d_{iqj1} = p^{-1/2} \sum_{j=1}^n u_{qj}  \mE_j \bz_j^T \bA_j(\lambda_i) \bZ_j v_{q}  \beta^{\tr}_j(\lambda_i) + O_{L_2}(r_q).\]
To show \eqref{eq:simplification_diqj1}, we only need to replace $\beta^{\tr}_j(\lambda_i)$ with $\beta^{\mE}(\lambda_i)$ and show that the induced difference is $O_{L_2}(r_q)$. To this end, observe that 
\[ \beta^{\tr}_j(\lambda_i) - \beta^{\mE}(\lambda_i) = \beta_j^{\tr}(\lambda_i) \beta^{\mE}(\lambda_i) (\frac{1}{n} \mE \tr(\bA_j(\lambda_i)) - \frac{1}{n}\tr(\bA_j(\lambda_i))).  \]
So, applying Lemma \ref{lemma:converg_A} and Lemma \ref{lemma:bound_z1_zalpha_b}, we conclude that the difference induced by replacing $\beta^{\tr}_j(\lambda_i)$ is such that 
\begin{align*} 
&\mE \Big| p^{-1/2} \sum_{j=1}^n u_{qj} \mE_j \bz_j^T \bA_j(\lambda_i) \bZ_j v_q \beta^{\tr}_j(\lambda_i)\beta^{\mE}(\lambda_i)(\frac{1}{n} \mE \tr(\bA_j(\lambda_i)) - \frac{1}{n}\tr(\bA_j(\lambda_i)))  \Big|^2 \\
&\leq \calK \sum_{j=1}^n |u_{qj}|^2 \mE\left[\Big|n^{-1/2} \bz_j^T \bA_j(\lambda_i)\bZ_j v_q \Big|^2 \Big|\frac{1}{n}\tr(\bA_j(\lambda_i)) - \frac{1}{n} \mE \tr(\bA_j(\lambda_i)) \Big|^2\right]\\
&\leq \calK n^{-1} \|u_q\|_2^2 \|v_q\|_2^2.
\end{align*}
The claim in \eqref{eq:simplification_diqj1} follows. 

Secondly, we show the proof of \eqref{eq:simplification_diqj2} by separately checking $d_{iqj2}^{(1)}$, $d_{iqj2}^{(2)}$, and $d_{iqj2}^{(3)}$. 

As for $d_{iqj2}^{(1)}$, 
\begin{align*}
&p^{1/2} \sum_{j=1}^n (\mE_{j} - \mE_{j-1}) \frac{1}{p} v_{qj} \beta_j(\lambda_i) \bz_j^T \bA_j(\lambda_i)\bZ_j u_q\\
&= p^{1/2} \sum_{j=1}^n (\mE_j - \mE_{j-1})  \frac{1}{p} v_{qj} \beta^{\mE}(\lambda_i) \bz_j^T \bA_j(\lambda_i)\bZ_j u_q \\
&\quad - p^{1/2} \sum_{j=1}^n (\mE_j - \mE_{j-1})  \frac{1}{p} v_{qj} \beta_j(\lambda_i) \beta^{\mE}(\lambda_i) \theta_j(\lambda_i) \bz_j^T \bA_j(\lambda_i)\bZ_j u_q\\
& \quad - p^{1/2} \sum_{j=1}^n (\mE_j -\mE_{j-1}) \frac{1}{p} v_{qj} \beta_j(\lambda_i) \beta^{\mE}(\lambda_i) \Big(\frac{1}{n}\tr[\bA_j(\lambda_i)] - \frac{1}{n}\mE \tr[\bA_j(\lambda_i)] \Big) \bz_j^T \bA_j(\lambda_i) \bZ_j u_q.
\end{align*}
Using Lemma \ref{lemma:converg_A}, Lemma \ref{lemma:bound_z1_zalpha_b}, and Lemma \ref{lemma:comprehensive}, the second and third terms are such that 
\begin{align*}
    &\mE \Big| p^{1/2} \sum_{j=1}^n (\mE_j - \mE_{j-1})  \frac{1}{p} v_{qj} \beta_j(\lambda_i) \beta^{\mE}(\lambda_i) \theta_j(\lambda_i) \bz_j^T \bA_j(\lambda_i)\bZ_j u_q \Big|^2\\
    &\leq \calK \sum_{j=1}^n v_{qj}^2 \mE \Big| \theta_j(\lambda_i) \Big(\frac{1}{\sqrt{n}} \bz_j^T \bA_j(\lambda_i) \bZ_j u_q \Big)  \Big|^2 \leq \calK n^{-1/2} \sum_{j=1}^n |v_{qj}|^2 \|u_q\|_2^2 \leq \calK n^{-1/2} \|u_q\|_2^2 \|v_q\|_2^2. \\
    & \mE \Big|  p^{1/2} \sum_{j=1}^n (\mE_j -\mE_{j-1}) \frac{1}{p} v_{qj} \beta_j(\lambda_i) \beta^{\mE}(\lambda_i) \Big(\frac{1}{n}\tr[\bA_j(\lambda_i)] - \frac{1}{n}\mE \tr[\bA_j(\lambda_i)] \Big) \bz_j^T \bA_j(\lambda_i) \bZ_j u_q\Big|^2 \\
    & \leq \calK \sum_{j=1}^n  |v_{qj}|^2 \mE \Big| \Big( \frac{1}{n} \tr[\bA_j(\lambda_i)] - \frac{1}{n}\mE \tr[\bA_j(\lambda_i)] \Big) \Big(\frac{1}{\sqrt{n}} \bz_j^T \bA_j(\lambda_i)\bZ_j u_q \Big)  \Big|^2  \\
    &\leq \calK n^{-1/2} \|u_q\|_2^2 \|v_{q}\|_2^2.
\end{align*}
It follows then
\begin{align*}
&\sqrt{p} \sum_{j=1}^n (\mE_j -\mE_{j-1}) d_{iqj2}^{(1)} = \sqrt{p} \sum_{j=1}^n (\mE_j - \mE_{j-1}) \frac{1}{p} v_{qj}[\beta^{\mE}(\lambda_i) -1] \bz_j^T \bA_j(\lambda_i)\bZ_j u_q  + O_{L_2}(r_q)\\
&= [\beta^{\mE}(\lambda_i) -1]  \sum_{j=1}^n  \calH_{iqj}^{(2)} +O_{L_2}(r_q). 
\end{align*}

As for $d_{iqj2}^{(2)}$, using analogous arguments, 
\begin{align*}
    &\sqrt{p} \sum_{j=1}^n (\mE_j-\mE_{j-1}) d_{iqj2}^{(2)} \\
    &=- \sqrt{p} \sum_{j=1}^n (\mE_j - \mE_{j-1}) \frac{1}{p}\varrho_{iqj}\beta_j(\lambda_i)\\
    &= -\sqrt{p} \sum_{j=1}^n (\mE_j - \mE_{j-1}) \frac{1}{p} \varrho_{iqj} \beta^{\mE}(\lambda_i) - \sqrt{p} \sum_{j=1}^n (\mE_j-\mE_{j-1}) p^{-1} \varrho_{iqj} [\beta_j(\lambda_i) - \beta^{\mE}(\lambda_i) ]\\
    & = -\beta^{\mE}(\lambda_i) \sum_{j=1}^n (\mE_j - \mE_{j-1}) p^{-1/2} \varrho_{iqj} + O_{L_2}(r_q)\\
    &= \beta^{\mE}(\lambda_i) \sum_{j=1}^n \calH_{iqj}^{(3)} + O_{L_2}(r_q). 
\end{align*}

As for $d_{iqj2}^{(3)}$, 
\begin{align*}
&p^{1/2} \sum_{j=1}^n (\mE_{j} -\mE_{j-1}) \frac{1}{pn} u_q^T \bZ_j^T \bA_j^2(\lambda_i) \bZ_j v_q \beta_j(\lambda_i) \\
& = p^{1/2} \sum_{j=1}^n (\mE_{j} -\mE_{j-1}) \frac{1}{pn} u_q^T \bZ_j^T \bA_j^2(\lambda_i) \bZ_j v_q \beta_j^{\tr}(\lambda_i)\\
&\qquad - p^{1/2}\sum_{j=1}^n (\mE_j -\mE_{j-1}) \frac{1}{pn} u_q^T \bZ_j^T \bA_j^2(\lambda_i) \bZ_j v_q \beta_j(\lambda_i)\beta^{\tr}_j(\lambda_i) \theta_j(\lambda_i)\\
& = p^{1/2} \sum_{j=1}^n (\mE_j -\mE_{j-1})  \frac{1}{pn} u_q^T \bZ_j^T \bA_j^2(\lambda_i) \bZ_j v_q \beta_j(\lambda_i)\beta^{\tr}_j(\lambda_i) \theta_j(\lambda_i).
\end{align*}
Here, we are using the fact that $(\mE_{j} -\mE_{j-1}) \frac{1}{pn} u_q^T \bZ_j^T \bA_j^2(\lambda_i) \bZ_j v_q \beta_j^{\tr}(\lambda_i) = 0$.

Using Lemma \ref{lemma:concentration_quadrat} and Lemma \ref{lemma:zalpha_a_zalpha_a},
\begin{align*}
    &\mE \Big| p^{1/2} \sum_{j=1}^n (\mE_j -\mE_{j-1}) \frac{1}{pn} u_q^T \bZ_j^T \bA_j^2(\lambda_i) \bZ_j v_q   \beta_j(\lambda_i)\beta^{\tr}_j(\lambda_i) \theta_j(\lambda_i) \Big|^2\\
    &\leq \calK p^{-1}\sum_{j=1}^n \mE \Big|\frac{1}{n}u_q^T \bZ_j^T \bA_j^2(\lambda_i) \bZ_j v_q   \theta_j(\lambda_i)  \Big|^2\\
    &\leq \calK \Big(\mE |\theta_j(\lambda_i)|^4\Big)^{1/2}\Big( \mE \Big| \frac{1}{n} u_q^T \bZ_j^T \bA_j^2(\lambda_i) \bZ_j u_q \Big|^4 \mE \Big| \frac{1}{n} v_q^T \bZ_j^T \bA_j^2(\lambda_i) \bZ_j v_q \Big|^4 \Big)^{1/4}\\
    &\leq \calK n^{-1/2} \|u_q\|_2^2 \|v_q\|_2^2.
\end{align*}
It follows then
\[ \sqrt{p} \sum_{j=1}^n (\mE_j -\mE_{j-1})d_{iqj2}^{(3)} = O_{L_2}(r_q). \]
It completes the proof of \eqref{eq:simplification_diqj2}.

Combining the analysis of $d_{iqj1}$, $d_{iqj2}$, and $d_{iqj3}$, we can conclude that 
\[
\sqrt{p}\bigl(\calH-\mE\calH\bigr)
=
\sum_{j=1}^n
\sum_{i=1}^M
\sum_{q=1}^m
a_{iq}\beta^{\mE}(\lambda_i)
\Bigl[
\calH_{iqj}^{(1)}
+
\calH_{iqj}^{(2)}
+
\calH_{iqj}^{(3)}
\Bigr]
+ O_{L_2}\Big( \sum_{q=1}^m  r_{q} \Big).
\]

\subsection{Convergence of the Mean} \label{subsec:convergence_mean}

In this subsection, we show the convergence of the expectation $\mE \calH$. Recall the definition of $\Omega(z)$ in the manuscript. The target is to show that for any fixed $\lambda>0$ and regular vectors $u_q$ and $v_q$, 
\begin{equation}\label{eq:mean_convergence_bound}
\sqrt{p} \left| \mE p^{-1} \Big[ u_q^T \bZ^T \bA(\lambda) \bZ v_q \Big] -   u_q^T v_q \frac{\Omega(-\lambda) -1}{\gamma \Omega(-\lambda)} \right|  = o(\|u_q\|_2 \|v_q\|_2).
\end{equation}
Following the results, we can then conclude that 
\[ \sqrt{p} \Big(\mE \calH - \sum_{i=1}^M \sum_{q=1}^m a_{iq} u_q^T v_q \frac{\Omega(-\lambda_i)-1}{\gamma \Omega(-\lambda_i)}  \Big) = o(1). \]

We make use of existing results in \citet{li2020high}. In particular, it is shown in Section S.3.4 of \citet{li2020high} that
\[ \sqrt{p}\left| \frac{ p^{-1} \mE \tr\tr[\bA_1(\lambda)]}{ 1+ n^{-1} \mE \tr[\bA_1(\lambda)]} -  \frac{\Omega(-\lambda) -1}{\gamma\Omega(-\lambda) } \right| = o(1). \]
Therefore, it suffices to show 
\[ \sqrt{p} \left| \mE p^{-1} \Big[ u_q^T \bZ^T \bA(\lambda) \bZ v_q \Big] -   u_q^T v_q \frac{ p^{-1} \mE \tr\tr[\bA_1(\lambda)]}{ 1+ n^{-1} \mE \tr[\bA_1(\lambda)]} \right|  = o(\|u_q\|_2 \|v_q\|_2). \]

To this end, we decompose the expectation as follows. 
\begin{align*}
&~\sqrt{p} \mE p^{-1} u_q^T\bZ^T \bA(\lambda) \bZ v_q \\  
&=p^{-1/2} \sum\limits_{j=1}^n \mE u_{qj} v_{qj} \bz_j^T \bA_j(\lambda) \bz_j \beta_j(\lambda)+ p^{-1/2} \sum\limits_{j=1}^n \mE u_{qj}  \bz_j^T \bA_j(\lambda) \bZ_j v_{q} \beta_j(\lambda)\\
&=p^{-1/2} \sum\limits_{j=1}^n \mE u_{qj} v_{qj} \bz_j^T \bA_j(\lambda) \bz_j \beta^{\mE}(\lambda)\\
&~~-p^{-1/2} \sum\limits_{j=1}^n \mE u_{qj} v_{qj} \bz_j^T \bA_j(\lambda) \bz_j \beta_j(\lambda) \beta^{\mE}(\lambda)\{\frac{1}{n}\bz_j^T \bA_j(\lambda) \bz_j - \frac{1}{n}\mE \tr \bA_j(\lambda)\} \\
&~~-p^{-1/2} \sum\limits_{j=1}^n \mE u_{qj} \bz_j^T \bA_j(\lambda) \bZ_j v_{q} \beta_j(\lambda)\beta^{\mE}(\lambda) \{\frac{1}{n}\bz_j^T \bA_j(\lambda) \bz_j - \frac{1}{n}\mE \tr \bA_j(\lambda)\}\\
&= p^{1/2} u_q^Tv_q\frac{ p^{-1}\mE\tr\bA_1(\lambda) }{1+ n^{-1}\mE \tr\bA_1(\lambda) }\\  
&~~+ n p^{-1/2}u_q^Tv_q\beta^{\mE}(\lambda)  \mE \beta_1(\lambda) \{\frac{1}{n}\bz_1^T \bA_1(\lambda) \bz_1 - \frac{1}{n}\mE \tr \bA_1(\lambda)\}\\  
&~~-p^{-1/2} \sum\limits_{j=1}^n \mE u_{qj} \bz_j^T \bA_j(\lambda) \bZ_jv_{q} (\beta^{\mE}(\lambda))^2 \{\frac{1}{n}\bz_j^T \bA_j (\lambda) \bz_j - \frac{1}{n}\mE \tr \bA_j(\lambda)\}\\  
&~~+ p^{-1/2} \sum\limits_{j=1}^n \mE u_{qj} \bz_j^T \bA_j(\lambda) \bZ_jv_{q}\beta_j(\lambda) (\beta^{\mE}(\lambda))^2 \{\frac{1}{n}\bz_1^T \bA_1(\lambda) \bz_1 - \frac{1}{n}\mE \tr \bA_1(\lambda)\}^2\\  
&= D_{1} + D_{2} + D_{3} + D_{4}, \mbox{ say}. 
\end{align*}
Therefore, it suffices to show that 
\[ |D_2| + |D_3| + |D_4| = o(\|u_q\|_2 \|v_q\|_2), \]

We discuss $D_2$ and $D_4$ first. 
\begin{align*}
&D_{2} = n p^{-1/2} u_q^T v_q (\beta^{\mE}(\lambda))^2 \mE \beta_1(\lambda)\Big|\frac{1}{n}\bz_1^T \bA_1(\lambda) \bz_1 - \frac{1}{n}\mE \tr \bA_1(\lambda)\Big|^2,\\
&|D_{4}| \leq  p^{-1/2} |\beta^{\mE}(\lambda)|^2 \sum\limits_{j=1}^n  |u_{qj}| \mE \Big|\bz_j^T \bA_j(\lambda) \bZ_jv_{q}\beta_j(\lambda)\Big|\Big|\frac{1}{n}\bz_j^T \bA_j(\lambda) \bz_j - \frac{1}{n}\mE \tr \bA_j(\lambda)\Big|^2.
\end{align*}
Due to Lemma \ref{lemma:converg_A} and Lemma \ref{lemma:concentration_quadrat}, 
\begin{align}
&\mE \Big|\frac{1}{n}\bz_j^T \bA_j(\lambda) \bz_j - \frac{1}{n}\mE \tr \bA_j(\lambda)\Big|^4  = o(n^{-1}),\nonumber\\
&\mE \Big|p^{-1/2} \bz_j^T \bA_j(\lambda) \bZ_jv_{q}\beta_j(\lambda)\Big|^2 = O (\|v_q\|_2^2).  \label{eq:z1_A_Z_b_beta}
\end{align}
The bound is uniform over $j$. 
Thus,
\[|D_{2}| + |D_{4}| = o(\|u_q\|_2 \|v_q\|_2 ). \] 
Here, we bound $\sum_{j=1}^n |u_{qj}|$ by $\sqrt{n}\|u_{q}\|_2$. 

We next show that 
\[|D_{3}|  = o(\|u_q\|_2 \|v_q\|_2).  \]
By Lemma \ref{lemma:z_1_z1_z1_z1}, 
\begin{align*}
p^{-1/2}\mE \bz_j^T \bA_j(\lambda) \bZ_jv_{q}\{\frac{1}{n}\bz_j^T \bA_j(\lambda) \bz_j - \frac{1}{n}\mE \tr \bA_j(\lambda)\}= p^{-1/2} n^{-1}\mE z_{11}^3 \mE\sum\limits_{i=1}^p e_i^T\bA_j(\lambda)e_i e_i^T\bA_j(\lambda) \bZ_jv_{q}.
\end{align*}
We first show 
\begin{align}
&\sup_{1\leq i,i'\leq p}\mE \Big|e_i^T\bA_{j}(\lambda)e_{i'} - \mE e_i^T\bA_j(\lambda)e_{i'}\Big|^2 = O(n^{-1}),\label{eq:concentrate_diag} \\
&\sup_{1\leq i\leq p, 1\leq j\leq n}\mE \Big| p^{-1/2} e_i^T\bA_{j}(\lambda)\bZ_j v_{q}  - \mE  p^{-1/2} e_i^T\bA_j(\lambda)\bZ_jv_{q}\Big|^2 =O (\|v_q\|_2^2).\label{eq:concentrate_A_zalpha_a}  
\end{align}

We define
\[{\beta}_{\ell j}(\lambda) = \frac{1}{1+n^{-1}\bz^T_\ell {\bA}_{\ell j}(\lambda)\bz_\ell }.\]

For (\ref{eq:concentrate_diag}), with any $i,i'$,
\begin{align*}
&\mE \Big|e_i^T \bA_{j}(\lambda) e_{i'} - \mE e_i^T \bA_j(\lambda) e_{i'}\Big|^2 \\
&~= \mE \Big| \sum\limits_{\ell\neq j}^{n} (\mE_{\ell} - \mE_{\ell-1}) [e_i^T \bA_{j}(\lambda) e_{i'} - \mE e_i^T \bA_{\ell j}(\lambda) e_{i'} ]\Big |^2\\
&~= \mE \Big| \sum\limits_{\ell\neq j}^{n} (\mE_{\ell} - \mE_{\ell-1}) [\frac{1}{n}\bz_\ell^T \bA_{\ell j}(\lambda)e_{i'} e_i^T \bA_{\ell j}(\lambda) \bz_\ell  \beta_{\ell j}(\lambda)]\Big|^2\\
&~ \leq \calK \frac{1}{n^2}\sum\limits_{\ell\neq j}^n \mE \Big| \bz_\ell^T \bA_{\ell j}(\lambda)e_{i'} e_i^T \bA_{\ell j}(\lambda) \bz_\ell  \beta_{\ell j}(\lambda)\Big|^2.
\end{align*}
By Lemma \ref{lemma:bound_quad_rankone}, 
\[\mE |\bz_\ell^T \bA_{\ell j}(\lambda)e_{i'} e_i^T \bA_{\ell j}(\lambda) \bz_\ell|^2 \leq \calK \mE \|\bA_{\ell j}(\lambda)\|^4_2 = O(1). \]
It completes the proof of (\ref{eq:concentrate_diag}).

To show (\ref{eq:concentrate_A_zalpha_a}), 
\begin{align*}
& p^{-1}\mE |e_i^T \bA_j(\lambda)\bZ_jv_{q} - \mE e_i^T \bA_{j}(\lambda)\bZ_j v_{q} |^2\\
&= p^{-1}\mE\Big|\sum\limits_{\ell\neq j}^n (\mE_{\ell}-\mE_{\ell-1}) [e_i^T \bA_j(\lambda)\bZ_jv_{q} - e_i^T \bA_{\ell j}(\lambda)\bZ_{j\ell} v_{q}] \Big|^2\\
& \leq \calK p^{-1}\sum\limits_{\ell\neq j}^n \mE \Big| v_{q\ell} \beta_{\ell j}(\lambda) e_i^T \bA_{\ell j}(\lambda) \bz_\ell - \frac{1}{n} e_i^T \bA_{\ell j}(\lambda) \bz_\ell \bz_{\ell}^T \bA_{\ell j}(\lambda)\bZ_{j\ell}v_{q} \beta_{\ell j}(\lambda)\Big|^2\\ 
&\leq \calK p^{-1} \sum_{\ell\neq j}^n |v_{q\ell}|^2 \mE \Big| e_i^T \bA_{\ell j} (\lambda) \bz_{\ell} \Big|^2  + \calK p^{-1} \sum_{\ell \neq j}^n \mE \Big|\frac{1}{n} e_i^T \bA_{\ell j}(\lambda) \bz_\ell \bz_{\ell}^T \bA_{\ell j}(\lambda)\bZ_{j\ell}v_{q} \beta_{\ell j}(\lambda) \Big|^2
\end{align*}
By Lemma \ref{lemma:bound_quad_rankone}, 
\begin{align*}
\sup_i \mE|e_i^T \bA_{\ell j}(\lambda) \bz_\ell |^4 \leq \calK \mE \|\bA_{\ell j}(\lambda)\|^4 = O(1).  
\end{align*}
Also, 
\[\mE |p^{-1/2} e_i^T \bA_{\ell j}(\lambda) \bz_\ell \bz_{\ell}^T \bA_{\ell j}(\lambda)\bZ_{j\ell}v_{q} \beta_{\ell j}(\lambda)|^2 =O(\|v_q\|_2^2).   \]
The proof of (\ref{eq:concentrate_A_zalpha_a}) is complete. 

Following from (\ref{eq:concentrate_diag}) and \eqref{eq:concentrate_A_zalpha_a}, 
\begin{align*}
D_{3} &= - p^{-1/2}(\beta^{\mE}(\lambda))^2 \sum\limits_{j=1}^n u_{qj} \mE \bz_j^T \bA_j(\lambda)\bZ_j v_{q} \{\frac{1}{n} \bz_j^T \bA_j(\lambda)\bz_j - \frac{1}{n}\mE\tr\bA_j(\lambda)\}\\
& = - n^{-1} p^{-1/2}(\beta^{\mE}(\lambda))^2 \sum\limits_{j=1}^n u_{qj} \mE z_{11}^3 \mE \sum\limits_{i=1}^p e_i^T \bA_j(\lambda)e_i e_i^T \bA_j(\lambda)\bZ_jv_{q}\\ 
& = -n^{-1} p^{-1/2}(\beta^{\mE}(\lambda))^2 \sum\limits_{j=1}^n u_{qj} \mE z_{11}^3  \sum\limits_{i=1}^p \mE e_i^T \bA_j(\lambda)e_i \mE e_i^T \bA_j(\lambda)\bZ_jv_{q} +o(\|u_q\|_2\|v_q\|_2 )\\
& = - n^{-1}p^{-1/2}(\beta^{\mE}(\lambda))^2 \mE z_{11}^3   \sum\limits_{j=1}^n u_{qj} \sum\limits_{\ell\neq j}^n v_{q\ell}  \sum\limits_{i=1}^p \mE e_i^T \bA_1(\lambda)e_i  \mE e_i^T \bA_1(\lambda)\bz_2 + o(\|u_q\|_2\|v_q\|_2 ).
\end{align*}
Note 
\begin{align*}
&\mE e_i^T\bA_1(\lambda) \bz_2\\
&=  \mE e_i^T\bA_{12}(\lambda)\bz_2 \beta_{21}(\lambda)\\
&= -\mE e_i^T\bA_{12}(\lambda)\bz_2 \beta_{21}(\lambda) \beta^{\mE}(\lambda) (\frac{1}{n}\bz_2^T\bA_{12}(\lambda)\bz_2 - \frac{1}{n} \mE\tr\bA_1(\lambda) )\\
&=  -\mE e_i^T\bA_{12}(\lambda)\bz_2 (\beta^{\mE}(\lambda))^2 (\frac{1}{n}\bz_2^T\bA_{12}(\lambda)\bz_2 - \frac{1}{n} \mE\tr\bA_1(\lambda) ) \\
&~~ +\mE e_i^T\bA_{12}(\lambda)\bz_2 (\beta^{\mE}(\lambda))^2\beta_{21}(\lambda) \{\frac{1}{n}\bz_2^T\bA_{12}(\lambda)\bz_2 - \frac{1}{n} \mE\tr\bA_1(\lambda) \}^2 \\ 
& = -\frac{1}{n} (\beta^{\mE}(\lambda))^2 \mE z_{11}^3 \sum\limits_{\ell=1}^p \mE e_\ell^T \bA_{12}(\lambda) e_\ell e_\ell^T \bA_{12}(\lambda) e_i\\ 
&~~ +\mE e_i^T\bA_{12}(\lambda)\bz_2 (\beta^{\mE}(\lambda))^2\beta_{21}(\lambda) \{\frac{1}{n}\bz_2^T\bA_{12}(\lambda)\bz_2 - \frac{1}{n} \mE\tr\bA_1(\lambda) \}^2.
\end{align*}
By Lemma \ref{lemma:concentration_quadrat},  Lemma \ref{lemma:bound_quad_rankone}, and Lemma \ref{lemma:bound_beta},
\[ \sup_i\mE |e_i^T\bA_{12}(\lambda)\bz_2 (\beta^{\mE}(\lambda))^2\beta_{21}(\lambda) \{\frac{1}{n}\bz_2^T\bA_{12}(\lambda)\bz_2 - \frac{1}{n} \mE\tr\bA_1(\lambda) \}^2| = o(n^{-1/2}).\]
Therefore, again using (\ref{eq:concentrate_diag}), 
\begin{align*}
&D_{3}=  n^{-2} p^{-1/2}(\beta^{\mE}(\lambda))^4 (\mE z_{11}^3)^2   \sum\limits_{j=1}^n u_{qj} \sum\limits_{\ell\neq j}^n v_{q\ell}  \sum\limits_{i=1}^p \mE e_i^T \bA_j(\lambda)e_i\times\\
&\qquad  \sum\limits_{\ell=1}^p \mE e_\ell^T \bA_{12}(\lambda) e_\ell e_\ell^T \bA_{12}(\lambda) e_i  + o(\|u_q\|_2 \|v_q\|_2 )\\
&=  n^{-2} p^{-1/2}(\beta^{\mE}(\lambda))^4 (\mE z_{11}^3)^2   \sum\limits_{j=1}^n u_{qj} \sum\limits_{\ell\neq j}^n v_{q\ell}  \sum\limits_{i=1}^p \sum\limits_{k=1}^p \mE e_i^T \bA_j(\lambda)e_i \mE e_k^T \bA_{12}(\lambda) e_k~\times \\ 
&\qquad\qquad\mE e_k^T \bA_{12}(\lambda) e_i  + o(\|u_q\|_2 \|v_q\|_2)\\
&=  n^{-2} p^{-1/2} (\beta^{\mE}(\lambda))^4 (\mE z_{11}^3)^2  \sum_{j=1}^n u_{qj} \sum_{\ell\neq j} v_{q\ell}
 \sum\limits_{i=1}^p \sum\limits_{k=1}^p \mE e_i^T \bA_{1}(\lambda)e_i \mE e_k^T \bA_{12}(\lambda) e_k \mE e_k^T \bA_{12}(\lambda) e_i\\
 & \quad + o(\|u_q\|\|v_q\|). 
\end{align*}
It can be verified that 
\[ \sum\limits_{i=1}^p \sum\limits_{k=1}^p \mE e_i^T \bA_1(\lambda)e_i \mE e_k^T \bA_{12}(\lambda) e_k \mE e_k^T \bA_{12}(\lambda) e_i =\alpha_1^T (\lambda) \mE \bA_{12} (\lambda) \alpha_2(\lambda), \]
where $\alpha_1(\lambda)$ is the diagonal of $\mE \bA_{1}(\lambda)$ and $\alpha_2(\lambda)$ is the diagonal of $\mE \bA_{12}(\lambda)$. 
\begin{align*}
 |\alpha_1^T (\lambda)\mE \bA_{12} (\lambda) \alpha_2(\lambda)| & \leq  \|\mE\bA_{12}(\lambda)\|_2 \|\alpha_1(\lambda)\|_2 \|\alpha_2(\lambda)\|_2 \\
&\leq n \|\mE\bA_{12}(\lambda)\|^2_2 \|\mE\bA_{1}(\lambda)\|_2 \|\mE \bA_{12}(\lambda) \|_2  = O(n). 
\end{align*}
It follows that 
\[  |D_{3}| = O(n^{-1/2}\|u_q\|_2 \|v_q\|_2).\] 

\subsection{Conditional Variance of the Martingale}
\label{subsec:variance_martingale}

In this subsection, we characterize the limit of the conditional variance
\[
\sum_{j=1}^n
\mE_{j-1}
\left[
\left(
\sum_{i=1}^M
\sum_{q=1}^m
a_{iq}\beta^{\mE}(\lambda_i)
\Big[
\calH_{iqj}^{(1)}
+
\calH_{iqj}^{(2)}
+
\calH_{iqj}^{(3)}
\Big]
\right)^2
\right].
\]
To this end, it suffices to study all cross-product terms of the form $\calH_{iqj}^{(g)} \calH_{i'q'j}^{(g')}$. Since $\calH_{iqj}^{(2)}$ is the counterpart of
$\calH_{iqj}^{(1)}$ obtained by interchanging $u_q$ and $v_q$,
it suffices to determine the limits of $\beta^{\mE}(\lambda)$ and of terms of the following four types:
\begin{align*}
&\sum_{j=1}^n
\mE_{j-1}
\calH_{iqj}^{(1)}
\calH_{i'q'j}^{(1)},
\quad \sum_{j=1}^n
\mE_{j-1}
\calH_{iqj}^{(1)}
\calH_{i'q'j}^{(2)},
\quad\sum_{j=1}^n
\mE_{j-1}
\calH_{iqj}^{(1)}
\calH_{i'q'j}^{(3)},
\quad\sum_{j=1}^n
\mE_{j-1}
\calH_{iqj}^{(3)}
\calH_{i'q'j}^{(3)},
\end{align*}
for all
$i,i'\in\{1,\ldots,M\}$ and
$q,q'\in\{1,\ldots,m\}$.
The remaining cross terms follow from these four cases by interchanging the corresponding $u$- and $v$-vectors.

We claim the following results. Define
\[ \eta_{q,q'} = n^{-3/16} \|u_q\|_2 \|u_{q'}\|_2 \|v_q \|_2 \|v_{q'}\|_2. \]
Note that $\eta_{q,q'} \asymp n^{-3/16}$ under the regularity condition of the vectors. 

\begin{align}
&\sum_{j=1}^n
\mE_{j-1}
\calH^{(1)}_{iqj}
\calH^{(1)}_{i'q'j}=
p^{-1}
\sum_{j=1}^n
u_{qj}u_{q'j}
\mE_{j-1}
\Big\{
\mE_j[\bz_j^T\bA_j(\lambda_i)\bZ_jv_q]
\,
\mE_j[\bz_j^T\bA_j(\lambda_{i'})\bZ_jv_{q'}]
\Big\}  \label{eq:limit_H1H1}\\
&=
p^{-1}
\beta^{\mE}(\lambda_i)
\beta^{\mE}(\lambda_{i'})
\sum_{j=1}^n
u_{qj}u_{q'j}
\tr\!\Big[
\mE_j\bA_j(\lambda_i)
\mE_j\bA_j(\lambda_{i'})
\Big]
\sum_{r=1}^{j-1}
v_{qr}v_{q'r}+ O_{L_1}\Big( \eta_{q,q'}\Big).\nonumber\\
&\sum_{j=1}^n
\mE_{j-1}
\calH^{(1)}_{iqj}
\calH^{(2)}_{i'q'j}
=
p^{-1}
\sum_{j=1}^n
u_{qj}v_{q'j}
\mE_{j-1}
\Big\{
\mE_j[\bz_j^T\bA_j(\lambda_i)\bZ_jv_q]
\,
\mE_j[\bz_j^T\bA_j(\lambda_{i'})\bZ_ju_{q'}]
\Big\}\label{eq:limit_H1H2} \\
&\qquad=
p^{-1}
\beta^{\mE}(\lambda_i)
\beta^{\mE}(\lambda_{i'})
\sum_{j=1}^n
u_{qj}v_{q'j}
\tr\!\Big[
\mE_j\bA_j(\lambda_i)
\mE_j\bA_j(\lambda_{i'})
\Big]
\sum_{r=1}^{j-1}
v_{qr}u_{q'r} + O_{L_1}\Big(\eta_{q,q'}\Big).\nonumber \\
&\sum_{j=1}^n
\mE_{j-1}
\calH^{(1)}_{iqj}
\calH^{(3)}_{i'q'j}
= -
p^{-1}
\sum_{j=1}^n
u_{qj}
\mE_{j-1}
\Big\{
\mE_j[\bz_j^T\bA_j(\lambda_i)\bZ_jv_q]
\,
\mE_j[\varrho_{i'q'j}]
\Big\} =  O_{L_1}\Big(\eta_{q,q'}\Big).  \label{eq:limit_H1H3}\\
&\sum_{j=1}^n
\mE_{j-1}
\calH^{(3)}_{iqj}
\calH^{(3)}_{i'q'j}
=
p^{-1}
\sum_{j=1}^n
\mE_{j-1}
\Big\{
\mE_j\varrho_{iqj}\,
\mE_j\varrho_{i'q'j}
\Big\} \nonumber \\
&\qquad=
\frac{1}{pn^2}
\sum_{j=1}^n
\Big[
\tr\!\big(
\mE_j\bA_j(\lambda_i)
\mE_j\bA_j(\lambda_{i'})
\big)
\Big]^2 \nonumber\\
&\qquad\qquad\times
\Bigg[
\left(
\sum_{r=1}^{j-1}u_{qr}v_{q'r}
\right)
\left(
\sum_{r=1}^{j-1}v_{qr}u_{q'r}
\right)
+
\left(
\sum_{r=1}^{j-1}u_{qr}u_{q'r}
\right)
\left(
\sum_{r=1}^{j-1}v_{qr}v_{q'r}
\right)
\Bigg] + O_{L_1}\Big(\eta_{q,q'}\Big).\label{eq:limit_H3H3}
\end{align}

We provide the proof only for \eqref{eq:limit_H1H1}. The remaining results follow from analogous arguments. Closely related calculations can be found in Sections~S.3.1.3--S.3.1.6 of \citet{li2020high}, and we therefore omit the corresponding details.

We introduce 
\[\underline{\bZ}_j = [\bz_1,\dots, \bz_{j-1}, 0, \underline{\bz}_{j+1}, \dots,\underline{\bz}_{n}]\] 
for $j=2,3,\dots,n$, where $\underline{\bz}_{j+1}$, $\dots,$ $\underline{\bz}_n$ are i.i.d. copies of $\bz_1$ and independent with $\bz_1,\dots,\bz_{j-1}$. What's more, introduce  $\underline{\bA}_j(\lambda)$ like $\bA_j(\lambda)$, but $\underline{\bA}_j(\lambda)$ is now defined on $\underline{\bZ}_j$ instead of $\bZ_j$. Note, conditional on $\bz_1,\dots,\bz_{j-1}$, $(\bZ_j,\bA_j)$ is independent with $(\underline{\bZ}_j,\underline{\bA}_j)$. Therefore, we can express
\begin{align*}
&\mE_{j-1} \Big[\mE_j [ \bz_j^T \bA_j(\lambda_i) \bZ_j v_q] \mE_j[\bz_j^T \bA_j(\lambda_{i'}) \bZ_j v_{q'}]\Big] = \mE_{j-1}\bz_j^T  \mE_{j} \Big[\bA_j(\lambda_i) \bZ_j v_q v^T_{q'} \underline{\bZ}^T_j \underline{\bA}_j(\lambda_{i'}) \Big]\bz_j\\
&=  \mE_{j-1} \tr\Big[ \bA_j(\lambda_i) \bZ_j v_q v_{q'}^T \underline{\bZ}_j^T \underline{\bA}_j (\lambda_{i'}) \Big].
\end{align*}

Further, we introduce $\underline{\bZ}_{rj}, r<j$ to be $[\bz_1, \dots, \bz_{r-1}, 0, \bz_{r+1}, \dots, \bz_{j-1}, 0 , \underline{\bz}_{j+1}, \dots,\underline{\bz}_{n}]$.  With these notations, for $j\geq 2$, 
\begin{align*}
&\mE_{j-1} \calH_{iqj}^{(1)} \calH_{i'q'j}^{(1)} = \mE_{j-1} \Big[ \mE_j[p^{-1/2} u_{qj} \bz_j^T \bA_j(\lambda_i) \bZ_{j} v_q ] \mE_j [p^{-1/2} u_{q'j} \bz_j^T \bA_j(\lambda_{i'}) \bZ_j v_{q'}]   \Big]\\
&=p^{-1} u_{qj}u_{q'j} \mE_{j-1}\Big[ \bz_j^T [\mE_j \bA_j(\lambda_i)\bZ_j v_q ] \mE_j [v_{q'}^T \bZ_j^T \bA_j(\lambda_{i'}) ] \bz_j  \Big] \\
&=p^{-1} u_{qj} u_{q'j}  \mE_{j-1}\tr\Big[ [\mE_j \bA_j(\lambda_i)\bZ_j v_q ] \mE_j [v_{q'}^T \bZ_j^T \bA_j(\lambda_{i'}) ]  \Big]  \\
&=p^{-1} u_{qj} u_{q'j} \mE_{j-1} \tr\Big[ \bA_j(\lambda_i) \bZ_j v_q v_{q'}^T \underline{\bZ}_j^T \underline{\bA}_j (\lambda_{i'}) \Big]\\
&=p^{-1} u_{qj} u_{q'j} \mE_{j-1} v_q^T \bZ_j^T \bA_j(\lambda_i) \underline{\bA}_j(\lambda_{i'}) \underline{\bZ}_j v_{q'}\\
&=p^{-1} u_{qj} u_{q'j} \sum_{r=1}^{j-1} v_{qr}v_{q'r} \mE_{j-1} \bz_r^T \bA_j(\lambda_i) \underline{\bA}_j(\lambda_{i'}) \bz_r\\
&\quad + p^{-1} u_{qj}u_{q'j} \sum_{r=1}^{j-1} v_{qr} \mE_{j-1} \bz_r^T \bA_j(\lambda_i) \underline{\bA}_j(\lambda_{i'}) \underline{\bZ}_{rj} v_{q'}\\
&\quad + p^{-1} u_{qj} u_{q'j} \sum_{r=j+1}^n v_{qr} \mE_{j-1}\bz_r^T \bA_j(\lambda_i) \underline{\bA}_{j}(\lambda_{i'})\underline{\bZ}_j v_{q'} \\
& = u_{qj} u_{q'j} \Big( \calJ_j^{(1)} + \calJ_j^{(2)} + \calJ_j^{(3)}\Big), \mbox{ say}. 
\end{align*}

We aim to show 
\begin{align*}
    &\sup_{j\geq2} \mE \Big| \calJ_j^{(1)} - \beta^{\mE}(\lambda_i) \beta^{\mE}(\lambda_{i'}) \Big(\sum_{r=1}^{j-1} v_{qr}v_{q'r} \Big) p^{-1}\tr\mE_{j-1} \Big[\bA_j(\lambda_i)\underline{\bA}_{j}(\lambda_{i'}) \Big]  \Big| = O(n^{-1/2} \|v_q\|_2 \|v_{q'}\|_2 ), \\ 
    &\sup_{j\geq2} \mE |\calJ_j^{(2)}| = O(n^{-3/16} \|v_q\|_2 \|v_{q'} \|_2 ),\\
    &\sup_{j\geq2} \mE |\calJ_j^{(3)}| = O(n^{-1/2} \|v_q\|_2 \|v_{q'} \|_2 ).\\
\end{align*}

Recall the definition of $\beta_{rj}(\lambda)$. Similarly, define
\begin{align*}
\underline{\beta}_{rj}(\lambda) = \frac{1}{1+n^{-1}\bz^T_r\underline{\bA}_{rj}(\lambda)\bz_r}, 
\end{align*}
where $\underline{\bA}_{rj}(\lambda)$ is defined in the same way as $\bA_{rj}(\lambda)$, but with $\underline{\bZ}_{rj}$ instead of $\bZ_{rj}$.

As for $\calJ_j^{(1)}$, 
\begin{align*}
&\mE \Big| \calJ_j^{(1)} -  \beta^{\mE}(\lambda_i)\beta^{\mE}(\lambda_{i'}) \Big(\sum\limits_{r=1}^{j-1} v_{qr} v_{q'r}\Big) p^{-1} \tr \left[\mE_{j-1}[\bA_j(\lambda_i)\underline{\bA}_j(\lambda_{i'})]\right]\Big| \\  
&=\mE \Big| \sum\limits_{r=1}^{j-1} v_{qr} v_{q'r} \mE_{j-1} \beta_{rj}(\lambda_i)\underline{\beta}_{rj}(\lambda_{i'}) 
   p^{-1} \bz_r^T \bA_{rj}(\lambda_i)\underline{\bA}_{rj}(\lambda_{i'})\bz_r -\\ 
&\qquad  \beta^{\mE}(\lambda_i)\beta^{\mE}(\lambda_{i'})\Big(\sum\limits_{r=1}^{j-1} v_{qr} v_{q'r}\Big) p^{-1} \tr \left[\mE_{j}[\bA_j(\lambda_i)\underline{\bA}_j(\lambda_{i'})]\right]\Big|  \\ 
&\leq \sum_{r=1}^{j-1} |v_{qr}||v_{q'r}| \mE \Big| \beta_{rj}(\lambda_i)\underline{\beta}_{rj}(\lambda_{i'}) 
  \frac{1}{p}\bz_r^T \bA_{rj}(\lambda_i)\underline{\bA}_{rj}(\lambda_{i'})\bz_r - \\ 
&\qquad \beta^{\mE}(\lambda_i)\beta^{\mE}(\lambda_{i'})  \frac{1}{p}\tr\{\bA_j(\lambda_i)\underline{\bA}_j(\lambda_{i'})\}\Big|\\ 
&\leq  \sum_{r=1}^{n} |v_{qr}||v_{q'r}| \mE \Big| (\beta_{rj}(\lambda_i)-\beta^{\mE}(\lambda_i))\underline{\beta}_{rj}(\lambda_{i'}) \frac{1}{p}\bz_r^T \bA_{rj}(\lambda_i)\underline{\bA}_{rj}(\lambda_{i'})\bz_r \Big| \\ 
&+ \sum_{r=1}^{n} |v_{qr}||v_{q'r}| \mE \Big| {\beta}^{\mE}(\lambda_i)(\underline{\beta}_{rj}(\lambda_{i'})-{\beta}^{\mE}(\lambda_{i'})) \frac{1}{p}\bz_r^T \bA_{rj}(\lambda_i)\underline{\bA}_{rj}(\lambda_{i'})\bz_r \Big| \\ 
&  + \sum_{r=1}^{n} |v_{qr}||v_{q'r}|  |{\beta}^{\mE}(\lambda_i)||{\beta}^{\mE}(\lambda_{i'})| \mE \Big|\frac{1}{p}\bz_r^T \bA_{rj}(\lambda_i)\underline{\bA}_{rj}(\lambda_{i'})\bz_r -\frac{1}{p}\tr\{\bA_j(\lambda_i)\underline{\bA}_j(\lambda_{i'})\} \Big| \\ 
&= O( n^{-1/2} \|v_q\|_2\|v_{q'}\|_2).
\end{align*}
For the last line, we need the following arguments that are direct consequences of Lemma \ref{lemma:bound_beta}, Lemma \ref{lemma:nonran_bound_resol}, Lemma \ref{lemma:converg_A}, and Lemma \ref{lemma:concentration_quadrat}. 
\begin{align*}
&|\beta^{\mE}(\lambda)|\leq 1, \\  
& \mE \Big| \frac{1}{p} \bz_r^T \bA_{rj}(\lambda_i)\underline{\bA}_{rj}(\lambda_{i'})\bz_r\Big|^2 = O(1),\\  
& \mE \Big|\frac{1}{p} \bz_r^T \bA_{rj}(\lambda_i)\underline{\bA}_{rj}(\lambda_{i'})\bz_r - \frac{1}{p} \tr\big[\bA_2(\lambda_i)\underline{\bA}_2(\lambda_{i'})\big] \Big|^2  = O(n^{-1}),\\  
& \mE|\beta_{rj}(\lambda)-\beta^{\mE}(\lambda)|^2\leq \mE\Big|\beta_{rj}(\lambda)\beta^{\mE}(\lambda)(\frac{1}{n}\bz_r^T \bA_{rj}\bz_r - \frac{1}{n}\mE \tr\bA_{rj})\Big|^2\\
 &\qquad + \mE\Big|\beta_{rj}(\lambda)\beta^{\mE}(\lambda)(\frac{1}{n}\mE \tr\bA_{rj}- \frac{1}{n}\mE \tr\bA_{j})\Big|^2 = O(n^{-1}). 
\end{align*}

As for $\calJ_j^{(2)}$,  due to 
\begin{align*}
&\sup_{r<j}\mE \Big|(\beta_{rj}(\lambda_i)-\beta^{\mE}(\lambda_i))p^{-1/2} \bz_r^T\bA_{rj}(\lambda_i)\underline{\bA}_{rj}(\lambda_{i'})\underline{\bZ}_{rj} v_{q'} \Big| = O\Big(\frac{1}{\sqrt{n}}\|v_{q'}\|_2\Big), \\
&\sup_{r<j}\mE \Big|\frac{1}{p}(\beta_{rj}(\lambda_i)- \beta^{\mE}(\lambda_i))\tr\big[\bA_{rj}(\lambda_i)\underline{\bA}_{rj}(\lambda_{i'}) \big] p^{-1/2} \bz_r^T\underline{\bA}_{rj}(\lambda_{i'})\underline{\bZ}_{rj} v_{q'} \Big| = O\Big(\frac{1}{\sqrt{n}}\|v_{q'}\|_2\Big),\\
&\sup_{r<j}\mE \Big| \beta_{rj}(\lambda_i)\underline{\beta}_{rj}(\lambda_{i'})\frac{1}{n}[\bz_r^T\bA_{rj}(\lambda_i)\underline{\bA}_{rj}(\lambda_{i'})\bz_r- \tr\{\bA_{rj}(\lambda_i)\underline{\bA}_{rj}(\lambda_{i'}) \}]
 p^{-1/2} \bz_r^T\underline{\bA}_{rj}(\lambda_{i'})\underline{\bZ}_{rj}v_{q'} \Big|\\
 &\quad= O\Big(\frac{1}{\sqrt{n}}\|v_{q'}\|_2\Big),
\end{align*}
which are consequences of Lemma \ref{lemma:bound_beta}, Lemma \ref{lemma:converg_A}, Lemma \ref{lemma:concentration_quadrat}, and Lemma \ref{lemma:bound_z1_zalpha_b}, 
\begin{align*}
\calJ_j^{(2)} &= p^{-1} \mE_{j-1}\sum\limits_{r=1}^{j-1} v_{qr}\beta_{rj}(\lambda_i)\bz_r^T \bA_{rj}(\lambda_i)\underline{\bA}_{j}(\lambda_{i'})\underline{\bZ}_{rj} v_{q'} \\
& = p^{-1}\mE_{j-1} \sum\limits_{r=1}^{j-1} v_{qr}\beta_{rj}(\lambda_i)\bz_r^T\bA_{rj}(\lambda_i)\underline{\bA}_{rj}(\lambda_{i'})\underline{\bZ}_{rj}v_{q'} \\
&~~ ~- p^{-1}\mE_{j-1}\sum\limits_{r=1}^{j-1} \frac{1}{n} v_{qr}\beta_{rj}(\lambda_i)\underline{\beta}_{rj}(\lambda_{i'})\bz_r^T\bA_{rj}(\lambda_i)\underline{\bA}_{rj}(\lambda_{i'})\bz_r\bz_r^T\underline{\bA}_{rj}(\lambda_{i'})\underline{\bZ}_{rj}v_{q'} \\
& =  \sum\limits_{r=1}^{j-1} v_{qr} \beta^{\mE}(\lambda_i) p^{-1} \mE_{j-1}\bz_r^T\bA_{rj}(\lambda_i)\underline{\bA}_{rj}(\lambda_{i'})\underline{\bZ}_{rj}v_{q'}\\
&- \sum\limits_{r=1}^{j-1} \frac{1}{n} v_{qr}\beta^{\mE}(\lambda_i)\beta^{\mE}(\lambda_{i'})\mE_{j-1} \frac{1}{p}\tr\{\bA_{rj}(\lambda_i)\underline{\bA}_{rj}(\lambda_{i'}) \}\bz_r^T\underline{\bA}_{rj}(\lambda_{i'})\underline{\bZ}_{rj}v_{q'} \\
&\qquad + O_{L_1}( n^{-1/2} \|v_q\|_2\|v_{q'}\|_2 ). 
\end{align*}
The residual term above is uniform over $j$. 

We claim that the first two terms are also negligible. To verify the claim, we first show  
\begin{align}
&\sup_j\mE\Big| p^{-1}\sum\limits_{r=1}^{j-1} v_{qr} \bz_r^T \bA_{rj}(\lambda_i) \underline{\bA}_{rj}(\lambda_{i'}) \underline{\bZ}_{rj}v_{q'} \Big|^2 = O(n^{-3/8}\|v_q\|_2^2 \|v_{q'}\|_2^2 ),\label{eq:exp_Jj2_1} \\ 
&\sup_j\mE\Big| p^{-1}\sum\limits_{r=1}^{j-1} v_{qr} \bz_r^T \underline{\bA}_{rj}(\lambda_{i'}) \underline{\bZ}_{rj}v_{q'} \Big|^2 = O(n^{-1}\|v_q\|_2^2 \|v_{q'}\|_2^2).\label{eq:exp_Jj2_2}
\end{align}
The proofs of the two are very similar. Therefore, we shall only present proof of (\ref{eq:exp_Jj2_1}). 

Like $\bZ_{rj}$, we define $\bZ_{rr'j}$ to be $\bZ$ with $\bz_r,\bz_{r'}, \bz_{j}$ replaced by 0. Similarly, define $\underline{\bZ}_{rr'j}$,  $r< j, r'<j$ by replacing the $r$-th and $r'$-th column of $\underline{\bZ}_{j}$ by the 0 vectors. Further, define $\bA_{rr'j}$ and $\underline{\bA}_{rr'j}$ to be the counterparts of $\bA_{rj}$ and $\underline{\bA}_{rj}$ with $\bZ_{rj}$ replaced by $\bZ_{rr'j}$, or with $\underline{\bZ}_{rj}$ replaced by $\underline{\bZ}_{rr'j}$. Accordingly, define
\begin{align*}
&\beta_{rj}^{\tr}(\lambda) = \frac{1}{1+ n^{-1}\tr\bA_{rj}(\lambda)},\\
&\theta_{rj}(\lambda)  =\frac{1}{n} \bz_r^T \bA_{rj}(\lambda) \bz_r - \frac{1}{n}\tr \bA_{rj}(\lambda),\\
&\beta_{rr'j}(\lambda) = \frac{1}{1+ n^{-1}\bz_{r'}^T\bA_{rr'j}(\lambda)\bz_{r'} },\\
&\underline{\beta}_{rr'j}(\lambda) = \frac{1}{1+ n^{-1}\bz_{r'}^T\underline{\bA}_{rr'j}(\lambda)\bz_{r'} }.
\end{align*}  

By Lemma \ref{lemma:bound_z1_zalpha_b}, for squared-terms in the expansion of (\ref{eq:exp_Jj2_1}), 
\begin{align*}
&\sum\limits_{r=1}^{j-1} v_{qr}^2 \mE\Big|p^{-1}\bz_r^T \bA_{rj}(\lambda_i)\underline{\bA}_{rj}(\lambda_{i'})\underline{\bZ}_{rj}v_{q'} \Big|^2
\leq p^{-1}\sum\limits_{r=1}^{n} v_{qr}^2 \mE\Big|p^{-1/2}\bz_r^T \bA_{rj}(\lambda_i)\underline{\bA}_{rj}(\lambda_{i'})\underline{\bZ}_{rj}v_{q'} \Big|^2\\
&=O\Big(n^{-1}\|v_q\|_2^2 \|v_{q'}\|_2^2 \Big).
\end{align*}

For crossed-terms with $r\neq r'$, using analogous arguments to those in the proof of (3.35) in \citet{Pan2011central}, we can show that 
\[ p^{-1}\mE \Big|\bz_r^T\bA_{rj}(\lambda_i)\underline{\bA}_{rj}(\lambda_{i'})\underline{\bZ}_{rj}v_{q'} \bz_{r'}^T\bA_{r'j}(\lambda_i)\underline{\bA}_{r'j}(\lambda_{i'})\underline{\bZ}_{r'j}v_{q'}\Big| \leq \calK n^{-3/8} \|v_{q'}\|_2^2. \]
It follows then 
\[ \sum_{r\neq r'} |v_{qr}||v_{qr'}| p^{-2}\mE \Big|\bz_r^T\bA_{rj}(\lambda_i)\underline{\bA}_{rj}(\lambda_{i'})\underline{\bZ}_{rj}v_{q'} \bz_{r'}^T\bA_{r'j}(\lambda_i)\underline{\bA}_{r'j}(\lambda_{i'})\underline{\bZ}_{r'j}v_{q'}\Big| \leq \calK n^{-3/8} \|v_{q}\|_2^2 \|v_{q'}\|_2^2. \]
Here, we bound  $ \sum_{r,r'}|v_{qr}||v_{qr'}|$ by $n\|v_q\|_2^2$.

The second term in the expansion of  $\calJ_j^{(2)}$ is such that, 
 \begin{align*}
 \mE\Big| p^{-1/2}\Big[ &\tr\{\bA_{rj}(\lambda_i)\underline{\bA}_{rj}(\lambda_{i'}) \} - 
\tr\{\bA_{j}(\lambda_i)\underline{\bA}_{j}(\lambda_{i'}) \} 
 \Big]\bz_r^T\underline{\bA}_{rj}(\lambda_{i'})\underline{\bZ}_{rj}v_{q'} \Big|^2\\
 &\leq \calK \mE \Big| p^{-1/2}\bz_r^T\underline{\bA}_{rj}(\lambda_{i'})\underline{\bZ}_{rj}v_{q'}  \Big|^2 = O(\|v_{q'}\|_2^2),
 \end{align*}
 due to an inequality similar to Lemma \ref{lemma:nonran_bound_resol}.

 Therefore,
 \begin{align*}
&\sum\limits_{r=1}^{j-1} \frac{1}{n}v_{qr}\beta^{\mE}(\lambda_i)\beta^{\mE}(\lambda_{i'})\frac{1}{p} \mE_{j-1}\tr\{\bA_{rj}(\lambda_i)\underline{\bA}_{rj}(\lambda_{i'}) \}\bz_r^T\underline{\bA}_{rj}(\lambda_{i'})\underline{\bZ}_{rj}v_{q'}  \\
&= \beta^{\mE}(\lambda_i)\beta^{\mE}(\lambda_{i'})\mE_{j-1}\Big[\frac{1}{np}\tr\{\bA_{j}(\lambda_i)\underline{\bA}_{j}(\lambda_{i'}) \}\sum\limits_{r=1}^{j-1}  v_{qr}\bz_r^T\underline{\bA}_{rj}(\lambda_{i'})\underline{\bZ}_{rj}v_{q'}\Big] + o_{L_1}\Big(n^{-1/2}\|v_q\|_2 \|v_{q'}\|_2 \Big).
 \end{align*}
By (\ref{eq:exp_Jj2_1}), (\ref{eq:exp_Jj2_2}), and Lemma \ref{lemma:nonran_bound_resol}, 
\begin{align*}
&\mE\Big[\frac{1}{np}\tr\{\bA_{j}(\lambda_i)\underline{\bA}_{j}(\lambda_{i'}) \}\sum\limits_{r=1}^{j-1}  v_{qr}\bz_r^T\underline{\bA}_{rj}(\lambda_{i'})\underline{\bZ}_{rj}v_{q'}\Big]^2 \leq \calK \mE\Big[p^{-1}\sum\limits_{r=1}^{n}  v_{qr}\bz_r^T\underline{\bA}_{rj}(\lambda_{i'})\underline{\bZ}_{rj}v_{q'}\Big]^2 \\
&=O(n^{-1} \|v_q\|_2^2 \|v_{q'}\|_2^2). 
\end{align*}
It implies that the second term in the expansion of  $\calJ_j^{(2)}$ is also $O_{L_1}(n^{-1/2} \|v_q\|_2 \|v_{q'}\|_2)$ uniformly for all $j$. 

As for $\calJ_j^{(3)}$,  due to
\begin{align*}
&\mE|\theta_{rj}(\lambda)|^2 = O(n^{-1}),\\
&\mE|p^{-1/2}\bz_r^T \bA_{rj}(\lambda_i)\underline{\bA}_{j}(\lambda_{i'})\underline{\bZ}_j v_{q'}|^2 = O(\|v_{q'}\|_2^2),
\end{align*}
we get
\begin{align*}
\calJ_j^{(3)}& = p^{-1}\mE_{j-1} \sum\limits_{r=j+1}^n v_{qr} \bz_r^T \bA_j(\lambda_i) \underline{\bA}_j(\lambda_{i'}) \underline{\bZ}_j v_{q'} \\
& = p^{-1}\mE_{j-1} \sum\limits_{r=j+1}^n v_{qr} \beta_{rj}(\lambda_i) \bz_r^T \bA_{rj}(\lambda_i) \underline{\bA}_j(\lambda_{i'})\underline{\bZ}_jv_{q'} \\ 
& = p^{-1}\mE_{j-1} \sum\limits_{r=j+1}^n v_{qr} \beta_{rj}^{\tr}(\lambda_i) \bz_r^T \bA_{rj}(\lambda_i)\underline{\bA}_{j}(\lambda_{i'})\underline{\bZ}_j v_{q'} \\
& - p^{-1}\mE_{j-1} \sum\limits_{r=j+1}^n v_{qr} \beta_{rj}(\lambda_i) \beta_{rj}^{\tr}(\lambda_i) \theta_{rj}(\lambda_i) \bz_r^T \bA_{rj}(\lambda_i)\underline{\bA}_{j}(\lambda_{i'})\underline{\bZ}_j v_{q'}\\
&= - p^{-1}\mE_{j-1} \sum\limits_{r=j+1}^n v_{qr} \beta_{rj}(\lambda_i) \beta_{rj}^{\tr}(\lambda_i) \theta_{rj}(\lambda_i) \bz_r^T \bA_{rj}(\lambda_i)\underline{\bA}_{j}(\lambda_{i'})\underline{\bZ}_j v_{q'} \\
&= O_{L_1}\Big(n^{-1/2} \|v_q\|_2 \|v_{q'}\|_2\Big). 
\end{align*}

Moreover, when $j=1$, it is straightforward to verify that 
\[ \mE \calH_{iq1}^{(1)}\calH_{i'q'1}^{(1)} = O_{L_1}\Big(n^{-3/16}\|u_q\|_2 \|u_{q'}\|_2 \|v_q\|_2 \|v_{q'}\|_2 \Big).\]

It follows then
\begin{align*}
   \sum_{j=1}^n \mE_{j-1} \calH_{iqj}^{(1)} \calH_{i'q'j}^{(1)}  &=\sum_{j=1}^n u_{qj} u_{q'j} \beta^{\mE}(\lambda_i) \beta^{\mE}(\lambda_{i'}) \Big( \sum_{r=1}^{j-1} v_{qr}v_{q'r} \Big) p^{-1} \tr\mE_{j-1} \Big[ \bA_j(\lambda_i)\underline{\bA}_{j}(\lambda_{i'}) \Big] \\
    &+ O_{L_1} \Big( n^{-3/16} \|u_q\|_2\|u_{q'}\|_2 \|v_{q}\|_2 \|v_{q'}\|_2 \Big)\\
    &=\sum_{j=1}^n u_{qj} u_{q'j} \beta^{\mE}(\lambda_i) \beta^{\mE}(\lambda_{i'}) \Big( \sum_{r=1}^{j-1} v_{qr}v_{q'r} \Big) p^{-1} \tr\mE_{j-1} \Big[ \mE\bA_j(\lambda_i)\mE_j{\bA}_{j}(\lambda_{i'}) \Big] \\
    &+ O_{L_1} \Big( n^{-3/16} \|u_q\|_2\|u_{q'}\|_2 \|v_{q}\|_2 \|v_{q'}\|_2 \Big).
\end{align*}
\eqref{eq:limit_H1H1} has been proved. 

\subsection{A High-order Moment Bound for Increments}
\label{subsec:increment_bound}
In this subsection, we fix $(m=M=1)$ and $(a_{11}=1)$, and consider the increment
\[
\Big\{
H_{\lambda}(u_1,v)-\mE H_{\lambda}(u_1,v)
\Big\}-
\Big\{
H_{\lambda}(u_2,v)-\mE H_{\lambda}(u_2,v)
\Big\},
\]
obtained by holding \(v\) fixed and varying \(u\) from \(u_1\) to \(u_2\). By the bilinearity of \(H_\lambda\), this increment can be written as
\[
H_{\lambda}(u_1-u_2,v)-\mE H_{\lambda}(u_1-u_2,v).
\]
Here, $u_1$, $u_2$, and $v$ are all regular.

We show the following increment bound
\begin{align}\label{eq:eighth_moment_bound}
    \mE \Big| \sqrt{p}\Big(H_{\lambda}(u_1- u_2, v) - \mE H_{\lambda}(u_1-u_2, v) \Big) \Big|^r = O(\|u_1- u_2\|_2^r ), \quad r\geq 8 \text{ and }r/2\in\mathbb{N}.
\end{align}
 For convenience, set $\tilde{u} = u_1 - u_2$ and $\calK_{\max} = \max_{j} \sqrt{n} |v_j|$.  

Recall the definitions of \(d_{iqj1}\), \(d_{iqj2}\), and \(d_{iqj3}\), together with all subsequent quantities appearing in their decompositions. Replacing \((u_q,v_q)\) by \((\tilde{u},v)\), we define the corresponding quantities and denote them by \(d_{j1}\), \(d_{j2}\), and \(d_{j3}\), respectively.

Then, $\sqrt{p} (H_\lambda(\tilde{u}, v) -\mE H_\lambda(\tilde{u}, v))$ is written as the following martingale difference sequence
\[ \sqrt{p} (H_{\lambda}(\tilde{u}, v) - \mE H_\lambda(\tilde{u}, v)) =  \sqrt{p} \sum_{j=1}^n (\mE_j - \mE_{j-1}) [d_{j1} + d_{j2} + d_{j3}].\]

We consider the $r$-th moment of $\sqrt{p} |d_{j1}|$, $\sqrt{p} |d_{j2}|$, and $\sqrt{p} |d_{j3}|$. Recall the decomposition of $d_{iqj1}$ in Subsection \ref{subsec:martingale_construct}. We similarly have 
\begin{align*}
    \mE\Big| \sqrt{p} d_{j1}\Big|^r &\leq \calK  |\tilde{u}_j|^r \mE \Big|p^{-1/2} \bz_j^T \bA_j(\lambda) \bZ_j v_q\Big|^r  + \calK p^{r/2} |\tilde{u}_j|^r |v_j|^r  \mE \Big|p^{-1} \bz_j^T \bA_j(\lambda) \bz_j\Big|^r \\
    & + \calK |\tilde{u}_j|^r \mE \Big| \Big(n^{-1}\bz_j^T \bA_j(\lambda)\bz_j \Big) \Big( p^{-1/2} \bz_j^T \bA_j(\lambda) \bZ_j v_q\Big)\beta_j(\lambda)\Big|^r\\
    &+ \calK p^{r/2} |\tilde{u}_{j}|^r |v_j|^r \mE  \Big| p^{-1}n^{-1} (\bz_j^T \bA_j(\lambda)\bz_j)^2 \beta_j(\lambda)\Big|^{r}. 
\end{align*}
By Lemma \ref{lemma:bound_beta}, Lemma \ref{lemma:concentration_quadrat}, and Lemma \ref{lemma:bound_z1_zalpha_b}, for any $r\geq 4$,
\begin{align*}
    &\sup_{j} \mE \Big| p^{-1/2} \bz_j^T \bA_j(\lambda) \bZ_j v_q\Big|^r \leq \calK \|v_q\|_2^r =  O(1), \\
    &\sup_{j} \mE \Big| p^{-1}\bz_j^T \bA_j(\lambda) \bz_j \Big|^r \leq  \calK = O(1).
\end{align*}
It follows then
\[ \mE \Big| \sqrt{p} d_{j1} \Big|^r \leq \calK \calK_{\max}^r |\tilde{u}_j|^r.\]

Recall the decomposition of $d_{iqj2}$ in Subsection \ref{subsec:martingale_construct}. We similarly have 
\begin{align*}
    \mE \Big| \sqrt{p} d_{j2} \Big|^r & \leq |v_{j}|^r \mE \Big| p^{-1/2} \bz_j^T \bA_j(\lambda) \bZ_j \tilde{u} \Big|^r + n^{-r/2} \mE \Big| \Big(p^{-1/2}\bz_j^T \bA_j(\lambda) \bZ_j v \Big) \Big(p^{-1/2}\bz_j^T \bA_j(\lambda) \bZ_j \tilde{u}   \Big)   \Big|^r \\
    &\leq \Big(|v_j|^r + n^{-r/2} \|v\|_2^r\Big) \|\tilde{u}\|_2^r\leq \calK \calK^r_{\max} n^{-r/2} \|\tilde{u}\|_2^r. 
\end{align*}

Lastly, 
\[\mE \Big| \sqrt{p} d_{j3} \Big|^r = |v_{j}|^r \mE \Big|p^{-1/2} \bz_j^T \bA_j(\lambda) \bZ_j \tilde{u} \Big|^r \leq \calK |v_j|^{r} \|\tilde{u}\|_2^r \leq \calK \calK^r_{\max} n^{-r/2} \|\tilde{u}\|_2^r. \]

Therefore, by Lemma \ref{lemma:Burkholder_martingale_moments}, setting $k = r/2$,
\begin{align*} 
&\mE \Big|\sqrt{p} \Big( H_\lambda(u_1 - u_2 , v) - \mE H_\lambda(u_1 - u_2, v) \Big) \Big|^{2k} \\
&\leq \calK \Big(\sum_{j_1, \cdots,j_k=1}^n  \Big( \mE  |\sqrt{p}d_{j_11}|^{2k} \times \cdots \times \mE |\sqrt{p}d_{j_k1}|^{2k}  \Big)^{1/k}  \Big)\\
&\quad + \calK \Big(\sum_{j_1, \cdots,j_k=1}^n \Big( \mE |\sqrt{p}d_{j_12}|^{2k} \times \cdots \times \mE |\sqrt{}d_{j_k2}|^{2k}  \Big)^{1/k}  \Big)\\
&\quad + \calK \Big(\sum_{j_1, \cdots,j_k=1}^n \Big( \mE |\sqrt{p}d_{j_13}|^{2k} \times \cdots \times \mE |\sqrt{}d_{j_k3}|^{2k}  \Big)^{1/k}  \Big)\\
& \quad + \calK \sum_{j=1}^n \Big[\mE |\sqrt{p} d_{j1}|^{2k} + \mE |\sqrt{p} d_{j2}|^{2k} + \mE |\sqrt{p} d_{j3}|^{2k}\Big]\\
&\leq \calK \sum_{j_1, \cdots,j_k=1}^n |\tilde{u}_{j_1}\times \cdots \times \tilde{u}_{j_k}|^{2}  + \calK \sum_{j=1}^n |\tilde{u}_j|^{2k} \\
&\leq \calK \| \tilde{u}\|_2^{2k} + \calK \sum_{j=1}^n |\tilde{u}_j|^{2k} \leq \calK \|\tilde{u}\|_2^{2k}.
\end{align*}

\subsection{Asymptotic normality of $\calH$}\label{subsec:asymptotic_normality_calH}

In this section, we establish the following asymptotic normality result.

\begin{theorem}
\label{thm:asymptotic_normality_general_form}
Suppose that Conditions~\ref{enum:moment_conditions}--\ref{enum:convergence_rate} hold. Let $\calH$ be as in \eqref{eq:def_calH} for some fixed collections of regular vectors $\{u_q\}$ and $\{v_q\}$. Then,   
\[\sqrt{p}\frac{\displaystyle \calH -  
\mathfrak{M}}{\mathfrak{V}^{1/2}}\stackrel{D}{\longrightarrow}
N(0,1),
\]
where 
\[\mathfrak{M} = \sum_{q=1}^m\sum_{i=1}^M a_{iq} u_q^T v_q \frac{\Omega(-\lambda_i) -1}{\gamma \Omega(-\lambda_i)}, \]
\[\mathfrak{V} = \sum_{i,i'}\sum_{q,q'} a_{iq}a_{i'q'} \Big[ (u_q^T u_{q'}) (v_q^T v_{q'} ) + (u_q^T v_{q'})(v_q^T u_{q'})\Big] \Gamma(-\lambda_i, -\lambda_{i'}) .\] 
\end{theorem}

First of all, results in Section \ref{subsec:convergence_mean} show that 
\[ \sqrt{p} \Big| \mE \calH - \mathfrak{M} \Big| = o(1).\]
Moreover, results in Section \ref{subsec:martingale_construct} show that 
\[ \sqrt{p} \Big(\calH - \mE\calH\Big) = \sum_{j=1}^n \sum_{i=1}^M \sum_{q=1}^m a_i \beta^{\mE}(\lambda_i) \Big[ \calH_{iqj}^{(1)} + \calH_{iqj}^{(2)} + \calH_{iqj}^{(3)} \Big] + o_p(1). \]
Therefore, it is enough to show that 
\[ \frac{ \sum_{j=1}^n \sum_{i=1}^M \sum_{q=1}^m a_i \beta^{\mE}(\lambda_i) \Big[ \calH_{iqj}^{(1)} + \calH_{iqj}^{(2)} + \calH_{iqj}^{(3)} \Big]}{\mathfrak{B}^{1/2}}  \stackrel{D}{\longrightarrow}N(0,1). \]

To this end, we use the following martingale central limit theorem. 
\begin{theorem}(Theorem 35.12 of \citet{billingsley1995probability}).
\label{thm:martingale_central_limit_theorem}
 Suppose $Y_{1},\dots,Y_{n}$ is a martinagle difference sequence with respsect to the increasing $\sigma$-field $\mathscr{F}_1,\dots,\mathscr{F}_n$, with finite second moments. If as $n\to\infty$, 
\begin{itemize}
\item[(i)] $\sum\limits_{j=1}^n \mE (Y_j^2\mid \mathscr{F}_{j-1})\to \sigma^2$ in probability where $\sigma^2$ is a positive constant,
\item[(ii)] $\sum\limits_{j=1}^n \mE\{Y^2_j \mathbbm{1}(|Y_j|\geq\epsilon)\}\to0 $ for each $\epsilon>0$,
\end{itemize}
then, 
\[\sum\limits_{j=1}^n Y_j \to \mathcal{N}(0,\sigma^2),\quad \mbox{ in distribution}.\]
\end{theorem}

We first verify Condition~(ii). Since $m$ and $M$ are fixed, it suffices to show that, for every $\epsilon>0$,
\begin{align*}
&
\sum_{j=1}^n
\mE\left\{
\left(
\sum_{i=1}^M\sum_{q=1}^m \sum_{g=1,2,3}
a_i\beta^{\mE}(\lambda_i)\calH_{iqj}^{(g)}
\right)^2
\mathbbm{1}
\left(
\left|
\sum_{i=1}^M\sum_{q=1}^m \sum_{g=1,2,3}
a_i\beta^{\mE}(\lambda_i)\calH_{iqj}^{(g)}
\right|>\epsilon
\right)
\right\}
\\
&\qquad\leq
\frac{\calK}{\epsilon^2}
\max_{1\le q\le m}\max_{g=1,2,3}
\sum_{j=1}^n
\mE
\left(
\sum_{i=1}^M
a_i\beta^{\mE}(\lambda_i)\calH_{iqj}^{(g)}
\right)^4
\longrightarrow0.
\end{align*}

For each fixed $i$ and $q$, 
\[ \mE \Big| p^{-1/2} u_{qj} \bz_j^T \bA_j(\lambda_i) \bZ_j v_q\Big|^4  = O(n^{-2}).\]
\[ \mE \Big| p^{-1/2} v_{qj} \bz_j^T \bA_j(\lambda_{i}) \bZ_j u_q \Big|^4 = O(n^{-2}).\]
\begin{align*}
&\mE \Big| p^{-1/2} \Big[\frac{1}{n} \bz_j^T \bA_j(\lambda_i)\bZ_j u_q v_q^T \bZ_j^T \bA_j(\lambda_i)\bz_j - \frac{1}{n}u_q^T \bZ_j^T \bA_j^2(\lambda_i)\bZ_j v_q \Big] \Big|^4 \\
&\leq \calK n^{-2} \mE \Big| n^{-1} u_q^T \bZ_j^T \bA^2_j(\lambda_i) \bZ_j v_q\Big|^4\\
&\leq \calK n^{-2} \Big(\mE\Big|n^{-1} u_q^T \bZ^T_j \bA^2_j(\lambda_i) \bZ_j u_q \Big|^4  \mE\Big|n^{-1} v_q^T \bZ^T_j \bA^2_j(\lambda_i) \bZ_j v_q \Big|^4   \Big)^{1/2}\\
&\leq \calK n^{-2} \Big(\mE\Big|n^{-1} u_q^T \bZ^T_j \bZ_j u_q \Big|^4  \mE\Big|n^{-1} v_q^T \bZ^T_j \bZ_j v_q \Big|^4   \Big)^{1/2} = O(n^{-2}).
\end{align*}
It follows then 
\[\sum_{j=1}^n  \mE \Big|\sum_{i=1}^M \sum_{q=1}^m a_i \beta^{\mE}(\lambda_i) \calH_{iqj}^{(1)}  \Big|^4 = O(n^{-1}),\]
\[\sum_{j=1}^n  \mE \Big|\sum_{i=1}^M \sum_{q=1}^m a_i \beta^{\mE}(\lambda_i) \calH_{iqj}^{(2)}  \Big|^4 = O(n^{-1}).\]
\[\sum_{j=1}^n  \mE \Big|\sum_{i=1}^M \sum_{q=1}^m a_i \beta^{\mE}(\lambda_i) \calH_{iqj}^{(3)}  \Big|^4 = O(n^{-1}).\]
It completes the proof of Condition (ii).

It remains to verify Condition (i). To this end, we need to find the limit in probability of the condition variance
\[ \sum_{j=1}^n \mE_{j-1} \Big(\sum_{g=1,2,3} \sum_{i=1}^M \sum_{q=1}^m a_i \beta^{\mE}(\lambda_i) \calH^{(g)}_{iqj} \Big)^2,\]
which is already studied in Subsection \ref{subsec:variance_martingale}. In particular, we showed \eqref{eq:limit_H1H1}--\eqref{eq:limit_H3H3}. It remains to deduce the limit of 
\[\beta^{\mE}(\lambda)\qquad \text{and}\qquad  n^{-1}\tr[\mE_j \bA_j(\lambda_i) \mE_j \bA_{j}(\lambda_{i'})].\]

Using arguments from \citet{bai2004clt}, it is established in \citet{li2020high} that for any fixed $\lambda, \lambda_1,\lambda_2 >0$,
\[ \beta^{\mE}(\lambda) \longrightarrow \Omega^{-1}(-\lambda),\]
\[ \sup_{j} \mE \Big|\beta^{\mE}(\lambda_i) \beta^{\mE}(\lambda_{i'}) \frac{1}{n} \tr [\mE_j \bA_j(\lambda_i) \mE_j \bA_j (\lambda_{i'})] - \frac{\calB(-\lambda_1, -\lambda_2)}{1- n^{-1}(j-1) \calB(-\lambda_1,-\lambda_2)} \Big| \longrightarrow 0,\]
where 
\[\displaystyle \calB(z_1, z_2) =  1+ \frac{z_1- z_2}{ \big[z_1^{-1} (\gamma-1) + \gamma \varphi(z_1) \big]^{-1}\big[z_2^{-1} (\gamma-1) + \gamma \varphi(z_2) \big]^{-1} }.\]
See Section S.2 and S.3.1.7 of \citet{li2020high} for details. 

Combining the preceding results with the deterministic limits of
$\beta^{\mE}(\lambda)$ and
\[
\frac{1}{n}
\beta^{\mE}(\lambda_i)
\beta^{\mE}(\lambda_{i'})
\tr\left[
\mE_j\bA_j(\lambda_i)
\mE_j\bA_j(\lambda_{i'})
\right],
\]
we obtain
\[
\sum_{g,g'=1}^3
\sum_{j=1}^n
\beta^{\mE}(\lambda_i)
\beta^{\mE}(\lambda_{i'})
\mE_{j-1}
\left[
\calH_{iqj}^{(g)}
\calH_{i'q'j}^{(g')}
\right]
\stackrel{P}{\longrightarrow}
\Gamma(-\lambda_i,-\lambda_{i'})
\left[
(u_q^Tu_{q'})(v_q^Tv_{q'})
+
(u_q^Tv_{q'})(v_q^Tu_{q'})
\right].
\]
The derivation of this limit is lengthy but otherwise routine, following the same arguments as those in Section~S.3.1.7 of \citet{li2020high}. We therefore omit the details.

Summing over all pairs of $i,i'$ and $q,q'$, we obtain that the asymptotic variance of $\sqrt{p}(\calH -\mE \calH)$ is 
\begin{align*}
\mathfrak{B} = \sum_{i,i'=1}^M 
\sum_{q,q'=1}^m a_{iq} a_{i'q'} \left[
(u_q^Tu_{q'})(v_q^Tv_{q'})
+
(u_q^Tv_{q'})(v_q^Tu_{q'})\right] \Gamma(-\lambda_i,-\lambda_{i'}).
\end{align*}

\subsection{Weak Convergence of the Processes}
\label{subsec:weak_convergence_process}

In this subsection, we establish the weak convergence of the processes
\[
\big\{H_\lambda(u(t)):\lambda\in\Lambda,\ t\in\calT_{\rm mc}(\varepsilon)\big\}
\qquad\text{and}\qquad
\big\{H_\lambda(u(t),h):\lambda\in\Lambda,\ t\in\calT_{\rm mc}(\varepsilon)\big\}.
\]

We first establish finite-dimensional convergence. Fix \(M,m\in\mathbb{N}_+\), \(\lambda_1,\ldots,\lambda_M\in\Lambda\), and \(t_1,\ldots,t_m\in\calT_{\rm mc}(\varepsilon)\). By the Cram\'{e}r--Wold device, it suffices to study, for an arbitrary deterministic set of coefficients \(\{a_{iq}\}\), the linear combination
\begin{align*}
\sum_{i=1}^M\sum_{q=1}^m
a_{iq}H_{\lambda_i}(u(t_q))
&=
p^{-1}
\sum_{i=1}^M\sum_{q=1}^m
a_{iq}\,
u^T(t_q)\bZ^T\bA(\lambda_i)\bZ u(t_q).
\end{align*}
Similarly, for the second process, it suffices to consider
\begin{align*}
\sum_{i=1}^M\sum_{q=1}^m
a_{iq}H_{\lambda_i}(u(t_q),h)
&=
p^{-1}
\sum_{i=1}^M\sum_{q=1}^m
a_{iq}\,
u^T(t_q)\bZ^T\bA(\lambda_i)\bZ h.
\end{align*}
Both expressions are in the form of the superposition \(\calH\) considered in Subsection~\ref{subsec:asymptotic_normality_calH}. The asymptotic normality established in Theorem \ref{thm:asymptotic_normality_general_form} therefore yields the required finite-dimensional convergence.

Second, we establish the stochastic equicontinuity of the standardized and scaled processes
\[
\{H_{\lambda}(u(t)):\lambda\in\Lambda,\ t\in\calT_{\rm mc}(\varepsilon)\}
\qquad\text{and}\qquad
\{H_{\lambda}(u(t),h):\lambda\in\Lambda,\ t\in\calT_{\rm mc}(\varepsilon)\}.
\]
Since $\Lambda$ is finite, tightness with respect to
$(\lambda,t)\in\Lambda\times\calT_{\rm mc}(\varepsilon)$ follows once
tightness is established for each fixed $\lambda\in\Lambda$. We therefore
fix $\lambda\in\Lambda$ and consider the processes
\[
\left\{
\sqrt{p}\left[
H_{\lambda}(u(t))
-u(t)^Tu(t)
\frac{\Omega(-\lambda)-1}{\gamma\Omega(-\lambda)}
\right]:
t\in\calT_{\rm mc}(\varepsilon)
\right\}
\]
and
\[
\left\{
\sqrt{p}\,H_{\lambda}(u(t),h):
t\in\calT_{\rm mc}(\varepsilon)
\right\},
\]
where we have used the fact that $u(t)^Th=0$ for every
$t\in\calT_{\rm mc}(\varepsilon)$.

Note that $\calT_{\rm mc}(\varepsilon)\subset[0,1]^3$. Since the indexing
set is three-dimensional, by the multi-parameter Kolmogorov--Chentsov
criterion, it suffices to establish tightness at a fixed point
$t_0\in\calT_{\rm mc}(\varepsilon)$, together with an increment moment
bound whose exponent in $\|t-t'\|_2$ is strictly greater than three (see for example, \citet[Theorem 2.5.1]{khoshnevisan2002multiparameter}).
Thus, we first establish tightness of
\[
\sqrt{p}\left[
H_{\lambda}(u(t_0))
-u(t_0)^Tu(t_0)
\frac{\Omega(-\lambda)-1}{\gamma\Omega(-\lambda)}
\right]
\qquad\text{and}\qquad
\sqrt{p}\,H_{\lambda}(u(t_0),h).
\]
It directly follows from the finite-dimensional asymptotic normality of the processes. 

It remains to establish the following eighth-order increment bounds:
\[
\begin{aligned}
&\mE\Bigg|
\sqrt{p}\left[
H_{\lambda}(u(t))
-u(t)^Tu(t)
\frac{\Omega(-\lambda)-1}{\gamma\Omega(-\lambda)}
\right]
-
\sqrt{p}\left[
H_{\lambda}(u(t'))
-u(t')^Tu(t')
\frac{\Omega(-\lambda)-1}{\gamma\Omega(-\lambda)}
\right]
\Bigg|^8\\
&\qquad\leq
\calK\left(
\|t-t'\|_2+\frac{1}{n}
\right)^4,
\qquad t,t'\in\calT_{\rm mc}(\varepsilon),
\end{aligned}
\]
and
\[
\mE\left|
\sqrt{p}\,H_{\lambda}(u(t),h)
-\sqrt{p}\,H_{\lambda}(u(t'),h)
\right|^8
\leq
\calK\left(
\|t-t'\|_2+\frac{1}{n}
\right)^4,
\qquad t,t'\in\calT_{\rm mc}(\varepsilon).
\]

These bounds follow from \eqref{eq:mean_convergence_bound} and
\eqref{eq:eighth_moment_bound}. Indeed, by the definition
$k(t)=\lfloor nt\rfloor+1$, we have
\[
|k(t_i)-k(t_i')|
\leq n|t_i-t_i'|+1,
\qquad i=1,2,3.
\]
Since all segment lengths are bounded below by a constant multiple of
$n$ on $\calT_{\rm mc}(\varepsilon)$, it follows that, for some constant
$\calK>0$,
\[
\|u(t)-u(t')\|_2^2
\leq
\calK\left(
\|t-t'\|_2+\frac{1}{n}
\right),
\qquad
t,t'\in\calT_{\rm mc}(\varepsilon).
\]
Moreover,
\[
u(t)^Th=0,
\qquad t\in\calT_{\rm mc}(\varepsilon).
\]

Finally, since $u(t)$ depends on $t$ only through
$(k(t_1),k(t_2),k(t_3))$, it is sufficient to consider the effective
$n^{-1}$-grid of distinct values of the process. For any two distinct
points $t$ and $t'$ on this grid,
\[
\|t-t'\|_2\geq \frac{1}{n},
\]
and hence
\[
\|t-t'\|_2+\frac{1}{n}
\leq 2\|t-t'\|_2.
\]
Consequently, the preceding increment bounds imply
\[
\begin{aligned}
&\mE\Bigg|
\sqrt{p}\left[
H_{\lambda}(u(t))
-u(t)^Tu(t)
\frac{\Omega(-\lambda)-1}{\gamma\Omega(-\lambda)}
\right]\\
&\qquad\qquad-
\sqrt{p}\left[
H_{\lambda}(u(t'))
-u(t')^Tu(t')
\frac{\Omega(-\lambda)-1}{\gamma\Omega(-\lambda)}
\right]
\Bigg|^8
\leq
\calK\|t-t'\|_2^4,
\end{aligned}
\]
and
\[
\mE\left|
\sqrt{p}\,H_{\lambda}(u(t),h)
-\sqrt{p}\,H_{\lambda}(u(t'),h)
\right|^8
\leq
\calK\|t-t'\|_2^4.
\]
Since $4>3$, the required stochastic equicontinuity follows from the multi-parameter Kolmogorov--Chentsov criterion.

In summary, we have established that
\[
\left\{
\sqrt{p}\left[
\|u(t)\|_2^{-2}H_\lambda(u(t))
-
\frac{\Omega(-\lambda)-1}
{\gamma\Omega(-\lambda)}
\right]
:
\lambda\in\Lambda,\ t\in\calT_{\rm mc}(\varepsilon)
\right\}
\rightsquigarrow
\left\{
G_\lambda(t):
\lambda\in\Lambda,\ t\in\calT_{\rm mc}(\varepsilon)
\right\}
\]
in $\ell^\infty(\Lambda\times\calT_{\rm mc}(\varepsilon))$. The asymptotic covariance kernel of the limiting Gaussian process can be obtained from the covariance formula established above, together with the normalization
$\mathfrak{B}^{1/2}$. The resulting calculation is straightforward but lengthy, and we therefore omit the details.

Moreover,
\[
\sup_{\substack{t\in\calT_{\rm mc}(\varepsilon)\\ \lambda\in\Lambda}}
\sqrt{p}\,
\bigl|H_\lambda(u(t),h)\bigr|
=
O_p(1).
\]

Lastly, we transform back to the process $p^{-1}V_{\lambda}(t)$. Recall the relationship in \eqref{eq:relation_V_H}. By the asymptotic normality of $H_{\lambda}(h)$ established in Subsection~\ref{subsec:asymptotic_normality_calH}, and since $\Lambda$ is finite,
\[
\max_{\lambda\in\Lambda}
\left|
H_\lambda(h)
-
h^Th\frac{\Omega(-\lambda)-1}
{\gamma\Omega(-\lambda)}
\right|
=
O_p(p^{-1/2}).
\]
Since $h^Th=1$ and $p/n = \gamma +o(p^{-1/2})$, it follows that, uniformly over
$\lambda\in\Lambda$,
\[
1-\frac{p}{n}H_\lambda(h)
=
\frac{1}{\Omega(-\lambda)}+o_p(1).
\]

Recall that it is demonstrated in Lemma \ref{lemma:estimate_Stieltjes} that for any
$\lambda\in\mathbb{R}_+$,
\[
\inf_{\lambda\in \Lambda }|\Omega^{-1}(-\lambda)| = \inf_{\lambda\in \Lambda }|1-\gamma + \gamma \lambda \varphi(-\lambda)| > c, \quad \text{for some constant }c>0.
\]
Consequently,
\[
\mP\left(
\min_{\lambda\in\Lambda}
\left|1-\frac{p}{n}H_\lambda(h)\right|>c/2
\right)
\longrightarrow 1.
\]

Moreover, from the stochastic boundedness established above,
\[
\sup_{\substack{t\in\calT_{\rm mc}(\varepsilon)\\\lambda\in\Lambda}}
\sqrt{p}\,
|H_\lambda(u(t),h)|
=
O_p(1),
\]
and hence
\[
\sup_{\substack{t\in\calT_{\rm mc}(\varepsilon)\\\lambda\in\Lambda}}
\sqrt{p}\,
\bigl(H_\lambda(u(t),h)\bigr)^2
=
o_p(1).
\]
Together with \eqref{eq:relation_V_H} and the fact that
$\|u(t)\|_2$ is uniformly bounded away from zero over
$t\in\calT_{\rm mc}(\varepsilon)$, this yields
\[
\sup_{\substack{t\in\calT_{\rm mc}(\varepsilon)\\\lambda\in\Lambda}}
\sqrt{p}\left|
p^{-1}V_\lambda(t)
-
\frac{H_\lambda(u(t))}{\|u(t)\|_2^2}
\right|
=o_p(1).
\]
Equivalently,
\[
p^{-1}V_\lambda(t)
=
\frac{H_\lambda(u(t))}{\|u(t)\|_2^2}
+
o_p(p^{-1/2}),
\]
uniformly over $(\lambda,t)\in\Lambda\times\calT_{\rm mc}(\varepsilon)$.

Therefore, by Slutsky's theorem,
\[
\left\{
\sqrt{p}\left[
p^{-1}V_\lambda(t)
-
\frac{\Omega(-\lambda)-1}
{\gamma\Omega(-\lambda)}
\right]
:
\lambda\in\Lambda,\ t\in\calT_{\rm mc}(\varepsilon)
\right\}
\rightsquigarrow
\left\{
G_\lambda(t):
\lambda\in\Lambda,\ t\in\calT_{\rm mc}(\varepsilon)
\right\}
\]
in $\ell^\infty(\Lambda\times\calT_{\rm mc}(\varepsilon))$.

\section{Proof of Lemma \ref{lemma:Stieltjes_under_alternative}}
\label{sec:proof_stieltjes_under_alternative}

Under $H_a$, recall the definition
\begin{equation}
\label{eq:def_calU}
\calU
=
\Big(
\mu_1,\mu_{k_1+1},\mu_{k_2+1},\dots,\mu_{k_s+1}
\Big),
\end{equation}
which collects the $s+1$ distinct mean vectors. Let
\begin{equation}
\label{eq:def_calV}
\calV
=
\Big(
\underbrace{e_{1},\dots,e_{1}}_{k_1},
\underbrace{e_{2},\dots,e_{2}}_{k_2-k_1},
\dots,
\underbrace{e_{s+1},\dots,e_{s+1}}_{n-k_s}
\Big).
\end{equation}
Here, we recall that $e_i$ is the $(s+1)$-dimensional canonical vector whose $i$th entry is one and all other entries are zero. With these definitions,
the observation matrix $X=(X_1,\dots,X_n)$ can be written as
\[
X=\calU\calV+\Sigma_p^{1/2}\bZ,
\]
 In what follows, we treat $\bZ$ as
fixed, since the argument relies only on a deterministic rank bound that holds uniformly for any fixed $\bZ$.

Let
\[
P_0=I_n-n^{-1}\mathbbm{1}_n\mathbbm{1}_n^T.
\]
The pooled sample covariance matrix can then be decomposed as
\begin{align*}
S_n
&=
\frac{1}{n}\sum_{j=1}^n
(X_j-\bar X)(X_j-\bar X)^T
=
\frac{1}{n}X P_0X^T\\
&=
\frac{1}{n}
\Sigma_p^{1/2}\bZ P_0\bZ^T\Sigma_p^{1/2}
+
\frac{1}{n}
\Sigma_p^{1/2}\bZ P_0\calV^T\calU^T\\
&\qquad
+
\frac{1}{n}
\calU\calV P_0\bZ^T\Sigma_p^{1/2}
+
\frac{1}{n}
\calU\calV P_0\calV^T\calU^T\\
&=
\tilde S_n
+
\frac{1}{n}
\Sigma_p^{1/2}\bZ P_0\calV^T\calU^T
+
\frac{1}{n}
\calU\calV P_0\bZ^T\Sigma_p^{1/2}
+
\frac{1}{n}
\calU\calV P_0\calV^T\calU^T.
\end{align*}
This decomposition shows that
\[
R_n\coloneqq S_n-\tilde S_n
\]
has rank at most $3(s+1)$, since each of the three perturbation terms
has rank at most $\operatorname{rank}(\calU)\leq s+1$. Hence,
\[
\operatorname{rank}(R_n)\leq 3(s+1).
\]

Recall that $F_p$ denotes the empirical spectral distribution of
$S_n$, and let $\tilde F_p$ denote that of $\tilde S_n$.
By Lemma~\ref{lemma:rank_one_perturbation},
\[
\|F_p-\tilde F_p\|_\infty
\leq
\frac{1}{p}\operatorname{rank}(R_n)
\leq
\frac{3(s+1)}{p}
\longrightarrow 0.
\]

For any $\lambda>0$, let
\[
\Delta(\tau)=F_p(\tau)-\tilde F_p(\tau).
\]
Then
\[
m_n(-\lambda)-\tilde m_n(-\lambda)
=
\int\frac{1}{\tau+\lambda}\,d\Delta(\tau).
\]
Since both $S_n$ and $\tilde S_n$ are positive semidefinite,
$\Delta(\tau)=0$ for $\tau<0$, while
$\Delta(\infty)=0$. Integration by parts therefore gives
\[
\begin{aligned}
\left|
m_n(-\lambda)-\tilde m_n(-\lambda)
\right|
=
\left|
\int_0^\infty
\frac{\Delta(\tau)}{(\tau+\lambda)^2}\,d\tau
\right|
\leq
\|\Delta\|_\infty
\int_0^\infty\frac{1}{(\tau+\lambda)^2}\,d\tau\leq
\frac{3(s+1)}{p\lambda}.
\end{aligned}
\]
Similarly, since
\[
m_n'(-\lambda)
=
\int\frac{1}{(\tau+\lambda)^2}\,dF_p(\tau),
\]
another integration by parts yields
\[
\left|
m_n'(-\lambda)-\tilde m_n'(-\lambda)
\right|
\leq
2\|\Delta\|_\infty
\int_0^\infty\frac{1}{(\tau+\lambda)^3}\,d\tau
\leq
\frac{3(s+1)}{p\lambda^2}.
\]
Consequently, for $\lambda$ bounded away from zero,
\[
m_n(-\lambda)-\tilde m_n(-\lambda)=O(p^{-1}),
\qquad
m_n'(-\lambda)-\tilde m_n'(-\lambda)=O(p^{-1}).
\]
The corresponding bounds for
$\widehat\Theta(\lambda)-\tilde\Theta(\lambda)$ and
$\widehat\Gamma(\lambda)-\tilde\Gamma(\lambda)$ then follow
from their definitions as smooth functions of
$m_n(-\lambda)$ and $m_n'(-\lambda)$.

Lastly, let $A$ be any deterministic matrix with bounded spectral
norm. By the resolvent identity,
\[
(S_n+\lambda I_p)^{-1}
-
(\tilde S_n+\lambda I_p)^{-1}
=
-(S_n+\lambda I_p)^{-1}
R_n
(\tilde S_n+\lambda I_p)^{-1},
\]
and hence
\[
\operatorname{rank}\left(
(S_n+\lambda I_p)^{-1}
-
(\tilde S_n+\lambda I_p)^{-1}
\right)
\leq
\operatorname{rank}(R_n)
\leq 3(s+1).
\]
Therefore,
\begin{align*}
&\left|
p^{-1}\tr\big[(S_n+\lambda I_p)^{-1}A \big]
-
p^{-1}\tr\big[(\tilde S_n+\lambda I_p)^{-1}A \big]
\right|\\
&\qquad\leq
\frac{1}{p}
\left\|
(S_n+\lambda I_p)^{-1}
-
(\tilde S_n+\lambda I_p)^{-1}
\right\|_2
\|A\|_2\,
\operatorname{rank}(R_n)
\leq
\frac{6(s+1)}{p\lambda}\|A\|_2.
\end{align*}
It follows that the convergence results in
Lemma~\ref{lemma:estimate_Stieltjes} continue to hold under $H_a$.

\section{Proof of Theorems \ref{thm:power_single_deterministic} and \ref{thm:power_multiple_deterministic}}
\label{sec:proof_power_single_deterministic}

In this section, we prove Theorems~\ref{thm:power_single_deterministic} and~\ref{thm:power_multiple_deterministic}. Since the former is a special case of the latter with $s=1$, it suffices to prove
Theorem~\ref{thm:power_multiple_deterministic}.

Recall the decomposition
\[
X=\calU\calV+\Sigma_p^{1/2}\bZ
\]
established in Section~\ref{sec:proof_stieltjes_under_alternative}. Also,
recall that $\tilde{S}_n$ denotes the pooled sample covariance matrix
formed from the noise component $\Sigma_p^{1/2}\bZ$, namely,
\[
\tilde{S}_n
=
\frac{1}{n}\Sigma_p^{1/2}\bZ P_0\bZ^T\Sigma_p^{1/2}
=
\frac{1}{n}\Sigma_p^{1/2}\bZ\bZ^T\Sigma_p^{1/2}
-
\Sigma_p^{1/2}\bar{\bz}\bar{\bz}^T\Sigma_p^{1/2}.
\]
Recall also the decomposition of $S_n$ given in
Section~\ref{sec:proof_stieltjes_under_alternative}. In particular,
\[
R_n
\coloneqq
S_n-\tilde{S}_n
=
\frac{1}{n}\Sigma_p^{1/2}\bZ P_0\calV^T\calU^T
+
\frac{1}{n}\calU\calV P_0\bZ^T\Sigma_p^{1/2}
+
\frac{1}{n}\calU\calV P_0\calV^T\calU^T.
\]
Thus, $\tilde{S}_n$ is the counterpart of $S_n$ under the null
hypothesis and is the covariance matrix analyzed in
Section~\ref{subsec:weak_convergence_process}. Therefore, the results
established in Section~\ref{sec:proof_weak_convergence_process} apply
to processes constructed from $\tilde{S}_n$. 
In particular,
\begin{align}
&\left\{
\frac{\sqrt{p}}{\sqrt{\Gamma(-\lambda)}}
\left[
\frac{1}{p\|u(t)\|_2^2}
u(t)^T\bZ^T\Sigma_p^{1/2}
(\tilde S_n+\lambda I_p)^{-1}
\Sigma_p^{1/2}\bZ u(t)
-\Theta(-\lambda)
\right]
\right\}
\rightsquigarrow
\{G_\lambda(t)\},
\label{eq:proof_alternative1}
\end{align}

We first characterize the tightness of several processes involving $\calU$ and $\calV$. Under the investigated class of local alternatives, 
\[\sqrt{p}\, \big\|\calU^T\calD_p(-\lambda)\calU\big\|_2 \asymp 1, \qquad \lambda\in\Lambda,\]
Since the deterministic equivalent $\calD_p(-\lambda)$ is such that 
\[\frac{1}{(1+\sqrt{\gamma})^2\limsup_p\alpha_{\max}(\Sigma_p)+\lambda}\leq \alpha_{\min}\bigl(\calD_p(-\lambda)\bigr) \leq \alpha_{\max}\bigl(\calD_p(-\lambda)\bigr) \leq\frac{1}{\lambda}.\]
Therefore, under the assumed regime
we have
\[p^{1/4}\,\|\calU\|_2\asymp 1.\]
Since the columns of $\calU$ consist of all distinct mean vectors under
the alternative, this further implies
\[\sup_{1\leq j\leq n} \sqrt{p}\,\|\mu_j\|_2^2 \asymp 1.\]

For $q=1,\dots, s+1$, call the $q$-th column of $p^{-1/2} P_0\calV^T$ as $v_q$. It is straightforward that $v_q$ is regular. Therefore, for any regular vector $u$ 
\[  \mE \Big| p^{-1} u^T \bZ^T \Sigma_p^{1/2} (\tilde{S}_n + \lambda I_p)^{-1} \Sigma_p^{1/2} \bZ v_q \Big|^{r}  = O(\|u\|_2^{r} \|v_q\|_2^r), \quad  r \geq 2. \]
The result is obtained by directly applying Lemma \ref{lemma:zalpha_a_zalpha_a}. It implies that the process
\begin{equation}\label{eq:tight_process_1}
\Big\{p^{-3/2} u(t)^T \bZ^T \Sigma_p^{1/2} (\tilde{S}_n + \lambda I_p)^{-1} \Sigma_p^{1/2} \bZ P_0 \calV^T, ~\lambda \in \Lambda,~  t\in \calT_{\rm mc}(\varepsilon) \Big\}  
\end{equation}
is tight. For each $(\lambda,t)\in\Lambda\times\calT_{\rm mc}(\varepsilon)$, the process takes values in $\mathbb R^{s+1}$. Since $s$ is fixed, tightness of the vector-valued process follows from tightness of each of its $s+1$ coordinate processes in $\ell^\infty(\Lambda\times\calT_{\rm mc}(\varepsilon))$.

Secondly, using analogous arguments as those in Subsection \ref{subsec:increment_bound}, we can show that for any regular vector $u$
\begin{align*}
\mE \Big\| u^T \bZ^T \Sigma_p^{1/2} (\tilde{S}_n + \lambda I_p)^{-1} \calU   \Big\|_2^{r} \leq \|u\|_2^r  \|\calU\|_2^r, \quad r\geq 2. 
\end{align*}
We omit the details in the proof. It implies that the process
\begin{equation}\label{eq:tight_process_2}
\Big\{ n^{1/4} u(t)^T \bZ^T \Sigma_p^{1/2} (\tilde{S}_n + \lambda I_p)^{-1} \calU, ~ \lambda \in \Lambda, ~t\in \calT_{\rm mc}(\varepsilon)\Big\}  
\end{equation}
is tight. Again, the process is $(s+1)$-dimensional for each $(\lambda,t)\in\Lambda\times\calT_{\rm mc}(\varepsilon)$.

Thirdly, Lemma \ref{lemma:zalpha_a_zalpha_a} implies that 
\[ \Big\{p^{-1} u(t)^T \bZ^T \bZ^T u(t), ~t\in \calT_{\rm mc}(\varepsilon) \Big\} \]
is tight. Therefore, we conclude that the two processes
\begin{align}
\Big\{p^{-1} u(t)^T \bZ^T (S_n+\lambda I_p)^{-1} \bZ u(t), ~t\in \calT_{\rm mc}(\varepsilon) \Big\} \label{eq:tight_process_3} \\
\Big\{p^{-1} u(t)^T \bZ^T (\tilde{S}_n+\lambda I_p)^{-1} \bZ u(t), ~t\in \calT_{\rm mc}(\varepsilon) \Big\} \label{eq:tight_process_4} 
\end{align} 
are tight. Here, we use the deterministic bound $\| (S_n +\lambda I_p)^{-1} \|_2 \leq 1/\lambda$ and $\| (\tilde{S}_n +\lambda I_p)^{-1} \|_2 \leq 1/\lambda$.

Lastly, we discuss the order of the spectral norm of $R_n$. For the first and the second term in the definition of $R_n$, 
\[\mE\Big\|\frac{1}{n} \Sigma_p^{1/2} \bZ P_0 \calV^T \calU^T\Big\|^r_2 \leq \|\calU\|_2^{r} \mE \Big\| \frac{1}{n^2} \calV P_0 \bZ^T \Sigma_p \bZ P_0 \calV^T \Big\|_2^{r/2} = O(\|\calU\|_2^r) = O(n^{-r/4}), \quad r\geq 2.\]
The third term is such that 
\[ \Big\| \frac{1}{n} \calU \calV P_0 \calV^T \calU^T\Big\|_2 \leq \calK \|\calU^T\calU\|_2 = O(n^{-1/2}). \]

We write $\tilde{G} = (\tilde{S}_n+ \lambda I_p)^{-1}$ and $G = (S_n + \lambda I)^{-1}$. Then, we can expand 
\[ G = \tilde{G} - \tilde{G} R_n \tilde{G} + \tilde{G} R_n \tilde{G} R_n \tilde{G} - \tilde{G} R_n \tilde{G} R_n \tilde{G} R_n G.\]
Meanwhile, as $X = \Sigma_p^{1/2} \bZ + \calU \calV$. We can decompose 
\begin{align*}
p^{-1} \|u(t)\|_2^2  V_\lambda(t) = T_0 + T_1 +T_2 + T_3,
\end{align*}
where 
\begin{align*}
    T_0 &= p^{-1} u(t)^T \bZ^T \Sigma_p^{1/2} \tilde{G} \Sigma_p^{1/2} \bZ u(t)\\
        &\quad + 2 p^{-1} u(t)^T \calV^T \calU^T \tilde{G} \Sigma_p^{1/2} \bZ u(t)\\
        &\quad + p^{-1} u(t)^T \calV^T \calU^T \tilde{G} \calU \calV u(t); 
\end{align*}
\begin{align*}
    T_1 &= - p^{-1} u(t)^T \bZ^T \Sigma_p^{1/2} \tilde{G} R_n \tilde{G} \Sigma^{1/2}_p \bZ u(t)\\
        &\quad -2 p^{-1} u(t)^T \calV^T \calU^T \tilde{G} R_n \tilde{G} \Sigma^{1/2}_p \bZ u(t)\\
        &\quad - p^{-1} u(t)^T \calV^T \calU^T \tilde{G} R_n \tilde{G} \calU \calV u(t). 
\end{align*}
\begin{align*}
    T_2 &= p^{-1} u(t)^T \bZ^T \Sigma_p^{1/2} \tilde{G} R_n \tilde{G} R_n \tilde{G} \Sigma^{1/2}_p \bZ u(t)\\
        & \quad + 2p^{-1} u(t)^T \calV^T \calU^T \tilde{G} R_n \tilde{G} R_n \tilde{G} \Sigma_p^{1/2} \bZ u(t)\\
        & \quad + p^{-1} u(t)^T \calV^T \calU^T \tilde{G} R_n \tilde{G} R_n \tilde{G} \calU \calV u(t);
\end{align*}
\begin{align*}
T_3 &= -p^{-1} u(t)^T \bZ^T \Sigma_p^{1/2} \tilde{G} R_n \tilde{G} R_n \tilde{G} R_n G \Sigma_p^{1/2} \bZ u(t)\\
    &\quad - p^{-1} u(t)^T \bZ^T \Sigma_p^{1/2} \tilde{G} R_n \tilde{G} R_n \tilde{G} R_n G \calU \calV u(t)\\
    &\quad - p^{-1} u(t)^T \calV^T \calU^T  \tilde{G} R_n \tilde{G} R_n \tilde{G} R_n G \Sigma_p^{1/2} \bZ u(t)\\
    &\quad - p^{-1} u(t)^T \calV^T \calU^T  \tilde{G} R_n \tilde{G} R_n \tilde{G} R_n G  \calU \calV u(t).
\end{align*}

Due to the tightness shown in \eqref{eq:tight_process_1}--\eqref{eq:tight_process_4} and the quantification of the spectral norm of $R_n$, it is straightforward to verify that
\[\max_{\lambda \in \Lambda} \sup_{t\in \calT_{\rm mc}(\varepsilon)} |T_3| = o_p(p^{-1/2}). \]
\[T_0 = p^{-1} u(t)^T \bZ^T \Sigma_p^{1/2} \tilde{G} \Sigma_p^{1/2} \bZ u(t) + p^{-1} u(t)^T \calV^T \calU^T \tilde{G} \calU \calV u(t) + o_p(p^{-1/2}),\]
where the above residual is uniform on $\Lambda \times \calT_{\rm mc} (\varepsilon)$.

Next, we analyze $T_1$ and $T_2$, by further expanding $R_n$. For $T_1$, it can be verified that the third term in its definition is $o_p(p^{-1/2})$ uniformly on $\Lambda \times \calT_{\rm mc} (\varepsilon)$. Again, using \eqref{eq:tight_process_1}--\eqref{eq:tight_process_4} and the convergence of the quadratic terms shown in Section \ref{sec:proof_weak_convergence_process}, we therefore have 
\begin{align*}
    T_1 &= -\frac{1}{pn} u(t)^T \bZ^T \Sigma_p^{1/2} \tilde{G}  \Sigma_p^{1/2} \bZ P_0 \calV^T \calU^T \tilde{G} \Sigma_p^{1/2} \bZ u(t)\\
        &~~~ -\frac{1}{pn} u(t)^T \bZ^T \Sigma_p^{1/2} \tilde{G}  \calU \calV P_0 \bZ^T \Sigma_p^{1/2} \tilde{G} \Sigma_p^{1/2} \bZ u(t)\\
        &~~~ -\frac{1}{pn} u(t)^T \bZ^T \Sigma_p^{1/2} \tilde{G} \calU \calV P_0 \calV^T \calU^T \tilde{G} \Sigma_p^{1/2} \bZ u(t)\\
        &~~~ - 2 \frac{1}{pn} u(t)^T \calV^T \calU^T \tilde{G}  \Sigma_p^{1/2} \bZ P_0 \calV^T \calU^T  \tilde{G} \Sigma^{1/2}_p \bZ u(t) \\
        &~~~ - 2 \frac{1}{pn} u(t)^T \calV^T \calU^T \tilde{G}  \calU \calV P_0 \bZ^T \Sigma_p^{1/2}  \tilde{G} \Sigma^{1/2}_p \bZ u(t) \\
        &~~~ - 2 \frac{1}{pn} u(t)^T \calV^T \calU^T \tilde{G} \calU \calV P_0 \calV^T \calU^T \tilde{G} \Sigma^{1/2}_p \bZ u(t)  + o_p(p^{-1/2})\\
        &= -\frac{2}{pn}  u(t)^T \calV^T \calU^T \tilde{G}  \calU \calV P_0 \bZ^T \Sigma_p^{1/2}  \tilde{G} \Sigma^{1/2}_p \bZ u(t) + o_p(p^{-1/2})\\
        &= - \frac{2\gamma}{p} u(t)^T \calV^T \calU^T \tilde{G} \calU \calV u(t)  \Theta(-\lambda) + o_p(p^{-1/2}).
\end{align*}
Here, the residuals are all uniform on $\Lambda \times \calT_{\rm mc}(\varepsilon)$. In particular, for the leading term, we use the approximation
\[\frac{1}{n} \calV P_0 \bZ^T \Sigma_p^{1/2} \tilde{G} \Sigma_p^{1/2} \bZ u(t) = \gamma \Theta(-\lambda) \calV P_0 u(t) + O_p(p^{-1/2}) = \gamma \Theta(-\lambda) \calV u(t) + O_p(p^{-1/2}).   \]

As for $T_2$, it can verified that the second and the third terms in its definition is $o_p(p^{-1/2})$ uniformly on $\Lambda \times \calT_{\rm mc}(\varepsilon)$. We expand the first term and obtain
\begin{align*}
    T_2 &= p^{-1} u(t)^T \bZ^T \Sigma_p^{1/2} \tilde{G} R_n \tilde{G} R_n \tilde{G} \Sigma^{1/2}_p \bZ u(t) + o_p(p^{-1/2})\\
        &=\frac{1}{pn^2} u(t)^T\bZ^T\Sigma_p^{1/2}\tilde G \Sigma_p^{1/2}\bZ P_0\calV^T\calU^T \tilde G \Sigma_p^{1/2}\bZ P_0\calV^T\calU^T \tilde G\Sigma_p^{1/2}\bZ u(t)\\
        &+\frac{1}{pn^2} u(t)^T\bZ^T\Sigma_p^{1/2}\tilde G \Sigma_p^{1/2}\bZ P_0\calV^T\calU^T \tilde G \calU\calV P_0\bZ^T\Sigma_p^{1/2} \tilde G\Sigma_p^{1/2}\bZ u(t)\\ 
        &+\frac{1}{pn^2}
u(t)^T\bZ^T\Sigma_p^{1/2}\tilde G
\Sigma_p^{1/2}\bZ P_0\calV^T\calU^T
\tilde G
\calU\calV P_0\calV^T\calU^T
\tilde G\Sigma_p^{1/2}\bZ u(t)\\
&+\frac{1}{pn^2}
u(t)^T\bZ^T\Sigma_p^{1/2}\tilde G
\calU\calV P_0\bZ^T\Sigma_p^{1/2}
\tilde G
\Sigma_p^{1/2}\bZ P_0\calV^T\calU^T
\tilde G\Sigma_p^{1/2}\bZ u(t)\\
&+\frac{1}{pn^2}
u(t)^T\bZ^T\Sigma_p^{1/2}\tilde G
\calU\calV P_0\bZ^T\Sigma_p^{1/2}
\tilde G
\calU\calV P_0\bZ^T\Sigma_p^{1/2}
\tilde G\Sigma_p^{1/2}\bZ u(t)\\
&+\frac{1}{pn^2}
u(t)^T\bZ^T\Sigma_p^{1/2}\tilde G
\calU\calV P_0\bZ^T\Sigma_p^{1/2}
\tilde G
\calU\calV P_0\calV^T\calU^T
\tilde G\Sigma_p^{1/2}\bZ u(t)\\
&+\frac{1}{pn^2}
u(t)^T\bZ^T\Sigma_p^{1/2}\tilde G
\calU\calV P_0\calV^T\calU^T
\tilde G
\Sigma_p^{1/2}\bZ P_0\calV^T\calU^T
\tilde G\Sigma_p^{1/2}\bZ u(t)\\
&+\frac{1}{pn^2}
u(t)^T\bZ^T\Sigma_p^{1/2}\tilde G
\calU\calV P_0\calV^T\calU^T
\tilde G
\calU\calV P_0\bZ^T\Sigma_p^{1/2}
\tilde G\Sigma_p^{1/2}\bZ u(t)\\
&+\frac{1}{pn^2}
u(t)^T\bZ^T\Sigma_p^{1/2}\tilde G
\calU\calV P_0\calV^T\calU^T
\tilde G
\calU\calV P_0\calV^T\calU^T
\tilde G\Sigma_p^{1/2}\bZ u(t) + o_p(p^{-1/2})\\
& = \gamma^2 \frac{1}{p} u(t)^T \calV^T \calU^T \tilde{G} \calU \calV u(t)\Theta^2(-\lambda) +o_p(p^{-1/2}).
\end{align*}
Here, the residual is uniform on $\Lambda \times \calT_{\rm mc}(\varepsilon)$. 

Combining the analysis of $T_0$, $T_1$, $T_2$ and  $T_3$, we conclude that 
\begin{align*}
p^{-1} V_\lambda(t) &=\frac{1}{p\|u(t)\|_2^2 } u(t)^T \bZ^T \Sigma_p^{1/2} (\tilde{S}_n + \lambda I_p)^{-1} \Sigma_p^{1/2} \bZ u(t)\\
&+ \frac{1}{p \|u(t)\|_2^2 } [1-\gamma \Theta(-\lambda)]^2 u(t)^T \calV^T \calU^T (\tilde{S}_n + \lambda I_p)^{-1}\calU \calV u(t)   + o_p(p^{-1/2}).
\end{align*}
Here, the convergence is uniform on $\Lambda \times \calT_{\rm mc}(\varepsilon)$. 

Results (2) of Lemma \ref{lemma:estimate_Stieltjes} implies that uniformly over $\lambda\in \Lambda $, the occurrence of $(\tilde{S}_n + \lambda I_p)^{-1}$ in the leading term can be replaced by its deterministic equivalent $\calD(-\lambda)$, with an $o_p(p^{-1/2})$ error. That is, 
\begin{align*}
p^{-1} V_\lambda(t) &=\frac{1}{p\|u(t)\|_2^2 } u(t)^T \bZ^T \Sigma_p^{1/2} (\tilde{S}_n + \lambda I_p)^{-1} \Sigma_p^{1/2} \bZ u(t)\\
&+ \frac{1}{p \|u(t)\|_2^2 } [1-\gamma \Theta(-\lambda)]^2 u(t)^T \calV^T \calU^T  \calD(-\lambda) \calU \calV u(t)   + o_p(p^{-1/2}).
\end{align*}
Moreover, it is straightforward to verify that, uniformly for
$t\in\calT_{\rm mc}(\varepsilon)$,
\begin{align*}
\frac{1}{\sqrt{n}}\calV u(t)
&=
\left[
\int_0^1 u(x;t)v_j(x)\,dx
\right]_{j=1}^{s+1}
+O\left(\frac{1}{n}\right)\\
&=
\sqrt{\kappa(t)}\,\psi(t)
+O\left(\frac{1}{n}\right)=
\psi(t)
\sqrt{
\|u(t)\|_2^2+O\left(\frac{1}{n}\right)
}
+O\left(\frac{1}{n}\right).
\end{align*}
We conclude that uniformly on $\Lambda \times \calT_{\rm mc}(\varepsilon)$,
\begin{align*}
p^{-1} V_{\lambda} (t) &=  \frac{1}{p\|u(t)\|_2^2 } u(t)^T \bZ^T \Sigma_p^{1/2} (\tilde{S}_n + \lambda I_p)^{-1} \Sigma_p^{1/2} \bZ u(t)  \\
    & \quad + \frac{1}{\gamma} [1-\gamma \Theta(-\lambda)]^2 \psi^T(t) \calU^T \calD(-\lambda) \calU \psi(t) + o_p(p^{-1/2}). 
\end{align*}
After standardization, 
\begin{align*}
&\sqrt{p} \frac{\Big(p^{-1}V_\lambda(t) - \Theta(-\lambda)\Big)}{\sqrt{ \Gamma(-\lambda)}} \\
&= \sqrt{p} \frac{ p^{-1} \|u(t)\|_2^{-2} u(t)^T \bZ^T \Sigma_p^{1/2} (\tilde{S}_n + \lambda I_p)^{-1} \Sigma_p^{1/2} \bZ u(t) -\Theta(-\lambda) }{\sqrt{\Gamma(-\lambda)}} +\calS_\lambda(t; \calU)  + o_p(1)\\
& \rightsquigarrow G_\lambda (t) + \calS_{\lambda}(t; \calU).
\end{align*}

\section{Proof of Theorem \ref{thm:power_multiple_PA}}\label{sec:proof_thm_power_PA}
In this section, we prove Theorem \ref{thm:power_multiple_PA}. Throughout the section, $\mP_{\calU}$ is the prior probability measure of $\calU$ under \textbf{PA}, and $\mP(\cdot \mid \calU)$ is the probability measure of the observations conditional on $\calU$, and $\mP_{\bZ}$ is the probability measure of $\bZ$. 

Under \textbf{PA}, using Lemma \ref{lemma:concentration_quadrat}, we can show that 
\[ \left\|\sqrt{p}\calU^T \calU - p^{-1} \tr(\calG\calG^T) \Xi\Xi^T \right\|_2 \stackrel{\mP_{\calU}}{\longrightarrow} 0.  \]
and also for any fixed $\lambda >0$,
\begin{equation}\label{eq:convergence_PA_quadratic_form}
\left\|\sqrt{p} \calU^T \calD_p(-\lambda) \calU - \calQ_p(\lambda, \calG) \Xi\Xi^T\right\|_2 \stackrel{\mP_{\calU}}{\longrightarrow} 0.    
\end{equation}
Since $\|\calG\|_2$ is bounded and $\Xi$ is fixed,  for any  $\zeta>0$,  there exists a sufficiently large constant $\calK_\zeta$ and $N_1$ such that when $n>N_1$, 
\[\mP_{\calU}\Big(K^{(1)}(\zeta) \Big) >1-\zeta/4,\] 
where
\[  K^{(1)} (\zeta) = \Big\{\calU ~:~ \left\|\sqrt{p}\calU^T \calU\right\|_2 \leq \calK_{\zeta} \Big\}.\]
Moreover, for any $\varphi>0$ and $\zeta>0$, there exists a sufficiently large $N_2$ such that when $n>N_2$, 
\[\mP_{\calU}\Big(K^{(2)}(\zeta, \varphi) \Big) >1-\zeta/4,\] 
where
{
\[  K^{(2)}(\zeta, \varphi) = \Big\{\calU~:~\max_{\lambda\in \Lambda}\sup_{t\in\calT_{\rm mc}(\varepsilon)} \Big| \calS_{\lambda}(t;\calU) - \calS_{\lambda}^{\rm PA}(t;\calG,\Xi) \Big| \leq \varphi/4 \Big\}.  \]
}
It implies that  when $n>\max(N_1,N_2)$, 
\begin{equation}\label{eq:bound_K1K2}
\mP_{\calU} \Big( K^{(1)}(\zeta) \cap K^{(2)}(\zeta, \varphi) \Big) \geq 1-\zeta/2.    
\end{equation}

For convenience, define 
\[ \Upsilon_\infty = \mP\left(   \max_{\lambda \in \Lambda} \sup_{t\in \calT_{\rm mc}(\varepsilon)} \Big\{ G_{\lambda}(t) + \calS_\lambda^{\rm PA}\big(t; \mathcal{G}, \Xi\big) \Big\} > \tilde{\xi}_{\rm mc}(1-\alpha; \varepsilon,\Lambda)  \right). \]

Therefore, to show the convergence of $\Upsilon_{\rm mc}(\calU, \varepsilon, \Lambda)$ in probability to $\Upsilon_\infty$ with respect to $\mP_{\calU}$, it suffices to assume that $\calU \in  K^{(1)}(\zeta) \cap K^{(2)}(\zeta, \varphi)$. Namely, we only need to show that for any fixed $\varphi>0$ and $\zeta>0$, we can find a sufficiently large constant $N_0$ such that when $n>N_0$,
\begin{equation}\label{eq:sufficient_condition}
\mP_{\calU} \Big( \Big\{| \Upsilon_{\rm mc}(\calU, \varepsilon, \Lambda) - \Upsilon_\infty |>2 \varphi\Big\}  \cap \Big\{ K^{(1)}(\zeta) \cap K^{(2)}(\zeta, \varphi) \Big\}  \Big) < \zeta/2.    
\end{equation}
It is because that, combining \eqref{eq:bound_K1K2} and \eqref{eq:sufficient_condition}, when $n>\max(N_1,N_2, N_0)$, we have 
\begin{align*}
\mP_{\calU}& \Big( | \Upsilon_{\rm mc}(\calU, \varepsilon, \Lambda) - \Upsilon_\infty | > 2\varphi\Big) \leq \mP_{\calU} \Big( \Big\{ | \Upsilon_{\rm mc}(\calU, \varepsilon, \Lambda) - \Upsilon_\infty | > 2\varphi\Big\}  \cap \Big\{ K^{(1)}(\zeta) \cap K^{(2)}(\zeta, \varphi) \Big\}  \Big)\\
&+ \mP_{\calU}\left(\Big\{ K^{(1)}(\zeta) \cap K^{(2)}(\zeta, \varphi) \Big\}^\complement\right) \leq \zeta.
\end{align*} 
Therefore, in the following analysis, without further mentioning, we assume that $\calU \in  K^{(1)}(\zeta) \cap K^{(2)}(\zeta, \varphi)$.

For convenience, denote 
\begin{align*}
   \tilde{D}_{\lambda}(t) = & \sqrt{p} \frac{ p^{-1} \|u(t)\|_2^{-2} u(t)^T \bZ^T \Sigma_p^{1/2} (\tilde{S}_n + \lambda I_p)^{-1} \Sigma_p^{1/2} \bZ u(t) -\Theta(-\lambda)}{\sqrt{\Gamma(-\lambda)}} +\calS^{\rm PA}_\lambda(t; \calG, \Xi).
\end{align*}
In Section \ref{sec:proof_power_single_deterministic}, we showed that conditional on $\calU$, the standardized statistic 
\[ \sqrt{p} \frac{p^{-1} V_\lambda(t) - \Theta(-\lambda)}{\Gamma^{1/2}(-\lambda)}  ,\]
is such that 
\[ \max_{\lambda \in \Lambda}\sup_{t\in \calT_{\rm mc}(\varepsilon)} \left|\sqrt{p} \frac{p^{-1} V_\lambda(t) - \Theta(-\lambda)}{\Gamma^{1/2}(-\lambda)} -  \tilde{D}_{\lambda}(t) - \Big(\calS_{\lambda}(t;\calU) - \calS_{\lambda}^{\rm PA}(t; \calG, \Xi) \Big) \right| \xrightarrow{\mP(\cdot \mid \calU)} 0\]
with a tail bound only depends on $\|\calU^T\calU\|_2$.
Moreover, Lemma \ref{lemma:estimate_Stieltjes} and Lemma \ref{lemma:Stieltjes_under_alternative} indicate that with a tail bound independent of $\calU$, we have
\[ \hat{\Theta}(-\lambda) - \Theta(-\lambda) \xrightarrow{\mP(\cdot \mid \calU)} 0 ,\quad   \hat{\Gamma}(-\lambda) - \Gamma(-\lambda) \xrightarrow{\mP(\cdot \mid \calU)} 0.\]

Combining the results, we conclude that for any $\varphi>0$ and $\zeta>0$, we can find a sufficiently large $N_3$ such that when $ n>N_3$,
\[ \inf_{\calU \in K^{(1)} \cap K^{(2)}} \mP_{\bZ} \Big( K^{(3)}(\calU) \mid \calU \Big) > 1-\varphi/2,\]
where
\[ K^{(3)}(\calU) = \Big\{ \bZ ~:~ \max_{\lambda \in \Lambda} \sup_{t \in \calT_{\rm mc}(\varepsilon) } |D_\lambda(t) - \tilde{D}_\lambda(t)| <\varphi/2\Big\}. \]
It indicates that we can approximate the distribution function of $D_\lambda(t)$ using that of $\tilde{D}_\lambda(t)$ giving $\calU \in K^{(1)} \cap K^{(2)}$.

Since 
\begin{align*}
\Upsilon_{\rm mc}(\calU, \varepsilon, \Lambda)  & = \mP\Big( \max_{\lambda\in\Lambda}\sup_{t\in\calT_{\rm mc}(\varepsilon)} D_\lambda(t) > \tilde{\xi}_{\rm mc}(1-\alpha; \varepsilon,\Lambda) \mid \calU \Big)\\
 &= \mP \Big(\Big\{ \max_{\lambda\in\Lambda}\sup_{t\in\calT_{\rm mc}(\varepsilon)} D_\lambda(t) > \tilde{\xi}_{\rm mc}(1-\alpha; \varepsilon,\Lambda)\Big\}  \cap K^{(3)}  \mid \calU \Big)  \\
& \qquad + \mP \Big(\Big\{ \max_{\lambda\in\Lambda}\sup_{t\in\calT_{\rm mc}(\varepsilon)} D_\lambda(t) > \tilde{\xi}_{\rm mc}(1-\alpha; \varepsilon,\Lambda)\Big\}  \cap \{K^{(3)}\}^\complement \mid \calU \Big),
\end{align*}
it follows that when $n>\max(N_1,N_2, N_3)$ and $\calU \in K^{(1)}(\zeta) \cap K^{(2)}(\zeta, \varphi)$,
\begin{align*}
   \Upsilon_{\rm mc}(\calU, \varepsilon, \Lambda) \leq \varphi/2 + \mP_{\bZ}\Big( \max_{\lambda \in \Lambda}\sup_{t\in\calT_{\rm mc}(\varepsilon)}\tilde{D}_\lambda(t) > \tilde{\xi}_{\rm mc}(1-\alpha; \varepsilon,\Lambda)  - \varphi/2  \Big) ,
\end{align*}
\begin{align*}
   \Upsilon_{\rm mc}(\calU, \varepsilon, \Lambda) \geq -\varphi/2 + \mP_Z\Big( \max_{\lambda \in \Lambda}\sup_{t\in\calT_{\rm mc}(\varepsilon)}\tilde{D}_\lambda(t) > \tilde{\xi}_{\rm mc}(1-\alpha; \varepsilon,\Lambda)  + \varphi/2 \Big) .
\end{align*}

Since the distribution of
\[
\max_{\lambda\in\Lambda}
\sup_{t\in\calT_{\rm mc}(\varepsilon)}
\Big\{
G_\lambda(t)
+
\calS_\lambda^{\rm PA}(t;\calG,\Xi)
\Big\}
\]
is continuous at $\tilde{\xi}_{\rm mc}(1-\alpha;\varepsilon,\Lambda)$, for any $\varphi>0$ there exists
$\delta>0$ sufficiently small such that
\[
\left|
\Upsilon_\infty(\tilde{\xi}_{\rm mc}(1-\alpha;\varepsilon,\Lambda)\pm\delta)
-
\Upsilon_\infty(\tilde{\xi}_{\rm mc}(1-\alpha;\varepsilon,\Lambda))
\right|
<
\frac{\varphi}{2}.
\]

Using the arguments in Section~\ref{sec:proof_power_single_deterministic},
there exists $N_4$ such that, for $n>N_4$,
\[
\left|
\mP_{\bZ}\left(
\max_{\lambda\in\Lambda}
\sup_{t\in\calT_{\rm mc}(\varepsilon)}
\tilde D_\lambda(t)
>
\tilde{\xi}_{\rm mc}(1-\alpha;\varepsilon,\Lambda)-\delta
\right)
-
\Upsilon_\infty(\tilde{\xi}_{\rm mc}(1-\alpha;\varepsilon,\Lambda)-\delta)
\right|
<
\frac{\varphi}{2},
\]
and
\[
\left|
\mP_{\bZ}\left(
\max_{\lambda\in\Lambda}
\sup_{t\in\calT_{\rm mc}(\varepsilon)}
\tilde{D}_\lambda(t)
>
\tilde{\xi}_{\rm mc}(1-\alpha;\varepsilon,\Lambda)+\delta
\right)
-
\Upsilon_\infty(\tilde{\xi}_{\rm mc}(1-\alpha;\varepsilon,\Lambda)+\delta)
\right|
<
\frac{\varphi}{2}.
\]
Here, $\Upsilon_\infty(\tilde{\xi}_{\rm mc}(1-\alpha;\varepsilon,\Lambda) \pm \delta)$ is $\Upsilon_\infty$ but at the critical value $\tilde{\xi}_{\rm mc}(1-\alpha;\varepsilon,\Lambda) \pm \delta$.

Combining these bounds with the approximation of $D_\lambda(t)$ by
$\tilde D_\lambda(t)$, we obtain, for all sufficiently large $n$ and
$\calU\in K^{(1)}(\zeta)\cap K^{(2)}(\zeta,\delta)$,
\[
\Upsilon_{\rm mc}(\calU,\varepsilon,\Lambda)
\le
\varphi
+
\Upsilon_\infty(\tilde{\xi}_{\rm mc}(1-\alpha;\varepsilon,\Lambda)-\delta),
\]
and
\[
\Upsilon_{\rm mc}(\calU,\varepsilon,\Lambda)
\ge
-\varphi
+
\Upsilon_\infty(\tilde{\xi}_{\rm mc}(1-\alpha;\varepsilon,\Lambda)+\delta).
\]
Therefore,
\[
\left|
\Upsilon_{\rm mc}(\calU,\varepsilon,\Lambda)
-
\Upsilon_\infty
\right|
<
\frac{3\varphi}{2}.
\]
After relabeling $\varphi$ if necessary, this establishes
\eqref{eq:sufficient_condition} and completes the proof.

\section{Technical Lemmas}\label{s_sec:technical_lemma}

We present a collection of technical lemmas used throughout the proofs. Proofs of some of the lemmas are omitted, as closely related results are available in the literature; see, for example, Section S.10 of \citet{li2020high} and (3.10)--(3.14) of \citet{Pan2011central}. The matrices appearing in the following lemmas may be either real- or complex-valued.

It is worth noting that \citet{Pan2011central} employs a truncation scheme to control higher-order moments of the random variables, and their moment bounds are therefore formulated under the corresponding truncation condition. In the present work, the bounds are instead derived directly under the moment condition in \ref{enum:moment_conditions}, and their rates are adjusted accordingly.

\begin{lemma}[Theorem A.43 of \cite{bai2010spectral}]
\label{lemma:rank_one_perturbation}
    Let $A$ ane $B$ be two $p\times p$ Hermitian matrices. Then,
    \[ \|F^A - F^B \|_\infty \leq \frac{1}{p} \operatorname{rank}(A-B).\]
\end{lemma}

\begin{lemma}[Woodbury Matrix Identity]\label{lemma:woodbury}
The following identity holds:
\[
(A + UCV)^{-1} = A^{-1} - A^{-1} U \left( C^{-1} + V A^{-1} U \right)^{-1} V A^{-1},
\]
for matrices \( A, U, C, V \) of conformable sizes, assuming all inverses exist and are well-defined.
\end{lemma}

\begin{lemma}(\citet{fan1951maximum})
\label{lemma:singular_value_ineq}
Let $A$ and $C$ be two $p\times n$ complex matrices. Then, for any nonnegative integers $i$ and $j$, we have 
\[s_{i+j+1}(A+C) \leq s_{i+1}(A) + s_{j+1} (C),\]
where $s_{i}(\cdot)$ is the $i$-th largest singular value of a matrix. 
\end{lemma}

\begin{lemma}(Burkholder)
\label{lemma:Burkholder_martingale_moments}
Let $\{Y_i\}$ be a complex martingale difference sequence with respect to the increasing $\sigma$-field $\{\sigma_i\}$. Then for $m\geq 2$
\[\mE \Big|\sum\limits_{i}Y_i \Big|^m \leq \calK_m \mE(\sum\limits_{i} \mE(|Y_i|^2\mid \sigma_{i-1}) )^{m/2} + \calK_m \mE (\sum\limits_{i} |Y_i|^m).\]
\end{lemma}

\begin{lemma}
\label{lemma:Burkholder}
(Lemma 2.7 of \citet{bai1998no}) Let $\bY = (Y_1,\dots, Y_p)^T$, where $Y_i$'s are i.i.d. real r.v.'s with mean 0 and variance $1$. Let $\bB = (b_{ij})_{p\times p}$, a deterministic complex matrix. Then for any $m\geq 2$, we have 
\[\mE|\bY^T \bB \bY -\tr\bB|^m \leq \calK_m (\mE Y_1^4\tr\bB\bB^*)^{m/2} + \calK_m \mE Y_1^{2m} \tr[(\bB\bB^*)^{m/2}],\]
where $\bB^*$ denotes the complex conjugate transpose of $\bB$, and $\calK_m$ is a constant only depending on $m$.   
\end{lemma}

\begin{lemma}
\label{lemma:bound_beta}
For any $\lambda >0$,
\[ \|\bA(\lambda)\|_2\leq \frac{\|\Sigma_p\|}{\lambda},  \qquad |\beta_1(\lambda)|  \leq 1, \qquad |\beta_1^{\tr}(\lambda)|  \leq 1. \]
This result is modified from (3.4) of \citet{bai1998no}.
\end{lemma}

\begin{lemma}
\label{lemma:nonran_bound_resol}
(Lemma 2.10 of \citet{bai1998no}) For any matrix $\bD$ and $\lambda > 0$, 
\[\Big| \tr\{\bA(\lambda)\bD - \bA_{j}(\lambda)\bD\} \Big| \leq \frac{\|\bD\Sigma_p\|_2}{\lambda}.\]
\end{lemma}

\begin{lemma}
\label{lemma:converg_A}
For $m\geq 2$ and any fixed $\lambda>0$,
\[n^{-m}\mE\Big|\tr\bA(\lambda) -\mE \tr\bA(\lambda) \Big|^m =O(n^{-m/2}). \]
The result is modified from results in Section 4 of \citet{bai1998no}. 
\end{lemma}

\begin{lemma}
\label{lemma:concentration_quadrat}
For a sequence of deterministic matrices $\bD$ such that $\|\bD\|_2 < \infty$, $m\geq2$,
\begin{align*}
n^{-m} \mE \Big| \bz_1^T \bD \bz_1 - \tr(\bD) \Big|^m\leq \calK_m n^{-m} \Big\{ [\tr (\bD\bD^*)]^{m/2} + \tr[(\bD\bD^*)^{m/2}] \Big\} = O (n^{-m/2}),
\end{align*}
for some constant $\calK_m$. 
The result follows directly from Lemma \ref{lemma:Burkholder}.
\end{lemma}

\begin{lemma}
\label{lemma:bound_quad_rankone}
For sequences of deterministic matrices $\bD$ and $\bG$ such that $\|\bD\|_2 <\infty$ and $\|\bG\|_2<\infty$, for $m\geq2$,
\[ n^{-m}\mE \Big|\bz_1^T \bD e_i e_j^T \bG \bz_1 \Big|^m\leq \calK_m n^{-m} \|\bD\|_2^m \|\bG\|_2^m  = O(n^{-m}).\]
for some constant $\calK_m$. The result is again shown using Lemma \ref{lemma:Burkholder}.
\end{lemma}

\begin{lemma}
\label{lemma:zalpha_a_zalpha_a}
For a sequence of deterministic matrices $\bD$ such that $\|\bD\|_2 < \infty$ and a sequence of vector $\bm{u}$ such that $\limsup_{n\to\infty} \sqrt{n}\|\bm{u}\|_{\infty}\leq \calK_{\max}<\infty$,
\[\mE | n^{-1} \bm{u}^T \bZ^T \bD \bZ\bm{u} |^m \leq \calK_m  \|\bm{u}\|_2^{2m} \|\bD\|^m_2, \]
for $ m\geq 2$ and some constant $\calK_m>0$.
\end{lemma}
The proof follows from repeated applications of Lemma~\ref{lemma:Burkholder_martingale_moments}. The argument is adapted from the proof of (3.17) in \citet{Pan2011central}. The only modification is to replace the vectors $\bm{s}_i$ in their proof by $\bm{u}_i\bz_i$. We therefore omit the routine details.

\begin{lemma}
\label{lemma:bound_z1_zalpha_b}
For a sequence of deterministic matrices $\bD$ such that $\|\bD\|_2 < \infty$ and a sequence of vector $\bm{u}$ such that $\limsup_{n\to\infty} \sqrt{n}\|\bm{u}\|_{\infty}\leq \calK_{\max}<\infty$, 
\[\mE \Big|n^{-1/2} \bz_1^T \bD \bZ_1\bm{u}\Big|^m  \leq \calK_m  \|\bm{u}\|_2^m \|\bD\|^m_2, \]
for  $ m\geq 4$ and some constant $\calK_m>0$.
\end{lemma}

The proof is based on Lemmas~\ref{lemma:Burkholder} and
\ref{lemma:zalpha_a_zalpha_a}. Indeed, by Lemma~\ref{lemma:Burkholder},
\begin{align*}
&\mE\Big|n^{-1/2}\bz_1^T\bD\bZ_1\bm{u}\Big|^m  \\
&\le
\calK
\mE\Big|
n^{-1}\bz_1^T\bD\bZ_1\bm{u}\bm{u}^T\bZ_1^T\bD^T\bz_1
-
n^{-1}\bm{u}^T\bZ_1^T\bD^T\bD\bZ_1\bm{u}
\Big|^{m/2} +
\calK
\mE\Big|
n^{-1}\bm{u}^T\bZ_1^T\bD^T\bD\bZ_1\bm{u}
\Big|^{m/2}.
\end{align*}
The first term is bounded by Lemma~\ref{lemma:Burkholder}, while the second follows directly from Lemma~\ref{lemma:zalpha_a_zalpha_a} with $\bD^T\bD$ in place of $\bD$. Combining the two bounds yields the desired result.


\begin{lemma}
\label{lemma:z1_z2}
For a sequence of deterministic matrices $\bD$ such that $\|\bD\|_2 < \infty$,
\[\mE\big| \bz_1^T \bD \bz_2 \big|^m \leq \calK_m \|\bD\|_2^m n^{m/2},\]
for  $m\geq4$ and some constant $\calK_m>0$.
\end{lemma}
The proof follows from the same type of moment argument as that used in Lemma~\ref{lemma:bound_z1_zalpha_b}, and is therefore omitted.


\begin{lemma} 
\label{lemma:comprehensive}
For sequences of deterministic matrices $\bD_1,\dots,\bD_m$, $\bG_1,\dots, \bG_s$, and $\bJ$ such that $\|\bD_j\|_2 <\infty$, $\|\bG_j\|_2<\infty$, and $\|\bJ\|_2<\infty$, and a sequence of vector $\bm{u}$ such that $\limsup_{n\to\infty}\sqrt{n}\|\bm{u}\|_{\infty}\leq \calK_{\max}<\infty$,
\begin{align*}
&\mE \Big|\prod\limits_{i=1}^m \frac{1}{n} \bz_1^T \bD_i \bz_1 \prod\limits_{j=1}^s \frac{1}{n} 
(\bz_1^T \bG_j \bz_1 -\tr \bG_j) \big(n^{-1/2}\bz_1^T \bJ\bZ_1\bm{u}\big)^{t} \Big|\\
&\leq \calK_{m,s,t} \prod_{i=1}^m\|\bD_i\|_2 \prod_{j=1}^s\|\bG_j\| \|\bJ\|_2^t \|\bm{u}\|_2^t n^{-s/2}, 
\end{align*}
 where $ m\geq 0, s\geq1, t\geq 0$ and some constant $\calK_{m,s,t}>0$.
\end{lemma}
The result follows from Lemma \ref{lemma:concentration_quadrat}, Lemma \ref{lemma:bound_z1_zalpha_b} and the Cauchy-Schwarz inequality.


\begin{lemma}
\label{lemma:z1_z1_z1_a}
For a vector $\mathbf{r}$ and deterministic matrices $\bD=(d_{ij})$ and $\bG$,   
\[\mE[(\bz_1^T \bD \bz_1 -\tr\bD) \bz_1^T \bG \mathbf{r} ] = \mE z_{11}^3 \sum\limits_{i=1}^p d_{ii} e_i^T \bG\mathbf{r}, \]
where $e_i$ is the canonical vector with the $i$th entry $1$. 
\end{lemma}

\begin{lemma}
\label{lemma:z_1_z1_z1_z1}
For deterministic matrices $\bD=(d_{ij})$ and $\bG=(g_{ij})$,
\begin{align*}
\mE[(\bz_1^T \bD \bz_1 -\tr\bD)(\bz_1^T \bG \bz_1 -\tr \bG)] =  |\mE z_{11}^2|^2 \tr \bD\bG^T + \tr\bD\bG + (\mE z_{11}^4 - |\mE z_{11}^2|^2 -2) \sum\limits_{i=1}^p d_{ii}g_{ii}.
\end{align*}
\end{lemma}